\documentclass[sigplan,screen]{acmart}

\usepackage{textgreek}
\usepackage{mathpartir}
\usepackage{stmaryrd}
\usepackage{mathpartir}
\usepackage[capitalise,noabbrev]{cleveref}
\usepackage{tikz}
\usepackage{tikz-cd}

\usepackage{agda}
\usepackage{agdadimmed}
\usepackage{newunicodechar}

\crefname{section}{\S}{\S\S}

\newcommand{\bnfdef}{\mathrel{:\kern-.3em:\kern-.15em=}}

\usetikzlibrary{bending,arrows,arrows.meta,shapes,positioning,fit,shadows,calc,decorations.markings}

\newcommand{\sledom}{\mathrel{\rotatebox[origin=c]{180}{$\vDash$}}}
\MakeRobust{\sledom}

\newcommand{\af}{\AgdaFunction}

\newcommand{\ad}{\AgdaDatatype}
\newcommand{\ab}{\AgdaBound}
\newcommand{\ac}{\AgdaInductiveConstructor}
\newcommand{\aF}{\AgdaField}
\newcommand{\as}{\AgdaSymbol}
\newcommand{\ak}{\AgdaKeyword}

\newcommand{\am}{\AgdaModule}

\newunicodechar{◂}{} 

\newcommand\superequiv{\mathrel{\rlap{\raisebox{\fontdimen22\textfont2}{$=$}}\raisebox{-0.5\fontdimen22\textfont2}{$ = $}}}

\newunicodechar{×}{\ensuremath{\mathnormal\times}}
\newunicodechar{→}{\ensuremath{\mathnormal\to}}
\newunicodechar{←}{\ensuremath{\mathnormal\leftarrow}}
\newunicodechar{⟦}{\ensuremath{\mathnormal\llbracket}}
\newunicodechar{⟧}{\ensuremath{\mathnormal\rrbracket}}
\newunicodechar{ℕ}{\ensuremath{\mathbb{N}}}
\newunicodechar{⊕}{\ensuremath{\mathnormal\oplus}}
\newunicodechar{∔}{\ensuremath{\mathnormal\dotplus}}
\newunicodechar{⋎}{\ensuremath{\mathnormal\curlyvee}}
\newunicodechar{⊞}{\ensuremath{\mathnormal\boxplus}}
\newunicodechar{⇒}{\ensuremath{\mathnormal\Rightarrow}}
\newunicodechar{⇛}{\ensuremath{\mathnormal\Rrightarrow}}
\newunicodechar{⟨}{\ensuremath{\mathnormal\langle}}
\newunicodechar{⟩}{\ensuremath{\mathnormal\rangle}}
\newunicodechar{∪}{\ensuremath{\mathnormal\cup}}
\newunicodechar{⦃}{\ensuremath{\mathnormal\{\mskip-4.5mu\mid}}
\newunicodechar{⦄}{\ensuremath{\mathnormal\mid\mskip-4.5mu\}}}
\newunicodechar{⊎}{\ensuremath{\mathnormal\uplus}}
\newunicodechar{✴}{\ensuremath{\mathnormal\ast}}
\newunicodechar{↦}{\ensuremath{\mathnormal\mapsto}}
\newunicodechar{≡}{\ensuremath{\mathnormal\equiv}}
\newunicodechar{∀}{\ensuremath{\mathnormal\forall}}
\newunicodechar{∙}{\ensuremath{\mathnormal\bullet}}
\newunicodechar{≣}{\ensuremath{\mathnormal\superequiv}}
\newunicodechar{▿}{\ensuremath{\mathnormal\triangledown}}
\newunicodechar{▸}{\raisebox{0.25ex}{\scaleobj{0.65}{\ensuremath{\mathnormal\blacktriangleright}}}}
\newunicodechar{∼}{\ensuremath{\mathnormal\sim}}
\newunicodechar{≤}{\ensuremath{\mathnormal\leq}}
\newunicodechar{↔}{\ensuremath{\mathnormal\leftrightarrow}}
\newunicodechar{⊂}{\ensuremath{\mathnormal\subset}}
\newunicodechar{∘}{\ensuremath{\mathnormal\circ}}
\newunicodechar{∃}{\ensuremath{\mathnormal\exists}}
\newunicodechar{↓}{\ensuremath{\mathnormal\downarrow}}
\newunicodechar{↑}{\ensuremath{\mathnormal\uparrow}}
\newunicodechar{⊔}{\ensuremath{\mathnormal\sqcup}}
\newunicodechar{⊢}{\ensuremath{\mathnormal\vdash}}
\newunicodechar{⊨}{\ensuremath{\mathnormal\vDash}}
\newunicodechar{◇}{\ensuremath{\mathnormal\diamond}}
\newunicodechar{⊙}{\ensuremath{\mathnormal\odot}}
\newunicodechar{⊤}{\ensuremath{\mathnormal\top}}
\newunicodechar{⊥}{\ensuremath{\mathnormal\bot}}
\newunicodechar{∣}{\ensuremath{\mathnormal\mid}}
\newunicodechar{‵}{\ensuremath{^\backprime}}
\newunicodechar{′}{\ensuremath{^\prime}}
\newunicodechar{″}{\ensuremath{^{\prime\prime}}}
\newunicodechar{‴}{\ensuremath{^{\prime\prime\prime}}}
\newunicodechar{⅋}{}
\newunicodechar{∅}{\ensuremath{\emptyset}}
\newunicodechar{≺}{\ensuremath{\mathnormal{\prec}}}
\newunicodechar{≼}{\ensuremath{\mathnormal{\preceq}}}
\newunicodechar{∩}{\ensuremath{\mathnormal{\cap}}}
\newunicodechar{⟪}{\ensuremath{\langle\kern-.2em\langle}}
\newunicodechar{⟫}{\ensuremath{\rangle\kern-.2em\rangle}}
\newunicodechar{⦉}{\ensuremath{(\kern-.18em\lvert}}
\newunicodechar{⦊}{\ensuremath{\rvert\kern-.18em)}}
\newunicodechar{【}{\ensuremath{\langle\kern-.41em[}}
\newunicodechar{】}{\ensuremath{]\kern-.39em\rangle}}
\newunicodechar{⊆}{\ensuremath{\mathnormal{\subseteq}}}
\newunicodechar{⊇}{\ensuremath{\mathnormal{\supseteq}}}
\newunicodechar{▻}{\ensuremath{\mathnormal\vartriangleright}}
\newunicodechar{∷}{\ensuremath{:\kern-.3em:}}
\newunicodechar{►}{\ensuremath{\mathnormal{\blacktriangleright}}}
\newunicodechar{▹}{\ensuremath{\mathnormal{\triangleright}}}
\newunicodechar{□}{\ensuremath{\mathnormal{\square}}}
\newunicodechar{⋯}{\ensuremath{\mathnormal{\cdots}}}
\newunicodechar{▣}{\ensuremath{\mathnormal{\ldots}}}
\newunicodechar{⋮}{\ensuremath{\mathnormal{\quad\quad\vdots}}}
\newunicodechar{∈}{\ensuremath{\mathnormal{\in}}}
\newunicodechar{⊑}{\ensuremath{\mathbin{\sqsubseteq}}}
\newunicodechar{𝓑}{\ensuremath{\mathnormal{\gg\!\!=}}}
\newunicodechar{𝓒}{\ensuremath{\mathbf{\mathscr{C}}}}
\newunicodechar{𝓓}{\ensuremath{\mathbf{\mathscr{D}}}}
\newunicodechar{♯}{\ensuremath{\sharp}}
\newunicodechar{∎}{\ensuremath{\qed}}
\newunicodechar{↣}{\ensuremath{\rightarrowtail}}
\newunicodechar{⟶}{\ensuremath{\longrightarrow}}
\newunicodechar{ẋ}{\ensuremath{\dot{\times}}}
\newunicodechar{∐}{\ensuremath{\dot{\Sigma}}}
\newunicodechar{̂}{\ensuremath{^{\prime}}}

\newunicodechar{φ}{\ensuremath{\phi}}
\newunicodechar{Φ}{\ensuremath{\Phi}}
\newunicodechar{ψ}{\ensuremath{\psi}}
\newunicodechar{α}{\ensuremath{\alpha}}
\newunicodechar{β}{\ensuremath{\beta}}
\newunicodechar{σ}{\ensuremath{\sigma}}
\newunicodechar{ξ}{\ensuremath{\xi}}
\newunicodechar{Ξ}{\ensuremath{\Xi}}
\newunicodechar{λ}{\ensuremath{\lambda}}
\newunicodechar{ε}{\ensuremath{\epsilon}}
\newunicodechar{γ}{\ensuremath{\gamma}}
\newunicodechar{Σ}{\ensuremath{\Sigma}}
\newunicodechar{Δ}{\ensuremath{\Delta}}
\newunicodechar{Π}{\ensuremath{\Pi}}
\newunicodechar{Γ}{\ensuremath{\Gamma}}
\newunicodechar{η}{\ensuremath{\eta}}
\newunicodechar{ζ}{\ensuremath{\zeta}}
\newunicodechar{∞}{\ensuremath{\infty}}
\newunicodechar{≈}{\ensuremath{\approx}}
\newunicodechar{∂}{\ensuremath{\partial}}

\newunicodechar{₀}{\ensuremath{_{0}}}
\newunicodechar{₁}{\ensuremath{_{1}}}
\newunicodechar{₂}{\ensuremath{_{2}}}
\newunicodechar{₃}{\ensuremath{_{3}}}
\newunicodechar{₄}{\ensuremath{_{4}}}
\newunicodechar{₅}{\ensuremath{_{5}}}
\newunicodechar{₆}{\ensuremath{_{6}}}
\newunicodechar{₇}{\ensuremath{_{7}}}
\newunicodechar{₈}{\ensuremath{_{8}}}
\newunicodechar{₉}{\ensuremath{_{9}}}
\newunicodechar{¹}{\ensuremath{^{1}}}
\newunicodechar{⁻}{$^{-}$}
\newunicodechar{ᴬ}{$^{A}$}
\newunicodechar{ᴮ}{$^{B}$}
\newunicodechar{ᴱ}{$^{E}$}
\newunicodechar{ᴴ}{$^{H}$}
\newunicodechar{ˣ}{$^{×}$}
\newunicodechar{ᵈ}{$^{d}$}
\newunicodechar{ᵘ}{$^{u}$}
\newunicodechar{ᶠ}{$^{f}$}
\newunicodechar{ᵖ}{$^{p}$}
\newunicodechar{ⁱ}{$^{i}$}
\newunicodechar{ᵒ}{$^{o}$}
\newunicodechar{ˢ}{$^{s}$}
\newunicodechar{ˡ}{$^{l}$}
\newunicodechar{ʳ}{$^{r}$}
\newunicodechar{ᴰ}{$^{D}$}
\newunicodechar{ᵇ}{$^{b}$}
\newunicodechar{ᶜ}{$^{c}$}
\newunicodechar{ᵗ}{$^{t}$}
\newunicodechar{ᵐ}{$_{m}$}
\newunicodechar{ⁿ}{$_{n}$}
\newunicodechar{ₚ}{$_{p}$}
\newunicodechar{ₒ}{$_{o}$}
\newunicodechar{ₙ}{$_{n}$}
\newunicodechar{ₘ}{$_{m}$}
\newunicodechar{ᵛ}{$_{v}$}
\newunicodechar{ᵍ}{$_{f}$}
\newunicodechar{ᵢ}{$_{i}$}
\newunicodechar{ₗ}{$_{l}$}
\newunicodechar{ᵣ}{$_{r}$}
\newunicodechar{ₛ}{$_{s}$}
\newunicodechar{ₜ}{$_{t}$}
\newunicodechar{ₐ}{$_{a}$}
\newunicodechar{∶}{$:$}
\newunicodechar{̅}{$^{\textit{d}}$}

\newunicodechar{𝑡}{\textit{t}}
\newunicodechar{ℎ}{\textit{h}}
\newunicodechar{𝑟}{\textit{r}}
\newunicodechar{𝑜}{\textit{o}}
\newunicodechar{𝑤}{\textit{w}}
\newunicodechar{𝑐}{\textit{c}}
\newunicodechar{𝑎}{\textit{a}}
\newunicodechar{𝑡}{\textit{t}}

\newunicodechar{𝓑}{\ensuremath{\mathnormal{\gg\!\!=}}}

\makeatletter
\newsavebox{\@brx}
\newcommand{\lLangle}[1][]{\savebox{\@brx}{\(\m@th{#1\langle}\)}%
  \mathopen{\copy\@brx\mkern2mu\kern-0.9\wd\@brx\usebox{\@brx}}}
\newcommand{\rRangle}[1][]{\savebox{\@brx}{\(\m@th{#1\rangle}\)}%
  \mathclose{\copy\@brx\mkern2mu\kern-0.9\wd\@brx\usebox{\@brx}}}
\makeatother

\newunicodechar{⟪}{\ensuremath{\mathnormal\lLangle}}
\newunicodechar{⟫}{\ensuremath{\mathnormal\rRangle}}

\newunicodechar{⋆}{*}
\newunicodechar{〔}{(}
\newunicodechar{〕}{)}
\newunicodechar{．}{.}

\AtBeginDocument{%
  }

\begin{document}

\title{Generic Reduction-Based Interpreters}


\author{Casper Bach}
\email{casperbach@imada.sdu.dk}
\affiliation{%
  \institution{University of Southern Denmark}
  \city{Odense}
  \country{Denmark}
}

\renewcommand{\shortauthors}{Bach}

\begin{abstract}
  Reduction-based interpreters are traditionally defined in terms of a one-step reduction function which systematically decomposes a term into a potential redex and context, contracts the redex, and recomposes it to construct the new term to be further reduced.
  While implementing such interpreters follows a systematic recipe, they often require interpreter engineers to write a substantial amount of code---much of it boilerplate.
  In this paper, we apply well-known techniques from generic programming to reduce boilerplate code in reduction-based interpreters.
\end{abstract}



\received{20 February 2007}
\received[revised]{12 March 2009}
\received[accepted]{5 June 2009}

\maketitle

\begin{code}[hide]%
\>[0]\AgdaKeyword{open}\AgdaSpace{}%
\AgdaKeyword{import}\AgdaSpace{}%
\AgdaModule{Data.Nat}\<%
\\
\>[0]\AgdaKeyword{open}\AgdaSpace{}%
\AgdaKeyword{import}\AgdaSpace{}%
\AgdaModule{Data.Maybe}\<%
\\
\>[0]\AgdaKeyword{open}\AgdaSpace{}%
\AgdaKeyword{import}\AgdaSpace{}%
\AgdaModule{Data.Product}\<%
\\
\\[\AgdaEmptyExtraSkip]%
\>[0]\AgdaKeyword{module}\AgdaSpace{}%
\AgdaModule{sections.01-intro}\AgdaSpace{}%
\AgdaKeyword{where}\<%
\end{code}

\section{Introduction}

Reduction semantics~\cite{FelleisenH92} is a framework for operational semantics based on rewriting in an evaluation context.
The framework is intuitive and expressive; for example, reduction semantics is often used to concisely define the semantics of \emph{control effects} such as Reynolds' escape~\cite{reynolds98definitional}, Scheme's call/cc, prompt/control~\cite{Felleisen88}, or shift/reset~\cite{DanvyF90}.
Danvy and coauthors have demonstrated how the structure of reduction semantics and evaluation contexts can be naturally implemented in functional languages, using recursive functions, pattern matching, and algebraic data types---and even how to mechanically derive other semantic formalisms from such implementations of reduction semantics~\cite{BiernackaD07,AgerDM05,Danvy08,Danvy08afp,DanvyJZ11}.

Reduction semantic interpreters \'{a} la Danvy follow a fixed recipe
which provides a direct way to implement and explain reduction semantics.
However, they also involve quite a bit of manually written code, much of it boilerplate.
What if, instead of instructive and direct implementations, we want implementations without boilerplate code?

In this paper we use known generic programming techniques to provide an algebraic characterization of reduction semantics with deterministic, standard reduction strategies---i.e., left/right-most inner/outer-most.
This characterization affords deriving implementations of reduction semantics from only a syntax specification and a contraction function.
The rest of the semantic artifacts of the reduction semantics (e.g., \ad{Context}, \af{decompose}, \af{recompose}, and \af{drive}) are generic, meaning we can define them once and for all, in a \emph{typed}, \emph{compositional}, and \emph{correct by construction} manner.

The key observation that underpins our algebraic characterization is that evaluation contexts are \emph{zippers}~\cite{Huet97}.
Following McBride~\cite{mcbride2001derivative,McBride08}, such zippers are given by \emph{derivatives} of data types; and such derivatives and their corresponding ``plugging'' (or \af{recompose}) functions can be defined and computed generically from a syntax description.
Furthermore, we demonstrate that, for deterministic, standard reduction strategies such as left/right-most inner/outer-most reduction, standard generic recursion schemes---specifically, \emph{paramorphisms}~\cite{MeijerFP91,Meertens92}---suffice to define context decomposition functions (\af{decompose}).

We illustrate our approach by using it to define and implement a reduction semantics for $\lambda$ expressions and shift/reset.


We make the following technical contributions:

\begin{itemize}
\item Building on the generic programming techniques of McBride~\cite{mcbride2001derivative}, we present a framework for implementing reduction semantics without boilerplate code.
Language designers need only specify a syntax and contraction function for that syntax.
\item We demonstrate our approach by applying it to calculi with $\lambda$ expressions and control effects, to obtain reduction semantic interpreters for them.
\end{itemize}

In \cref{sec:02-redsem} we recall how to implement a reduction semantics using recursive functions and algebraic data types in Agda~\cite{norell2008dependently}.
\cref{sec:04-redsem-gen} generalizes this implementation using the generic programming techniques recalled in \cref{sec:03-prelude}.
\Cref{sec:07-rel-work} discusses related and future work. 
\cref{sec:08-conclusion} concludes.

The techniques we use rely on dependent types, and we use Agda our implementation language.
The framework should be possible to implement in any language with dependent types, such as Idris, Coq, or Lean.
The techniques should also be portable to languages without dependent types, by sacrificing some static guarantees.

We assume familiarity with basic dependently typed programming and basic category theory.
Agda-specific notation is explained throughout.
The paper is a literate Agda file, and interested readers are invited to consult the source of the paper for the definitions we sometimes elide for brevity.


\begin{code}[hide]%
\>[0]\AgdaSymbol{\{-\#}\AgdaSpace{}%
\AgdaKeyword{OPTIONS}\AgdaSpace{}%
\AgdaPragma{--guardedness}\AgdaSpace{}%
\AgdaSymbol{\#-\}}\<%
\\
\\[\AgdaEmptyExtraSkip]%
\>[0]\AgdaKeyword{open}\AgdaSpace{}%
\AgdaKeyword{import}\AgdaSpace{}%
\AgdaModule{Function}\<%
\\
\\[\AgdaEmptyExtraSkip]%
\>[0]\AgdaKeyword{open}\AgdaSpace{}%
\AgdaKeyword{import}\AgdaSpace{}%
\AgdaModule{Data.Empty}\<%
\\
\>[0]\AgdaKeyword{open}\AgdaSpace{}%
\AgdaKeyword{import}\AgdaSpace{}%
\AgdaModule{Data.Unit}\<%
\\
\>[0]\AgdaKeyword{open}\AgdaSpace{}%
\AgdaKeyword{import}\AgdaSpace{}%
\AgdaModule{Data.Bool}\<%
\\
\>[0]\AgdaKeyword{open}\AgdaSpace{}%
\AgdaKeyword{import}\AgdaSpace{}%
\AgdaModule{Data.Nat}\<%
\\
\>[0]\AgdaKeyword{open}\AgdaSpace{}%
\AgdaKeyword{import}\AgdaSpace{}%
\AgdaModule{Data.Maybe}\AgdaSpace{}%
\AgdaSymbol{as}\AgdaSpace{}%
\AgdaModule{Maybe}\<%
\\
\>[0]\AgdaKeyword{open}\AgdaSpace{}%
\AgdaKeyword{import}\AgdaSpace{}%
\AgdaModule{Data.Product}\<%
\\
\>[0]\AgdaKeyword{open}\AgdaSpace{}%
\AgdaKeyword{import}\AgdaSpace{}%
\AgdaModule{Data.Sum}\AgdaSpace{}%
\AgdaKeyword{hiding}\AgdaSpace{}%
\AgdaSymbol{(}\AgdaFunction{reduce}\AgdaSymbol{)}\AgdaSpace{}%
\AgdaKeyword{renaming}\AgdaSpace{}%
\AgdaSymbol{(}\AgdaOperator{\AgdaFunction{[\AgdaUnderscore{},\AgdaUnderscore{}]}}\AgdaSpace{}%
\AgdaSymbol{to}\AgdaSpace{}%
\AgdaOperator{\AgdaFunction{\AgdaUnderscore{}∇\AgdaUnderscore{}}}\AgdaSymbol{)}\<%
\\
\>[0]\AgdaKeyword{open}\AgdaSpace{}%
\AgdaKeyword{import}\AgdaSpace{}%
\AgdaModule{Data.List}\<%
\\
\\[\AgdaEmptyExtraSkip]%
\>[0]\AgdaKeyword{open}\AgdaSpace{}%
\AgdaKeyword{import}\AgdaSpace{}%
\AgdaModule{Relation.Nullary}\<%
\\
\>[0]\AgdaKeyword{open}\AgdaSpace{}%
\AgdaKeyword{import}\AgdaSpace{}%
\AgdaModule{Relation.Binary.PropositionalEquality}\<%
\\
\\[\AgdaEmptyExtraSkip]%
\>[0]\AgdaKeyword{module}\AgdaSpace{}%
\AgdaModule{sections.02-redsem}\AgdaSpace{}%
\AgdaKeyword{where}\<%
\end{code}

\section{A Primer on Reduction Semantics and Their Implementation}
\label{sec:02-redsem}

We recall what a reduction semantics is, and how to implement it following a similar \emph{modus operandi} as Danvy~\cite{Danvy08afp}:
\begin{enumerate}
\item Abstract syntax
\item Notion of contraction
\item Reduction strategy: reduction contexts, decomposition, and recomposition
\item One-step reduction
\item Reduction-based normalization
\end{enumerate}
The reduction semantics we will implement is shown in \cref{fig:redsem-arith}.
The abstract syntax has two kinds of expressions (ranged over by $e,e',e_0,\ldots$): addition expressions and numbers.
Numbers are values (ranged over by $v,v',v_0,\ldots$).

The contraction rule in the figure contracts addition expressions whose immediate subterms are numbers into a number expression (or value).
The one-step reduction rule in the figure lets us reduce nested addition expressions, by decomposing it into into a \emph{reduction context} and a redex; e.g., $E[e]$ which represents the recomposition of the expression $e$ into the evaluation context $E$.
Thus the rule says, for any input expression $e_0 = E[e]$, if $e \longmapsto e'$, then the result of a single reduction step is $E[e']$.
The \textsc{[Trans]} rule transitively applies the one-step reduction relation, and the \textsc{[Refl]} rule applies when the left-hand side of the relation is a value $v$.

By convention, we require that languages can be uniquely decomposed; i.e., for any $e, e', e'', E, E'$, if $e = E[e']$ and $e = E'[e'']$ then $E = E'$ and $e' = e''$.
Languages that do not have this property are beyond the scope of this work.


\begin{figure}
\raggedright
\textbf{Abstract Syntax}
\begin{mathpar}
  e \bnfdef \mathsf{add}\ e\ e \mid \mathsf{num}\ n
  \and
  v \bnfdef \mathsf{num}\ n
\end{mathpar}
\textbf{Notion of Contraction}
\begin{mathpar}
\inferrule{
}{
  \mathsf{add}\ (\mathsf{num}\ n_0)\ (\mathsf{num}\ n_1)
  \longmapsto
  \mathsf{num}\ (n_0 + n_1)
}
\end{mathpar}
\textbf{Reduction Contexts}
\begin{mathpar}
  E \bnfdef [] \mid \mathsf{add}\ E\ e \mid \mathsf{add}\ v\ E
\end{mathpar}
\textbf{One-Step Reduction}
\begin{mathpar}
\inferrule{
  e \longmapsto e'
}{
  E[e] \longrightarrow E[e']
}
\end{mathpar}
\textbf{Reduction-Based Normalization}
\begin{mathpar}
\inferrule[Trans]{
  e_0 \longrightarrow e_1
  \and
  e_1 \longrightarrow^{*} e_2
}{
  e_0 \longrightarrow^{*} e_2
}
\and
\inferrule[Refl]{}{
  v \longrightarrow^{*} v
}
\end{mathpar}
\caption{Reduction semantics for simple arithmetic}
\label{fig:redsem-arith}
\end{figure}

\subsection{Implementing a Reduction Semantics for Simple Arithmetic Expressions}
\label{sec:simple-arith}

\begin{code}[hide]%
\>[0]\AgdaKeyword{module}\AgdaSpace{}%
\AgdaModule{SimpleArith}\AgdaSpace{}%
\AgdaKeyword{where}\<%
\end{code}

The semantics in \cref{fig:redsem-arith} can be implemented using the modus operandi from the beginning of the section.
While the modus operandi is borrowed from Danvy, our types differ slightly from Danvy's.
\cref{sec:relation-to-danvy} discusses differences.

\subsubsection{Abstract Syntax}
The syntax of arithmetic expressions is given below where \as{(}\ab{t₁~t₂}~\as{:}~\ad{Term}\as{)~→}~\ad{Term} is a two-argument term constructor.
\begin{code}%
\>[0][@{}l@{\AgdaIndent{1}}]%
\>[2]\AgdaKeyword{data}\AgdaSpace{}%
\AgdaDatatype{Term}\AgdaSpace{}%
\AgdaSymbol{:}\AgdaSpace{}%
\AgdaPrimitive{Set}\AgdaSpace{}%
\AgdaKeyword{where}\<%
\\
\>[2][@{}l@{\AgdaIndent{0}}]%
\>[4]\AgdaInductiveConstructor{add}%
\>[9]\AgdaSymbol{:}\AgdaSpace{}%
\AgdaSymbol{(}\AgdaBound{t₀}\AgdaSpace{}%
\AgdaBound{t₁}\AgdaSpace{}%
\AgdaSymbol{:}\AgdaSpace{}%
\AgdaDatatype{Term}\AgdaSymbol{)}%
\>[27]\AgdaSymbol{→}\AgdaSpace{}%
\AgdaDatatype{Term}\<%
\\
\>[4]\AgdaInductiveConstructor{num}%
\>[9]\AgdaSymbol{:}\AgdaSpace{}%
\AgdaDatatype{ℕ}%
\>[27]\AgdaSymbol{→}\AgdaSpace{}%
\AgdaDatatype{Term}\<%
\end{code}
Values are numbers.
\begin{code}%
\>[2]\AgdaKeyword{data}\AgdaSpace{}%
\AgdaDatatype{Val}\AgdaSpace{}%
\AgdaSymbol{:}\AgdaSpace{}%
\AgdaPrimitive{Set}\AgdaSpace{}%
\AgdaKeyword{where}\AgdaSpace{}%
\AgdaInductiveConstructor{num}\AgdaSpace{}%
\AgdaSymbol{:}\AgdaSpace{}%
\AgdaDatatype{ℕ}\AgdaSpace{}%
\AgdaSymbol{→}\AgdaSpace{}%
\AgdaDatatype{Val}\<%
\end{code}
Note that Agda lets us overload constructor names, so the \ac{num} constructors of \ad{Term} and \ad{Val} construct values of distinct types.
However, we can convert a value to a term:
\begin{code}%
\>[2]\AgdaFunction{v2t}\AgdaSpace{}%
\AgdaSymbol{:}\AgdaSpace{}%
\AgdaDatatype{Val}\AgdaSpace{}%
\AgdaSymbol{→}\AgdaSpace{}%
\AgdaDatatype{Term}\<%
\\
\>[2]\AgdaFunction{v2t}\AgdaSpace{}%
\AgdaSymbol{(}\AgdaInductiveConstructor{num}\AgdaSpace{}%
\AgdaBound{n}\AgdaSymbol{)}\AgdaSpace{}%
\AgdaSymbol{=}\AgdaSpace{}%
\AgdaInductiveConstructor{num}\AgdaSpace{}%
\AgdaBound{n}\<%
\end{code}
The inverse is a partial function:
\begin{code}%
\>[2]\AgdaFunction{t2v}\AgdaSpace{}%
\AgdaSymbol{:}\AgdaSpace{}%
\AgdaDatatype{Term}\AgdaSpace{}%
\AgdaSymbol{→}\AgdaSpace{}%
\AgdaDatatype{Maybe}\AgdaSpace{}%
\AgdaDatatype{Val}\<%
\\
\>[2]\AgdaFunction{t2v}\AgdaSpace{}%
\AgdaSymbol{(}\AgdaInductiveConstructor{num}\AgdaSpace{}%
\AgdaBound{n}\AgdaSymbol{)}%
\>[15]\AgdaSymbol{=}\AgdaSpace{}%
\AgdaInductiveConstructor{just}\AgdaSpace{}%
\AgdaSymbol{(}\AgdaInductiveConstructor{num}\AgdaSpace{}%
\AgdaBound{n}\AgdaSymbol{)}\<%
\\
\>[2]\AgdaCatchallClause{\AgdaFunction{t2v}}\AgdaSpace{}%
\AgdaCatchallClause{\AgdaSymbol{\AgdaUnderscore{}}}%
\>[15]\AgdaSymbol{=}\AgdaSpace{}%
\AgdaInductiveConstructor{nothing}\<%
\end{code}

\subsubsection{Notion of Contraction}
For simple arithmetic expressions, redexes are contracted as follows.
\begin{code}%
\>[2]\AgdaFunction{contract}\AgdaSpace{}%
\AgdaSymbol{:}\AgdaSpace{}%
\AgdaDatatype{Term}\AgdaSpace{}%
\AgdaSymbol{→}\AgdaSpace{}%
\AgdaDatatype{Maybe}\AgdaSpace{}%
\AgdaDatatype{Term}\<%
\\
\>[2]\AgdaFunction{contract}\AgdaSpace{}%
\AgdaSymbol{(}\AgdaInductiveConstructor{add}\AgdaSpace{}%
\AgdaSymbol{(}\AgdaInductiveConstructor{num}\AgdaSpace{}%
\AgdaBound{n₀}\AgdaSymbol{)}\AgdaSpace{}%
\AgdaSymbol{(}\AgdaInductiveConstructor{num}\AgdaSpace{}%
\AgdaBound{n₁}\AgdaSymbol{))}%
\>[36]\AgdaSymbol{=}\AgdaSpace{}%
\AgdaInductiveConstructor{just}\AgdaSpace{}%
\AgdaSymbol{(}\AgdaInductiveConstructor{num}\AgdaSpace{}%
\AgdaSymbol{(}\AgdaBound{n₀}\AgdaSpace{}%
\AgdaOperator{\AgdaPrimitive{+}}\AgdaSpace{}%
\AgdaBound{n₁}\AgdaSymbol{))}\<%
\\
\>[2]\AgdaCatchallClause{\AgdaFunction{contract}}\AgdaSpace{}%
\AgdaCatchallClause{\AgdaSymbol{\AgdaUnderscore{}}}%
\>[36]\AgdaSymbol{=}\AgdaSpace{}%
\AgdaInductiveConstructor{nothing}\<%
\end{code}
If the input term is an addition expression with numbers as immediate sub-terms, we contract it.
Otherwise, it is not possible to contract the term.

In order to decompose a term we will also make use of the following helper function for recognizing a redex.
\begin{code}%
\>[2]\AgdaFunction{is-redex}\AgdaSpace{}%
\AgdaSymbol{:}\AgdaSpace{}%
\AgdaDatatype{Term}\AgdaSpace{}%
\AgdaSymbol{→}\AgdaSpace{}%
\AgdaDatatype{Bool}\<%
\\
\>[2]\AgdaFunction{is-redex}\AgdaSpace{}%
\AgdaSymbol{(}\AgdaInductiveConstructor{add}\AgdaSpace{}%
\AgdaSymbol{(}\AgdaInductiveConstructor{num}\AgdaSpace{}%
\AgdaBound{n₀}\AgdaSymbol{)}\AgdaSpace{}%
\AgdaSymbol{(}\AgdaInductiveConstructor{num}\AgdaSpace{}%
\AgdaBound{n₁}\AgdaSymbol{))}%
\>[36]\AgdaSymbol{=}\AgdaSpace{}%
\AgdaInductiveConstructor{true}\<%
\\
\>[2]\AgdaCatchallClause{\AgdaFunction{is-redex}}\AgdaSpace{}%
\AgdaCatchallClause{\AgdaSymbol{\AgdaUnderscore{}}}%
\>[36]\AgdaSymbol{=}\AgdaSpace{}%
\AgdaInductiveConstructor{false}\<%
\end{code}

The type of \af{contract} above differs from \ad{contract} function from the introduction which also accepted a \ad{Context} argument.
The type with the \ad{Context} parameter is generally preferable as that supports \emph{context-sensitive contraction}, as we will discuss in \cref{sec:lambda-shift-reset}.
However, for our simple arithmetic language, the simpler type suffices.

\subsubsection{Reduction Strategy}
\label{sec:reduction-strategy-arith}

The reduction strategy is determined by the notion of \emph{reduction context} and its \emph{decomposition} and \emph{recomposition} functions.
The following data type defines reduction contexts corresponding to \cref{fig:redsem-arith}.
\begin{code}%
\>[2]\AgdaKeyword{data}\AgdaSpace{}%
\AgdaDatatype{Context₀}\AgdaSpace{}%
\AgdaSymbol{:}\AgdaSpace{}%
\AgdaPrimitive{Set}\AgdaSpace{}%
\AgdaKeyword{where}\<%
\\
\>[2][@{}l@{\AgdaIndent{0}}]%
\>[4]\AgdaInductiveConstructor{top}%
\>[11]\AgdaSymbol{:}%
\>[38]\AgdaDatatype{Context₀}\<%
\\
\>[4]\AgdaInductiveConstructor{add-l}%
\>[11]\AgdaSymbol{:}\AgdaSpace{}%
\AgdaDatatype{Context₀}%
\>[23]\AgdaSymbol{→}\AgdaSpace{}%
\AgdaDatatype{Term}%
\>[35]\AgdaSymbol{→}%
\>[38]\AgdaDatatype{Context₀}\<%
\\
\>[4]\AgdaInductiveConstructor{add-r}%
\>[11]\AgdaSymbol{:}\AgdaSpace{}%
\AgdaDatatype{Val}%
\>[23]\AgdaSymbol{→}\AgdaSpace{}%
\AgdaDatatype{Context₀}%
\>[35]\AgdaSymbol{→}%
\>[38]\AgdaDatatype{Context₀}\<%
\end{code}
An isomorphic---and slightly more generic---encoding of contexts is given by a list of \emph{context frames}.
\begin{code}%
\>[2]\AgdaKeyword{data}\AgdaSpace{}%
\AgdaDatatype{CtxFrame}\AgdaSpace{}%
\AgdaSymbol{:}\AgdaSpace{}%
\AgdaPrimitive{Set}\AgdaSpace{}%
\AgdaKeyword{where}\<%
\\
\>[2][@{}l@{\AgdaIndent{0}}]%
\>[4]\AgdaInductiveConstructor{add-l}\AgdaSpace{}%
\AgdaSymbol{:}\AgdaSpace{}%
\AgdaDatatype{Term}%
\>[18]\AgdaSymbol{→}\AgdaSpace{}%
\AgdaDatatype{CtxFrame}\<%
\\
\>[4]\AgdaInductiveConstructor{add-r}\AgdaSpace{}%
\AgdaSymbol{:}\AgdaSpace{}%
\AgdaDatatype{Val}%
\>[18]\AgdaSymbol{→}\AgdaSpace{}%
\AgdaDatatype{CtxFrame}\<%
\\
\\[\AgdaEmptyExtraSkip]%
\>[2]\AgdaFunction{Context}\AgdaSpace{}%
\AgdaSymbol{=}\AgdaSpace{}%
\AgdaDatatype{List}\AgdaSpace{}%
\AgdaDatatype{CtxFrame}\<%
\end{code}
This context representation affords a simple definition of context composition:\footnote{\af{\_++\_} is list concatenation, as defined in Agda's standard library~\cite{agda-stdlib}.}
\begin{code}%
\>[2]\AgdaFunction{compose}\AgdaSpace{}%
\AgdaSymbol{:}\AgdaSpace{}%
\AgdaFunction{Context}\AgdaSpace{}%
\AgdaSymbol{→}\AgdaSpace{}%
\AgdaFunction{Context}\AgdaSpace{}%
\AgdaSymbol{→}\AgdaSpace{}%
\AgdaFunction{Context}\<%
\\
\>[2]\AgdaFunction{compose}\AgdaSpace{}%
\AgdaSymbol{=}\AgdaSpace{}%
\AgdaOperator{\AgdaFunction{\AgdaUnderscore{}++\AgdaUnderscore{}}}\<%
\end{code}
For the purpose of this paper we will order lists of context frames from left-to-right in increasing distance from the context hole.

\begin{example}
  The following context represents the context $\mathsf{add}\ (\mathsf{add}\ (\mathsf{num}\ 1)\ [])\ (\mathsf{num}\ 2)$.
  \begin{code}%
\>[2]\AgdaFunction{add-[add-[]-1]-2}\AgdaSpace{}%
\AgdaSymbol{:}\AgdaSpace{}%
\AgdaFunction{Context}\<%
\\
\>[2]\AgdaFunction{add-[add-[]-1]-2}\AgdaSpace{}%
\AgdaSymbol{=}\AgdaSpace{}%
\AgdaInductiveConstructor{add-r}\AgdaSpace{}%
\AgdaSymbol{(}\AgdaInductiveConstructor{num}\AgdaSpace{}%
\AgdaNumber{1}\AgdaSymbol{)}\AgdaSpace{}%
\AgdaOperator{\AgdaInductiveConstructor{∷}}\AgdaSpace{}%
\AgdaInductiveConstructor{add-l}\AgdaSpace{}%
\AgdaSymbol{(}\AgdaInductiveConstructor{num}\AgdaSpace{}%
\AgdaNumber{2}\AgdaSymbol{)}\AgdaSpace{}%
\AgdaOperator{\AgdaInductiveConstructor{∷}}\AgdaSpace{}%
\AgdaInductiveConstructor{[]}\<%
\end{code}
\end{example}

Next, we need a decomposition function which decomposes a term into a potential redex (i.e., a term) and context, if possible:\footnote{The syntax \ab{t}\as{@(}\ac{add}~\ab{t₀}~\ab{t₁}\as{)} matches the case where the input term is an \ac{add} expression, binding \ab{t} to the input term.}
\begin{code}%
\>[2]\AgdaFunction{decompose}\AgdaSpace{}%
\AgdaSymbol{:}\AgdaSpace{}%
\AgdaDatatype{Term}\AgdaSpace{}%
\AgdaSymbol{→}\AgdaSpace{}%
\AgdaFunction{Context}\AgdaSpace{}%
\AgdaSymbol{→}\AgdaSpace{}%
\AgdaDatatype{Maybe}\AgdaSpace{}%
\AgdaSymbol{(}\AgdaDatatype{Term}\AgdaSpace{}%
\AgdaOperator{\AgdaFunction{×}}\AgdaSpace{}%
\AgdaFunction{Context}\AgdaSymbol{)}\<%
\\
\>[2]\AgdaFunction{decompose}\AgdaSpace{}%
\AgdaSymbol{(}\AgdaInductiveConstructor{num}\AgdaSpace{}%
\AgdaBound{x}\AgdaSymbol{)}%
\>[27]\AgdaBound{c}\AgdaSpace{}%
\AgdaSymbol{=}\AgdaSpace{}%
\AgdaInductiveConstructor{nothing}\<%
\\
\>[2]\AgdaFunction{decompose}\AgdaSpace{}%
\AgdaBound{t}\AgdaSymbol{@(}\AgdaInductiveConstructor{add}\AgdaSpace{}%
\AgdaBound{t₀}\AgdaSpace{}%
\AgdaBound{t₁}\AgdaSymbol{)}%
\>[27]\AgdaBound{c}\AgdaSpace{}%
\AgdaSymbol{=}\AgdaSpace{}%
\AgdaFunction{maybe}\AgdaSpace{}%
\AgdaInductiveConstructor{just}\<%
\\
\>[2][@{}l@{\AgdaIndent{0}}]%
\>[4]\AgdaSymbol{(}\AgdaOperator{\AgdaFunction{if}}\AgdaSpace{}%
\AgdaFunction{is-redex}\AgdaSpace{}%
\AgdaBound{t}\AgdaSpace{}%
\AgdaOperator{\AgdaFunction{then}}\AgdaSpace{}%
\AgdaInductiveConstructor{just}\AgdaSpace{}%
\AgdaSymbol{(}\AgdaBound{t}\AgdaSpace{}%
\AgdaOperator{\AgdaInductiveConstructor{,}}\AgdaSpace{}%
\AgdaBound{c}\AgdaSymbol{)}\AgdaSpace{}%
\AgdaOperator{\AgdaFunction{else}}\AgdaSpace{}%
\AgdaInductiveConstructor{nothing}\AgdaSymbol{)}\<%
\\
\>[4]\AgdaSymbol{(}\AgdaFunction{find-left-most}\<%
\\
\>[4][@{}l@{\AgdaIndent{0}}]%
\>[6]\AgdaSymbol{(}%
\>[9]\AgdaFunction{decompose}\AgdaSpace{}%
\AgdaBound{t₀}\AgdaSpace{}%
\AgdaSymbol{(}\AgdaInductiveConstructor{add-l}\AgdaSpace{}%
\AgdaBound{t₁}\AgdaSpace{}%
\AgdaOperator{\AgdaInductiveConstructor{∷}}\AgdaSpace{}%
\AgdaBound{c}\AgdaSymbol{)}\<%
\\
\>[6]\AgdaOperator{\AgdaInductiveConstructor{∷}}%
\>[9]\AgdaFunction{maybe}%
\>[16]\AgdaSymbol{(λ}\AgdaSpace{}%
\AgdaBound{v}\AgdaSpace{}%
\AgdaSymbol{→}\AgdaSpace{}%
\AgdaFunction{decompose}\AgdaSpace{}%
\AgdaBound{t₁}\AgdaSpace{}%
\AgdaSymbol{(}\AgdaInductiveConstructor{add-r}\AgdaSpace{}%
\AgdaBound{v}\AgdaSpace{}%
\AgdaOperator{\AgdaInductiveConstructor{∷}}\AgdaSpace{}%
\AgdaBound{c}\AgdaSymbol{))}\<%
\\
\>[16]\AgdaInductiveConstructor{nothing}\<%
\\
\>[16]\AgdaSymbol{(}\AgdaFunction{t2v}\AgdaSpace{}%
\AgdaBound{t₀}\AgdaSymbol{)}\<%
\\
\>[6]\AgdaOperator{\AgdaInductiveConstructor{∷}}%
\>[9]\AgdaInductiveConstructor{[]}\AgdaSpace{}%
\AgdaSymbol{))}\<%
\end{code}%
\begin{code}[hide]%
\>[4]\AgdaKeyword{where}\<%
\\
\>[4][@{}l@{\AgdaIndent{0}}]%
\>[6]\AgdaFunction{find-left-most}%
\>[22]\AgdaSymbol{:}\AgdaSpace{}%
\AgdaDatatype{List}\AgdaSpace{}%
\AgdaSymbol{(}\AgdaDatatype{Maybe}\AgdaSpace{}%
\AgdaSymbol{(}\AgdaDatatype{Term}\AgdaSpace{}%
\AgdaOperator{\AgdaFunction{×}}\AgdaSpace{}%
\AgdaFunction{Context}\AgdaSymbol{))}\<%
\\
\>[22]\AgdaSymbol{→}\AgdaSpace{}%
\AgdaDatatype{Maybe}\AgdaSpace{}%
\AgdaSymbol{(}\AgdaDatatype{Term}\AgdaSpace{}%
\AgdaOperator{\AgdaFunction{×}}\AgdaSpace{}%
\AgdaFunction{Context}\AgdaSymbol{)}\<%
\\
\>[6]\AgdaFunction{find-left-most}\AgdaSpace{}%
\AgdaInductiveConstructor{[]}%
\>[30]\AgdaSymbol{=}\AgdaSpace{}%
\AgdaInductiveConstructor{nothing}\<%
\\
\>[6]\AgdaFunction{find-left-most}\AgdaSpace{}%
\AgdaSymbol{(}\AgdaBound{m}\AgdaSpace{}%
\AgdaOperator{\AgdaInductiveConstructor{∷}}\AgdaSpace{}%
\AgdaBound{r}\AgdaSymbol{)}%
\>[30]\AgdaSymbol{=}\AgdaSpace{}%
\AgdaFunction{maybe}\AgdaSpace{}%
\AgdaInductiveConstructor{just}\<%
\\
\>[6][@{}l@{\AgdaIndent{0}}]%
\>[8]\AgdaSymbol{(}\AgdaFunction{find-left-most}\AgdaSpace{}%
\AgdaBound{r}\AgdaSymbol{)}\<%
\\
\>[8]\AgdaBound{m}\<%
\end{code}%
The \af{maybe} function is the eliminator for the \ad{Maybe} type---i.e.,
\begin{code}[hide]%
\>[2]\AgdaKeyword{postulate}\<%
\end{code}%
\begin{code}[inline]%
\>[2][@{}l@{\AgdaIndent{1}}]%
\>[4]\AgdaPostulate{⅋maybe}\AgdaSpace{}%
\AgdaSymbol{:}\AgdaSpace{}%
\AgdaSymbol{\{}\AgdaBound{A}\AgdaSpace{}%
\AgdaBound{B}\AgdaSpace{}%
\AgdaSymbol{:}\AgdaSpace{}%
\AgdaPrimitive{Set}\AgdaSymbol{\}}\AgdaSpace{}%
\AgdaSymbol{→}\AgdaSpace{}%
\AgdaSymbol{(}\AgdaBound{A}\AgdaSpace{}%
\AgdaSymbol{→}\AgdaSpace{}%
\AgdaBound{B}\AgdaSymbol{)}\AgdaSpace{}%
\AgdaSymbol{→}\AgdaSpace{}%
\AgdaBound{B}\AgdaSpace{}%
\AgdaSymbol{→}\AgdaSpace{}%
\AgdaDatatype{Maybe}\AgdaSpace{}%
\AgdaBound{A}\AgdaSpace{}%
\AgdaSymbol{→}\AgdaSpace{}%
\AgdaBound{B}\<%
\end{code}%
.\footnote{The curly braces represent implicitly-bound parameters which Agda will automatically infer for us at function application sites.}
The function constructs a list of possible decompositions in left-to-right order, and uses a function \af{find-left-most} to yield the left-most inner-most redex.

Recomposition ``plugs'' the hole of a one-hole context with a term:
\begin{code}%
\>[2]\AgdaFunction{recompose-frame}\AgdaSpace{}%
\AgdaSymbol{:}\AgdaSpace{}%
\AgdaDatatype{CtxFrame}\AgdaSpace{}%
\AgdaSymbol{→}\AgdaSpace{}%
\AgdaDatatype{Term}\AgdaSpace{}%
\AgdaSymbol{→}\AgdaSpace{}%
\AgdaDatatype{Term}\<%
\\
\>[2]\AgdaFunction{recompose-frame}\AgdaSpace{}%
\AgdaSymbol{(}\AgdaInductiveConstructor{add-l}\AgdaSpace{}%
\AgdaBound{t₁}\AgdaSymbol{)}%
\>[30]\AgdaBound{t}\AgdaSpace{}%
\AgdaSymbol{=}\AgdaSpace{}%
\AgdaInductiveConstructor{add}\AgdaSpace{}%
\AgdaBound{t}\AgdaSpace{}%
\AgdaBound{t₁}\<%
\\
\>[2]\AgdaFunction{recompose-frame}\AgdaSpace{}%
\AgdaSymbol{(}\AgdaInductiveConstructor{add-r}\AgdaSpace{}%
\AgdaBound{v}\AgdaSymbol{)}%
\>[30]\AgdaBound{t}%
\>[33]\AgdaSymbol{=}\AgdaSpace{}%
\AgdaInductiveConstructor{add}\AgdaSpace{}%
\AgdaSymbol{(}\AgdaFunction{v2t}\AgdaSpace{}%
\AgdaBound{v}\AgdaSymbol{)}\AgdaSpace{}%
\AgdaBound{t}\<%
\\
\\[\AgdaEmptyExtraSkip]%
\>[2]\AgdaFunction{recompose}\AgdaSpace{}%
\AgdaSymbol{:}\AgdaSpace{}%
\AgdaFunction{Context}\AgdaSpace{}%
\AgdaSymbol{→}\AgdaSpace{}%
\AgdaDatatype{Term}\AgdaSpace{}%
\AgdaSymbol{→}\AgdaSpace{}%
\AgdaDatatype{Term}\<%
\\
\>[2]\AgdaFunction{recompose}\AgdaSpace{}%
\AgdaInductiveConstructor{[]}%
\>[23]\AgdaBound{t}\AgdaSpace{}%
\AgdaSymbol{=}\AgdaSpace{}%
\AgdaBound{t}\<%
\\
\>[2]\AgdaFunction{recompose}\AgdaSpace{}%
\AgdaSymbol{(}\AgdaBound{frm}\AgdaSpace{}%
\AgdaOperator{\AgdaInductiveConstructor{∷}}\AgdaSpace{}%
\AgdaBound{c}\AgdaSymbol{)}%
\>[23]\AgdaBound{t}\AgdaSpace{}%
\AgdaSymbol{=}\<%
\\
\>[2][@{}l@{\AgdaIndent{0}}]%
\>[4]\AgdaFunction{recompose}\AgdaSpace{}%
\AgdaBound{c}\AgdaSpace{}%
\AgdaSymbol{(}\AgdaFunction{recompose-frame}\AgdaSpace{}%
\AgdaBound{frm}\AgdaSpace{}%
\AgdaBound{t}\AgdaSymbol{)}\<%
\end{code}

Decomposition and recomposition satisfy the following properties:
\begin{code}%
\>[2]\AgdaFunction{recompose-compose}\AgdaSpace{}%
\AgdaSymbol{:}\<%
\\
\>[2][@{}l@{\AgdaIndent{0}}]%
\>[4]\AgdaSymbol{∀}\AgdaSpace{}%
\AgdaSymbol{\{}\AgdaBound{t}\AgdaSpace{}%
\AgdaBound{c₀}\AgdaSpace{}%
\AgdaBound{c₁}\AgdaSymbol{\}}\<%
\\
\>[4]\AgdaSymbol{→}%
\>[263I]\AgdaFunction{recompose}\AgdaSpace{}%
\AgdaSymbol{(}\AgdaFunction{compose}\AgdaSpace{}%
\AgdaBound{c₀}\AgdaSpace{}%
\AgdaBound{c₁}\AgdaSymbol{)}\AgdaSpace{}%
\AgdaBound{t}\<%
\\
\>[.][@{}l@{}]\<[263I]%
\>[6]\AgdaOperator{\AgdaDatatype{≡}}\AgdaSpace{}%
\AgdaFunction{recompose}\AgdaSpace{}%
\AgdaBound{c₁}\AgdaSpace{}%
\AgdaSymbol{(}\AgdaFunction{recompose}\AgdaSpace{}%
\AgdaBound{c₀}\AgdaSpace{}%
\AgdaBound{t}\AgdaSymbol{)}\<%
\\
\\[\AgdaEmptyExtraSkip]%
\>[2]\AgdaFunction{decompose-recompose}\AgdaSpace{}%
\AgdaSymbol{:}\<%
\\
\>[2][@{}l@{\AgdaIndent{0}}]%
\>[4]\AgdaSymbol{∀}\AgdaSpace{}%
\AgdaSymbol{\{}\AgdaBound{t₀}\AgdaSpace{}%
\AgdaBound{t}\AgdaSpace{}%
\AgdaBound{c₀}\AgdaSpace{}%
\AgdaBound{c}\AgdaSymbol{\}}\<%
\\
\>[4]\AgdaSymbol{→}\AgdaSpace{}%
\AgdaFunction{decompose}\AgdaSpace{}%
\AgdaBound{t₀}\AgdaSpace{}%
\AgdaBound{c₀}\AgdaSpace{}%
\AgdaOperator{\AgdaDatatype{≡}}\AgdaSpace{}%
\AgdaInductiveConstructor{just}\AgdaSpace{}%
\AgdaSymbol{(}\AgdaBound{t}\AgdaSpace{}%
\AgdaOperator{\AgdaInductiveConstructor{,}}\AgdaSpace{}%
\AgdaBound{c}\AgdaSymbol{)}\<%
\\
\>[4]\AgdaSymbol{→}\AgdaSpace{}%
\AgdaFunction{recompose}\AgdaSpace{}%
\AgdaBound{c}\AgdaSpace{}%
\AgdaBound{t}\AgdaSpace{}%
\AgdaOperator{\AgdaDatatype{≡}}\AgdaSpace{}%
\AgdaFunction{recompose}\AgdaSpace{}%
\AgdaBound{c₀}\AgdaSpace{}%
\AgdaBound{t₀}\<%
\end{code}
\begin{code}[hide]%
\>[2]\AgdaFunction{rc}%
\>[293I]\AgdaSymbol{:}\AgdaSpace{}%
\AgdaSymbol{(}\AgdaBound{t}\AgdaSpace{}%
\AgdaSymbol{:}\AgdaSpace{}%
\AgdaDatatype{Term}\AgdaSymbol{)}\AgdaSpace{}%
\AgdaSymbol{(}\AgdaBound{c₀}\AgdaSpace{}%
\AgdaBound{c₁}\AgdaSpace{}%
\AgdaSymbol{:}\AgdaSpace{}%
\AgdaFunction{Context}\AgdaSymbol{)}\<%
\\
\>[.][@{}l@{}]\<[293I]%
\>[5]\AgdaSymbol{→}\AgdaSpace{}%
\AgdaFunction{recompose}\AgdaSpace{}%
\AgdaSymbol{(}\AgdaBound{c₀}\AgdaSpace{}%
\AgdaOperator{\AgdaFunction{++}}\AgdaSpace{}%
\AgdaBound{c₁}\AgdaSymbol{)}\AgdaSpace{}%
\AgdaBound{t}\AgdaSpace{}%
\AgdaOperator{\AgdaDatatype{≡}}\AgdaSpace{}%
\AgdaFunction{recompose}\AgdaSpace{}%
\AgdaBound{c₁}\AgdaSpace{}%
\AgdaSymbol{(}\AgdaFunction{recompose}\AgdaSpace{}%
\AgdaBound{c₀}\AgdaSpace{}%
\AgdaBound{t}\AgdaSymbol{)}\<%
\\
\>[2]\AgdaFunction{dr}%
\>[312I]\AgdaSymbol{:}\AgdaSpace{}%
\AgdaSymbol{(}\AgdaBound{t₀}\AgdaSpace{}%
\AgdaBound{t}\AgdaSpace{}%
\AgdaSymbol{:}\AgdaSpace{}%
\AgdaDatatype{Term}\AgdaSymbol{)}\AgdaSpace{}%
\AgdaSymbol{(}\AgdaBound{c₀}\AgdaSpace{}%
\AgdaBound{c}\AgdaSpace{}%
\AgdaSymbol{:}\AgdaSpace{}%
\AgdaFunction{Context}\AgdaSymbol{)}\<%
\\
\>[.][@{}l@{}]\<[312I]%
\>[5]\AgdaSymbol{→}\AgdaSpace{}%
\AgdaFunction{decompose}\AgdaSpace{}%
\AgdaBound{t₀}\AgdaSpace{}%
\AgdaBound{c₀}\AgdaSpace{}%
\AgdaOperator{\AgdaDatatype{≡}}\AgdaSpace{}%
\AgdaInductiveConstructor{just}\AgdaSpace{}%
\AgdaSymbol{(}\AgdaBound{t}\AgdaSpace{}%
\AgdaOperator{\AgdaInductiveConstructor{,}}\AgdaSpace{}%
\AgdaBound{c}\AgdaSymbol{)}\<%
\\
\>[5]\AgdaSymbol{→}\AgdaSpace{}%
\AgdaFunction{recompose}\AgdaSpace{}%
\AgdaBound{c}\AgdaSpace{}%
\AgdaBound{t}\AgdaSpace{}%
\AgdaOperator{\AgdaDatatype{≡}}\AgdaSpace{}%
\AgdaFunction{recompose}\AgdaSpace{}%
\AgdaBound{c₀}\AgdaSpace{}%
\AgdaBound{t₀}\<%
\\
\>[0]\<%
\\
\>[2]\AgdaFunction{recompose-compose}\AgdaSpace{}%
\AgdaSymbol{\{}\AgdaBound{t}\AgdaSymbol{\}}\AgdaSpace{}%
\AgdaSymbol{\{}\AgdaBound{c₀}\AgdaSymbol{\}}\AgdaSpace{}%
\AgdaSymbol{\{}\AgdaBound{c₁}\AgdaSymbol{\}}\AgdaSpace{}%
\AgdaSymbol{=}\AgdaSpace{}%
\AgdaFunction{rc}\AgdaSpace{}%
\AgdaBound{t}\AgdaSpace{}%
\AgdaBound{c₀}\AgdaSpace{}%
\AgdaBound{c₁}\<%
\\
\>[2]\AgdaFunction{decompose-recompose}\AgdaSpace{}%
\AgdaSymbol{\{}\AgdaBound{t₀}\AgdaSymbol{\}}\AgdaSpace{}%
\AgdaSymbol{\{}\AgdaBound{t}\AgdaSymbol{\}}\AgdaSpace{}%
\AgdaSymbol{\{}\AgdaBound{c₀}\AgdaSymbol{\}}\AgdaSpace{}%
\AgdaSymbol{\{}\AgdaBound{c}\AgdaSymbol{\}}\AgdaSpace{}%
\AgdaSymbol{=}\AgdaSpace{}%
\AgdaFunction{dr}\AgdaSpace{}%
\AgdaBound{t₀}\AgdaSpace{}%
\AgdaBound{t}\AgdaSpace{}%
\AgdaBound{c₀}\AgdaSpace{}%
\AgdaBound{c}\<%
\\
\\[\AgdaEmptyExtraSkip]%
\>[2]\AgdaFunction{rc}\AgdaSpace{}%
\AgdaBound{t}\AgdaSpace{}%
\AgdaInductiveConstructor{[]}\AgdaSpace{}%
\AgdaBound{c₁}\AgdaSpace{}%
\AgdaSymbol{=}\AgdaSpace{}%
\AgdaInductiveConstructor{refl}\<%
\\
\>[2]\AgdaFunction{rc}\AgdaSpace{}%
\AgdaBound{t}\AgdaSpace{}%
\AgdaSymbol{(}\AgdaInductiveConstructor{add-l}\AgdaSpace{}%
\AgdaBound{x}\AgdaSpace{}%
\AgdaOperator{\AgdaInductiveConstructor{∷}}\AgdaSpace{}%
\AgdaBound{c₀}\AgdaSymbol{)}\AgdaSpace{}%
\AgdaBound{c₁}\AgdaSpace{}%
\AgdaSymbol{=}\AgdaSpace{}%
\AgdaFunction{rc}\AgdaSpace{}%
\AgdaSymbol{(}\AgdaInductiveConstructor{add}\AgdaSpace{}%
\AgdaBound{t}\AgdaSpace{}%
\AgdaBound{x}\AgdaSymbol{)}\AgdaSpace{}%
\AgdaBound{c₀}\AgdaSpace{}%
\AgdaBound{c₁}\<%
\\
\>[2]\AgdaFunction{rc}\AgdaSpace{}%
\AgdaBound{t}\AgdaSpace{}%
\AgdaSymbol{(}\AgdaInductiveConstructor{add-r}\AgdaSpace{}%
\AgdaBound{x}\AgdaSpace{}%
\AgdaOperator{\AgdaInductiveConstructor{∷}}\AgdaSpace{}%
\AgdaBound{c₀}\AgdaSymbol{)}\AgdaSpace{}%
\AgdaBound{c₁}\AgdaSpace{}%
\AgdaSymbol{=}\AgdaSpace{}%
\AgdaFunction{rc}\AgdaSpace{}%
\AgdaSymbol{(}\AgdaInductiveConstructor{add}\AgdaSpace{}%
\AgdaSymbol{(}\AgdaFunction{v2t}\AgdaSpace{}%
\AgdaBound{x}\AgdaSymbol{)}\AgdaSpace{}%
\AgdaBound{t}\AgdaSymbol{)}\AgdaSpace{}%
\AgdaBound{c₀}\AgdaSpace{}%
\AgdaBound{c₁}\<%
\\
\\[\AgdaEmptyExtraSkip]%
\\[\AgdaEmptyExtraSkip]%
\>[2]\AgdaFunction{dr}\AgdaSpace{}%
\AgdaSymbol{(}\AgdaInductiveConstructor{add}\AgdaSpace{}%
\AgdaBound{t₀}\AgdaSpace{}%
\AgdaBound{t₁}\AgdaSymbol{)}\AgdaSpace{}%
\AgdaBound{t}\AgdaSpace{}%
\AgdaBound{c₀}\AgdaSpace{}%
\AgdaBound{c}\AgdaSpace{}%
\AgdaBound{eq}%
\>[393I]\AgdaKeyword{with}\AgdaSpace{}%
\AgdaFunction{decompose}\AgdaSpace{}%
\AgdaBound{t₀}\AgdaSpace{}%
\AgdaSymbol{(}\AgdaInductiveConstructor{add-l}\AgdaSpace{}%
\AgdaBound{t₁}\AgdaSpace{}%
\AgdaOperator{\AgdaInductiveConstructor{∷}}\AgdaSpace{}%
\AgdaBound{c₀}\AgdaSymbol{)}\<%
\\
\>[393I][@{}l@{\AgdaIndent{0}}]%
\>[30]\AgdaSymbol{|}\AgdaSpace{}%
\AgdaFunction{inspect}\AgdaSpace{}%
\AgdaSymbol{(}\AgdaFunction{decompose}\AgdaSpace{}%
\AgdaBound{t₀}\AgdaSymbol{)}\AgdaSpace{}%
\AgdaSymbol{(}\AgdaInductiveConstructor{add-l}\AgdaSpace{}%
\AgdaBound{t₁}\AgdaSpace{}%
\AgdaOperator{\AgdaInductiveConstructor{∷}}\AgdaSpace{}%
\AgdaBound{c₀}\AgdaSymbol{)}\<%
\\
\>[2]\AgdaFunction{dr}\AgdaSpace{}%
\AgdaSymbol{(}\AgdaInductiveConstructor{add}\AgdaSpace{}%
\AgdaBound{t₀}\AgdaSpace{}%
\AgdaBound{t₁}\AgdaSymbol{)}\AgdaSpace{}%
\AgdaBound{t}\AgdaSpace{}%
\AgdaBound{c₀}\AgdaSpace{}%
\AgdaBound{c}\AgdaSpace{}%
\AgdaInductiveConstructor{refl}\AgdaSpace{}%
\AgdaSymbol{|}\AgdaSpace{}%
\AgdaInductiveConstructor{just}\AgdaSpace{}%
\AgdaDottedPattern{\AgdaSymbol{.(}}\AgdaDottedPattern{\AgdaBound{t}}\AgdaSpace{}%
\AgdaDottedPattern{\AgdaOperator{\AgdaInductiveConstructor{,}}}\AgdaSpace{}%
\AgdaDottedPattern{\AgdaBound{c}}\AgdaDottedPattern{\AgdaSymbol{)}}\AgdaSpace{}%
\AgdaSymbol{|}\AgdaSpace{}%
\AgdaOperator{\AgdaInductiveConstructor{[}}\AgdaSpace{}%
\AgdaBound{eq′}\AgdaSpace{}%
\AgdaOperator{\AgdaInductiveConstructor{]}}\AgdaSpace{}%
\AgdaSymbol{=}\AgdaSpace{}%
\AgdaFunction{dr}\AgdaSpace{}%
\AgdaBound{t₀}\AgdaSpace{}%
\AgdaBound{t}\AgdaSpace{}%
\AgdaSymbol{(}\AgdaInductiveConstructor{add-l}\AgdaSpace{}%
\AgdaBound{t₁}\AgdaSpace{}%
\AgdaOperator{\AgdaInductiveConstructor{∷}}\AgdaSpace{}%
\AgdaBound{c₀}\AgdaSymbol{)}\AgdaSpace{}%
\AgdaBound{c}\AgdaSpace{}%
\AgdaBound{eq′}\<%
\\
\>[2]\AgdaFunction{dr}\AgdaSpace{}%
\AgdaSymbol{(}\AgdaInductiveConstructor{add}\AgdaSpace{}%
\AgdaBound{t₀}\AgdaSpace{}%
\AgdaBound{t₁}\AgdaSymbol{)}\AgdaSpace{}%
\AgdaBound{t}\AgdaSpace{}%
\AgdaBound{c₀}\AgdaSpace{}%
\AgdaBound{c}\AgdaSpace{}%
\AgdaBound{eq}\AgdaSpace{}%
\AgdaSymbol{|}\AgdaSpace{}%
\AgdaInductiveConstructor{nothing}\AgdaSpace{}%
\AgdaSymbol{|}\AgdaSpace{}%
\AgdaOperator{\AgdaInductiveConstructor{[}}\AgdaSpace{}%
\AgdaBound{eq′}\AgdaSpace{}%
\AgdaOperator{\AgdaInductiveConstructor{]}}\AgdaSpace{}%
\AgdaKeyword{with}\AgdaSpace{}%
\AgdaFunction{t2v}\AgdaSpace{}%
\AgdaBound{t₀}\AgdaSpace{}%
\AgdaSymbol{|}\AgdaSpace{}%
\AgdaFunction{inspect}\AgdaSpace{}%
\AgdaFunction{t2v}\AgdaSpace{}%
\AgdaBound{t₀}\<%
\\
\>[2]\AgdaFunction{dr}\AgdaSpace{}%
\AgdaSymbol{(}\AgdaInductiveConstructor{add}\AgdaSpace{}%
\AgdaBound{t₀}\AgdaSpace{}%
\AgdaBound{t₁}\AgdaSymbol{)}\AgdaSpace{}%
\AgdaBound{t}\AgdaSpace{}%
\AgdaBound{c₀}\AgdaSpace{}%
\AgdaBound{c}\AgdaSpace{}%
\AgdaBound{eq}%
\>[460I]\AgdaSymbol{|}\AgdaSpace{}%
\AgdaInductiveConstructor{nothing}\AgdaSpace{}%
\AgdaSymbol{|}\AgdaSpace{}%
\AgdaOperator{\AgdaInductiveConstructor{[}}\AgdaSpace{}%
\AgdaBound{eq′}\AgdaSpace{}%
\AgdaOperator{\AgdaInductiveConstructor{]}}\<%
\\
\>[.][@{}l@{}]\<[460I]%
\>[27]\AgdaSymbol{|}\AgdaSpace{}%
\AgdaInductiveConstructor{just}\AgdaSpace{}%
\AgdaBound{x}\AgdaSpace{}%
\AgdaSymbol{|}\AgdaSpace{}%
\AgdaOperator{\AgdaInductiveConstructor{[}}\AgdaSpace{}%
\AgdaBound{eq₀}\AgdaSpace{}%
\AgdaOperator{\AgdaInductiveConstructor{]}}%
\>[472I]\AgdaKeyword{with}\AgdaSpace{}%
\AgdaFunction{decompose}\AgdaSpace{}%
\AgdaBound{t₁}\AgdaSpace{}%
\AgdaSymbol{(}\AgdaInductiveConstructor{add-r}\AgdaSpace{}%
\AgdaBound{x}\AgdaSpace{}%
\AgdaOperator{\AgdaInductiveConstructor{∷}}\AgdaSpace{}%
\AgdaBound{c₀}\AgdaSymbol{)}\<%
\\
\>[472I][@{}l@{\AgdaIndent{0}}]%
\>[49]\AgdaSymbol{|}\AgdaSpace{}%
\AgdaFunction{inspect}\AgdaSpace{}%
\AgdaSymbol{(}\AgdaFunction{decompose}\AgdaSpace{}%
\AgdaBound{t₁}\AgdaSymbol{)}\AgdaSpace{}%
\AgdaSymbol{(}\AgdaInductiveConstructor{add-r}\AgdaSpace{}%
\AgdaBound{x}\AgdaSpace{}%
\AgdaOperator{\AgdaInductiveConstructor{∷}}\AgdaSpace{}%
\AgdaBound{c₀}\AgdaSymbol{)}\<%
\\
\>[2]\AgdaFunction{dr}\AgdaSpace{}%
\AgdaSymbol{(}\AgdaInductiveConstructor{add}\AgdaSpace{}%
\AgdaSymbol{(}\AgdaInductiveConstructor{add}\AgdaSpace{}%
\AgdaBound{t₀}\AgdaSpace{}%
\AgdaBound{t₂}\AgdaSymbol{)}\AgdaSpace{}%
\AgdaBound{t₁}\AgdaSymbol{)}\AgdaSpace{}%
\AgdaBound{t}\AgdaSpace{}%
\AgdaBound{c₀}\AgdaSpace{}%
\AgdaBound{c}\AgdaSpace{}%
\AgdaBound{eq}%
\>[495I]\AgdaSymbol{|}\AgdaSpace{}%
\AgdaInductiveConstructor{nothing}\AgdaSpace{}%
\AgdaSymbol{|}\AgdaSpace{}%
\AgdaOperator{\AgdaInductiveConstructor{[}}\AgdaSpace{}%
\AgdaBound{eq′}\AgdaSpace{}%
\AgdaOperator{\AgdaInductiveConstructor{]}}\<%
\\
\>[.][@{}l@{}]\<[495I]%
\>[36]\AgdaSymbol{|}\AgdaSpace{}%
\AgdaInductiveConstructor{just}\AgdaSpace{}%
\AgdaBound{x}\AgdaSpace{}%
\AgdaSymbol{|}\AgdaSpace{}%
\AgdaOperator{\AgdaInductiveConstructor{[}}\AgdaSpace{}%
\AgdaSymbol{()}\AgdaSpace{}%
\AgdaOperator{\AgdaInductiveConstructor{]}}\<%
\\
\>[36]\AgdaSymbol{|}\AgdaSpace{}%
\AgdaInductiveConstructor{nothing}\AgdaSpace{}%
\AgdaSymbol{|}\AgdaSpace{}%
\AgdaOperator{\AgdaInductiveConstructor{[}}\AgdaSpace{}%
\AgdaBound{eq₁}\AgdaSpace{}%
\AgdaOperator{\AgdaInductiveConstructor{]}}\<%
\\
\>[2]\AgdaFunction{dr}\AgdaSpace{}%
\AgdaSymbol{(}\AgdaInductiveConstructor{add}\AgdaSpace{}%
\AgdaSymbol{(}\AgdaInductiveConstructor{num}\AgdaSpace{}%
\AgdaBound{x}\AgdaSymbol{)}\AgdaSpace{}%
\AgdaSymbol{(}\AgdaInductiveConstructor{num}\AgdaSpace{}%
\AgdaBound{x₁}\AgdaSymbol{))}\AgdaSpace{}%
\AgdaDottedPattern{\AgdaSymbol{.(}}\AgdaDottedPattern{\AgdaInductiveConstructor{add}}%
\>[518I]\AgdaDottedPattern{\AgdaSymbol{(}}\AgdaDottedPattern{\AgdaInductiveConstructor{num}}\AgdaSpace{}%
\AgdaDottedPattern{\AgdaBound{x}}\AgdaDottedPattern{\AgdaSymbol{)}}\AgdaSpace{}%
\AgdaDottedPattern{\AgdaSymbol{(}}\AgdaDottedPattern{\AgdaInductiveConstructor{num}}\AgdaSpace{}%
\AgdaDottedPattern{\AgdaBound{x₁}}\AgdaDottedPattern{\AgdaSymbol{))}}\AgdaSpace{}%
\AgdaBound{c₀}\AgdaSpace{}%
\AgdaDottedPattern{\AgdaSymbol{.}}\AgdaDottedPattern{\AgdaBound{c₀}}\AgdaSpace{}%
\AgdaInductiveConstructor{refl}\AgdaSpace{}%
\AgdaSymbol{|}\AgdaSpace{}%
\AgdaInductiveConstructor{nothing}\AgdaSpace{}%
\AgdaSymbol{|}\AgdaSpace{}%
\AgdaOperator{\AgdaInductiveConstructor{[}}\AgdaSpace{}%
\AgdaBound{eq′}\AgdaSpace{}%
\AgdaOperator{\AgdaInductiveConstructor{]}}\<%
\\
\>[518I][@{}l@{\AgdaIndent{0}}]%
\>[36]\AgdaSymbol{|}\AgdaSpace{}%
\AgdaInductiveConstructor{just}\AgdaSpace{}%
\AgdaDottedPattern{\AgdaSymbol{.(}}\AgdaDottedPattern{\AgdaInductiveConstructor{num}}\AgdaSpace{}%
\AgdaDottedPattern{\AgdaBound{x}}\AgdaDottedPattern{\AgdaSymbol{)}}\AgdaSpace{}%
\AgdaSymbol{|}\AgdaSpace{}%
\AgdaOperator{\AgdaInductiveConstructor{[}}\AgdaSpace{}%
\AgdaInductiveConstructor{refl}\AgdaSpace{}%
\AgdaOperator{\AgdaInductiveConstructor{]}}\<%
\\
\>[36]\AgdaSymbol{|}\AgdaSpace{}%
\AgdaInductiveConstructor{nothing}\AgdaSpace{}%
\AgdaSymbol{|}\AgdaSpace{}%
\AgdaOperator{\AgdaInductiveConstructor{[}}\AgdaSpace{}%
\AgdaBound{eq₁}\AgdaSpace{}%
\AgdaOperator{\AgdaInductiveConstructor{]}}\AgdaSpace{}%
\AgdaSymbol{=}\AgdaSpace{}%
\AgdaInductiveConstructor{refl}\<%
\\
\>[2]\AgdaFunction{dr}%
\>[545I]\AgdaSymbol{(}\AgdaInductiveConstructor{add}\AgdaSpace{}%
\AgdaSymbol{(}\AgdaInductiveConstructor{num}\AgdaSpace{}%
\AgdaBound{x}\AgdaSymbol{)}\AgdaSpace{}%
\AgdaBound{t₁}\AgdaSymbol{)}\AgdaSpace{}%
\AgdaDottedPattern{\AgdaSymbol{.}}\AgdaDottedPattern{\AgdaBound{t₂}}\AgdaSpace{}%
\AgdaBound{c₀}\AgdaSpace{}%
\AgdaDottedPattern{\AgdaSymbol{.}}\AgdaDottedPattern{\AgdaBound{c₂}}%
\>[552I]\AgdaInductiveConstructor{refl}\AgdaSpace{}%
\AgdaSymbol{|}\AgdaSpace{}%
\AgdaInductiveConstructor{nothing}\AgdaSpace{}%
\AgdaSymbol{|}\AgdaSpace{}%
\AgdaOperator{\AgdaInductiveConstructor{[}}\AgdaSpace{}%
\AgdaBound{eq′}\AgdaSpace{}%
\AgdaOperator{\AgdaInductiveConstructor{]}}\<%
\\
\>[.][@{}l@{}]\<[552I]%
\>[33]\AgdaSymbol{|}\AgdaSpace{}%
\AgdaInductiveConstructor{just}\AgdaSpace{}%
\AgdaDottedPattern{\AgdaSymbol{.(}}\AgdaDottedPattern{\AgdaInductiveConstructor{num}}\AgdaSpace{}%
\AgdaDottedPattern{\AgdaBound{x}}\AgdaDottedPattern{\AgdaSymbol{)}}\AgdaSpace{}%
\AgdaSymbol{|}\AgdaSpace{}%
\AgdaOperator{\AgdaInductiveConstructor{[}}\AgdaSpace{}%
\AgdaInductiveConstructor{refl}\AgdaSpace{}%
\AgdaOperator{\AgdaInductiveConstructor{]}}\<%
\\
\>[33]\AgdaSymbol{|}\AgdaSpace{}%
\AgdaInductiveConstructor{just}\AgdaSpace{}%
\AgdaSymbol{(}\AgdaBound{t₂}\AgdaSpace{}%
\AgdaOperator{\AgdaInductiveConstructor{,}}\AgdaSpace{}%
\AgdaBound{c₂}\AgdaSymbol{)}\AgdaSpace{}%
\AgdaSymbol{|}\AgdaSpace{}%
\AgdaOperator{\AgdaInductiveConstructor{[}}\AgdaSpace{}%
\AgdaBound{eq₁}\AgdaSpace{}%
\AgdaOperator{\AgdaInductiveConstructor{]}}\AgdaSpace{}%
\AgdaSymbol{=}\<%
\\
\>[.][@{}l@{}]\<[545I]%
\>[5]\AgdaFunction{dr}\AgdaSpace{}%
\AgdaBound{t₁}\AgdaSpace{}%
\AgdaBound{t₂}\AgdaSpace{}%
\AgdaSymbol{(}\AgdaInductiveConstructor{add-r}\AgdaSpace{}%
\AgdaSymbol{(}\AgdaInductiveConstructor{num}\AgdaSpace{}%
\AgdaBound{x}\AgdaSymbol{)}\AgdaSpace{}%
\AgdaOperator{\AgdaInductiveConstructor{∷}}\AgdaSpace{}%
\AgdaBound{c₀}\AgdaSymbol{)}\AgdaSpace{}%
\AgdaBound{c₂}\AgdaSpace{}%
\AgdaBound{eq₁}\<%
\\
\>[2]\AgdaFunction{dr}\AgdaSpace{}%
\AgdaSymbol{(}\AgdaInductiveConstructor{add}\AgdaSpace{}%
\AgdaSymbol{(}\AgdaInductiveConstructor{add}\AgdaSpace{}%
\AgdaBound{t₀}\AgdaSpace{}%
\AgdaBound{t₂}\AgdaSymbol{)}\AgdaSpace{}%
\AgdaBound{t₁}\AgdaSymbol{)}\AgdaSpace{}%
\AgdaBound{t}\AgdaSpace{}%
\AgdaBound{c₀}\AgdaSpace{}%
\AgdaBound{c}\AgdaSpace{}%
\AgdaSymbol{()}\AgdaSpace{}%
\AgdaSymbol{|}\AgdaSpace{}%
\AgdaInductiveConstructor{nothing}\AgdaSpace{}%
\AgdaSymbol{|}\AgdaSpace{}%
\AgdaOperator{\AgdaInductiveConstructor{[}}\AgdaSpace{}%
\AgdaBound{eq′}\AgdaSpace{}%
\AgdaOperator{\AgdaInductiveConstructor{]}}\AgdaSpace{}%
\AgdaSymbol{|}\AgdaSpace{}%
\AgdaInductiveConstructor{nothing}\AgdaSpace{}%
\AgdaSymbol{|}\AgdaSpace{}%
\AgdaOperator{\AgdaInductiveConstructor{[}}\AgdaSpace{}%
\AgdaBound{eq₀}\AgdaSpace{}%
\AgdaOperator{\AgdaInductiveConstructor{]}}\<%
\\
\>[2]\AgdaFunction{dr}\AgdaSpace{}%
\AgdaSymbol{(}\AgdaInductiveConstructor{add}\AgdaSpace{}%
\AgdaSymbol{(}\AgdaInductiveConstructor{num}\AgdaSpace{}%
\AgdaBound{x}\AgdaSymbol{)}\AgdaSpace{}%
\AgdaBound{t₁}\AgdaSymbol{)}\AgdaSpace{}%
\AgdaBound{t}\AgdaSpace{}%
\AgdaBound{c₀}\AgdaSpace{}%
\AgdaBound{c}\AgdaSpace{}%
\AgdaBound{eq}\AgdaSpace{}%
\AgdaSymbol{|}\AgdaSpace{}%
\AgdaInductiveConstructor{nothing}\AgdaSpace{}%
\AgdaSymbol{|}\AgdaSpace{}%
\AgdaOperator{\AgdaInductiveConstructor{[}}\AgdaSpace{}%
\AgdaBound{eq′}\AgdaSpace{}%
\AgdaOperator{\AgdaInductiveConstructor{]}}\AgdaSpace{}%
\AgdaSymbol{|}\AgdaSpace{}%
\AgdaInductiveConstructor{nothing}\AgdaSpace{}%
\AgdaSymbol{|}\AgdaSpace{}%
\AgdaOperator{\AgdaInductiveConstructor{[}}\AgdaSpace{}%
\AgdaSymbol{()}\AgdaSpace{}%
\AgdaOperator{\AgdaInductiveConstructor{]}}\<%
\end{code}
The first property says that recomposing \ab{t} with the composed context \af{compose}~\ab{c₀~c₁} is the same as recomposing \ab{t} into \ab{c₀} first, and then recomposing that result with \ab{c₁}.
The second property ensures that recomposing a decomposition yields the original term.
%
\begin{code}[hide]%
\>[2]\AgdaFunction{decompose-recompose-top}%
\>[625I]\AgdaSymbol{:}\AgdaSpace{}%
\AgdaSymbol{∀}\AgdaSpace{}%
\AgdaSymbol{\{}\AgdaBound{t₀}\AgdaSpace{}%
\AgdaBound{t}\AgdaSpace{}%
\AgdaBound{c}\AgdaSymbol{\}}\<%
\\
\>[.][@{}l@{}]\<[625I]%
\>[26]\AgdaSymbol{→}\AgdaSpace{}%
\AgdaFunction{decompose}\AgdaSpace{}%
\AgdaBound{t₀}\AgdaSpace{}%
\AgdaInductiveConstructor{[]}\AgdaSpace{}%
\AgdaOperator{\AgdaDatatype{≡}}\AgdaSpace{}%
\AgdaInductiveConstructor{just}\AgdaSpace{}%
\AgdaSymbol{(}\AgdaBound{t}\AgdaSpace{}%
\AgdaOperator{\AgdaInductiveConstructor{,}}\AgdaSpace{}%
\AgdaBound{c}\AgdaSymbol{)}\<%
\\
\>[26]\AgdaSymbol{→}\AgdaSpace{}%
\AgdaFunction{recompose}\AgdaSpace{}%
\AgdaBound{c}\AgdaSpace{}%
\AgdaBound{t}\AgdaSpace{}%
\AgdaOperator{\AgdaDatatype{≡}}\AgdaSpace{}%
\AgdaBound{t₀}\<%
\\
\>[2]\AgdaFunction{decompose-recompose-top}\AgdaSpace{}%
\AgdaSymbol{\{}\AgdaBound{t₀}\AgdaSymbol{\}}\AgdaSpace{}%
\AgdaSymbol{\{}\AgdaBound{t}\AgdaSymbol{\}}\AgdaSpace{}%
\AgdaSymbol{=}\<%
\\
\>[2][@{}l@{\AgdaIndent{0}}]%
\>[4]\AgdaFunction{decompose-recompose}\AgdaSpace{}%
\AgdaSymbol{\{}\AgdaBound{t₀}\AgdaSymbol{\}}\AgdaSpace{}%
\AgdaSymbol{\{}\AgdaBound{t}\AgdaSymbol{\}}\AgdaSpace{}%
\AgdaSymbol{\{}\AgdaInductiveConstructor{[]}\AgdaSymbol{\}}\<%
\end{code}

\subsubsection{One-Step Reduction}

One-step reduction is given by the following function which decomposes a term into a redex and context, contracts the result, and recomposes the result, if possible:
\begin{code}%
\>[2]\AgdaFunction{reduce}\AgdaSpace{}%
\AgdaSymbol{:}\AgdaSpace{}%
\AgdaDatatype{Term}\AgdaSpace{}%
\AgdaSymbol{→}\AgdaSpace{}%
\AgdaDatatype{Maybe}\AgdaSpace{}%
\AgdaDatatype{Term}\<%
\\
\>[2]\AgdaFunction{reduce}\AgdaSpace{}%
\AgdaBound{t}\AgdaSpace{}%
\AgdaSymbol{=}\AgdaSpace{}%
\AgdaKeyword{do}\<%
\\
\>[2][@{}l@{\AgdaIndent{0}}]%
\>[4]\AgdaSymbol{(}\AgdaBound{t₀}\AgdaSpace{}%
\AgdaOperator{\AgdaInductiveConstructor{,}}\AgdaSpace{}%
\AgdaBound{c}\AgdaSymbol{)}%
\>[14]\AgdaOperator{\AgdaFunction{←}}\AgdaSpace{}%
\AgdaFunction{decompose}\AgdaSpace{}%
\AgdaBound{t}\AgdaSpace{}%
\AgdaInductiveConstructor{[]}\<%
\\
\>[4]\AgdaBound{t₁}%
\>[14]\AgdaOperator{\AgdaFunction{←}}\AgdaSpace{}%
\AgdaFunction{contract}\AgdaSpace{}%
\AgdaBound{t₀}\<%
\\
\>[4]\AgdaInductiveConstructor{just}\AgdaSpace{}%
\AgdaSymbol{(}\AgdaFunction{recompose}\AgdaSpace{}%
\AgdaBound{c}\AgdaSpace{}%
\AgdaBound{t₁}\AgdaSymbol{)}\<%
\end{code}
The code uses the standard monadic bind for the maybe monad to propagate errors arising from either decomposition or contraction.

\subsubsection{Reduction-Based Normalization using Coinduction in Agda}

Now, all that remains is to implement a driver loop that iterates a term towards a final value, if it exists.
While it is possible for a human to see that continued one-step reduction is guaranteed to terminate, Agda needs proof that it is before it will let us define a driver loop as a recursive function.
However, since we are often interested in expressing languages where reduction \emph{may} non-terminate, we define a driver-loop that tells Agda this instead.
We define our driver loop as a \emph{guarded corecursive function}.
That is, a function which, at each reduction step, emits an output term that reduction has progressed.
To this end, we use the following single-field \ad{Delay} record---usually called the \emph{delay monad}~\cite{Capretta05,Danielsson12,AbelC14}.\footnote{Here \ad{\_⊎\_} is the sum type from Agda's standard library, whose constructors are \ac{inj₁}~\as{:}~\ab{A}~\as{→}~\ab{A}~\ad{⊎}~\ab{B} and \ac{inj₂}~\as{:}~\ab{B}~\as{→}~\ab{A}~\ad{⊎}~\ab{B}.}
\begin{code}%
\>[2]\AgdaKeyword{record}\AgdaSpace{}%
\AgdaRecord{Delay}\AgdaSpace{}%
\AgdaSymbol{(}\AgdaBound{A}\AgdaSpace{}%
\AgdaSymbol{:}\AgdaSpace{}%
\AgdaPrimitive{Set}\AgdaSymbol{)}\AgdaSpace{}%
\AgdaSymbol{:}\AgdaSpace{}%
\AgdaPrimitive{Set}\AgdaSpace{}%
\AgdaKeyword{where}\<%
\\
\>[2][@{}l@{\AgdaIndent{0}}]%
\>[4]\AgdaKeyword{constructor}\AgdaSpace{}%
\AgdaOperator{\AgdaCoinductiveConstructor{D⟨\AgdaUnderscore{}⟩}}\<%
\\
\>[4]\AgdaKeyword{coinductive}\<%
\\
\>[4]\AgdaKeyword{field}\AgdaSpace{}%
\AgdaField{insp}\AgdaSpace{}%
\AgdaSymbol{:}\AgdaSpace{}%
\AgdaBound{A}\AgdaSpace{}%
\AgdaOperator{\AgdaDatatype{⊎}}\AgdaSpace{}%
\AgdaRecord{Delay}\AgdaSpace{}%
\AgdaBound{A}\<%
\end{code}
\begin{code}[hide]%
\>[2]\AgdaKeyword{open}\AgdaSpace{}%
\AgdaModule{Delay}\<%
\end{code}
The keyword \ak{coinductive} declares the record type as coinductive.\footnote{See also \url{https://agda.readthedocs.io/en/latest/language/coinduction.html}}
This lets us use Agda's support for \emph{guarded corecursion} to define corecursive functions where each recursive call is guarded by one or more constructors of type \ad{Delay}.
More concretely, the type \ad{Delay}~\ab{A} represents a potentially infinitely delayed value of type \ab{A}.
\begin{example}
  The following represents a value without delay:\footnote{The definition uses Agda copatterns; see \url{https://agda.readthedocs.io/en/latest/language/copatterns.html}}
  \begin{code}%
\>[2]\AgdaFunction{value₀}\AgdaSpace{}%
\AgdaSymbol{:}\AgdaSpace{}%
\AgdaRecord{Delay}\AgdaSpace{}%
\AgdaDatatype{ℕ}\<%
\\
\>[2]\AgdaField{insp}\AgdaSpace{}%
\AgdaFunction{value₀}\AgdaSpace{}%
\AgdaSymbol{=}\AgdaSpace{}%
\AgdaInductiveConstructor{inj₁}\AgdaSpace{}%
\AgdaNumber{0}\<%
\end{code}
  In contrast, the following two values are once delayed and infinitely delayed, respectively:
  \begin{code}%
\>[2]\AgdaFunction{value₁}\AgdaSpace{}%
\AgdaSymbol{:}\AgdaSpace{}%
\AgdaRecord{Delay}\AgdaSpace{}%
\AgdaDatatype{ℕ}\<%
\\
\>[2]\AgdaField{insp}\AgdaSpace{}%
\AgdaFunction{value₁}\AgdaSpace{}%
\AgdaSymbol{=}\AgdaSpace{}%
\AgdaInductiveConstructor{inj₂}\AgdaSpace{}%
\AgdaFunction{value₀}\<%
\\
\\[\AgdaEmptyExtraSkip]%
\>[2]\AgdaFunction{value∞}\AgdaSpace{}%
\AgdaSymbol{:}\AgdaSpace{}%
\AgdaRecord{Delay}\AgdaSpace{}%
\AgdaDatatype{ℕ}\<%
\\
\>[2]\AgdaField{insp}\AgdaSpace{}%
\AgdaFunction{value∞}\AgdaSpace{}%
\AgdaSymbol{=}\AgdaSpace{}%
\AgdaInductiveConstructor{inj₂}\AgdaSpace{}%
\AgdaFunction{value∞}\<%
\end{code}
\end{example}

We can define a driver loop as a guarded corecursive function which witnesses that progress (i.e., reduction) happens in each step.
The driver loop below terminates (using the \ac{inj₁} constructor) if the input is a value or a stuck term, and otherwise witnesses (using the \ac{inj₂} constructor) that a single reduction step has happened before recursively applying the driver loop.\footnote{The definition uses Agda's \ak{with}-abstraction, which lets us pattern match on the result of intermediate computations: \url{https://agda.readthedocs.io/en/latest/language/with-abstraction.html}}
\begin{code}%
\>[2]\AgdaFunction{drive}\AgdaSpace{}%
\AgdaSymbol{:}\AgdaSpace{}%
\AgdaDatatype{Term}\AgdaSpace{}%
\AgdaSymbol{→}\AgdaSpace{}%
\AgdaRecord{Delay}\AgdaSpace{}%
\AgdaSymbol{(}\AgdaDatatype{Maybe}\AgdaSpace{}%
\AgdaDatatype{Val}\AgdaSymbol{)}\<%
\\
\>[2]\AgdaField{insp}\AgdaSpace{}%
\AgdaSymbol{(}\AgdaFunction{drive}\AgdaSpace{}%
\AgdaBound{t}\AgdaSymbol{)}\AgdaSpace{}%
\AgdaKeyword{with}\AgdaSpace{}%
\AgdaFunction{t2v}\AgdaSpace{}%
\AgdaBound{t}\<%
\\
\>[2]\AgdaField{insp}\AgdaSpace{}%
\AgdaSymbol{(}\AgdaFunction{drive}\AgdaSpace{}%
\AgdaBound{t}\AgdaSymbol{)}\AgdaSpace{}%
\AgdaSymbol{|}\AgdaSpace{}%
\AgdaInductiveConstructor{just}\AgdaSpace{}%
\AgdaBound{x}\AgdaSpace{}%
\AgdaSymbol{=}\AgdaSpace{}%
\AgdaInductiveConstructor{inj₁}\AgdaSpace{}%
\AgdaSymbol{(}\AgdaInductiveConstructor{just}\AgdaSpace{}%
\AgdaBound{x}\AgdaSymbol{)}\<%
\\
\>[2]\AgdaField{insp}\AgdaSpace{}%
\AgdaSymbol{(}\AgdaFunction{drive}\AgdaSpace{}%
\AgdaBound{t}\AgdaSymbol{)}\AgdaSpace{}%
\AgdaSymbol{|}\AgdaSpace{}%
\AgdaInductiveConstructor{nothing}\AgdaSpace{}%
\AgdaKeyword{with}\AgdaSpace{}%
\AgdaFunction{reduce}\AgdaSpace{}%
\AgdaBound{t}\<%
\\
\>[2]\AgdaField{insp}\AgdaSpace{}%
\AgdaSymbol{(}\AgdaFunction{drive}\AgdaSpace{}%
\AgdaBound{t}\AgdaSymbol{)}\AgdaSpace{}%
\AgdaSymbol{|}\AgdaSpace{}%
\AgdaInductiveConstructor{nothing}\AgdaSpace{}%
\AgdaSymbol{|}\AgdaSpace{}%
\AgdaInductiveConstructor{just}\AgdaSpace{}%
\AgdaBound{x}%
\>[38]\AgdaSymbol{=}\AgdaSpace{}%
\AgdaInductiveConstructor{inj₂}\AgdaSpace{}%
\AgdaSymbol{(}\AgdaFunction{drive}\AgdaSpace{}%
\AgdaBound{x}\AgdaSymbol{)}\<%
\\
\>[2]\AgdaField{insp}\AgdaSpace{}%
\AgdaSymbol{(}\AgdaFunction{drive}\AgdaSpace{}%
\AgdaBound{t}\AgdaSymbol{)}\AgdaSpace{}%
\AgdaSymbol{|}\AgdaSpace{}%
\AgdaInductiveConstructor{nothing}\AgdaSpace{}%
\AgdaSymbol{|}\AgdaSpace{}%
\AgdaInductiveConstructor{nothing}%
\>[38]\AgdaSymbol{=}\AgdaSpace{}%
\AgdaInductiveConstructor{inj₁}\AgdaSpace{}%
\AgdaInductiveConstructor{nothing}\<%
\end{code}

Values of type \ad{Delay} may be a potentially infinite stream.
So, since \af{drive} returns a value of this type, Agda is careful about unfolding recursive calls to it.
In order to compare values of type \ad{Delay}, and thus \af{drive} calls, it is useful to define a bisimilarity relation.
To this end, we (mutually) define a bisimulation step relation \ad{\_≈₀\_} and a coinductive bisimilarity relation \ad{\_≈\_} defined in terms of it:
\begin{code}%
\>[2]\AgdaKeyword{mutual}\<%
\\
\>[2][@{}l@{\AgdaIndent{0}}]%
\>[4]\AgdaKeyword{data}\AgdaSpace{}%
\AgdaOperator{\AgdaDatatype{\AgdaUnderscore{}≈₀\AgdaUnderscore{}}}\AgdaSpace{}%
\AgdaSymbol{\{}\AgdaBound{A}\AgdaSymbol{\}}\AgdaSpace{}%
\AgdaSymbol{:}\AgdaSpace{}%
\AgdaRecord{Delay}\AgdaSpace{}%
\AgdaBound{A}\AgdaSpace{}%
\AgdaSymbol{→}\AgdaSpace{}%
\AgdaRecord{Delay}\AgdaSpace{}%
\AgdaBound{A}\AgdaSpace{}%
\AgdaSymbol{→}\AgdaSpace{}%
\AgdaPrimitive{Set}\AgdaSpace{}%
\AgdaKeyword{where}\<%
\\
\>[4][@{}l@{\AgdaIndent{0}}]%
\>[6]\AgdaInductiveConstructor{val}%
\>[759I]\AgdaSymbol{:}\AgdaSpace{}%
\AgdaSymbol{\{}\AgdaBound{d₀}\AgdaSpace{}%
\AgdaBound{d₁}\AgdaSpace{}%
\AgdaSymbol{:}\AgdaSpace{}%
\AgdaRecord{Delay}\AgdaSpace{}%
\AgdaBound{A}\AgdaSymbol{\}}\AgdaSpace{}%
\AgdaSymbol{\{}\AgdaBound{x}\AgdaSpace{}%
\AgdaSymbol{:}\AgdaSpace{}%
\AgdaBound{A}\AgdaSymbol{\}}\<%
\\
\>[.][@{}l@{}]\<[759I]%
\>[10]\AgdaSymbol{→}\AgdaSpace{}%
\AgdaField{insp}\AgdaSpace{}%
\AgdaBound{d₀}\AgdaSpace{}%
\AgdaOperator{\AgdaDatatype{≡}}\AgdaSpace{}%
\AgdaInductiveConstructor{inj₁}\AgdaSpace{}%
\AgdaBound{x}\AgdaSpace{}%
\AgdaSymbol{→}\AgdaSpace{}%
\AgdaField{insp}\AgdaSpace{}%
\AgdaBound{d₁}\AgdaSpace{}%
\AgdaOperator{\AgdaDatatype{≡}}\AgdaSpace{}%
\AgdaInductiveConstructor{inj₁}\AgdaSpace{}%
\AgdaBound{x}\<%
\\
\>[10]\AgdaSymbol{→}\AgdaSpace{}%
\AgdaBound{d₀}\AgdaSpace{}%
\AgdaOperator{\AgdaDatatype{≈₀}}\AgdaSpace{}%
\AgdaBound{d₁}\<%
\\
\>[6]\AgdaInductiveConstructor{tau}%
\>[782I]\AgdaSymbol{:}\AgdaSpace{}%
\AgdaSymbol{\{}\AgdaBound{d₀}\AgdaSpace{}%
\AgdaBound{d₀₁}\AgdaSpace{}%
\AgdaBound{d₁}\AgdaSpace{}%
\AgdaBound{d₁₁}\AgdaSpace{}%
\AgdaSymbol{:}\AgdaSpace{}%
\AgdaRecord{Delay}\AgdaSpace{}%
\AgdaBound{A}\AgdaSymbol{\}}\<%
\\
\>[.][@{}l@{}]\<[782I]%
\>[10]\AgdaSymbol{→}\AgdaSpace{}%
\AgdaField{insp}\AgdaSpace{}%
\AgdaBound{d₀}\AgdaSpace{}%
\AgdaOperator{\AgdaDatatype{≡}}\AgdaSpace{}%
\AgdaInductiveConstructor{inj₂}\AgdaSpace{}%
\AgdaBound{d₀₁}\AgdaSpace{}%
\AgdaSymbol{→}\AgdaSpace{}%
\AgdaField{insp}\AgdaSpace{}%
\AgdaBound{d₁}\AgdaSpace{}%
\AgdaOperator{\AgdaDatatype{≡}}\AgdaSpace{}%
\AgdaInductiveConstructor{inj₂}\AgdaSpace{}%
\AgdaBound{d₁₁}\<%
\\
\>[10]\AgdaSymbol{→}\AgdaSpace{}%
\AgdaBound{d₀₁}\AgdaSpace{}%
\AgdaOperator{\AgdaRecord{≈}}\AgdaSpace{}%
\AgdaBound{d₁₁}\AgdaSpace{}%
\AgdaSymbol{→}\AgdaSpace{}%
\AgdaBound{d₀}\AgdaSpace{}%
\AgdaOperator{\AgdaDatatype{≈₀}}\AgdaSpace{}%
\AgdaBound{d₁}\<%
\\
\\[\AgdaEmptyExtraSkip]%
\>[4]\AgdaKeyword{record}\AgdaSpace{}%
\AgdaOperator{\AgdaRecord{\AgdaUnderscore{}≈\AgdaUnderscore{}}}\AgdaSpace{}%
\AgdaSymbol{\{}\AgdaBound{A}\AgdaSymbol{\}}\AgdaSpace{}%
\AgdaSymbol{(}\AgdaBound{d₀}\AgdaSpace{}%
\AgdaBound{d₁}\AgdaSpace{}%
\AgdaSymbol{:}\AgdaSpace{}%
\AgdaRecord{Delay}\AgdaSpace{}%
\AgdaBound{A}\AgdaSymbol{)}\AgdaSpace{}%
\AgdaSymbol{:}\AgdaSpace{}%
\AgdaPrimitive{Set}\AgdaSpace{}%
\AgdaKeyword{where}\<%
\\
\>[4][@{}l@{\AgdaIndent{0}}]%
\>[6]\AgdaKeyword{coinductive}\<%
\\
\>[6]\AgdaKeyword{constructor}\AgdaSpace{}%
\AgdaOperator{\AgdaCoinductiveConstructor{≈⟨\AgdaUnderscore{}⟩}}\<%
\\
\>[6]\AgdaKeyword{field}\AgdaSpace{}%
\AgdaField{bisim}\AgdaSpace{}%
\AgdaSymbol{:}\AgdaSpace{}%
\AgdaBound{d₀}\AgdaSpace{}%
\AgdaOperator{\AgdaDatatype{≈₀}}\AgdaSpace{}%
\AgdaBound{d₁}\<%
\end{code}
\begin{code}[hide]%
\>[2]\AgdaFunction{delay}\AgdaSpace{}%
\AgdaSymbol{:}\AgdaSpace{}%
\AgdaSymbol{∀}\AgdaSpace{}%
\AgdaSymbol{\{}\AgdaBound{A}\AgdaSymbol{\}}\AgdaSpace{}%
\AgdaSymbol{→}\AgdaSpace{}%
\AgdaDatatype{ℕ}\AgdaSpace{}%
\AgdaSymbol{→}\AgdaSpace{}%
\AgdaBound{A}\AgdaSpace{}%
\AgdaSymbol{→}\AgdaSpace{}%
\AgdaRecord{Delay}\AgdaSpace{}%
\AgdaBound{A}\<%
\\
\>[2]\AgdaField{insp}\AgdaSpace{}%
\AgdaSymbol{(}\AgdaFunction{delay}\AgdaSpace{}%
\AgdaInductiveConstructor{zero}\AgdaSpace{}%
\AgdaBound{x}\AgdaSymbol{)}%
\>[25]\AgdaSymbol{=}\AgdaSpace{}%
\AgdaInductiveConstructor{inj₁}\AgdaSpace{}%
\AgdaBound{x}\<%
\\
\>[2]\AgdaField{insp}\AgdaSpace{}%
\AgdaSymbol{(}\AgdaFunction{delay}\AgdaSpace{}%
\AgdaSymbol{(}\AgdaInductiveConstructor{suc}\AgdaSpace{}%
\AgdaBound{n}\AgdaSymbol{)}\AgdaSpace{}%
\AgdaBound{x}\AgdaSymbol{)}\AgdaSpace{}%
\AgdaSymbol{=}\AgdaSpace{}%
\AgdaInductiveConstructor{inj₂}\AgdaSpace{}%
\AgdaSymbol{(}\AgdaFunction{delay}\AgdaSpace{}%
\AgdaBound{n}\AgdaSpace{}%
\AgdaBound{x}\AgdaSymbol{)}\<%
\\
\\[\AgdaEmptyExtraSkip]%
\>[2]\AgdaKeyword{open}\AgdaSpace{}%
\AgdaOperator{\AgdaModule{\AgdaUnderscore{}≈\AgdaUnderscore{}}}\<%
\end{code}
The \ac{val} constructor says that, if both sides of a bisimulation step are values, then they must be equal.
The \ac{tau} constructor says, if both sides are delayed values, then the results of peeling off the delay must yield equal values.

Using these relations, we can prove that our reduction semantics computes.
\begin{example}
The definition below witnesses that $(1 + 2) + 39 = 42$.\footnote{\ac{refl}~\as{:}~\ab{x}~\ad{≡}~\ab{x} is the constructor witnessing a propositional equality.}
\begin{code}%
\>[2]\AgdaFunction{test42}\AgdaSpace{}%
\AgdaSymbol{:}%
\>[850I]\AgdaFunction{drive}\AgdaSpace{}%
\AgdaSymbol{(}\AgdaInductiveConstructor{add}\AgdaSpace{}%
\AgdaSymbol{(}\AgdaInductiveConstructor{add}\AgdaSpace{}%
\AgdaSymbol{(}\AgdaInductiveConstructor{num}\AgdaSpace{}%
\AgdaNumber{1}\AgdaSymbol{)}\AgdaSpace{}%
\AgdaSymbol{(}\AgdaInductiveConstructor{num}\AgdaSpace{}%
\AgdaNumber{2}\AgdaSymbol{))}\AgdaSpace{}%
\AgdaSymbol{(}\AgdaInductiveConstructor{num}\AgdaSpace{}%
\AgdaNumber{39}\AgdaSymbol{))}\<%
\\
\>[.][@{}l@{}]\<[850I]%
\>[11]\AgdaOperator{\AgdaRecord{≈}}\AgdaSpace{}%
\AgdaFunction{delay}\AgdaSpace{}%
\AgdaNumber{2}\AgdaSpace{}%
\AgdaSymbol{(}\AgdaInductiveConstructor{just}\AgdaSpace{}%
\AgdaSymbol{(}\AgdaInductiveConstructor{num}\AgdaSpace{}%
\AgdaNumber{42}\AgdaSymbol{))}\<%
\\
\>[2]\AgdaField{bisim}\AgdaSpace{}%
\AgdaFunction{test42}\AgdaSpace{}%
\AgdaSymbol{=}\AgdaSpace{}%
\AgdaInductiveConstructor{tau}\AgdaSpace{}%
\AgdaInductiveConstructor{refl}\AgdaSpace{}%
\AgdaInductiveConstructor{refl}\<%
\\
\>[2][@{}l@{\AgdaIndent{0}}]%
\>[4]\AgdaOperator{\AgdaCoinductiveConstructor{≈⟨}}\AgdaSpace{}%
\AgdaInductiveConstructor{tau}\AgdaSpace{}%
\AgdaInductiveConstructor{refl}\AgdaSpace{}%
\AgdaInductiveConstructor{refl}\<%
\\
\>[4][@{}l@{\AgdaIndent{0}}]%
\>[6]\AgdaOperator{\AgdaCoinductiveConstructor{≈⟨}}\AgdaSpace{}%
\AgdaInductiveConstructor{val}\AgdaSpace{}%
\AgdaInductiveConstructor{refl}\AgdaSpace{}%
\AgdaInductiveConstructor{refl}\AgdaSpace{}%
\AgdaOperator{\AgdaCoinductiveConstructor{⟩}}\AgdaSpace{}%
\AgdaOperator{\AgdaCoinductiveConstructor{⟩}}\<%
\end{code}
Here, \af{delay}~\ab{n} is a helper that delays a value by \ab{n} steps.
\end{example}

\subsubsection{Relation to Danvy's Modus Operandi}
\label{sec:relation-to-danvy}

While we have followed the modus operandi of Danvy~\cite{Danvy08afp}, our definitions and types differ slightly from his.
These differences are motivated by our goal of providing a generic reduction semantics in the next section which factors boilerplate code into a library.
As we will see, our functions for decomposition, recomposition, one-step reduction, and reduction-based normalization are all boilerplate code.
Before turning our attention to genericizing the definitions from this section, we first provide a reduction semantics for a more interesting language.
Doing so will demonstrate both (1) that the modus operandi we have followed scales to more interesting reduction semantics and (2) that the resulting definitions contain repetitive code.

\subsection{Implementing a Reduction Semantics for Lambda Expressions and Shift/Reset}
\label{sec:lambda-shift-reset}

We follow the same modus operandi as in the previous section, except we will provide our reduction strategy and notion of contraction simultaneously.
The reason for this simultaneity is that the language we consider in this section has \emph{context-sensitive contraction}; i.e., reductions may \emph{unwind}~\cite{Danvy08} or otherwise manipulate the context.

The abstract syntax and reduction rules provided in \cref{fig:redsem-lam} extend \cref{fig:redsem-arith}.
The application rule is standard, and uses an elided notion of De Bruijn substitution.
The first rule for \textsf{reset} is for returning the value computed by a $\mathsf{reset}$ expression.
The final rule defines the semantics of Danvy and Filinski's shift/reset operators~\cite{DanvyF90}.\footnote{This formulation of the rule is borrowed from here \url{https://docs.racket-lang.org/reference/cont.html\#\%28form._\%28\%28lib._racket\%2Fcontrol..rkt\%29._reset\%29\%29}}
The rule says that when a $(\mathsf{shift}\ e)$ expression is encountered in a context, we unwind that context $E$ to the nearest \textsf{reset} expression.
That \textsf{reset} expression is then replaced by a new expression which applies $e$ to a function closure containing the context $E$ that we shifted from.
The operator $\uparrow$ is the standard operation for incrementing free De Bruijn variables by one.

\begin{figure}
\raggedright
\textbf{Abstract Syntax}
\begin{mathpar}
  e \bnfdef \cdots \mid \mathsf{lam}\ e
                   \mid \mathsf{var}\ n
                   \mid \mathsf{app}\ e\ e
                   \mid \mathsf{reset}\ e
                   \mid \mathsf{shift}\ e
  \and
  v \bnfdef \cdots \mid \mathsf{lam}\ e
\end{mathpar}
\textbf{Notion of Contraction}
\begin{mathpar}
\cdots
\\
\inferrule{
}{
  \mathsf{app}\ (\mathsf{lam}\ e_0)\ v
  \longmapsto
  e_0[ 0 \mapsto v ]
}
\and
\inferrule{
}{
  \mathsf{shift}\ v
  \longmapsto
  v
}
\and
\inferrule{
  \text{$E$ has no \textsf{reset}}
}{
  \mathsf{reset}\ E[\mathsf{shift}\ e]
  \longmapsto
  \mathsf{reset}\ (\mathsf{app}\ e\ (\mathsf{lam}\ (\mathsf{reset}\ (\uparrow E)[\mathsf{var}\ 0])))
}
\end{mathpar}
\textbf{Reduction Contexts}
\begin{mathpar}
  E \bnfdef \cdots \mid \mathsf{app}\ E\ e
            \mid \mathsf{app}\ v\ E
            \mid \mathsf{reset}\ E
\end{mathpar}
\caption{Reduction semantics for $\lambda$s and shift/reset}
\label{fig:redsem-lam}
\end{figure}

\begin{code}[hide]%
\>[0]\AgdaKeyword{module}\AgdaSpace{}%
\AgdaModule{LambdaControl}\AgdaSpace{}%
\AgdaKeyword{where}\<%
\end{code}

\subsubsection{Abstract Syntax}
The abstract syntax of the language we consider is given by the following type which uses De Bruijn indices for name binding.
\begin{code}%
\>[0][@{}l@{\AgdaIndent{1}}]%
\>[2]\AgdaKeyword{data}\AgdaSpace{}%
\AgdaDatatype{Term}\AgdaSpace{}%
\AgdaSymbol{:}\AgdaSpace{}%
\AgdaPrimitive{Set}\AgdaSpace{}%
\AgdaKeyword{where}\<%
\\
\>[2][@{}l@{\AgdaIndent{0}}]%
\>[4]\AgdaInductiveConstructor{add}%
\>[11]\AgdaSymbol{:}\AgdaSpace{}%
\AgdaSymbol{(}\AgdaBound{t₀}\AgdaSpace{}%
\AgdaBound{t₁}\AgdaSpace{}%
\AgdaSymbol{:}\AgdaSpace{}%
\AgdaDatatype{Term}\AgdaSymbol{)}%
\>[29]\AgdaSymbol{→}\AgdaSpace{}%
\AgdaDatatype{Term}\AgdaSpace{}%
\AgdaComment{--\ As\ before}\<%
\\
\>[4]\AgdaInductiveConstructor{num}%
\>[11]\AgdaSymbol{:}\AgdaSpace{}%
\AgdaDatatype{ℕ}%
\>[29]\AgdaSymbol{→}\AgdaSpace{}%
\AgdaDatatype{Term}\AgdaSpace{}%
\AgdaComment{--\ As\ before}\<%
\\
\>[4]\AgdaInductiveConstructor{lam}%
\>[11]\AgdaSymbol{:}\AgdaSpace{}%
\AgdaDatatype{Term}%
\>[29]\AgdaSymbol{→}\AgdaSpace{}%
\AgdaDatatype{Term}\<%
\\
\>[4]\AgdaInductiveConstructor{var}%
\>[11]\AgdaSymbol{:}\AgdaSpace{}%
\AgdaDatatype{ℕ}%
\>[29]\AgdaSymbol{→}\AgdaSpace{}%
\AgdaDatatype{Term}\<%
\\
\>[4]\AgdaInductiveConstructor{app}%
\>[11]\AgdaSymbol{:}\AgdaSpace{}%
\AgdaSymbol{(}\AgdaBound{t₀}\AgdaSpace{}%
\AgdaBound{t₁}\AgdaSpace{}%
\AgdaSymbol{:}\AgdaSpace{}%
\AgdaDatatype{Term}\AgdaSymbol{)}%
\>[29]\AgdaSymbol{→}\AgdaSpace{}%
\AgdaDatatype{Term}\<%
\\
\>[4]\AgdaInductiveConstructor{reset}%
\>[11]\AgdaSymbol{:}\AgdaSpace{}%
\AgdaDatatype{Term}%
\>[29]\AgdaSymbol{→}\AgdaSpace{}%
\AgdaDatatype{Term}\<%
\\
\>[4]\AgdaInductiveConstructor{shift}%
\>[11]\AgdaSymbol{:}\AgdaSpace{}%
\AgdaDatatype{Term}%
\>[29]\AgdaSymbol{→}\AgdaSpace{}%
\AgdaDatatype{Term}\<%
\end{code}
Our semantics will use a standard notion of De Bruijn substitution whose definition we elide for brevity.
\begin{code}%
\>[2]\AgdaFunction{Subst}\AgdaSpace{}%
\AgdaSymbol{:}\AgdaSpace{}%
\AgdaPrimitive{Set}\<%
\\
\>[2]\AgdaFunction{Subst}\AgdaSpace{}%
\AgdaSymbol{=}\AgdaSpace{}%
\AgdaDatatype{ℕ}\AgdaSpace{}%
\AgdaSymbol{→}\AgdaSpace{}%
\AgdaDatatype{Term}\<%
\\
\\[\AgdaEmptyExtraSkip]%
\>[2]\AgdaFunction{sb0}\AgdaSpace{}%
\AgdaSymbol{:}\AgdaSpace{}%
\AgdaDatatype{Term}\AgdaSpace{}%
\AgdaSymbol{→}\AgdaSpace{}%
\AgdaFunction{Subst}\AgdaSpace{}%
\AgdaComment{--\ Substitute\ 0th\ De\ Bruijn\ index}\<%
\\
\>[2]\AgdaFunction{sb0}\AgdaSpace{}%
\AgdaBound{t}\AgdaSpace{}%
\AgdaNumber{0}%
\>[17]\AgdaSymbol{=}\AgdaSpace{}%
\AgdaBound{t}\<%
\\
\>[2]\AgdaFunction{sb0}\AgdaSpace{}%
\AgdaSymbol{\AgdaUnderscore{}}\AgdaSpace{}%
\AgdaSymbol{(}\AgdaInductiveConstructor{suc}\AgdaSpace{}%
\AgdaBound{n}\AgdaSymbol{)}%
\>[17]\AgdaSymbol{=}\AgdaSpace{}%
\AgdaInductiveConstructor{var}\AgdaSpace{}%
\AgdaBound{n}\<%
\\
\\[\AgdaEmptyExtraSkip]%
\>[2]\AgdaFunction{sb}\AgdaSpace{}%
\AgdaSymbol{:}\AgdaSpace{}%
\AgdaDatatype{Term}\AgdaSpace{}%
\AgdaSymbol{→}\AgdaSpace{}%
\AgdaFunction{Subst}\AgdaSpace{}%
\AgdaSymbol{→}\AgdaSpace{}%
\AgdaDatatype{Term}\AgdaSpace{}%
\AgdaComment{--\ Apply\ substitution\ to\ term}\<%
\end{code}
\begin{code}[hide]%
\>[2]\AgdaFunction{Ren}\AgdaSpace{}%
\AgdaSymbol{:}\AgdaSpace{}%
\AgdaPrimitive{Set}\<%
\\
\>[2]\AgdaFunction{Ren}\AgdaSpace{}%
\AgdaSymbol{=}\AgdaSpace{}%
\AgdaDatatype{ℕ}\AgdaSpace{}%
\AgdaSymbol{→}\AgdaSpace{}%
\AgdaDatatype{ℕ}\<%
\\
\\[\AgdaEmptyExtraSkip]%
\>[2]\AgdaFunction{rn+}\AgdaSpace{}%
\AgdaSymbol{:}\AgdaSpace{}%
\AgdaFunction{Ren}\AgdaSpace{}%
\AgdaSymbol{→}\AgdaSpace{}%
\AgdaFunction{Ren}\<%
\\
\>[2]\AgdaFunction{rn+}\AgdaSpace{}%
\AgdaBound{r}\AgdaSpace{}%
\AgdaInductiveConstructor{zero}\AgdaSpace{}%
\AgdaSymbol{=}\AgdaSpace{}%
\AgdaInductiveConstructor{zero}\<%
\\
\>[2]\AgdaFunction{rn+}\AgdaSpace{}%
\AgdaBound{r}\AgdaSpace{}%
\AgdaSymbol{(}\AgdaInductiveConstructor{suc}\AgdaSpace{}%
\AgdaBound{n}\AgdaSymbol{)}\AgdaSpace{}%
\AgdaSymbol{=}\AgdaSpace{}%
\AgdaBound{r}\AgdaSpace{}%
\AgdaBound{n}\<%
\\
\>[0]\<%
\\
\>[2]\AgdaFunction{rn}\AgdaSpace{}%
\AgdaSymbol{:}\AgdaSpace{}%
\AgdaDatatype{Term}\AgdaSpace{}%
\AgdaSymbol{→}\AgdaSpace{}%
\AgdaFunction{Ren}\AgdaSpace{}%
\AgdaSymbol{→}\AgdaSpace{}%
\AgdaDatatype{Term}\<%
\\
\>[2]\AgdaFunction{rn}\AgdaSpace{}%
\AgdaSymbol{(}\AgdaInductiveConstructor{add}\AgdaSpace{}%
\AgdaBound{t₀}\AgdaSpace{}%
\AgdaBound{t₁}\AgdaSymbol{)}\AgdaSpace{}%
\AgdaBound{r}\AgdaSpace{}%
\AgdaSymbol{=}\AgdaSpace{}%
\AgdaInductiveConstructor{add}\AgdaSpace{}%
\AgdaSymbol{(}\AgdaFunction{rn}\AgdaSpace{}%
\AgdaBound{t₀}\AgdaSpace{}%
\AgdaBound{r}\AgdaSymbol{)}\AgdaSpace{}%
\AgdaSymbol{(}\AgdaFunction{rn}\AgdaSpace{}%
\AgdaBound{t₁}\AgdaSpace{}%
\AgdaBound{r}\AgdaSymbol{)}\<%
\\
\>[2]\AgdaFunction{rn}\AgdaSpace{}%
\AgdaSymbol{(}\AgdaInductiveConstructor{num}\AgdaSpace{}%
\AgdaBound{n}\AgdaSymbol{)}\AgdaSpace{}%
\AgdaBound{r}\AgdaSpace{}%
\AgdaSymbol{=}\AgdaSpace{}%
\AgdaInductiveConstructor{num}\AgdaSpace{}%
\AgdaBound{n}\<%
\\
\>[2]\AgdaFunction{rn}\AgdaSpace{}%
\AgdaSymbol{(}\AgdaInductiveConstructor{lam}\AgdaSpace{}%
\AgdaBound{t₀}\AgdaSymbol{)}\AgdaSpace{}%
\AgdaBound{r}\AgdaSpace{}%
\AgdaSymbol{=}\AgdaSpace{}%
\AgdaInductiveConstructor{lam}\AgdaSpace{}%
\AgdaSymbol{(}\AgdaFunction{rn}\AgdaSpace{}%
\AgdaBound{t₀}\AgdaSpace{}%
\AgdaSymbol{(}\AgdaFunction{rn+}\AgdaSpace{}%
\AgdaBound{r}\AgdaSymbol{))}\<%
\\
\>[2]\AgdaFunction{rn}\AgdaSpace{}%
\AgdaSymbol{(}\AgdaInductiveConstructor{var}\AgdaSpace{}%
\AgdaBound{n}\AgdaSymbol{)}\AgdaSpace{}%
\AgdaBound{r}\AgdaSpace{}%
\AgdaSymbol{=}\AgdaSpace{}%
\AgdaInductiveConstructor{var}\AgdaSpace{}%
\AgdaSymbol{(}\AgdaBound{r}\AgdaSpace{}%
\AgdaBound{n}\AgdaSymbol{)}\<%
\\
\>[2]\AgdaFunction{rn}\AgdaSpace{}%
\AgdaSymbol{(}\AgdaInductiveConstructor{app}\AgdaSpace{}%
\AgdaBound{t₀}\AgdaSpace{}%
\AgdaBound{t₁}\AgdaSymbol{)}\AgdaSpace{}%
\AgdaBound{r}\AgdaSpace{}%
\AgdaSymbol{=}\AgdaSpace{}%
\AgdaInductiveConstructor{app}\AgdaSpace{}%
\AgdaSymbol{(}\AgdaFunction{rn}\AgdaSpace{}%
\AgdaBound{t₀}\AgdaSpace{}%
\AgdaBound{r}\AgdaSymbol{)}\AgdaSpace{}%
\AgdaSymbol{(}\AgdaFunction{rn}\AgdaSpace{}%
\AgdaBound{t₁}\AgdaSpace{}%
\AgdaBound{r}\AgdaSymbol{)}\<%
\\
\>[2]\AgdaFunction{rn}\AgdaSpace{}%
\AgdaSymbol{(}\AgdaInductiveConstructor{reset}\AgdaSpace{}%
\AgdaBound{t₀}\AgdaSymbol{)}\AgdaSpace{}%
\AgdaBound{r}\AgdaSpace{}%
\AgdaSymbol{=}\AgdaSpace{}%
\AgdaInductiveConstructor{reset}\AgdaSpace{}%
\AgdaSymbol{(}\AgdaFunction{rn}\AgdaSpace{}%
\AgdaBound{t₀}\AgdaSpace{}%
\AgdaBound{r}\AgdaSymbol{)}\<%
\\
\>[2]\AgdaFunction{rn}\AgdaSpace{}%
\AgdaSymbol{(}\AgdaInductiveConstructor{shift}\AgdaSpace{}%
\AgdaBound{t₀}\AgdaSymbol{)}\AgdaSpace{}%
\AgdaBound{r}\AgdaSpace{}%
\AgdaSymbol{=}\AgdaSpace{}%
\AgdaInductiveConstructor{shift}\AgdaSpace{}%
\AgdaSymbol{(}\AgdaFunction{rn}\AgdaSpace{}%
\AgdaBound{t₀}\AgdaSpace{}%
\AgdaSymbol{(}\AgdaFunction{rn+}\AgdaSpace{}%
\AgdaBound{r}\AgdaSymbol{))}\<%
\\
\\[\AgdaEmptyExtraSkip]%
\>[2]\AgdaFunction{sb+}\AgdaSpace{}%
\AgdaSymbol{:}\AgdaSpace{}%
\AgdaFunction{Subst}\AgdaSpace{}%
\AgdaSymbol{→}\AgdaSpace{}%
\AgdaFunction{Subst}\<%
\\
\>[2]\AgdaFunction{sb+}\AgdaSpace{}%
\AgdaBound{s}\AgdaSpace{}%
\AgdaInductiveConstructor{zero}%
\>[16]\AgdaSymbol{=}\AgdaSpace{}%
\AgdaInductiveConstructor{var}\AgdaSpace{}%
\AgdaInductiveConstructor{zero}\<%
\\
\>[2]\AgdaFunction{sb+}\AgdaSpace{}%
\AgdaBound{s}\AgdaSpace{}%
\AgdaSymbol{(}\AgdaInductiveConstructor{suc}\AgdaSpace{}%
\AgdaBound{n}\AgdaSymbol{)}\AgdaSpace{}%
\AgdaSymbol{=}\AgdaSpace{}%
\AgdaFunction{rn}\AgdaSpace{}%
\AgdaSymbol{(}\AgdaBound{s}\AgdaSpace{}%
\AgdaBound{n}\AgdaSymbol{)}\AgdaSpace{}%
\AgdaInductiveConstructor{suc}\<%
\\
\\[\AgdaEmptyExtraSkip]%
\>[2]\AgdaFunction{sb}\AgdaSpace{}%
\AgdaSymbol{(}\AgdaInductiveConstructor{add}\AgdaSpace{}%
\AgdaBound{t₀}\AgdaSpace{}%
\AgdaBound{t₁}\AgdaSymbol{)}\AgdaSpace{}%
\AgdaBound{s}\AgdaSpace{}%
\AgdaSymbol{=}\AgdaSpace{}%
\AgdaInductiveConstructor{add}\AgdaSpace{}%
\AgdaSymbol{(}\AgdaFunction{sb}\AgdaSpace{}%
\AgdaBound{t₀}\AgdaSpace{}%
\AgdaBound{s}\AgdaSymbol{)}\AgdaSpace{}%
\AgdaSymbol{(}\AgdaFunction{sb}\AgdaSpace{}%
\AgdaBound{t₁}\AgdaSpace{}%
\AgdaBound{s}\AgdaSymbol{)}\<%
\\
\>[2]\AgdaFunction{sb}\AgdaSpace{}%
\AgdaSymbol{(}\AgdaInductiveConstructor{num}\AgdaSpace{}%
\AgdaBound{n}\AgdaSymbol{)}%
\>[17]\AgdaSymbol{\AgdaUnderscore{}}\AgdaSpace{}%
\AgdaSymbol{=}\AgdaSpace{}%
\AgdaInductiveConstructor{num}\AgdaSpace{}%
\AgdaBound{n}\<%
\\
\>[2]\AgdaFunction{sb}\AgdaSpace{}%
\AgdaSymbol{(}\AgdaInductiveConstructor{lam}\AgdaSpace{}%
\AgdaBound{t₀}\AgdaSymbol{)}%
\>[17]\AgdaBound{s}\AgdaSpace{}%
\AgdaSymbol{=}\AgdaSpace{}%
\AgdaInductiveConstructor{lam}\AgdaSpace{}%
\AgdaSymbol{(}\AgdaFunction{sb}\AgdaSpace{}%
\AgdaBound{t₀}\AgdaSpace{}%
\AgdaSymbol{(}\AgdaFunction{sb+}\AgdaSpace{}%
\AgdaBound{s}\AgdaSymbol{))}\<%
\\
\>[2]\AgdaFunction{sb}\AgdaSpace{}%
\AgdaSymbol{(}\AgdaInductiveConstructor{var}\AgdaSpace{}%
\AgdaBound{x}\AgdaSymbol{)}%
\>[17]\AgdaBound{s}\AgdaSpace{}%
\AgdaSymbol{=}\AgdaSpace{}%
\AgdaBound{s}\AgdaSpace{}%
\AgdaBound{x}\<%
\\
\>[2]\AgdaFunction{sb}\AgdaSpace{}%
\AgdaSymbol{(}\AgdaInductiveConstructor{app}\AgdaSpace{}%
\AgdaBound{t₀}\AgdaSpace{}%
\AgdaBound{t₁}\AgdaSymbol{)}\AgdaSpace{}%
\AgdaBound{s}\AgdaSpace{}%
\AgdaSymbol{=}\AgdaSpace{}%
\AgdaInductiveConstructor{app}\AgdaSpace{}%
\AgdaSymbol{(}\AgdaFunction{sb}\AgdaSpace{}%
\AgdaBound{t₀}\AgdaSpace{}%
\AgdaBound{s}\AgdaSymbol{)}\AgdaSpace{}%
\AgdaSymbol{(}\AgdaFunction{sb}\AgdaSpace{}%
\AgdaBound{t₁}\AgdaSpace{}%
\AgdaBound{s}\AgdaSymbol{)}\<%
\\
\>[2]\AgdaFunction{sb}\AgdaSpace{}%
\AgdaSymbol{(}\AgdaInductiveConstructor{reset}\AgdaSpace{}%
\AgdaBound{t₀}\AgdaSymbol{)}%
\>[17]\AgdaBound{s}\AgdaSpace{}%
\AgdaSymbol{=}\AgdaSpace{}%
\AgdaInductiveConstructor{reset}\AgdaSpace{}%
\AgdaSymbol{(}\AgdaFunction{sb}\AgdaSpace{}%
\AgdaBound{t₀}\AgdaSpace{}%
\AgdaBound{s}\AgdaSymbol{)}\<%
\\
\>[2]\AgdaFunction{sb}\AgdaSpace{}%
\AgdaSymbol{(}\AgdaInductiveConstructor{shift}\AgdaSpace{}%
\AgdaBound{t₀}\AgdaSymbol{)}%
\>[17]\AgdaBound{s}\AgdaSpace{}%
\AgdaSymbol{=}\AgdaSpace{}%
\AgdaInductiveConstructor{shift}\AgdaSpace{}%
\AgdaSymbol{(}\AgdaFunction{sb}\AgdaSpace{}%
\AgdaBound{t₀}\AgdaSpace{}%
\AgdaSymbol{(}\AgdaFunction{sb+}\AgdaSpace{}%
\AgdaBound{s}\AgdaSymbol{))}\<%
\end{code}
\begin{code}[hide]%
\>[2]\AgdaFunction{sb-test₀}\AgdaSpace{}%
\AgdaSymbol{:}%
\>[1094I]\AgdaFunction{sb}\AgdaSpace{}%
\AgdaSymbol{(}\AgdaInductiveConstructor{lam}\AgdaSpace{}%
\AgdaSymbol{(}\AgdaInductiveConstructor{lam}\AgdaSpace{}%
\AgdaSymbol{(}\AgdaInductiveConstructor{add}\AgdaSpace{}%
\AgdaSymbol{(}\AgdaInductiveConstructor{var}\AgdaSpace{}%
\AgdaNumber{0}\AgdaSymbol{)}\AgdaSpace{}%
\AgdaSymbol{(}\AgdaInductiveConstructor{var}\AgdaSpace{}%
\AgdaNumber{1}\AgdaSymbol{))))}\AgdaSpace{}%
\AgdaSymbol{(}\AgdaFunction{sb0}\AgdaSpace{}%
\AgdaSymbol{(}\AgdaInductiveConstructor{num}\AgdaSpace{}%
\AgdaNumber{42}\AgdaSymbol{))}\<%
\\
\>[.][@{}l@{}]\<[1094I]%
\>[13]\AgdaOperator{\AgdaDatatype{≡}}\AgdaSpace{}%
\AgdaInductiveConstructor{lam}\AgdaSpace{}%
\AgdaSymbol{(}\AgdaInductiveConstructor{lam}\AgdaSpace{}%
\AgdaSymbol{(}\AgdaInductiveConstructor{add}\AgdaSpace{}%
\AgdaSymbol{(}\AgdaInductiveConstructor{var}\AgdaSpace{}%
\AgdaNumber{0}\AgdaSymbol{)}\AgdaSpace{}%
\AgdaSymbol{(}\AgdaInductiveConstructor{var}\AgdaSpace{}%
\AgdaNumber{1}\AgdaSymbol{)))}\<%
\\
\>[2]\AgdaFunction{sb-test₀}\AgdaSpace{}%
\AgdaSymbol{=}\AgdaSpace{}%
\AgdaInductiveConstructor{refl}\<%
\\
\\[\AgdaEmptyExtraSkip]%
\>[2]\AgdaFunction{sb-test₁}\AgdaSpace{}%
\AgdaSymbol{:}%
\>[1115I]\AgdaFunction{sb}\AgdaSpace{}%
\AgdaSymbol{(}\AgdaInductiveConstructor{lam}\AgdaSpace{}%
\AgdaSymbol{(}\AgdaInductiveConstructor{lam}\AgdaSpace{}%
\AgdaSymbol{(}\AgdaInductiveConstructor{add}\AgdaSpace{}%
\AgdaSymbol{(}\AgdaInductiveConstructor{var}\AgdaSpace{}%
\AgdaNumber{0}\AgdaSymbol{)}\AgdaSpace{}%
\AgdaSymbol{(}\AgdaInductiveConstructor{var}\AgdaSpace{}%
\AgdaNumber{2}\AgdaSymbol{))))}\AgdaSpace{}%
\AgdaSymbol{(}\AgdaFunction{sb0}\AgdaSpace{}%
\AgdaSymbol{(}\AgdaInductiveConstructor{num}\AgdaSpace{}%
\AgdaNumber{42}\AgdaSymbol{))}\<%
\\
\>[.][@{}l@{}]\<[1115I]%
\>[13]\AgdaOperator{\AgdaDatatype{≡}}\AgdaSpace{}%
\AgdaInductiveConstructor{lam}\AgdaSpace{}%
\AgdaSymbol{(}\AgdaInductiveConstructor{lam}\AgdaSpace{}%
\AgdaSymbol{(}\AgdaInductiveConstructor{add}\AgdaSpace{}%
\AgdaSymbol{(}\AgdaInductiveConstructor{var}\AgdaSpace{}%
\AgdaNumber{0}\AgdaSymbol{)}\AgdaSpace{}%
\AgdaSymbol{(}\AgdaInductiveConstructor{num}\AgdaSpace{}%
\AgdaNumber{42}\AgdaSymbol{)))}\<%
\\
\>[2]\AgdaFunction{sb-test₁}\AgdaSpace{}%
\AgdaSymbol{=}\AgdaSpace{}%
\AgdaInductiveConstructor{refl}\<%
\\
\\[\AgdaEmptyExtraSkip]%
\>[2]\AgdaFunction{sb-test₂}\AgdaSpace{}%
\AgdaSymbol{:}%
\>[1136I]\AgdaFunction{sb}\AgdaSpace{}%
\AgdaSymbol{(}\AgdaInductiveConstructor{lam}\AgdaSpace{}%
\AgdaSymbol{(}\AgdaInductiveConstructor{lam}\AgdaSpace{}%
\AgdaSymbol{(}\AgdaInductiveConstructor{add}\AgdaSpace{}%
\AgdaSymbol{(}\AgdaInductiveConstructor{var}\AgdaSpace{}%
\AgdaNumber{0}\AgdaSymbol{)}\AgdaSpace{}%
\AgdaSymbol{(}\AgdaInductiveConstructor{var}\AgdaSpace{}%
\AgdaNumber{3}\AgdaSymbol{))))}\AgdaSpace{}%
\AgdaSymbol{(}\AgdaFunction{sb0}\AgdaSpace{}%
\AgdaSymbol{(}\AgdaInductiveConstructor{num}\AgdaSpace{}%
\AgdaNumber{42}\AgdaSymbol{))}\<%
\\
\>[.][@{}l@{}]\<[1136I]%
\>[13]\AgdaOperator{\AgdaDatatype{≡}}\AgdaSpace{}%
\AgdaInductiveConstructor{lam}\AgdaSpace{}%
\AgdaSymbol{(}\AgdaInductiveConstructor{lam}\AgdaSpace{}%
\AgdaSymbol{(}\AgdaInductiveConstructor{add}\AgdaSpace{}%
\AgdaSymbol{(}\AgdaInductiveConstructor{var}\AgdaSpace{}%
\AgdaNumber{0}\AgdaSymbol{)}\AgdaSpace{}%
\AgdaSymbol{(}\AgdaInductiveConstructor{var}\AgdaSpace{}%
\AgdaNumber{2}\AgdaSymbol{)))}\<%
\\
\>[2]\AgdaFunction{sb-test₂}\AgdaSpace{}%
\AgdaSymbol{=}\AgdaSpace{}%
\AgdaInductiveConstructor{refl}\<%
\end{code}
Values in our language are numbers and function closures:
\begin{code}%
\>[2]\AgdaKeyword{data}\AgdaSpace{}%
\AgdaDatatype{Val}\AgdaSpace{}%
\AgdaSymbol{:}\AgdaSpace{}%
\AgdaPrimitive{Set}\AgdaSpace{}%
\AgdaKeyword{where}\<%
\\
\>[2][@{}l@{\AgdaIndent{0}}]%
\>[4]\AgdaInductiveConstructor{num}%
\>[9]\AgdaSymbol{:}\AgdaSpace{}%
\AgdaDatatype{ℕ}%
\>[17]\AgdaSymbol{→}\AgdaSpace{}%
\AgdaDatatype{Val}\<%
\\
\>[4]\AgdaInductiveConstructor{clo}%
\>[9]\AgdaSymbol{:}\AgdaSpace{}%
\AgdaDatatype{Term}%
\>[17]\AgdaSymbol{→}\AgdaSpace{}%
\AgdaDatatype{Val}\<%
\end{code}
Values can be converted to terms; not always the other way around.
Furthermore, we use a function \af{is-val} for recognizing terms that are values:
\begin{code}%
\>[2]\AgdaFunction{v2t}%
\>[10]\AgdaSymbol{:}\AgdaSpace{}%
\AgdaDatatype{Val}\AgdaSpace{}%
\AgdaSymbol{→}\AgdaSpace{}%
\AgdaDatatype{Term}\<%
\\
\>[2]\AgdaFunction{t2v}%
\>[10]\AgdaSymbol{:}\AgdaSpace{}%
\AgdaDatatype{Term}\AgdaSpace{}%
\AgdaSymbol{→}\AgdaSpace{}%
\AgdaDatatype{Maybe}\AgdaSpace{}%
\AgdaDatatype{Val}\<%
\\
\>[2]\AgdaFunction{is-val}%
\>[10]\AgdaSymbol{:}\AgdaSpace{}%
\AgdaDatatype{Term}\AgdaSpace{}%
\AgdaSymbol{→}\AgdaSpace{}%
\AgdaDatatype{Bool}\<%
\end{code}
\begin{code}[hide]%
\>[2]\AgdaFunction{v2t}\AgdaSpace{}%
\AgdaSymbol{(}\AgdaInductiveConstructor{num}\AgdaSpace{}%
\AgdaBound{n}\AgdaSymbol{)}%
\>[15]\AgdaSymbol{=}\AgdaSpace{}%
\AgdaInductiveConstructor{num}\AgdaSpace{}%
\AgdaBound{n}\<%
\\
\>[2]\AgdaFunction{v2t}\AgdaSpace{}%
\AgdaSymbol{(}\AgdaInductiveConstructor{clo}\AgdaSpace{}%
\AgdaBound{t}\AgdaSymbol{)}%
\>[15]\AgdaSymbol{=}\AgdaSpace{}%
\AgdaInductiveConstructor{lam}\AgdaSpace{}%
\AgdaBound{t}\<%
\\
\\[\AgdaEmptyExtraSkip]%
\>[2]\AgdaFunction{t2v}\AgdaSpace{}%
\AgdaSymbol{(}\AgdaInductiveConstructor{num}\AgdaSpace{}%
\AgdaBound{n}\AgdaSymbol{)}\AgdaSpace{}%
\AgdaSymbol{=}\AgdaSpace{}%
\AgdaInductiveConstructor{just}\AgdaSpace{}%
\AgdaSymbol{(}\AgdaInductiveConstructor{num}\AgdaSpace{}%
\AgdaBound{n}\AgdaSymbol{)}\<%
\\
\>[2]\AgdaFunction{t2v}\AgdaSpace{}%
\AgdaSymbol{(}\AgdaInductiveConstructor{lam}\AgdaSpace{}%
\AgdaBound{t}\AgdaSymbol{)}\AgdaSpace{}%
\AgdaSymbol{=}\AgdaSpace{}%
\AgdaInductiveConstructor{just}\AgdaSpace{}%
\AgdaSymbol{(}\AgdaInductiveConstructor{clo}\AgdaSpace{}%
\AgdaBound{t}\AgdaSymbol{)}\<%
\\
\>[2]\AgdaCatchallClause{\AgdaFunction{t2v}}\AgdaSpace{}%
\AgdaCatchallClause{\AgdaSymbol{\AgdaUnderscore{}}}%
\>[14]\AgdaSymbol{=}\AgdaSpace{}%
\AgdaInductiveConstructor{nothing}\<%
\\
\\[\AgdaEmptyExtraSkip]%
\>[2]\AgdaFunction{is-val}\AgdaSpace{}%
\AgdaSymbol{(}\AgdaInductiveConstructor{num}\AgdaSpace{}%
\AgdaBound{n}\AgdaSymbol{)}\AgdaSpace{}%
\AgdaSymbol{=}\AgdaSpace{}%
\AgdaInductiveConstructor{true}\<%
\\
\>[2]\AgdaFunction{is-val}\AgdaSpace{}%
\AgdaSymbol{(}\AgdaInductiveConstructor{lam}\AgdaSpace{}%
\AgdaBound{t}\AgdaSymbol{)}\AgdaSpace{}%
\AgdaSymbol{=}\AgdaSpace{}%
\AgdaInductiveConstructor{true}\<%
\\
\>[2]\AgdaCatchallClause{\AgdaFunction{is-val}}\AgdaSpace{}%
\AgdaCatchallClause{\AgdaSymbol{\AgdaUnderscore{}}}%
\>[17]\AgdaSymbol{=}\AgdaSpace{}%
\AgdaInductiveConstructor{false}\<%
\end{code}

\subsubsection{Reduction Strategy and Contraction}

Contexts are defined similarly as earlier, and directly encode the notion of reduction context in \cref{fig:redsem-lam}.
\begin{code}%
\>[2]\AgdaKeyword{data}\AgdaSpace{}%
\AgdaDatatype{CtxFrame}\AgdaSpace{}%
\AgdaSymbol{:}\AgdaSpace{}%
\AgdaPrimitive{Set}\AgdaSpace{}%
\AgdaKeyword{where}\<%
\\
\>[2][@{}l@{\AgdaIndent{0}}]%
\>[4]\AgdaInductiveConstructor{add-l}\AgdaSpace{}%
\AgdaSymbol{:}\AgdaSpace{}%
\AgdaDatatype{Term}%
\>[18]\AgdaSymbol{→}\AgdaSpace{}%
\AgdaDatatype{CtxFrame}\AgdaSpace{}%
\AgdaComment{--\ As\ before}\<%
\\
\>[4]\AgdaInductiveConstructor{add-r}\AgdaSpace{}%
\AgdaSymbol{:}\AgdaSpace{}%
\AgdaDatatype{Val}%
\>[18]\AgdaSymbol{→}\AgdaSpace{}%
\AgdaDatatype{CtxFrame}\AgdaSpace{}%
\AgdaComment{--\ As\ before}\<%
\\
\>[4]\AgdaInductiveConstructor{app-l}\AgdaSpace{}%
\AgdaSymbol{:}\AgdaSpace{}%
\AgdaDatatype{Term}%
\>[18]\AgdaSymbol{→}\AgdaSpace{}%
\AgdaDatatype{CtxFrame}\<%
\\
\>[4]\AgdaInductiveConstructor{app-r}\AgdaSpace{}%
\AgdaSymbol{:}\AgdaSpace{}%
\AgdaDatatype{Val}%
\>[18]\AgdaSymbol{→}\AgdaSpace{}%
\AgdaDatatype{CtxFrame}\<%
\\
\>[4]\AgdaInductiveConstructor{reset}\AgdaSpace{}%
\AgdaSymbol{:}%
\>[20]\AgdaDatatype{CtxFrame}\<%
\\
\\[\AgdaEmptyExtraSkip]%
\>[2]\AgdaFunction{Context}\AgdaSpace{}%
\AgdaSymbol{=}\AgdaSpace{}%
\AgdaDatatype{List}\AgdaSpace{}%
\AgdaDatatype{CtxFrame}\<%
\end{code}
Here and in \cref{fig:redsem-lam} there is no context frame for either \ac{lam} or \ac{shift} constructors.
If there were, we might perform reduction under $\lambda$s which, for this language, we disallow.
While \ad{CtxFrame} has new cases, we will see in \cref{sec:04-redsem-gen} that contexts can be systematically derived from an abstract syntax definition.

\begin{code}[hide]%
\>[2]\AgdaFunction{is-redex}\AgdaSpace{}%
\AgdaSymbol{:}\AgdaSpace{}%
\AgdaDatatype{Term}\AgdaSpace{}%
\AgdaSymbol{→}\AgdaSpace{}%
\AgdaDatatype{Bool}\<%
\\
\>[2]\AgdaFunction{is-redex}\AgdaSpace{}%
\AgdaSymbol{(}\AgdaInductiveConstructor{add}\AgdaSpace{}%
\AgdaSymbol{(}\AgdaInductiveConstructor{num}\AgdaSpace{}%
\AgdaBound{n₀}\AgdaSymbol{)}\AgdaSpace{}%
\AgdaSymbol{(}\AgdaInductiveConstructor{num}\AgdaSpace{}%
\AgdaBound{n₁}\AgdaSymbol{))}%
\>[36]\AgdaSymbol{=}\AgdaSpace{}%
\AgdaInductiveConstructor{true}\<%
\\
\>[2]\AgdaFunction{is-redex}\AgdaSpace{}%
\AgdaSymbol{(}\AgdaInductiveConstructor{app}\AgdaSpace{}%
\AgdaSymbol{(}\AgdaInductiveConstructor{lam}\AgdaSpace{}%
\AgdaBound{t₀}\AgdaSymbol{)}\AgdaSpace{}%
\AgdaBound{t₁}\AgdaSymbol{)}\AgdaSpace{}%
\AgdaKeyword{with}\AgdaSpace{}%
\AgdaFunction{is-val}\AgdaSpace{}%
\AgdaBound{t₁}\<%
\\
\>[2]\AgdaSymbol{...}\AgdaSpace{}%
\AgdaSymbol{|}\AgdaSpace{}%
\AgdaInductiveConstructor{true}\AgdaSpace{}%
\AgdaSymbol{=}\AgdaSpace{}%
\AgdaInductiveConstructor{true}\<%
\\
\>[2]\AgdaSymbol{...}\AgdaSpace{}%
\AgdaSymbol{|}\AgdaSpace{}%
\AgdaInductiveConstructor{false}\AgdaSpace{}%
\AgdaSymbol{=}\AgdaSpace{}%
\AgdaInductiveConstructor{false}\<%
\\
\>[2]\AgdaFunction{is-redex}\AgdaSpace{}%
\AgdaSymbol{(}\AgdaInductiveConstructor{reset}\AgdaSpace{}%
\AgdaBound{t}\AgdaSymbol{)}\AgdaSpace{}%
\AgdaKeyword{with}\AgdaSpace{}%
\AgdaFunction{is-val}\AgdaSpace{}%
\AgdaBound{t}\<%
\\
\>[2]\AgdaSymbol{...}\AgdaSpace{}%
\AgdaSymbol{|}\AgdaSpace{}%
\AgdaInductiveConstructor{true}\AgdaSpace{}%
\AgdaSymbol{=}\AgdaSpace{}%
\AgdaInductiveConstructor{true}\<%
\\
\>[2]\AgdaSymbol{...}\AgdaSpace{}%
\AgdaSymbol{|}\AgdaSpace{}%
\AgdaInductiveConstructor{false}\AgdaSpace{}%
\AgdaSymbol{=}\AgdaSpace{}%
\AgdaInductiveConstructor{false}\<%
\\
\>[2]\AgdaFunction{is-redex}\AgdaSpace{}%
\AgdaSymbol{(}\AgdaInductiveConstructor{shift}\AgdaSpace{}%
\AgdaBound{t}\AgdaSymbol{)}\AgdaSpace{}%
\AgdaSymbol{=}\AgdaSpace{}%
\AgdaInductiveConstructor{true}\<%
\\
\>[2]\AgdaCatchallClause{\AgdaFunction{is-redex}}\AgdaSpace{}%
\AgdaCatchallClause{\AgdaSymbol{\AgdaUnderscore{}}}\AgdaSpace{}%
\AgdaSymbol{=}\AgdaSpace{}%
\AgdaInductiveConstructor{false}\<%
\\
\\[\AgdaEmptyExtraSkip]%
\>[2]\AgdaFunction{↑}\AgdaSpace{}%
\AgdaSymbol{:}\AgdaSpace{}%
\AgdaDatatype{Term}\AgdaSpace{}%
\AgdaSymbol{→}\AgdaSpace{}%
\AgdaDatatype{Term}\<%
\\
\>[2]\AgdaFunction{↑}\AgdaSpace{}%
\AgdaBound{t}\AgdaSpace{}%
\AgdaSymbol{=}\AgdaSpace{}%
\AgdaFunction{rn}\AgdaSpace{}%
\AgdaBound{t}\AgdaSpace{}%
\AgdaInductiveConstructor{suc}\<%
\end{code}

Redexes in our language are either addition expressions whose immediate sub-terms are numbers, application expressions whose first sub-term is a $\lambda$ and second sub-term is a value, reset expressions whose sub-term is a value, or shift expressions.
The following function contracts these.
\begin{code}%
\>[2]\AgdaFunction{contract}\AgdaSpace{}%
\AgdaSymbol{:}\AgdaSpace{}%
\AgdaDatatype{Term}\AgdaSpace{}%
\AgdaSymbol{→}\AgdaSpace{}%
\AgdaFunction{Context}\AgdaSpace{}%
\AgdaSymbol{→}\AgdaSpace{}%
\AgdaDatatype{Maybe}\AgdaSpace{}%
\AgdaSymbol{(}\AgdaDatatype{Term}\AgdaSpace{}%
\AgdaOperator{\AgdaFunction{×}}\AgdaSpace{}%
\AgdaFunction{Context}\AgdaSymbol{)}\<%
\end{code}
\begin{code}[hide]%
\>[2]\AgdaFunction{contract}\AgdaSpace{}%
\AgdaSymbol{(}\AgdaInductiveConstructor{add}\AgdaSpace{}%
\AgdaSymbol{(}\AgdaInductiveConstructor{num}\AgdaSpace{}%
\AgdaBound{n₀}\AgdaSymbol{)}\AgdaSpace{}%
\AgdaSymbol{(}\AgdaInductiveConstructor{num}\AgdaSpace{}%
\AgdaBound{n₁}\AgdaSymbol{))}\AgdaSpace{}%
\AgdaBound{c}\AgdaSpace{}%
\AgdaSymbol{=}\<%
\\
\>[2][@{}l@{\AgdaIndent{0}}]%
\>[4]\AgdaInductiveConstructor{just}\AgdaSpace{}%
\AgdaSymbol{(}\AgdaInductiveConstructor{num}\AgdaSpace{}%
\AgdaSymbol{(}\AgdaBound{n₀}\AgdaSpace{}%
\AgdaOperator{\AgdaPrimitive{+}}\AgdaSpace{}%
\AgdaBound{n₁}\AgdaSymbol{)}\AgdaSpace{}%
\AgdaOperator{\AgdaInductiveConstructor{,}}\AgdaSpace{}%
\AgdaBound{c}\AgdaSymbol{)}\<%
\\
\>[2]\AgdaFunction{contract}\AgdaSpace{}%
\AgdaSymbol{(}\AgdaInductiveConstructor{app}\AgdaSpace{}%
\AgdaSymbol{(}\AgdaInductiveConstructor{lam}\AgdaSpace{}%
\AgdaBound{t₀}\AgdaSymbol{)}\AgdaSpace{}%
\AgdaBound{t₁}\AgdaSymbol{)}%
\>[35]\AgdaBound{c}\AgdaSpace{}%
\AgdaKeyword{with}\AgdaSpace{}%
\AgdaFunction{is-val}\AgdaSpace{}%
\AgdaBound{t₁}\<%
\\
\>[2]\AgdaSymbol{...}\AgdaSpace{}%
\AgdaSymbol{|}\AgdaSpace{}%
\AgdaInductiveConstructor{false}%
\>[15]\AgdaSymbol{=}\AgdaSpace{}%
\AgdaInductiveConstructor{nothing}\<%
\\
\>[2]\AgdaSymbol{...}\AgdaSpace{}%
\AgdaSymbol{|}\AgdaSpace{}%
\AgdaInductiveConstructor{true}%
\>[15]\AgdaSymbol{=}\AgdaSpace{}%
\AgdaInductiveConstructor{just}\AgdaSpace{}%
\AgdaSymbol{(}\AgdaFunction{sb}\AgdaSpace{}%
\AgdaBound{t₀}\AgdaSpace{}%
\AgdaSymbol{(}\AgdaFunction{sb0}\AgdaSpace{}%
\AgdaBound{t₁}\AgdaSymbol{)}\AgdaSpace{}%
\AgdaOperator{\AgdaInductiveConstructor{,}}\AgdaSpace{}%
\AgdaBound{c}\AgdaSymbol{)}\<%
\\
\>[2]\AgdaFunction{contract}\AgdaSpace{}%
\AgdaSymbol{(}\AgdaInductiveConstructor{reset}\AgdaSpace{}%
\AgdaBound{t₀}\AgdaSymbol{)}%
\>[35]\AgdaBound{c}\AgdaSpace{}%
\AgdaKeyword{with}\AgdaSpace{}%
\AgdaFunction{is-val}\AgdaSpace{}%
\AgdaBound{t₀}\<%
\\
\>[2]\AgdaSymbol{...}\AgdaSpace{}%
\AgdaSymbol{|}\AgdaSpace{}%
\AgdaInductiveConstructor{false}\AgdaSpace{}%
\AgdaSymbol{=}\AgdaSpace{}%
\AgdaInductiveConstructor{nothing}\<%
\\
\>[2]\AgdaSymbol{...}\AgdaSpace{}%
\AgdaSymbol{|}\AgdaSpace{}%
\AgdaInductiveConstructor{true}%
\>[14]\AgdaSymbol{=}\AgdaSpace{}%
\AgdaInductiveConstructor{just}\AgdaSpace{}%
\AgdaSymbol{(}\AgdaBound{t₀}\AgdaSpace{}%
\AgdaOperator{\AgdaInductiveConstructor{,}}\AgdaSpace{}%
\AgdaBound{c}\AgdaSymbol{)}\<%
\end{code}
The most interesting case of this function is the case for \ac{shift}.
\begin{code}%
\>[2]\AgdaFunction{contract}\AgdaSpace{}%
\AgdaSymbol{(}\AgdaInductiveConstructor{shift}\AgdaSpace{}%
\AgdaBound{t₀}\AgdaSymbol{)}%
\>[35]\AgdaBound{c}\AgdaSpace{}%
\AgdaSymbol{=}\AgdaSpace{}%
\AgdaKeyword{do}\<%
\\
\>[2][@{}l@{\AgdaIndent{0}}]%
\>[4]\AgdaSymbol{(}\AgdaBound{c₁}\AgdaSpace{}%
\AgdaOperator{\AgdaInductiveConstructor{,}}\AgdaSpace{}%
\AgdaBound{t₁}\AgdaSymbol{)}\AgdaSpace{}%
\AgdaOperator{\AgdaFunction{←}}\AgdaSpace{}%
\AgdaFunction{unwind}\AgdaSpace{}%
\AgdaBound{c}\AgdaSpace{}%
\AgdaSymbol{(}\AgdaInductiveConstructor{var}\AgdaSpace{}%
\AgdaNumber{0}\AgdaSymbol{)}\<%
\\
\>[4]\AgdaInductiveConstructor{just}\AgdaSpace{}%
\AgdaSymbol{(}\AgdaInductiveConstructor{app}\AgdaSpace{}%
\AgdaSymbol{(}\AgdaInductiveConstructor{lam}\AgdaSpace{}%
\AgdaBound{t₀}\AgdaSymbol{)}\AgdaSpace{}%
\AgdaSymbol{(}\AgdaInductiveConstructor{lam}\AgdaSpace{}%
\AgdaSymbol{(}\AgdaInductiveConstructor{reset}\AgdaSpace{}%
\AgdaBound{t₁}\AgdaSymbol{))}\AgdaSpace{}%
\AgdaOperator{\AgdaInductiveConstructor{,}}\AgdaSpace{}%
\AgdaBound{c₁}\AgdaSymbol{)}\<%
\\
\>[4]\AgdaKeyword{where}\<%
\\
\>[4][@{}l@{\AgdaIndent{0}}]%
\>[6]\AgdaFunction{unwind}\AgdaSpace{}%
\AgdaSymbol{:}\AgdaSpace{}%
\AgdaFunction{Context}\AgdaSpace{}%
\AgdaSymbol{→}\AgdaSpace{}%
\AgdaDatatype{Term}\AgdaSpace{}%
\AgdaSymbol{→}\AgdaSpace{}%
\AgdaDatatype{Maybe}\AgdaSpace{}%
\AgdaSymbol{(}\AgdaFunction{Context}\AgdaSpace{}%
\AgdaOperator{\AgdaFunction{×}}\AgdaSpace{}%
\AgdaDatatype{Term}\AgdaSymbol{)}\<%
\\
\>[6]\AgdaFunction{unwind}\AgdaSpace{}%
\AgdaInductiveConstructor{[]}%
\>[29]\AgdaSymbol{\AgdaUnderscore{}}%
\>[33]\AgdaSymbol{=}\AgdaSpace{}%
\AgdaInductiveConstructor{nothing}\<%
\\
\>[6]\AgdaFunction{unwind}\AgdaSpace{}%
\AgdaSymbol{(}\AgdaInductiveConstructor{add-l}\AgdaSpace{}%
\AgdaBound{t₁}\AgdaSpace{}%
\AgdaOperator{\AgdaInductiveConstructor{∷}}\AgdaSpace{}%
\AgdaBound{c}\AgdaSymbol{)}%
\>[29]\AgdaBound{t₀}%
\>[33]\AgdaSymbol{=}\AgdaSpace{}%
\AgdaFunction{unwind}\AgdaSpace{}%
\AgdaBound{c}\AgdaSpace{}%
\AgdaSymbol{(}\AgdaInductiveConstructor{add}\AgdaSpace{}%
\AgdaBound{t₀}\AgdaSpace{}%
\AgdaSymbol{(}\AgdaFunction{↑}\AgdaSpace{}%
\AgdaBound{t₁}\AgdaSymbol{))}\<%
\\
\>[6]\AgdaFunction{unwind}\AgdaSpace{}%
\AgdaSymbol{(}\AgdaInductiveConstructor{add-r}\AgdaSpace{}%
\AgdaBound{v}\AgdaSpace{}%
\AgdaOperator{\AgdaInductiveConstructor{∷}}\AgdaSpace{}%
\AgdaBound{c}\AgdaSymbol{)}%
\>[29]\AgdaBound{t₁}%
\>[33]\AgdaSymbol{=}\AgdaSpace{}%
\AgdaFunction{unwind}\AgdaSpace{}%
\AgdaBound{c}\AgdaSpace{}%
\AgdaSymbol{(}\AgdaInductiveConstructor{add}\AgdaSpace{}%
\AgdaSymbol{(}\AgdaFunction{↑}\AgdaSpace{}%
\AgdaSymbol{(}\AgdaFunction{v2t}\AgdaSpace{}%
\AgdaBound{v}\AgdaSymbol{))}\AgdaSpace{}%
\AgdaBound{t₁}\AgdaSymbol{)}\<%
\\
\>[6]\AgdaFunction{unwind}\AgdaSpace{}%
\AgdaSymbol{(}\AgdaInductiveConstructor{app-l}\AgdaSpace{}%
\AgdaBound{t₁}\AgdaSpace{}%
\AgdaOperator{\AgdaInductiveConstructor{∷}}\AgdaSpace{}%
\AgdaBound{c}\AgdaSymbol{)}%
\>[29]\AgdaBound{t₀}%
\>[33]\AgdaSymbol{=}\AgdaSpace{}%
\AgdaFunction{unwind}\AgdaSpace{}%
\AgdaBound{c}\AgdaSpace{}%
\AgdaSymbol{(}\AgdaInductiveConstructor{app}\AgdaSpace{}%
\AgdaBound{t₀}\AgdaSpace{}%
\AgdaSymbol{(}\AgdaFunction{↑}\AgdaSpace{}%
\AgdaBound{t₁}\AgdaSymbol{))}\<%
\\
\>[6]\AgdaFunction{unwind}\AgdaSpace{}%
\AgdaSymbol{(}\AgdaInductiveConstructor{app-r}\AgdaSpace{}%
\AgdaBound{v}\AgdaSpace{}%
\AgdaOperator{\AgdaInductiveConstructor{∷}}\AgdaSpace{}%
\AgdaBound{c}\AgdaSymbol{)}%
\>[29]\AgdaBound{t₁}%
\>[33]\AgdaSymbol{=}\AgdaSpace{}%
\AgdaFunction{unwind}\AgdaSpace{}%
\AgdaBound{c}\AgdaSpace{}%
\AgdaSymbol{(}\AgdaInductiveConstructor{app}\AgdaSpace{}%
\AgdaSymbol{(}\AgdaFunction{↑}\AgdaSpace{}%
\AgdaSymbol{(}\AgdaFunction{v2t}\AgdaSpace{}%
\AgdaBound{v}\AgdaSymbol{))}\AgdaSpace{}%
\AgdaBound{t₁}\AgdaSymbol{)}\<%
\\
\>[6]\AgdaFunction{unwind}\AgdaSpace{}%
\AgdaSymbol{(}\AgdaInductiveConstructor{reset}\AgdaSpace{}%
\AgdaOperator{\AgdaInductiveConstructor{∷}}\AgdaSpace{}%
\AgdaBound{c}\AgdaSymbol{)}%
\>[29]\AgdaBound{t}%
\>[33]\AgdaSymbol{=}\AgdaSpace{}%
\AgdaInductiveConstructor{just}\AgdaSpace{}%
\AgdaSymbol{(}\AgdaInductiveConstructor{reset}\AgdaSpace{}%
\AgdaOperator{\AgdaInductiveConstructor{∷}}\AgdaSpace{}%
\AgdaBound{c}\AgdaSpace{}%
\AgdaOperator{\AgdaInductiveConstructor{,}}\AgdaSpace{}%
\AgdaBound{t}\AgdaSymbol{)}\<%
\end{code}
\begin{code}[hide]%
\>[2]\AgdaCatchallClause{\AgdaFunction{contract}}\AgdaSpace{}%
\AgdaCatchallClause{\AgdaSymbol{\AgdaUnderscore{}}}%
\>[35]\AgdaCatchallClause{\AgdaBound{c}}\AgdaSpace{}%
\AgdaSymbol{=}\AgdaSpace{}%
\AgdaInductiveConstructor{nothing}\<%
\end{code}
This case uses an auxiliary function to simultaneously (1) unwind the context to find the nearest \ac{reset} and (2) increment De Bruijn indices by one to prevent name capture.

Decomposition and recomposition all follow a similar pattern as earlier; we elide their full definitions for brevity.
\begin{code}%
\>[2]\AgdaFunction{decompose}\AgdaSpace{}%
\AgdaSymbol{:}\AgdaSpace{}%
\AgdaDatatype{Term}\AgdaSpace{}%
\AgdaSymbol{→}\AgdaSpace{}%
\AgdaFunction{Context}\AgdaSpace{}%
\AgdaSymbol{→}\AgdaSpace{}%
\AgdaDatatype{Maybe}\AgdaSpace{}%
\AgdaSymbol{(}\AgdaDatatype{Term}\AgdaSpace{}%
\AgdaOperator{\AgdaFunction{×}}\AgdaSpace{}%
\AgdaFunction{Context}\AgdaSymbol{)}\<%
\\
\>[2]\AgdaFunction{recompose-frame}\AgdaSpace{}%
\AgdaSymbol{:}\AgdaSpace{}%
\AgdaDatatype{Term}\AgdaSpace{}%
\AgdaSymbol{→}\AgdaSpace{}%
\AgdaDatatype{CtxFrame}\AgdaSpace{}%
\AgdaSymbol{→}\AgdaSpace{}%
\AgdaDatatype{Term}\<%
\\
\>[2]\AgdaFunction{recompose}\AgdaSpace{}%
\AgdaSymbol{:}\AgdaSpace{}%
\AgdaDatatype{Term}\AgdaSpace{}%
\AgdaSymbol{→}\AgdaSpace{}%
\AgdaFunction{Context}\AgdaSpace{}%
\AgdaSymbol{→}\AgdaSpace{}%
\AgdaDatatype{Term}\<%
\end{code}
\begin{code}[hide]%
\>[2]\AgdaFunction{recompose}\AgdaSpace{}%
\AgdaBound{t}\AgdaSpace{}%
\AgdaInductiveConstructor{[]}%
\>[25]\AgdaSymbol{=}\AgdaSpace{}%
\AgdaBound{t}%
\>[30]\AgdaComment{--\ As\ earlier}\<%
\\
\>[2]\AgdaFunction{recompose}\AgdaSpace{}%
\AgdaBound{t}\AgdaSpace{}%
\AgdaSymbol{(}\AgdaBound{frm}\AgdaSpace{}%
\AgdaOperator{\AgdaInductiveConstructor{∷}}\AgdaSpace{}%
\AgdaBound{c}\AgdaSymbol{)}%
\>[25]\AgdaSymbol{=}%
\>[30]\AgdaComment{--\ As\ earlier}\<%
\\
\>[2][@{}l@{\AgdaIndent{0}}]%
\>[4]\AgdaFunction{recompose}\AgdaSpace{}%
\AgdaSymbol{(}\AgdaFunction{recompose-frame}\AgdaSpace{}%
\AgdaBound{t}\AgdaSpace{}%
\AgdaBound{frm}\AgdaSymbol{)}\AgdaSpace{}%
\AgdaBound{c}\<%
\\
\\[\AgdaEmptyExtraSkip]%
\>[2]\AgdaFunction{decompose}\AgdaSpace{}%
\AgdaSymbol{(}\AgdaInductiveConstructor{lam}\AgdaSpace{}%
\AgdaBound{t}\AgdaSymbol{)}\AgdaSpace{}%
\AgdaBound{c}\AgdaSpace{}%
\AgdaSymbol{=}\AgdaSpace{}%
\AgdaInductiveConstructor{nothing}\<%
\\
\>[2]\AgdaFunction{decompose}\AgdaSpace{}%
\AgdaSymbol{(}\AgdaInductiveConstructor{var}\AgdaSpace{}%
\AgdaBound{x}\AgdaSymbol{)}\AgdaSpace{}%
\AgdaBound{c}\AgdaSpace{}%
\AgdaSymbol{=}\AgdaSpace{}%
\AgdaInductiveConstructor{nothing}\<%
\\
\>[2]\AgdaFunction{decompose}\AgdaSpace{}%
\AgdaBound{t}\AgdaSymbol{@(}\AgdaInductiveConstructor{app}\AgdaSpace{}%
\AgdaBound{t₀}\AgdaSpace{}%
\AgdaBound{t₁}\AgdaSymbol{)}\AgdaSpace{}%
\AgdaBound{c}\AgdaSpace{}%
\AgdaSymbol{=}\AgdaSpace{}%
\AgdaKeyword{let}%
\>[1470I]\AgdaComment{--\ TODO:\ this\ reads\ a\ little\ differently\ from}\<%
\\
\>[.][@{}l@{}]\<[1470I]%
\>[34]\AgdaComment{--\ earlier.\ \ It\ is\ easy\ to\ reformulate\ to\ make}\<%
\\
\>[34]\AgdaComment{--\ it\ read\ the\ same.\ \ This\ is\ left\ as\ an}\<%
\\
\>[34]\AgdaComment{--\ exercise\ to\ the\ reader.}\<%
\\
\>[2][@{}l@{\AgdaIndent{0}}]%
\>[8]\AgdaBound{subs}%
\>[1471I]\AgdaSymbol{=}\AgdaSpace{}%
\AgdaFunction{decompose}\AgdaSpace{}%
\AgdaBound{t₀}\AgdaSpace{}%
\AgdaSymbol{(}\AgdaInductiveConstructor{app-l}\AgdaSpace{}%
\AgdaBound{t₁}\AgdaSpace{}%
\AgdaOperator{\AgdaInductiveConstructor{∷}}\AgdaSpace{}%
\AgdaBound{c}\AgdaSymbol{)}\<%
\\
\>[.][@{}l@{}]\<[1471I]%
\>[13]\AgdaOperator{\AgdaInductiveConstructor{∷}}\AgdaSpace{}%
\AgdaSymbol{(}\AgdaFunction{maybe}%
\>[23]\AgdaSymbol{(λ}\AgdaSpace{}%
\AgdaBound{v}\AgdaSpace{}%
\AgdaSymbol{→}\AgdaSpace{}%
\AgdaFunction{decompose}\AgdaSpace{}%
\AgdaBound{t₁}\AgdaSpace{}%
\AgdaSymbol{(}\AgdaInductiveConstructor{app-r}\AgdaSpace{}%
\AgdaBound{v}\AgdaSpace{}%
\AgdaOperator{\AgdaInductiveConstructor{∷}}\AgdaSpace{}%
\AgdaBound{c}\AgdaSymbol{))}\<%
\\
\>[23]\AgdaInductiveConstructor{nothing}\<%
\\
\>[23]\AgdaSymbol{(}\AgdaFunction{t2v}\AgdaSpace{}%
\AgdaBound{t₀}\AgdaSymbol{))}\<%
\\
\>[13]\AgdaOperator{\AgdaInductiveConstructor{∷}}\AgdaSpace{}%
\AgdaInductiveConstructor{[]}\<%
\\
\>[2][@{}l@{\AgdaIndent{0}}]%
\>[4]\AgdaKeyword{in}%
\>[8]\AgdaFunction{foldr}%
\>[15]\AgdaSymbol{(λ}\AgdaSpace{}%
\AgdaBound{hd}\AgdaSpace{}%
\AgdaBound{tl}\AgdaSpace{}%
\AgdaSymbol{→}\AgdaSpace{}%
\AgdaFunction{maybe}\AgdaSpace{}%
\AgdaInductiveConstructor{just}\AgdaSpace{}%
\AgdaBound{tl}\AgdaSpace{}%
\AgdaBound{hd}\AgdaSymbol{)}\<%
\\
\>[15]\AgdaSymbol{(}\AgdaOperator{\AgdaFunction{if}}\AgdaSpace{}%
\AgdaSymbol{(}\AgdaFunction{is-redex}\AgdaSpace{}%
\AgdaBound{t}\AgdaSymbol{)}\<%
\\
\>[15][@{}l@{\AgdaIndent{0}}]%
\>[16]\AgdaOperator{\AgdaFunction{then}}\AgdaSpace{}%
\AgdaInductiveConstructor{just}\AgdaSpace{}%
\AgdaSymbol{(}\AgdaBound{t}\AgdaSpace{}%
\AgdaOperator{\AgdaInductiveConstructor{,}}\AgdaSpace{}%
\AgdaBound{c}\AgdaSymbol{)}\<%
\\
\>[16]\AgdaOperator{\AgdaFunction{else}}\AgdaSpace{}%
\AgdaInductiveConstructor{nothing}\AgdaSymbol{)}\<%
\\
\>[15]\AgdaBound{subs}\<%
\\
\>[2]\AgdaFunction{decompose}\AgdaSpace{}%
\AgdaBound{t}\AgdaSymbol{@(}\AgdaInductiveConstructor{reset}\AgdaSpace{}%
\AgdaBound{t₀}\AgdaSymbol{)}\AgdaSpace{}%
\AgdaBound{c}\AgdaSpace{}%
\AgdaSymbol{=}\AgdaSpace{}%
\AgdaFunction{maybe}\<%
\\
\>[2][@{}l@{\AgdaIndent{0}}]%
\>[4]\AgdaInductiveConstructor{just}\<%
\\
\>[4]\AgdaSymbol{(}\AgdaOperator{\AgdaFunction{if}}\AgdaSpace{}%
\AgdaSymbol{(}\AgdaFunction{is-redex}\AgdaSpace{}%
\AgdaBound{t}\AgdaSymbol{)}\AgdaSpace{}%
\AgdaOperator{\AgdaFunction{then}}\AgdaSpace{}%
\AgdaInductiveConstructor{just}\AgdaSpace{}%
\AgdaSymbol{(}\AgdaBound{t}\AgdaSpace{}%
\AgdaOperator{\AgdaInductiveConstructor{,}}\AgdaSpace{}%
\AgdaBound{c}\AgdaSymbol{)}\AgdaSpace{}%
\AgdaOperator{\AgdaFunction{else}}\AgdaSpace{}%
\AgdaInductiveConstructor{nothing}\AgdaSymbol{)}\<%
\\
\>[4]\AgdaSymbol{(}\AgdaFunction{decompose}\AgdaSpace{}%
\AgdaBound{t₀}\AgdaSpace{}%
\AgdaSymbol{(}\AgdaInductiveConstructor{reset}\AgdaSpace{}%
\AgdaOperator{\AgdaInductiveConstructor{∷}}\AgdaSpace{}%
\AgdaBound{c}\AgdaSymbol{))}\<%
\\
\>[2]\AgdaFunction{decompose}\AgdaSpace{}%
\AgdaBound{t}\AgdaSymbol{@(}\AgdaInductiveConstructor{shift}\AgdaSpace{}%
\AgdaBound{t₀}\AgdaSymbol{)}\AgdaSpace{}%
\AgdaBound{c}\AgdaSpace{}%
\AgdaSymbol{=}\AgdaSpace{}%
\AgdaInductiveConstructor{just}\AgdaSpace{}%
\AgdaSymbol{(}\AgdaBound{t}\AgdaSpace{}%
\AgdaOperator{\AgdaInductiveConstructor{,}}\AgdaSpace{}%
\AgdaBound{c}\AgdaSymbol{)}\<%
\\
\>[2]\AgdaFunction{decompose}\AgdaSpace{}%
\AgdaBound{t}\AgdaSymbol{@(}\AgdaInductiveConstructor{add}\AgdaSpace{}%
\AgdaBound{t₀}\AgdaSpace{}%
\AgdaBound{t₁}\AgdaSymbol{)}\AgdaSpace{}%
\AgdaBound{c}\AgdaSpace{}%
\AgdaSymbol{=}\AgdaSpace{}%
\AgdaKeyword{let}\<%
\\
\>[2][@{}l@{\AgdaIndent{0}}]%
\>[8]\AgdaBound{subs}%
\>[1535I]\AgdaSymbol{=}\AgdaSpace{}%
\AgdaFunction{decompose}\AgdaSpace{}%
\AgdaBound{t₀}\AgdaSpace{}%
\AgdaSymbol{(}\AgdaInductiveConstructor{add-l}\AgdaSpace{}%
\AgdaBound{t₁}\AgdaSpace{}%
\AgdaOperator{\AgdaInductiveConstructor{∷}}\AgdaSpace{}%
\AgdaBound{c}\AgdaSymbol{)}\<%
\\
\>[.][@{}l@{}]\<[1535I]%
\>[13]\AgdaOperator{\AgdaInductiveConstructor{∷}}\AgdaSpace{}%
\AgdaSymbol{(}\AgdaFunction{maybe}%
\>[23]\AgdaSymbol{(λ}\AgdaSpace{}%
\AgdaBound{v}\AgdaSpace{}%
\AgdaSymbol{→}\AgdaSpace{}%
\AgdaFunction{decompose}\AgdaSpace{}%
\AgdaBound{t₁}\AgdaSpace{}%
\AgdaSymbol{(}\AgdaInductiveConstructor{add-r}\AgdaSpace{}%
\AgdaBound{v}\AgdaSpace{}%
\AgdaOperator{\AgdaInductiveConstructor{∷}}\AgdaSpace{}%
\AgdaBound{c}\AgdaSymbol{))}\<%
\\
\>[23]\AgdaInductiveConstructor{nothing}\<%
\\
\>[23]\AgdaSymbol{(}\AgdaFunction{t2v}\AgdaSpace{}%
\AgdaBound{t₀}\AgdaSymbol{))}\<%
\\
\>[13]\AgdaOperator{\AgdaInductiveConstructor{∷}}\AgdaSpace{}%
\AgdaInductiveConstructor{[]}\<%
\\
\>[2][@{}l@{\AgdaIndent{0}}]%
\>[4]\AgdaKeyword{in}%
\>[8]\AgdaFunction{foldr}%
\>[15]\AgdaSymbol{(λ}\AgdaSpace{}%
\AgdaBound{hd}\AgdaSpace{}%
\AgdaBound{tl}\AgdaSpace{}%
\AgdaSymbol{→}\AgdaSpace{}%
\AgdaFunction{maybe}\AgdaSpace{}%
\AgdaInductiveConstructor{just}\AgdaSpace{}%
\AgdaBound{tl}\AgdaSpace{}%
\AgdaBound{hd}\AgdaSymbol{)}\<%
\\
\>[15]\AgdaSymbol{(}\AgdaOperator{\AgdaFunction{if}}\AgdaSpace{}%
\AgdaSymbol{(}\AgdaFunction{is-redex}\AgdaSpace{}%
\AgdaBound{t}\AgdaSymbol{)}\<%
\\
\>[15][@{}l@{\AgdaIndent{0}}]%
\>[16]\AgdaOperator{\AgdaFunction{then}}\AgdaSpace{}%
\AgdaInductiveConstructor{just}\AgdaSpace{}%
\AgdaSymbol{(}\AgdaBound{t}\AgdaSpace{}%
\AgdaOperator{\AgdaInductiveConstructor{,}}\AgdaSpace{}%
\AgdaBound{c}\AgdaSymbol{)}\<%
\\
\>[16]\AgdaOperator{\AgdaFunction{else}}\AgdaSpace{}%
\AgdaInductiveConstructor{nothing}\AgdaSymbol{)}\<%
\\
\>[15]\AgdaBound{subs}\<%
\\
\>[2]\AgdaFunction{decompose}\AgdaSpace{}%
\AgdaSymbol{(}\AgdaInductiveConstructor{num}\AgdaSpace{}%
\AgdaBound{x}\AgdaSymbol{)}\AgdaSpace{}%
\AgdaBound{c}\AgdaSpace{}%
\AgdaSymbol{=}\AgdaSpace{}%
\AgdaInductiveConstructor{nothing}\<%
\\
\\[\AgdaEmptyExtraSkip]%
\>[2]\AgdaFunction{recompose-frame}\AgdaSpace{}%
\AgdaBound{t₀}\AgdaSpace{}%
\AgdaSymbol{(}\AgdaInductiveConstructor{add-l}\AgdaSpace{}%
\AgdaBound{t₁}\AgdaSymbol{)}\AgdaSpace{}%
\AgdaSymbol{=}\AgdaSpace{}%
\AgdaInductiveConstructor{add}\AgdaSpace{}%
\AgdaBound{t₀}\AgdaSpace{}%
\AgdaBound{t₁}\<%
\\
\>[2]\AgdaFunction{recompose-frame}\AgdaSpace{}%
\AgdaBound{t₁}\AgdaSpace{}%
\AgdaSymbol{(}\AgdaInductiveConstructor{add-r}\AgdaSpace{}%
\AgdaBound{v}\AgdaSymbol{)}%
\>[32]\AgdaSymbol{=}\AgdaSpace{}%
\AgdaInductiveConstructor{add}\AgdaSpace{}%
\AgdaSymbol{(}\AgdaFunction{v2t}\AgdaSpace{}%
\AgdaBound{v}\AgdaSymbol{)}\AgdaSpace{}%
\AgdaBound{t₁}\<%
\\
\>[2]\AgdaFunction{recompose-frame}\AgdaSpace{}%
\AgdaBound{t₀}\AgdaSpace{}%
\AgdaSymbol{(}\AgdaInductiveConstructor{app-l}\AgdaSpace{}%
\AgdaBound{t₁}\AgdaSymbol{)}\AgdaSpace{}%
\AgdaSymbol{=}\AgdaSpace{}%
\AgdaInductiveConstructor{add}\AgdaSpace{}%
\AgdaBound{t₀}\AgdaSpace{}%
\AgdaBound{t₁}\<%
\\
\>[2]\AgdaFunction{recompose-frame}\AgdaSpace{}%
\AgdaBound{t₁}\AgdaSpace{}%
\AgdaSymbol{(}\AgdaInductiveConstructor{app-r}\AgdaSpace{}%
\AgdaBound{v}\AgdaSymbol{)}%
\>[32]\AgdaSymbol{=}\AgdaSpace{}%
\AgdaInductiveConstructor{app}\AgdaSpace{}%
\AgdaSymbol{(}\AgdaFunction{v2t}\AgdaSpace{}%
\AgdaBound{v}\AgdaSymbol{)}\AgdaSpace{}%
\AgdaBound{t₁}\<%
\\
\>[2]\AgdaFunction{recompose-frame}\AgdaSpace{}%
\AgdaBound{t₀}\AgdaSpace{}%
\AgdaInductiveConstructor{reset}%
\>[32]\AgdaSymbol{=}\AgdaSpace{}%
\AgdaInductiveConstructor{reset}\AgdaSpace{}%
\AgdaBound{t₀}\<%
\end{code}
We leave it as an exercise to prove that the notions of decomposition and recomposition satisfy the same properties as discussed in \cref{sec:reduction-strategy-arith}.

\subsubsection{One-Step Reduction and Reduction-Based Normalization}

Our one-step reduction function and driver loops are both the same as earlier.
\begin{code}[hide]%
\>[2]\AgdaKeyword{record}\AgdaSpace{}%
\AgdaRecord{Delay}\AgdaSpace{}%
\AgdaSymbol{(}\AgdaBound{A}\AgdaSpace{}%
\AgdaSymbol{:}\AgdaSpace{}%
\AgdaPrimitive{Set}\AgdaSymbol{)}\AgdaSpace{}%
\AgdaSymbol{:}\AgdaSpace{}%
\AgdaPrimitive{Set}\AgdaSpace{}%
\AgdaKeyword{where}\<%
\\
\>[2][@{}l@{\AgdaIndent{0}}]%
\>[4]\AgdaKeyword{constructor}\AgdaSpace{}%
\AgdaOperator{\AgdaCoinductiveConstructor{D⟨\AgdaUnderscore{}⟩}}\<%
\\
\>[4]\AgdaKeyword{coinductive}\<%
\\
\>[4]\AgdaKeyword{field}\AgdaSpace{}%
\AgdaField{insp}\AgdaSpace{}%
\AgdaSymbol{:}\AgdaSpace{}%
\AgdaBound{A}\AgdaSpace{}%
\AgdaOperator{\AgdaDatatype{⊎}}\AgdaSpace{}%
\AgdaRecord{Delay}\AgdaSpace{}%
\AgdaBound{A}\<%
\\
\\[\AgdaEmptyExtraSkip]%
\>[2]\AgdaKeyword{open}\AgdaSpace{}%
\AgdaModule{Delay}\<%
\end{code}%
\begin{code}%
\>[2]\AgdaFunction{reduce}\AgdaSpace{}%
\AgdaSymbol{:}\AgdaSpace{}%
\AgdaDatatype{Term}\AgdaSpace{}%
\AgdaSymbol{→}\AgdaSpace{}%
\AgdaDatatype{Maybe}\AgdaSpace{}%
\AgdaDatatype{Term}\<%
\\
\>[2]\AgdaFunction{drive}\AgdaSpace{}%
\AgdaSymbol{:}\AgdaSpace{}%
\AgdaDatatype{Term}\AgdaSpace{}%
\AgdaSymbol{→}\AgdaSpace{}%
\AgdaRecord{Delay}\AgdaSpace{}%
\AgdaSymbol{(}\AgdaDatatype{Maybe}\AgdaSpace{}%
\AgdaDatatype{Val}\AgdaSymbol{)}\<%
\end{code}

\begin{code}[hide]%
\>[2]\AgdaFunction{reduce}\AgdaSpace{}%
\AgdaBound{t}\AgdaSpace{}%
\AgdaSymbol{=}\AgdaSpace{}%
\AgdaKeyword{do}\<%
\\
\>[2][@{}l@{\AgdaIndent{0}}]%
\>[4]\AgdaSymbol{(}\AgdaBound{t₀}\AgdaSpace{}%
\AgdaOperator{\AgdaInductiveConstructor{,}}\AgdaSpace{}%
\AgdaBound{c}\AgdaSymbol{)}\AgdaSpace{}%
\AgdaOperator{\AgdaFunction{←}}\AgdaSpace{}%
\AgdaFunction{decompose}\AgdaSpace{}%
\AgdaBound{t}\AgdaSpace{}%
\AgdaInductiveConstructor{[]}\<%
\\
\>[4]\AgdaSymbol{(}\AgdaBound{t₁}\AgdaSpace{}%
\AgdaOperator{\AgdaInductiveConstructor{,}}\AgdaSpace{}%
\AgdaBound{c₁}\AgdaSymbol{)}\AgdaSpace{}%
\AgdaOperator{\AgdaFunction{←}}\AgdaSpace{}%
\AgdaFunction{contract}\AgdaSpace{}%
\AgdaBound{t₀}\AgdaSpace{}%
\AgdaBound{c}\<%
\\
\>[4]\AgdaInductiveConstructor{just}\AgdaSpace{}%
\AgdaSymbol{(}\AgdaFunction{recompose}\AgdaSpace{}%
\AgdaBound{t₁}\AgdaSpace{}%
\AgdaBound{c₁}\AgdaSymbol{)}\<%
\\
\>[0]\<%
\\
\>[2]\AgdaField{insp}\AgdaSpace{}%
\AgdaSymbol{(}\AgdaFunction{drive}\AgdaSpace{}%
\AgdaBound{t}\AgdaSymbol{)}\AgdaSpace{}%
\AgdaKeyword{with}\AgdaSpace{}%
\AgdaFunction{t2v}\AgdaSpace{}%
\AgdaBound{t}\<%
\\
\>[2]\AgdaField{insp}\AgdaSpace{}%
\AgdaSymbol{(}\AgdaFunction{drive}\AgdaSpace{}%
\AgdaBound{t}\AgdaSymbol{)}\AgdaSpace{}%
\AgdaSymbol{|}\AgdaSpace{}%
\AgdaInductiveConstructor{just}\AgdaSpace{}%
\AgdaBound{x}\AgdaSpace{}%
\AgdaSymbol{=}\AgdaSpace{}%
\AgdaInductiveConstructor{inj₁}\AgdaSpace{}%
\AgdaSymbol{(}\AgdaInductiveConstructor{just}\AgdaSpace{}%
\AgdaBound{x}\AgdaSymbol{)}\<%
\\
\>[2]\AgdaField{insp}\AgdaSpace{}%
\AgdaSymbol{(}\AgdaFunction{drive}\AgdaSpace{}%
\AgdaBound{t}\AgdaSymbol{)}\AgdaSpace{}%
\AgdaSymbol{|}\AgdaSpace{}%
\AgdaInductiveConstructor{nothing}\AgdaSpace{}%
\AgdaKeyword{with}\AgdaSpace{}%
\AgdaFunction{reduce}\AgdaSpace{}%
\AgdaBound{t}\<%
\\
\>[2]\AgdaField{insp}\AgdaSpace{}%
\AgdaSymbol{(}\AgdaFunction{drive}\AgdaSpace{}%
\AgdaBound{t}\AgdaSymbol{)}\AgdaSpace{}%
\AgdaSymbol{|}\AgdaSpace{}%
\AgdaInductiveConstructor{nothing}\AgdaSpace{}%
\AgdaSymbol{|}\AgdaSpace{}%
\AgdaInductiveConstructor{just}\AgdaSpace{}%
\AgdaBound{x}%
\>[38]\AgdaSymbol{=}\AgdaSpace{}%
\AgdaInductiveConstructor{inj₂}\AgdaSpace{}%
\AgdaSymbol{(}\AgdaFunction{drive}\AgdaSpace{}%
\AgdaBound{x}\AgdaSymbol{)}\<%
\\
\>[2]\AgdaField{insp}\AgdaSpace{}%
\AgdaSymbol{(}\AgdaFunction{drive}\AgdaSpace{}%
\AgdaBound{t}\AgdaSymbol{)}\AgdaSpace{}%
\AgdaSymbol{|}\AgdaSpace{}%
\AgdaInductiveConstructor{nothing}\AgdaSpace{}%
\AgdaSymbol{|}\AgdaSpace{}%
\AgdaInductiveConstructor{nothing}%
\>[38]\AgdaSymbol{=}\AgdaSpace{}%
\AgdaInductiveConstructor{inj₁}\AgdaSpace{}%
\AgdaInductiveConstructor{nothing}\<%
\\
\>[0]\<%
\\
\>[2]\AgdaKeyword{mutual}\<%
\\
\>[2][@{}l@{\AgdaIndent{0}}]%
\>[4]\AgdaKeyword{data}\AgdaSpace{}%
\AgdaOperator{\AgdaDatatype{\AgdaUnderscore{}≈₀\AgdaUnderscore{}}}\AgdaSpace{}%
\AgdaSymbol{\{}\AgdaBound{A}\AgdaSymbol{\}}\AgdaSpace{}%
\AgdaSymbol{:}\AgdaSpace{}%
\AgdaRecord{Delay}\AgdaSpace{}%
\AgdaBound{A}\AgdaSpace{}%
\AgdaSymbol{→}\AgdaSpace{}%
\AgdaRecord{Delay}\AgdaSpace{}%
\AgdaBound{A}\AgdaSpace{}%
\AgdaSymbol{→}\AgdaSpace{}%
\AgdaPrimitive{Set}\AgdaSpace{}%
\AgdaKeyword{where}\<%
\\
\>[4][@{}l@{\AgdaIndent{0}}]%
\>[6]\AgdaInductiveConstructor{val}%
\>[1698I]\AgdaSymbol{:}\AgdaSpace{}%
\AgdaSymbol{\{}\AgdaBound{d₀}\AgdaSpace{}%
\AgdaBound{d₁}\AgdaSpace{}%
\AgdaSymbol{:}\AgdaSpace{}%
\AgdaRecord{Delay}\AgdaSpace{}%
\AgdaBound{A}\AgdaSymbol{\}}\AgdaSpace{}%
\AgdaSymbol{\{}\AgdaBound{x}\AgdaSpace{}%
\AgdaSymbol{:}\AgdaSpace{}%
\AgdaBound{A}\AgdaSymbol{\}}\<%
\\
\>[.][@{}l@{}]\<[1698I]%
\>[10]\AgdaSymbol{→}\AgdaSpace{}%
\AgdaField{insp}\AgdaSpace{}%
\AgdaBound{d₀}\AgdaSpace{}%
\AgdaOperator{\AgdaDatatype{≡}}\AgdaSpace{}%
\AgdaInductiveConstructor{inj₁}\AgdaSpace{}%
\AgdaBound{x}\AgdaSpace{}%
\AgdaSymbol{→}\AgdaSpace{}%
\AgdaField{insp}\AgdaSpace{}%
\AgdaBound{d₁}\AgdaSpace{}%
\AgdaOperator{\AgdaDatatype{≡}}\AgdaSpace{}%
\AgdaInductiveConstructor{inj₁}\AgdaSpace{}%
\AgdaBound{x}\<%
\\
\>[10]\AgdaSymbol{→}\AgdaSpace{}%
\AgdaBound{d₀}\AgdaSpace{}%
\AgdaOperator{\AgdaDatatype{≈₀}}\AgdaSpace{}%
\AgdaBound{d₁}\<%
\\
\>[6]\AgdaInductiveConstructor{tau}%
\>[1721I]\AgdaSymbol{:}\AgdaSpace{}%
\AgdaSymbol{\{}\AgdaBound{d₀}\AgdaSpace{}%
\AgdaBound{d₀₁}\AgdaSpace{}%
\AgdaBound{d₁}\AgdaSpace{}%
\AgdaBound{d₁₁}\AgdaSpace{}%
\AgdaSymbol{:}\AgdaSpace{}%
\AgdaRecord{Delay}\AgdaSpace{}%
\AgdaBound{A}\AgdaSymbol{\}}\<%
\\
\>[.][@{}l@{}]\<[1721I]%
\>[10]\AgdaSymbol{→}\AgdaSpace{}%
\AgdaField{insp}\AgdaSpace{}%
\AgdaBound{d₀}\AgdaSpace{}%
\AgdaOperator{\AgdaDatatype{≡}}\AgdaSpace{}%
\AgdaInductiveConstructor{inj₂}\AgdaSpace{}%
\AgdaBound{d₀₁}\AgdaSpace{}%
\AgdaSymbol{→}\AgdaSpace{}%
\AgdaField{insp}\AgdaSpace{}%
\AgdaBound{d₁}\AgdaSpace{}%
\AgdaOperator{\AgdaDatatype{≡}}\AgdaSpace{}%
\AgdaInductiveConstructor{inj₂}\AgdaSpace{}%
\AgdaBound{d₁₁}\<%
\\
\>[10]\AgdaSymbol{→}\AgdaSpace{}%
\AgdaBound{d₀₁}\AgdaSpace{}%
\AgdaOperator{\AgdaRecord{≈}}\AgdaSpace{}%
\AgdaBound{d₁₁}\AgdaSpace{}%
\AgdaSymbol{→}\AgdaSpace{}%
\AgdaBound{d₀}\AgdaSpace{}%
\AgdaOperator{\AgdaDatatype{≈₀}}\AgdaSpace{}%
\AgdaBound{d₁}\<%
\\
\\[\AgdaEmptyExtraSkip]%
\>[4]\AgdaKeyword{record}\AgdaSpace{}%
\AgdaOperator{\AgdaRecord{\AgdaUnderscore{}≈\AgdaUnderscore{}}}\AgdaSpace{}%
\AgdaSymbol{\{}\AgdaBound{A}\AgdaSymbol{\}}\AgdaSpace{}%
\AgdaSymbol{(}\AgdaBound{d₀}\AgdaSpace{}%
\AgdaBound{d₁}\AgdaSpace{}%
\AgdaSymbol{:}\AgdaSpace{}%
\AgdaRecord{Delay}\AgdaSpace{}%
\AgdaBound{A}\AgdaSymbol{)}\AgdaSpace{}%
\AgdaSymbol{:}\AgdaSpace{}%
\AgdaPrimitive{Set}\AgdaSpace{}%
\AgdaKeyword{where}\<%
\\
\>[4][@{}l@{\AgdaIndent{0}}]%
\>[6]\AgdaKeyword{coinductive}\<%
\\
\>[6]\AgdaKeyword{constructor}\AgdaSpace{}%
\AgdaOperator{\AgdaCoinductiveConstructor{≈⟨\AgdaUnderscore{}⟩}}\<%
\\
\>[6]\AgdaKeyword{field}\AgdaSpace{}%
\AgdaField{bisim}\AgdaSpace{}%
\AgdaSymbol{:}\AgdaSpace{}%
\AgdaBound{d₀}\AgdaSpace{}%
\AgdaOperator{\AgdaDatatype{≈₀}}\AgdaSpace{}%
\AgdaBound{d₁}\<%
\\
\\[\AgdaEmptyExtraSkip]%
\>[2]\AgdaKeyword{open}\AgdaSpace{}%
\AgdaOperator{\AgdaModule{\AgdaUnderscore{}≈\AgdaUnderscore{}}}\<%
\\
\\[\AgdaEmptyExtraSkip]%
\>[2]\AgdaFunction{refl-≈}%
\>[1764I]\AgdaSymbol{:}\AgdaSpace{}%
\AgdaSymbol{∀}\AgdaSpace{}%
\AgdaSymbol{\{}\AgdaBound{A}\AgdaSymbol{\}}\AgdaSpace{}%
\AgdaSymbol{(}\AgdaBound{d}\AgdaSpace{}%
\AgdaSymbol{:}\AgdaSpace{}%
\AgdaRecord{Delay}\AgdaSpace{}%
\AgdaBound{A}\AgdaSymbol{)}\<%
\\
\>[.][@{}l@{}]\<[1764I]%
\>[9]\AgdaSymbol{→}\AgdaSpace{}%
\AgdaBound{d}\AgdaSpace{}%
\AgdaOperator{\AgdaRecord{≈}}\AgdaSpace{}%
\AgdaBound{d}\<%
\\
\>[2]\AgdaField{bisim}\AgdaSpace{}%
\AgdaSymbol{(}\AgdaFunction{refl-≈}\AgdaSpace{}%
\AgdaBound{d}\AgdaSymbol{)}\AgdaSpace{}%
\AgdaKeyword{with}\AgdaSpace{}%
\AgdaField{insp}\AgdaSpace{}%
\AgdaBound{d}\AgdaSpace{}%
\AgdaSymbol{|}\AgdaSpace{}%
\AgdaFunction{inspect}\AgdaSpace{}%
\AgdaField{insp}\AgdaSpace{}%
\AgdaBound{d}\<%
\\
\>[2]\AgdaSymbol{...}\AgdaSpace{}%
\AgdaSymbol{|}\AgdaSpace{}%
\AgdaInductiveConstructor{inj₁}\AgdaSpace{}%
\AgdaBound{x}\AgdaSpace{}%
\AgdaSymbol{|}\AgdaSpace{}%
\AgdaOperator{\AgdaInductiveConstructor{[}}\AgdaSpace{}%
\AgdaBound{eq}\AgdaSpace{}%
\AgdaOperator{\AgdaInductiveConstructor{]}}\AgdaSpace{}%
\AgdaSymbol{=}\AgdaSpace{}%
\AgdaInductiveConstructor{val}\AgdaSpace{}%
\AgdaBound{eq}\AgdaSpace{}%
\AgdaBound{eq}\<%
\\
\>[2]\AgdaSymbol{...}\AgdaSpace{}%
\AgdaSymbol{|}\AgdaSpace{}%
\AgdaInductiveConstructor{inj₂}\AgdaSpace{}%
\AgdaBound{y}\AgdaSpace{}%
\AgdaSymbol{|}\AgdaSpace{}%
\AgdaOperator{\AgdaInductiveConstructor{[}}\AgdaSpace{}%
\AgdaBound{eq}\AgdaSpace{}%
\AgdaOperator{\AgdaInductiveConstructor{]}}\AgdaSpace{}%
\AgdaSymbol{=}\AgdaSpace{}%
\AgdaInductiveConstructor{tau}\AgdaSpace{}%
\AgdaBound{eq}\AgdaSpace{}%
\AgdaBound{eq}\AgdaSpace{}%
\AgdaSymbol{(}\AgdaFunction{refl-≈}\AgdaSpace{}%
\AgdaBound{y}\AgdaSymbol{)}\<%
\\
\\[\AgdaEmptyExtraSkip]%
\>[2]\AgdaFunction{trans-≈}\<%
\\
\>[2][@{}l@{\AgdaIndent{0}}]%
\>[4]\AgdaSymbol{:}\AgdaSpace{}%
\AgdaSymbol{∀}\AgdaSpace{}%
\AgdaSymbol{\{}\AgdaBound{A}\AgdaSymbol{\}}\AgdaSpace{}%
\AgdaSymbol{\{}\AgdaBound{d₀}\AgdaSpace{}%
\AgdaBound{d₁}\AgdaSpace{}%
\AgdaBound{d₂}\AgdaSpace{}%
\AgdaSymbol{:}\AgdaSpace{}%
\AgdaRecord{Delay}\AgdaSpace{}%
\AgdaBound{A}\AgdaSymbol{\}}\<%
\\
\>[4]\AgdaSymbol{→}\AgdaSpace{}%
\AgdaBound{d₀}\AgdaSpace{}%
\AgdaOperator{\AgdaRecord{≈}}\AgdaSpace{}%
\AgdaBound{d₁}\<%
\\
\>[4]\AgdaSymbol{→}\AgdaSpace{}%
\AgdaBound{d₁}\AgdaSpace{}%
\AgdaOperator{\AgdaRecord{≈}}\AgdaSpace{}%
\AgdaBound{d₂}\<%
\\
\>[4]\AgdaSymbol{→}\AgdaSpace{}%
\AgdaBound{d₀}\AgdaSpace{}%
\AgdaOperator{\AgdaRecord{≈}}\AgdaSpace{}%
\AgdaBound{d₂}\<%
\\
\>[2]\AgdaField{bisim}\AgdaSpace{}%
\AgdaSymbol{(}\AgdaFunction{trans-≈}\AgdaSpace{}%
\AgdaBound{eq₀}\AgdaSpace{}%
\AgdaBound{eq₁}\AgdaSymbol{)}\AgdaSpace{}%
\AgdaKeyword{with}\AgdaSpace{}%
\AgdaField{bisim}\AgdaSpace{}%
\AgdaBound{eq₀}\AgdaSpace{}%
\AgdaSymbol{|}\AgdaSpace{}%
\AgdaField{bisim}\AgdaSpace{}%
\AgdaBound{eq₁}\<%
\\
\>[2]\AgdaField{bisim}\AgdaSpace{}%
\AgdaSymbol{(}\AgdaFunction{trans-≈}\AgdaSpace{}%
\AgdaBound{eq₀}\AgdaSpace{}%
\AgdaBound{eq₁}\AgdaSymbol{)}\AgdaSpace{}%
\AgdaSymbol{|}\AgdaSpace{}%
\AgdaInductiveConstructor{val}\AgdaSpace{}%
\AgdaBound{q₀}\AgdaSpace{}%
\AgdaBound{q₁}\AgdaSpace{}%
\AgdaSymbol{|}\AgdaSpace{}%
\AgdaInductiveConstructor{val}\AgdaSpace{}%
\AgdaBound{q₂}\AgdaSpace{}%
\AgdaBound{q₃}\AgdaSpace{}%
\AgdaKeyword{with}\AgdaSpace{}%
\AgdaFunction{trans}\AgdaSpace{}%
\AgdaSymbol{(}\AgdaFunction{sym}\AgdaSpace{}%
\AgdaBound{q₁}\AgdaSymbol{)}\AgdaSpace{}%
\AgdaBound{q₂}\<%
\\
\>[2]\AgdaSymbol{...}\AgdaSpace{}%
\AgdaSymbol{|}\AgdaSpace{}%
\AgdaInductiveConstructor{refl}\AgdaSpace{}%
\AgdaSymbol{=}\AgdaSpace{}%
\AgdaInductiveConstructor{val}\AgdaSpace{}%
\AgdaBound{q₀}\AgdaSpace{}%
\AgdaBound{q₃}\<%
\\
\>[2]\AgdaField{bisim}\AgdaSpace{}%
\AgdaSymbol{(}\AgdaFunction{trans-≈}\AgdaSpace{}%
\AgdaBound{eq₀}\AgdaSpace{}%
\AgdaBound{eq₁}\AgdaSymbol{)}\AgdaSpace{}%
\AgdaSymbol{|}\AgdaSpace{}%
\AgdaInductiveConstructor{val}\AgdaSpace{}%
\AgdaBound{q₀}\AgdaSpace{}%
\AgdaBound{q₁}\AgdaSpace{}%
\AgdaSymbol{|}\AgdaSpace{}%
\AgdaInductiveConstructor{tau}\AgdaSpace{}%
\AgdaBound{q₂}\AgdaSpace{}%
\AgdaBound{q₃}\AgdaSpace{}%
\AgdaBound{eq₄}\AgdaSpace{}%
\AgdaKeyword{with}\AgdaSpace{}%
\AgdaFunction{trans}\AgdaSpace{}%
\AgdaSymbol{(}\AgdaFunction{sym}\AgdaSpace{}%
\AgdaBound{q₁}\AgdaSymbol{)}\AgdaSpace{}%
\AgdaBound{q₂}\<%
\\
\>[2]\AgdaSymbol{...}\AgdaSpace{}%
\AgdaSymbol{|}\AgdaSpace{}%
\AgdaSymbol{()}\<%
\\
\>[2]\AgdaField{bisim}\AgdaSpace{}%
\AgdaSymbol{(}\AgdaFunction{trans-≈}\AgdaSpace{}%
\AgdaBound{eq₀}\AgdaSpace{}%
\AgdaBound{eq₁}\AgdaSymbol{)}\AgdaSpace{}%
\AgdaSymbol{|}\AgdaSpace{}%
\AgdaInductiveConstructor{tau}\AgdaSpace{}%
\AgdaBound{q₀}\AgdaSpace{}%
\AgdaBound{q₁}\AgdaSpace{}%
\AgdaBound{eq₂}\AgdaSpace{}%
\AgdaSymbol{|}\AgdaSpace{}%
\AgdaInductiveConstructor{val}\AgdaSpace{}%
\AgdaBound{q₂}\AgdaSpace{}%
\AgdaBound{q₃}\AgdaSpace{}%
\AgdaKeyword{with}\AgdaSpace{}%
\AgdaFunction{trans}\AgdaSpace{}%
\AgdaSymbol{(}\AgdaFunction{sym}\AgdaSpace{}%
\AgdaBound{q₁}\AgdaSymbol{)}\AgdaSpace{}%
\AgdaBound{q₂}\<%
\\
\>[2]\AgdaSymbol{...}\AgdaSpace{}%
\AgdaSymbol{|}\AgdaSpace{}%
\AgdaSymbol{()}\<%
\\
\>[2]\AgdaField{bisim}\AgdaSpace{}%
\AgdaSymbol{(}\AgdaFunction{trans-≈}\AgdaSpace{}%
\AgdaBound{eq₀}\AgdaSpace{}%
\AgdaBound{eq₁}\AgdaSymbol{)}\AgdaSpace{}%
\AgdaSymbol{|}\AgdaSpace{}%
\AgdaInductiveConstructor{tau}\AgdaSpace{}%
\AgdaBound{q₀}\AgdaSpace{}%
\AgdaBound{q₁}\AgdaSpace{}%
\AgdaBound{eq₂}\AgdaSpace{}%
\AgdaSymbol{|}\AgdaSpace{}%
\AgdaInductiveConstructor{tau}\AgdaSpace{}%
\AgdaBound{q₂}\AgdaSpace{}%
\AgdaBound{q₃}\AgdaSpace{}%
\AgdaBound{eq₃}\AgdaSpace{}%
\AgdaKeyword{with}\AgdaSpace{}%
\AgdaFunction{trans}\AgdaSpace{}%
\AgdaSymbol{(}\AgdaFunction{sym}\AgdaSpace{}%
\AgdaBound{q₁}\AgdaSymbol{)}\AgdaSpace{}%
\AgdaBound{q₂}\<%
\\
\>[2]\AgdaSymbol{...}\AgdaSpace{}%
\AgdaSymbol{|}\AgdaSpace{}%
\AgdaInductiveConstructor{refl}\AgdaSpace{}%
\AgdaSymbol{=}\AgdaSpace{}%
\AgdaInductiveConstructor{tau}\AgdaSpace{}%
\AgdaBound{q₀}\AgdaSpace{}%
\AgdaBound{q₃}\AgdaSpace{}%
\AgdaSymbol{(}\AgdaFunction{trans-≈}\AgdaSpace{}%
\AgdaBound{eq₂}\AgdaSpace{}%
\AgdaBound{eq₃}\AgdaSymbol{)}\<%
\\
\\[\AgdaEmptyExtraSkip]%
\\[\AgdaEmptyExtraSkip]%
\>[2]\AgdaFunction{step-≈}\AgdaSpace{}%
\AgdaSymbol{:}\AgdaSpace{}%
\AgdaSymbol{∀}\AgdaSpace{}%
\AgdaSymbol{\{}\AgdaBound{A}\AgdaSymbol{\}}\AgdaSpace{}%
\AgdaSymbol{(}\AgdaBound{d₀}\AgdaSpace{}%
\AgdaSymbol{:}\AgdaSpace{}%
\AgdaRecord{Delay}\AgdaSpace{}%
\AgdaBound{A}\AgdaSymbol{)}\AgdaSpace{}%
\AgdaSymbol{\{}\AgdaBound{d₁}\AgdaSpace{}%
\AgdaBound{d₂}\AgdaSymbol{\}}\AgdaSpace{}%
\AgdaSymbol{→}\AgdaSpace{}%
\AgdaBound{d₀}\AgdaSpace{}%
\AgdaOperator{\AgdaRecord{≈}}\AgdaSpace{}%
\AgdaBound{d₁}\AgdaSpace{}%
\AgdaSymbol{→}\AgdaSpace{}%
\AgdaBound{d₁}\AgdaSpace{}%
\AgdaOperator{\AgdaRecord{≈}}\AgdaSpace{}%
\AgdaBound{d₂}\AgdaSpace{}%
\AgdaSymbol{→}\AgdaSpace{}%
\AgdaBound{d₀}\AgdaSpace{}%
\AgdaOperator{\AgdaRecord{≈}}\AgdaSpace{}%
\AgdaBound{d₂}\<%
\\
\>[2]\AgdaFunction{step-≈}\AgdaSpace{}%
\AgdaSymbol{\AgdaUnderscore{}}\AgdaSpace{}%
\AgdaSymbol{=}\AgdaSpace{}%
\AgdaFunction{trans-≈}\<%
\\
\\[\AgdaEmptyExtraSkip]%
\>[2]\AgdaKeyword{infix}\AgdaSpace{}%
\AgdaNumber{1}\AgdaSpace{}%
\AgdaOperator{\AgdaFunction{begin\AgdaUnderscore{}}}\<%
\\
\>[2]\AgdaOperator{\AgdaFunction{begin\AgdaUnderscore{}}}\AgdaSpace{}%
\AgdaSymbol{=}\AgdaSpace{}%
\AgdaFunction{id}\<%
\\
\\[\AgdaEmptyExtraSkip]%
\>[2]\AgdaKeyword{infixr}\AgdaSpace{}%
\AgdaNumber{2}\AgdaSpace{}%
\AgdaFunction{step-≈}\<%
\\
\>[2]\AgdaKeyword{syntax}\AgdaSpace{}%
\AgdaFunction{step-≈}\AgdaSpace{}%
\AgdaBound{d₀}\AgdaSpace{}%
\AgdaBound{eq₀}\AgdaSpace{}%
\AgdaBound{eq₁}\AgdaSpace{}%
\AgdaSymbol{=}\AgdaSpace{}%
\AgdaBound{d₀}\AgdaSpace{}%
\AgdaFunction{≈[}\AgdaSpace{}%
\AgdaBound{eq₀}\AgdaSpace{}%
\AgdaFunction{]}\AgdaSpace{}%
\AgdaBound{eq₁}\<%
\\
\\[\AgdaEmptyExtraSkip]%
\>[2]\AgdaKeyword{infixr}\AgdaSpace{}%
\AgdaNumber{3}\AgdaSpace{}%
\AgdaOperator{\AgdaFunction{\AgdaUnderscore{}∎}}\<%
\\
\>[2]\AgdaOperator{\AgdaFunction{\AgdaUnderscore{}∎}}\AgdaSpace{}%
\AgdaSymbol{:}\AgdaSpace{}%
\AgdaSymbol{∀}\AgdaSpace{}%
\AgdaSymbol{\{}\AgdaBound{A}\AgdaSymbol{\}}\AgdaSpace{}%
\AgdaSymbol{(}\AgdaBound{d}\AgdaSpace{}%
\AgdaSymbol{:}\AgdaSpace{}%
\AgdaRecord{Delay}\AgdaSpace{}%
\AgdaBound{A}\AgdaSymbol{)}\AgdaSpace{}%
\AgdaSymbol{→}\AgdaSpace{}%
\AgdaBound{d}\AgdaSpace{}%
\AgdaOperator{\AgdaRecord{≈}}\AgdaSpace{}%
\AgdaBound{d}\<%
\\
\>[2]\AgdaOperator{\AgdaFunction{\AgdaUnderscore{}∎}}\AgdaSpace{}%
\AgdaSymbol{=}\AgdaSpace{}%
\AgdaFunction{refl-≈}\<%
\end{code}
\begin{code}[hide]%
\>[2]\AgdaFunction{delay}\AgdaSpace{}%
\AgdaSymbol{:}\AgdaSpace{}%
\AgdaSymbol{∀}\AgdaSpace{}%
\AgdaSymbol{\{}\AgdaBound{A}\AgdaSymbol{\}}\AgdaSpace{}%
\AgdaSymbol{→}\AgdaSpace{}%
\AgdaDatatype{ℕ}\AgdaSpace{}%
\AgdaSymbol{→}\AgdaSpace{}%
\AgdaBound{A}\AgdaSpace{}%
\AgdaSymbol{→}\AgdaSpace{}%
\AgdaRecord{Delay}\AgdaSpace{}%
\AgdaBound{A}\<%
\\
\>[2]\AgdaField{insp}\AgdaSpace{}%
\AgdaSymbol{(}\AgdaFunction{delay}\AgdaSpace{}%
\AgdaInductiveConstructor{zero}\AgdaSpace{}%
\AgdaBound{x}\AgdaSymbol{)}%
\>[25]\AgdaSymbol{=}\AgdaSpace{}%
\AgdaInductiveConstructor{inj₁}\AgdaSpace{}%
\AgdaBound{x}\<%
\\
\>[2]\AgdaField{insp}\AgdaSpace{}%
\AgdaSymbol{(}\AgdaFunction{delay}\AgdaSpace{}%
\AgdaSymbol{(}\AgdaInductiveConstructor{suc}\AgdaSpace{}%
\AgdaBound{n}\AgdaSymbol{)}\AgdaSpace{}%
\AgdaBound{x}\AgdaSymbol{)}\AgdaSpace{}%
\AgdaSymbol{=}\AgdaSpace{}%
\AgdaInductiveConstructor{inj₂}\AgdaSpace{}%
\AgdaSymbol{(}\AgdaFunction{delay}\AgdaSpace{}%
\AgdaBound{n}\AgdaSpace{}%
\AgdaBound{x}\AgdaSymbol{)}\<%
\\
\\[\AgdaEmptyExtraSkip]%
\>[2]\AgdaFunction{delay+}\AgdaSpace{}%
\AgdaSymbol{:}\AgdaSpace{}%
\AgdaSymbol{∀}\AgdaSpace{}%
\AgdaSymbol{\{}\AgdaBound{A}\AgdaSymbol{\}}\AgdaSpace{}%
\AgdaSymbol{→}\AgdaSpace{}%
\AgdaDatatype{ℕ}\AgdaSpace{}%
\AgdaSymbol{→}\AgdaSpace{}%
\AgdaRecord{Delay}\AgdaSpace{}%
\AgdaBound{A}\AgdaSpace{}%
\AgdaSymbol{→}\AgdaSpace{}%
\AgdaRecord{Delay}\AgdaSpace{}%
\AgdaBound{A}\<%
\\
\>[2]\AgdaFunction{delay+}\AgdaSpace{}%
\AgdaInductiveConstructor{zero}\AgdaSpace{}%
\AgdaBound{x}%
\>[19]\AgdaSymbol{=}\AgdaSpace{}%
\AgdaBound{x}\<%
\\
\>[2]\AgdaField{insp}\AgdaSpace{}%
\AgdaSymbol{(}\AgdaFunction{delay+}\AgdaSpace{}%
\AgdaSymbol{(}\AgdaInductiveConstructor{suc}\AgdaSpace{}%
\AgdaBound{n}\AgdaSymbol{)}\AgdaSpace{}%
\AgdaBound{x}\AgdaSymbol{)}\AgdaSpace{}%
\AgdaSymbol{=}\AgdaSpace{}%
\AgdaInductiveConstructor{inj₂}\AgdaSpace{}%
\AgdaSymbol{(}\AgdaFunction{delay+}\AgdaSpace{}%
\AgdaBound{n}\AgdaSpace{}%
\AgdaBound{x}\AgdaSymbol{)}\<%
\\
\\[\AgdaEmptyExtraSkip]%
\>[2]\AgdaKeyword{open}\AgdaSpace{}%
\AgdaOperator{\AgdaModule{\AgdaUnderscore{}≈\AgdaUnderscore{}}}\<%
\end{code}

Using these, we can now prove that the expression $1+\mathsf{reset}\ (2+\mathsf{shift}_k\ (k\ (k\ 3)))$ reduces to $1+2+2+3=8$.

\begin{code}%
\>[2]\AgdaFunction{test-shift}\AgdaSpace{}%
\AgdaSymbol{:}%
\>[2024I]\AgdaFunction{drive}%
\>[2025I]\AgdaSymbol{(}\AgdaInductiveConstructor{add}\AgdaSpace{}%
\AgdaSymbol{(}\AgdaInductiveConstructor{num}\AgdaSpace{}%
\AgdaNumber{1}\AgdaSymbol{)}\<%
\\
\>[2025I][@{}l@{\AgdaIndent{0}}]%
\>[23]\AgdaSymbol{(}\AgdaInductiveConstructor{reset}\<%
\\
\>[23][@{}l@{\AgdaIndent{0}}]%
\>[25]\AgdaSymbol{(}\AgdaInductiveConstructor{add}\AgdaSpace{}%
\AgdaSymbol{(}\AgdaInductiveConstructor{num}\AgdaSpace{}%
\AgdaNumber{2}\AgdaSymbol{)}\<%
\\
\>[25][@{}l@{\AgdaIndent{0}}]%
\>[27]\AgdaSymbol{(}\AgdaInductiveConstructor{shift}\<%
\\
\>[27][@{}l@{\AgdaIndent{0}}]%
\>[29]\AgdaSymbol{(}\AgdaInductiveConstructor{app}\AgdaSpace{}%
\AgdaSymbol{(}\AgdaInductiveConstructor{var}\AgdaSpace{}%
\AgdaNumber{0}\AgdaSymbol{)}\<%
\\
\>[29][@{}l@{\AgdaIndent{0}}]%
\>[31]\AgdaSymbol{(}\AgdaInductiveConstructor{app}\AgdaSpace{}%
\AgdaSymbol{(}\AgdaInductiveConstructor{var}\AgdaSpace{}%
\AgdaNumber{0}\AgdaSymbol{)}\AgdaSpace{}%
\AgdaSymbol{(}\AgdaInductiveConstructor{num}\AgdaSpace{}%
\AgdaNumber{3}\AgdaSymbol{)))))))}\<%
\\
\>[.][@{}l@{}]\<[2024I]%
\>[15]\AgdaOperator{\AgdaRecord{≈}}\AgdaSpace{}%
\AgdaFunction{delay}\AgdaSpace{}%
\AgdaNumber{10}\AgdaSpace{}%
\AgdaSymbol{(}\AgdaInductiveConstructor{just}\AgdaSpace{}%
\AgdaSymbol{(}\AgdaInductiveConstructor{num}\AgdaSpace{}%
\AgdaNumber{8}\AgdaSymbol{))}\<%
\end{code}
\begin{code}[hide]%
\>[2]\AgdaFunction{test-shift}\AgdaSpace{}%
\AgdaSymbol{=}\AgdaSpace{}%
\AgdaOperator{\AgdaFunction{begin}}\AgdaSpace{}%
\AgdaSymbol{\AgdaUnderscore{}}\<%
\\
\>[2][@{}l@{\AgdaIndent{0}}]%
\>[4]\AgdaFunction{≈[}\AgdaSpace{}%
\AgdaFunction{1-step}\AgdaSpace{}%
\AgdaFunction{]}\<%
\\
\>[4][@{}l@{\AgdaIndent{0}}]%
\>[6]\AgdaFunction{delay+}\AgdaSpace{}%
\AgdaNumber{1}\<%
\\
\>[6][@{}l@{\AgdaIndent{0}}]%
\>[8]\AgdaSymbol{(}\AgdaFunction{drive}\<%
\\
\>[8][@{}l@{\AgdaIndent{0}}]%
\>[10]\AgdaSymbol{(}\AgdaInductiveConstructor{add}\AgdaSpace{}%
\AgdaSymbol{(}\AgdaInductiveConstructor{num}\AgdaSpace{}%
\AgdaNumber{1}\AgdaSymbol{)}\<%
\\
\>[10][@{}l@{\AgdaIndent{0}}]%
\>[12]\AgdaSymbol{(}\AgdaInductiveConstructor{reset}\<%
\\
\>[12][@{}l@{\AgdaIndent{0}}]%
\>[14]\AgdaSymbol{(}\AgdaInductiveConstructor{app}\AgdaSpace{}%
\AgdaSymbol{(}\AgdaInductiveConstructor{lam}\AgdaSpace{}%
\AgdaSymbol{(}\AgdaInductiveConstructor{app}\AgdaSpace{}%
\AgdaSymbol{(}\AgdaInductiveConstructor{var}\AgdaSpace{}%
\AgdaNumber{0}\AgdaSymbol{)}\AgdaSpace{}%
\AgdaSymbol{(}\AgdaInductiveConstructor{app}\AgdaSpace{}%
\AgdaSymbol{(}\AgdaInductiveConstructor{var}\AgdaSpace{}%
\AgdaNumber{0}\AgdaSymbol{)}\AgdaSpace{}%
\AgdaSymbol{(}\AgdaInductiveConstructor{num}\AgdaSpace{}%
\AgdaNumber{3}\AgdaSymbol{))))}\<%
\\
\>[14][@{}l@{\AgdaIndent{0}}]%
\>[16]\AgdaSymbol{(}\AgdaInductiveConstructor{lam}\AgdaSpace{}%
\AgdaSymbol{(}\AgdaInductiveConstructor{reset}\AgdaSpace{}%
\AgdaSymbol{(}\AgdaInductiveConstructor{add}\AgdaSpace{}%
\AgdaSymbol{(}\AgdaInductiveConstructor{num}\AgdaSpace{}%
\AgdaNumber{2}\AgdaSymbol{)}\AgdaSpace{}%
\AgdaSymbol{(}\AgdaInductiveConstructor{var}\AgdaSpace{}%
\AgdaNumber{0}\AgdaSymbol{))))))))}\<%
\\
\>[4]\AgdaFunction{≈[}\AgdaSpace{}%
\AgdaFunction{5-steps}\AgdaSpace{}%
\AgdaFunction{]}\<%
\\
\>[4][@{}l@{\AgdaIndent{0}}]%
\>[6]\AgdaFunction{delay+}\AgdaSpace{}%
\AgdaNumber{5}\<%
\\
\>[6][@{}l@{\AgdaIndent{0}}]%
\>[8]\AgdaSymbol{(}\AgdaFunction{drive}\<%
\\
\>[8][@{}l@{\AgdaIndent{0}}]%
\>[10]\AgdaSymbol{(}\AgdaInductiveConstructor{add}\AgdaSpace{}%
\AgdaSymbol{(}\AgdaInductiveConstructor{num}\AgdaSpace{}%
\AgdaNumber{1}\AgdaSymbol{)}\<%
\\
\>[10][@{}l@{\AgdaIndent{0}}]%
\>[12]\AgdaSymbol{(}\AgdaInductiveConstructor{reset}\<%
\\
\>[12][@{}l@{\AgdaIndent{0}}]%
\>[14]\AgdaSymbol{(}\AgdaInductiveConstructor{app}\AgdaSpace{}%
\AgdaSymbol{(}\AgdaInductiveConstructor{lam}\AgdaSpace{}%
\AgdaSymbol{(}\AgdaInductiveConstructor{reset}\AgdaSpace{}%
\AgdaSymbol{(}\AgdaInductiveConstructor{add}\AgdaSpace{}%
\AgdaSymbol{(}\AgdaInductiveConstructor{num}\AgdaSpace{}%
\AgdaNumber{2}\AgdaSymbol{)}\AgdaSpace{}%
\AgdaSymbol{(}\AgdaInductiveConstructor{var}\AgdaSpace{}%
\AgdaNumber{0}\AgdaSymbol{))))}\<%
\\
\>[14][@{}l@{\AgdaIndent{0}}]%
\>[16]\AgdaSymbol{(}\AgdaInductiveConstructor{num}\AgdaSpace{}%
\AgdaNumber{5}\AgdaSymbol{)))))}\<%
\\
\>[4]\AgdaFunction{≈[}\AgdaSpace{}%
\AgdaFunction{10-steps}\AgdaSpace{}%
\AgdaFunction{]}\<%
\\
\>[4][@{}l@{\AgdaIndent{0}}]%
\>[6]\AgdaFunction{delay}\AgdaSpace{}%
\AgdaNumber{10}\AgdaSpace{}%
\AgdaSymbol{(}\AgdaInductiveConstructor{just}\AgdaSpace{}%
\AgdaSymbol{(}\AgdaInductiveConstructor{num}\AgdaSpace{}%
\AgdaNumber{8}\AgdaSymbol{))}\<%
\\
\>[4]\AgdaOperator{\AgdaFunction{∎}}\<%
\\
\>[4]\AgdaKeyword{where}\<%
\\
\>[4]\AgdaFunction{1-step}%
\>[12]\AgdaSymbol{=}\AgdaSpace{}%
\AgdaOperator{\AgdaCoinductiveConstructor{≈⟨}}\AgdaSpace{}%
\AgdaInductiveConstructor{tau}\AgdaSpace{}%
\AgdaInductiveConstructor{refl}\AgdaSpace{}%
\AgdaInductiveConstructor{refl}\AgdaSpace{}%
\AgdaSymbol{(}\AgdaFunction{refl-≈}\AgdaSpace{}%
\AgdaSymbol{\AgdaUnderscore{})}\AgdaSpace{}%
\AgdaOperator{\AgdaCoinductiveConstructor{⟩}}\<%
\\
\>[4]\AgdaFunction{5-steps}%
\>[13]\AgdaSymbol{=}\AgdaSpace{}%
\AgdaOperator{\AgdaCoinductiveConstructor{≈⟨}}\AgdaSpace{}%
\AgdaInductiveConstructor{tau}\AgdaSpace{}%
\AgdaInductiveConstructor{refl}\AgdaSpace{}%
\AgdaInductiveConstructor{refl}\<%
\\
\>[4][@{}l@{\AgdaIndent{0}}]%
\>[10]\AgdaOperator{\AgdaCoinductiveConstructor{≈⟨}}%
\>[2094I]\AgdaInductiveConstructor{tau}\AgdaSpace{}%
\AgdaInductiveConstructor{refl}\AgdaSpace{}%
\AgdaInductiveConstructor{refl}\<%
\\
\>[.][@{}l@{}]\<[2094I]%
\>[13]\AgdaOperator{\AgdaCoinductiveConstructor{≈⟨}}%
\>[2097I]\AgdaInductiveConstructor{tau}\AgdaSpace{}%
\AgdaInductiveConstructor{refl}\AgdaSpace{}%
\AgdaInductiveConstructor{refl}\<%
\\
\>[.][@{}l@{}]\<[2097I]%
\>[16]\AgdaOperator{\AgdaCoinductiveConstructor{≈⟨}}%
\>[2100I]\AgdaInductiveConstructor{tau}\AgdaSpace{}%
\AgdaInductiveConstructor{refl}\AgdaSpace{}%
\AgdaInductiveConstructor{refl}\<%
\\
\>[.][@{}l@{}]\<[2100I]%
\>[19]\AgdaOperator{\AgdaCoinductiveConstructor{≈⟨}}%
\>[2103I]\AgdaInductiveConstructor{tau}\AgdaSpace{}%
\AgdaInductiveConstructor{refl}\AgdaSpace{}%
\AgdaInductiveConstructor{refl}\<%
\\
\>[.][@{}l@{}]\<[2103I]%
\>[22]\AgdaSymbol{(}\AgdaFunction{refl-≈}\AgdaSpace{}%
\AgdaSymbol{\AgdaUnderscore{})}\AgdaSpace{}%
\AgdaOperator{\AgdaCoinductiveConstructor{⟩}}\AgdaSpace{}%
\AgdaOperator{\AgdaCoinductiveConstructor{⟩}}\AgdaSpace{}%
\AgdaOperator{\AgdaCoinductiveConstructor{⟩}}\AgdaSpace{}%
\AgdaOperator{\AgdaCoinductiveConstructor{⟩}}\AgdaSpace{}%
\AgdaOperator{\AgdaCoinductiveConstructor{⟩}}\<%
\\
\>[4]\AgdaFunction{10-steps}\AgdaSpace{}%
\AgdaSymbol{=}\AgdaSpace{}%
\AgdaOperator{\AgdaCoinductiveConstructor{≈⟨}}\AgdaSpace{}%
\AgdaInductiveConstructor{tau}\AgdaSpace{}%
\AgdaInductiveConstructor{refl}\AgdaSpace{}%
\AgdaInductiveConstructor{refl}\<%
\\
\>[4][@{}l@{\AgdaIndent{0}}]%
\>[10]\AgdaOperator{\AgdaCoinductiveConstructor{≈⟨}}%
\>[2117I]\AgdaInductiveConstructor{tau}\AgdaSpace{}%
\AgdaInductiveConstructor{refl}\AgdaSpace{}%
\AgdaInductiveConstructor{refl}\<%
\\
\>[.][@{}l@{}]\<[2117I]%
\>[13]\AgdaOperator{\AgdaCoinductiveConstructor{≈⟨}}%
\>[2120I]\AgdaInductiveConstructor{tau}\AgdaSpace{}%
\AgdaInductiveConstructor{refl}\AgdaSpace{}%
\AgdaInductiveConstructor{refl}\<%
\\
\>[.][@{}l@{}]\<[2120I]%
\>[16]\AgdaOperator{\AgdaCoinductiveConstructor{≈⟨}}%
\>[2123I]\AgdaInductiveConstructor{tau}\AgdaSpace{}%
\AgdaInductiveConstructor{refl}\AgdaSpace{}%
\AgdaInductiveConstructor{refl}\<%
\\
\>[.][@{}l@{}]\<[2123I]%
\>[19]\AgdaOperator{\AgdaCoinductiveConstructor{≈⟨}}%
\>[2126I]\AgdaInductiveConstructor{tau}\AgdaSpace{}%
\AgdaInductiveConstructor{refl}\AgdaSpace{}%
\AgdaInductiveConstructor{refl}\<%
\\
\>[.][@{}l@{}]\<[2126I]%
\>[22]\AgdaOperator{\AgdaCoinductiveConstructor{≈⟨}}%
\>[2129I]\AgdaInductiveConstructor{tau}\AgdaSpace{}%
\AgdaInductiveConstructor{refl}\AgdaSpace{}%
\AgdaInductiveConstructor{refl}\<%
\\
\>[.][@{}l@{}]\<[2129I]%
\>[25]\AgdaOperator{\AgdaCoinductiveConstructor{≈⟨}}%
\>[2132I]\AgdaInductiveConstructor{tau}\AgdaSpace{}%
\AgdaInductiveConstructor{refl}\AgdaSpace{}%
\AgdaInductiveConstructor{refl}\<%
\\
\>[.][@{}l@{}]\<[2132I]%
\>[28]\AgdaOperator{\AgdaCoinductiveConstructor{≈⟨}}%
\>[2135I]\AgdaInductiveConstructor{tau}\AgdaSpace{}%
\AgdaInductiveConstructor{refl}\AgdaSpace{}%
\AgdaInductiveConstructor{refl}\<%
\\
\>[.][@{}l@{}]\<[2135I]%
\>[31]\AgdaOperator{\AgdaCoinductiveConstructor{≈⟨}}%
\>[2138I]\AgdaInductiveConstructor{tau}\AgdaSpace{}%
\AgdaInductiveConstructor{refl}\AgdaSpace{}%
\AgdaInductiveConstructor{refl}\<%
\\
\>[.][@{}l@{}]\<[2138I]%
\>[34]\AgdaOperator{\AgdaCoinductiveConstructor{≈⟨}}%
\>[2141I]\AgdaInductiveConstructor{tau}\AgdaSpace{}%
\AgdaInductiveConstructor{refl}\AgdaSpace{}%
\AgdaInductiveConstructor{refl}\<%
\\
\>[.][@{}l@{}]\<[2141I]%
\>[37]\AgdaOperator{\AgdaCoinductiveConstructor{≈⟨}}\AgdaSpace{}%
\AgdaInductiveConstructor{val}\AgdaSpace{}%
\AgdaInductiveConstructor{refl}\AgdaSpace{}%
\AgdaInductiveConstructor{refl}\AgdaSpace{}%
\AgdaOperator{\AgdaCoinductiveConstructor{⟩}}\AgdaSpace{}%
\AgdaOperator{\AgdaCoinductiveConstructor{⟩}}\AgdaSpace{}%
\AgdaOperator{\AgdaCoinductiveConstructor{⟩}}\AgdaSpace{}%
\AgdaOperator{\AgdaCoinductiveConstructor{⟩}}\AgdaSpace{}%
\AgdaOperator{\AgdaCoinductiveConstructor{⟩}}\AgdaSpace{}%
\AgdaOperator{\AgdaCoinductiveConstructor{⟩}}\AgdaSpace{}%
\AgdaOperator{\AgdaCoinductiveConstructor{⟩}}\AgdaSpace{}%
\AgdaOperator{\AgdaCoinductiveConstructor{⟩}}\AgdaSpace{}%
\AgdaOperator{\AgdaCoinductiveConstructor{⟩}}\AgdaSpace{}%
\AgdaOperator{\AgdaCoinductiveConstructor{⟩}}\AgdaSpace{}%
\AgdaOperator{\AgdaCoinductiveConstructor{⟩}}\<%
\end{code}

\subsection{On Boilerplate Code}

We have demonstrated how to implement reduction semantics with deterministic reduction strategies, using functional programming in Agda.
We have also discussed how to use Agda to reason about the reduction semantics (e.g., that recomposing a decomposition yields the original term).
However, the code we have been writing for defining contexts, decomposition, and recomposition is repetitive, boilerplate code.


\begin{code}[hide]%
\>[0]\AgdaKeyword{open}\AgdaSpace{}%
\AgdaKeyword{import}\AgdaSpace{}%
\AgdaModule{Function}\<%
\\
\\[\AgdaEmptyExtraSkip]%
\>[0]\AgdaKeyword{open}\AgdaSpace{}%
\AgdaKeyword{import}\AgdaSpace{}%
\AgdaModule{Data.Empty}\<%
\\
\>[0]\AgdaKeyword{open}\AgdaSpace{}%
\AgdaKeyword{import}\AgdaSpace{}%
\AgdaModule{Data.Unit}\<%
\\
\>[0]\AgdaKeyword{open}\AgdaSpace{}%
\AgdaKeyword{import}\AgdaSpace{}%
\AgdaModule{Data.Nat}\<%
\\
\>[0]\AgdaKeyword{open}\AgdaSpace{}%
\AgdaKeyword{import}\AgdaSpace{}%
\AgdaModule{Data.Fin}\<%
\\
\>[0]\AgdaKeyword{open}\AgdaSpace{}%
\AgdaKeyword{import}\AgdaSpace{}%
\AgdaModule{Data.Sum}\<%
\\
\>[0]\AgdaKeyword{open}\AgdaSpace{}%
\AgdaKeyword{import}\AgdaSpace{}%
\AgdaModule{Data.Product}\<%
\\
\>[0]\AgdaKeyword{open}\AgdaSpace{}%
\AgdaKeyword{import}\AgdaSpace{}%
\AgdaModule{Data.List}\<%
\\
\\[\AgdaEmptyExtraSkip]%
\>[0]\AgdaKeyword{open}\AgdaSpace{}%
\AgdaKeyword{import}\AgdaSpace{}%
\AgdaModule{Relation.Unary}\<%
\\
\>[0]\AgdaKeyword{open}\AgdaSpace{}%
\AgdaKeyword{import}\AgdaSpace{}%
\AgdaModule{Relation.Binary.PropositionalEquality}\<%
\\
\\[\AgdaEmptyExtraSkip]%
\>[0]\AgdaKeyword{module}\AgdaSpace{}%
\AgdaModule{sections.03-prelude}\AgdaSpace{}%
\AgdaKeyword{where}\<%
\end{code}

\section{Generic Programming with Descriptions, Algebras, and Derivatives}
\label{sec:03-prelude}

Generic programming lets us factor boilerplate code into generic code that can be reused between languages.
The techniques we use to this end are borrowed from existing work.
This section embeds and applies the techniques in Agda, making use of dependent types to enforce similar typing guarantees as the interpreters in the previous section.

We first (\cref{sec:syntax-desc}) recall how to specify abstract syntax using \emph{descriptions}~\cite{chapman2010thegentle}.
Next, we consider how to interpret such syntax using \emph{F-algebras}.
With these basics in place, we recall how to take the \emph{derivative} of a syntax, and how this derivative gives rise to zippers and the familiar notion of context we saw in \cref{sec:02-redsem}.

While this section contains a substantial amount of code, we emphasize that the code is generic; i.e., written once and reused between different languages and semantics.

\subsection{Syntax Descriptions and Their Semantics}
\label{sec:syntax-desc}

In order to take the derivative of an abstract syntax, we must treat that syntax as a piece of data.
Plain Agda data types cannot easily be programmatically manipulated in Agda itself.
Therefore, we define a different data type of \emph{syntax descriptions}.
That is, a data type that represents an abstract description of the shape of a syntax.
For the intents and purposes of this paper, we use the following type of descriptions.
\begin{code}%
\>[0]\AgdaKeyword{data}\AgdaSpace{}%
\AgdaDatatype{Desc}\AgdaSpace{}%
\AgdaSymbol{:}\AgdaSpace{}%
\AgdaPrimitive{Set₁}\AgdaSpace{}%
\AgdaKeyword{where}\<%
\\
\>[0][@{}l@{\AgdaIndent{0}}]%
\>[2]\AgdaInductiveConstructor{I}%
\>[7]\AgdaSymbol{:}\AgdaSpace{}%
\AgdaDatatype{Desc}%
\>[42]\AgdaComment{--\ Identity}\<%
\\
\>[2]\AgdaInductiveConstructor{K}%
\>[7]\AgdaSymbol{:}\AgdaSpace{}%
\AgdaPrimitive{Set}\AgdaSpace{}%
\AgdaSymbol{→}\AgdaSpace{}%
\AgdaDatatype{Desc}%
\>[42]\AgdaComment{--\ Constant}\<%
\\
\>[2]\AgdaOperator{\AgdaInductiveConstructor{\AgdaUnderscore{}ẋ\AgdaUnderscore{}}}%
\>[7]\AgdaSymbol{:}\AgdaSpace{}%
\AgdaDatatype{Desc}\AgdaSpace{}%
\AgdaSymbol{→}\AgdaSpace{}%
\AgdaDatatype{Desc}\AgdaSpace{}%
\AgdaSymbol{→}\AgdaSpace{}%
\AgdaDatatype{Desc}%
\>[42]\AgdaComment{--\ Product}\<%
\\
\>[2]\AgdaInductiveConstructor{∐}%
\>[7]\AgdaSymbol{:}\AgdaSpace{}%
\AgdaSymbol{(}\AgdaBound{n}\AgdaSpace{}%
\AgdaSymbol{:}\AgdaSpace{}%
\AgdaDatatype{ℕ}\AgdaSymbol{)}\AgdaSpace{}%
\AgdaSymbol{→}\AgdaSpace{}%
\AgdaSymbol{(}\AgdaDatatype{Fin}\AgdaSpace{}%
\AgdaBound{n}\AgdaSpace{}%
\AgdaSymbol{→}\AgdaSpace{}%
\AgdaDatatype{Desc}\AgdaSymbol{)}\AgdaSpace{}%
\AgdaSymbol{→}\AgdaSpace{}%
\AgdaDatatype{Desc}%
\>[42]\AgdaComment{--\ Tagged\ sum}\<%
\end{code}
Starting from the bottom, the constructor \ac{∐}~\ab{n}~\ab{f} represents a \emph{sum} of possible descriptions.
The values of type \ad{Fin}~\ab{n} are the numbers between zero and $\ab{n}-1$.
So \ac{∐}~\ab{n}~\ab{f} is describing a choice between $n$ different abstract syntax constructors; e.g., $x \bnfdef x_0 \mid \cdots \mid x_{n-1}$ where each $x_{i\in\{0\ldots n\}}$ is a tagged grammar production described by $f~i$.
In contrast, \ab{d₀}~\ac{ẋ}~\ab{d₁} is describing a syntax with two adjacent abstract syntax constructors described by, respectively, \ab{d₀} and \ab{d₁}; e.g., $x \bnfdef x_0\ x_1$ where $x_{i\in\{0,1\}}$ is described by $d_i$.
\ac{K}~\ab{A} describes a sub-term of type \ab{A}, corresponding to a terminal grammar production; and \ac{I} describes a sub-term of the same type as the one we are currently defining, corresponding to a grammar non-terminal production.

\begin{example}
The following description describes the type of natural numbers.\footnote{The type \ad{⊤} is the unit type from Agda's standard library, where \ac{tt} is the unit value.}
\begin{code}%
\>[0]\AgdaFunction{NatDesc}\AgdaSpace{}%
\AgdaSymbol{:}\AgdaSpace{}%
\AgdaDatatype{Desc}\<%
\\
\>[0]\AgdaFunction{NatDesc}\AgdaSpace{}%
\AgdaSymbol{=}\AgdaSpace{}%
\AgdaInductiveConstructor{∐}\AgdaSpace{}%
\AgdaNumber{2}\AgdaSpace{}%
\AgdaSymbol{(λ}\AgdaSpace{}%
\AgdaKeyword{where}\<%
\\
\>[0][@{}l@{\AgdaIndent{0}}]%
\>[2]\AgdaInductiveConstructor{zero}\AgdaSpace{}%
\AgdaSymbol{→}\AgdaSpace{}%
\AgdaInductiveConstructor{K}\AgdaSpace{}%
\AgdaRecord{⊤}\<%
\\
\>[2]\AgdaSymbol{(}\AgdaInductiveConstructor{suc}\AgdaSpace{}%
\AgdaInductiveConstructor{zero}\AgdaSymbol{)}\AgdaSpace{}%
\AgdaSymbol{→}\AgdaSpace{}%
\AgdaInductiveConstructor{I}\AgdaSymbol{)}\<%
\end{code}
That is, we have a choice between two different syntax constructors, one that corresponds to a nullary terminal production (zero) and one that corresponds to a unary non-terminal production (successor of a natural number).
\end{example}

The semantics of descriptions is given by a function which maps a description to a \emph{signature functor}; specifically, an \emph{endofunctor} on \ad{Set}, of type \ad{Set}~\as{→}~\ad{Set} in Agda.\footnote{The type \ad{Σ[}~\ab{x}~\ad{∈}~\ab{Y}~\ad{]}~\as{(}\ab{P}~\ab{x}\as{)} is the type of a \emph{dependent sum}, also known as a $\Sigma$-type.}
\begin{code}%
\>[0]\AgdaOperator{\AgdaFunction{⟦\AgdaUnderscore{}⟧}}\AgdaSpace{}%
\AgdaSymbol{:}\AgdaSpace{}%
\AgdaDatatype{Desc}\AgdaSpace{}%
\AgdaSymbol{→}\AgdaSpace{}%
\AgdaPrimitive{Set}\AgdaSpace{}%
\AgdaSymbol{→}\AgdaSpace{}%
\AgdaPrimitive{Set}\<%
\\
\>[0]\AgdaOperator{\AgdaFunction{⟦}}\AgdaSpace{}%
\AgdaInductiveConstructor{I}%
\>[11]\AgdaOperator{\AgdaFunction{⟧}}\AgdaSpace{}%
\AgdaBound{X}%
\>[16]\AgdaSymbol{=}\AgdaSpace{}%
\AgdaBound{X}\<%
\\
\>[0]\AgdaOperator{\AgdaFunction{⟦}}\AgdaSpace{}%
\AgdaInductiveConstructor{K}\AgdaSpace{}%
\AgdaBound{A}%
\>[11]\AgdaOperator{\AgdaFunction{⟧}}\AgdaSpace{}%
\AgdaBound{X}%
\>[16]\AgdaSymbol{=}\AgdaSpace{}%
\AgdaBound{A}\<%
\\
\>[0]\AgdaOperator{\AgdaFunction{⟦}}\AgdaSpace{}%
\AgdaBound{d₁}\AgdaSpace{}%
\AgdaOperator{\AgdaInductiveConstructor{ẋ}}\AgdaSpace{}%
\AgdaBound{d₂}%
\>[11]\AgdaOperator{\AgdaFunction{⟧}}\AgdaSpace{}%
\AgdaBound{X}%
\>[16]\AgdaSymbol{=}\AgdaSpace{}%
\AgdaOperator{\AgdaFunction{⟦}}\AgdaSpace{}%
\AgdaBound{d₁}\AgdaSpace{}%
\AgdaOperator{\AgdaFunction{⟧}}\AgdaSpace{}%
\AgdaBound{X}\AgdaSpace{}%
\AgdaOperator{\AgdaFunction{×}}\AgdaSpace{}%
\AgdaOperator{\AgdaFunction{⟦}}\AgdaSpace{}%
\AgdaBound{d₂}\AgdaSpace{}%
\AgdaOperator{\AgdaFunction{⟧}}\AgdaSpace{}%
\AgdaBound{X}\<%
\\
\>[0]\AgdaOperator{\AgdaFunction{⟦}}\AgdaSpace{}%
\AgdaInductiveConstructor{∐}\AgdaSpace{}%
\AgdaBound{n}\AgdaSpace{}%
\AgdaBound{f}%
\>[11]\AgdaOperator{\AgdaFunction{⟧}}\AgdaSpace{}%
\AgdaBound{X}%
\>[16]\AgdaSymbol{=}\AgdaSpace{}%
\AgdaFunction{Σ[}\AgdaSpace{}%
\AgdaBound{m}\AgdaSpace{}%
\AgdaFunction{∈}\AgdaSpace{}%
\AgdaDatatype{Fin}\AgdaSpace{}%
\AgdaBound{n}\AgdaSpace{}%
\AgdaFunction{]}\AgdaSpace{}%
\AgdaSymbol{(}\AgdaOperator{\AgdaFunction{⟦}}\AgdaSpace{}%
\AgdaBound{f}\AgdaSpace{}%
\AgdaBound{m}\AgdaSpace{}%
\AgdaOperator{\AgdaFunction{⟧}}\AgdaSpace{}%
\AgdaBound{X}\AgdaSymbol{)}\<%
\end{code}
The following function defines the \emph{action on morphisms} of the functor \af{⟦}~\ab{d}~\af{⟧}.
That is, how morphisms $X \to Y$ map to morphisms \af{⟦}~\ab{d}~\af{⟧}~\ab{X}~\as{→}~\af{⟦}~\ab{d}~\af{⟧}~\ab{Y}, preserving the structure described by $d$:
\begin{code}%
\>[0]\AgdaFunction{fmap}%
\>[6]\AgdaSymbol{:}\AgdaSpace{}%
\AgdaSymbol{\{}\AgdaBound{d}\AgdaSpace{}%
\AgdaSymbol{:}\AgdaSpace{}%
\AgdaDatatype{Desc}\AgdaSymbol{\}}\AgdaSpace{}%
\AgdaSymbol{\{}\AgdaBound{X}\AgdaSpace{}%
\AgdaBound{Y}\AgdaSpace{}%
\AgdaSymbol{:}\AgdaSpace{}%
\AgdaPrimitive{Set}\AgdaSymbol{\}}\<%
\\
\>[6]\AgdaSymbol{→}\AgdaSpace{}%
\AgdaSymbol{(}\AgdaBound{X}\AgdaSpace{}%
\AgdaSymbol{→}\AgdaSpace{}%
\AgdaBound{Y}\AgdaSymbol{)}\AgdaSpace{}%
\AgdaSymbol{→}\AgdaSpace{}%
\AgdaOperator{\AgdaFunction{⟦}}\AgdaSpace{}%
\AgdaBound{d}\AgdaSpace{}%
\AgdaOperator{\AgdaFunction{⟧}}\AgdaSpace{}%
\AgdaBound{X}\AgdaSpace{}%
\AgdaSymbol{→}\AgdaSpace{}%
\AgdaOperator{\AgdaFunction{⟦}}\AgdaSpace{}%
\AgdaBound{d}\AgdaSpace{}%
\AgdaOperator{\AgdaFunction{⟧}}\AgdaSpace{}%
\AgdaBound{Y}\<%
\\
\>[0]\AgdaFunction{fmap}\AgdaSpace{}%
\AgdaSymbol{\{}\AgdaInductiveConstructor{I}%
\>[13]\AgdaSymbol{\}}\AgdaSpace{}%
\AgdaBound{f}\AgdaSpace{}%
\AgdaBound{x}%
\>[26]\AgdaSymbol{=}\AgdaSpace{}%
\AgdaBound{f}\AgdaSpace{}%
\AgdaBound{x}\<%
\\
\>[0]\AgdaFunction{fmap}\AgdaSpace{}%
\AgdaSymbol{\{}\AgdaInductiveConstructor{K}\AgdaSpace{}%
\AgdaBound{c}%
\>[13]\AgdaSymbol{\}}\AgdaSpace{}%
\AgdaBound{f}\AgdaSpace{}%
\AgdaBound{x}%
\>[26]\AgdaSymbol{=}\AgdaSpace{}%
\AgdaBound{x}\<%
\\
\>[0]\AgdaFunction{fmap}\AgdaSpace{}%
\AgdaSymbol{\{}\AgdaBound{d}\AgdaSpace{}%
\AgdaOperator{\AgdaInductiveConstructor{ẋ}}\AgdaSpace{}%
\AgdaBound{d₁}\AgdaSpace{}%
\AgdaSymbol{\}}\AgdaSpace{}%
\AgdaBound{f}\AgdaSpace{}%
\AgdaSymbol{(}\AgdaBound{x}\AgdaSpace{}%
\AgdaOperator{\AgdaInductiveConstructor{,}}\AgdaSpace{}%
\AgdaBound{y}\AgdaSymbol{)}%
\>[26]\AgdaSymbol{=}\AgdaSpace{}%
\AgdaFunction{fmap}\AgdaSpace{}%
\AgdaSymbol{\{}\AgdaBound{d}\AgdaSymbol{\}}\AgdaSpace{}%
\AgdaBound{f}\AgdaSpace{}%
\AgdaBound{x}\AgdaSpace{}%
\AgdaOperator{\AgdaInductiveConstructor{,}}\AgdaSpace{}%
\AgdaFunction{fmap}\AgdaSpace{}%
\AgdaSymbol{\{}\AgdaBound{d₁}\AgdaSymbol{\}}\AgdaSpace{}%
\AgdaBound{f}\AgdaSpace{}%
\AgdaBound{y}\<%
\\
\>[0]\AgdaFunction{fmap}\AgdaSpace{}%
\AgdaSymbol{\{}\AgdaInductiveConstructor{∐}\AgdaSpace{}%
\AgdaBound{n}\AgdaSpace{}%
\AgdaBound{g}%
\>[13]\AgdaSymbol{\}}\AgdaSpace{}%
\AgdaBound{f}\AgdaSpace{}%
\AgdaSymbol{(}\AgdaBound{m}\AgdaSpace{}%
\AgdaOperator{\AgdaInductiveConstructor{,}}\AgdaSpace{}%
\AgdaBound{x}\AgdaSymbol{)}%
\>[26]\AgdaSymbol{=}\AgdaSpace{}%
\AgdaBound{m}\AgdaSpace{}%
\AgdaOperator{\AgdaInductiveConstructor{,}}\AgdaSpace{}%
\AgdaFunction{fmap}\AgdaSpace{}%
\AgdaSymbol{\{}\AgdaBound{g}\AgdaSpace{}%
\AgdaBound{m}\AgdaSymbol{\}}\AgdaSpace{}%
\AgdaBound{f}\AgdaSpace{}%
\AgdaBound{x}\<%
\end{code}
The functorial action respects the \emph{functor laws}.
In other words, identity morphisms are mapped to identity morphisms.\footnote{The \as{\{}\ab{d}\as{\}} in \af{fmap}~\as{\{}\ab{d}\as{\}} makes explicit to Agda that the description to be used for the function application of \af{fmap} is the description value \ab{d}.  Making implicit arguments explicit in this way is sometimes necessary when Agda cannot uniquely determine what the implicit argument should be.}
\begin{code}%
\>[0]\AgdaFunction{fmap-id}%
\>[160I]\AgdaSymbol{:}\AgdaSpace{}%
\AgdaSymbol{∀}\AgdaSpace{}%
\AgdaBound{d}\AgdaSpace{}%
\AgdaSymbol{\{}\AgdaBound{X}\AgdaSymbol{\}}\AgdaSpace{}%
\AgdaSymbol{(}\AgdaBound{x}\AgdaSpace{}%
\AgdaSymbol{:}\AgdaSpace{}%
\AgdaOperator{\AgdaFunction{⟦}}\AgdaSpace{}%
\AgdaBound{d}\AgdaSpace{}%
\AgdaOperator{\AgdaFunction{⟧}}\AgdaSpace{}%
\AgdaBound{X}\AgdaSymbol{)}\<%
\\
\>[.][@{}l@{}]\<[160I]%
\>[8]\AgdaSymbol{→}\AgdaSpace{}%
\AgdaFunction{fmap}\AgdaSpace{}%
\AgdaSymbol{\{}\AgdaBound{d}\AgdaSymbol{\}}\AgdaSpace{}%
\AgdaFunction{id}\AgdaSpace{}%
\AgdaBound{x}\AgdaSpace{}%
\AgdaOperator{\AgdaDatatype{≡}}\AgdaSpace{}%
\AgdaFunction{id}\AgdaSpace{}%
\AgdaBound{x}\<%
\end{code}
\begin{code}[hide]%
\>[0]\AgdaFunction{fmap-id}\AgdaSpace{}%
\AgdaInductiveConstructor{I}\AgdaSpace{}%
\AgdaBound{x}\AgdaSpace{}%
\AgdaSymbol{=}\AgdaSpace{}%
\AgdaInductiveConstructor{refl}\<%
\\
\>[0]\AgdaFunction{fmap-id}\AgdaSpace{}%
\AgdaSymbol{(}\AgdaInductiveConstructor{K}\AgdaSpace{}%
\AgdaBound{x₁}\AgdaSymbol{)}\AgdaSpace{}%
\AgdaBound{x}\AgdaSpace{}%
\AgdaSymbol{=}\AgdaSpace{}%
\AgdaInductiveConstructor{refl}\<%
\\
\>[0]\AgdaFunction{fmap-id}\AgdaSpace{}%
\AgdaSymbol{(}\AgdaBound{d₀}\AgdaSpace{}%
\AgdaOperator{\AgdaInductiveConstructor{ẋ}}\AgdaSpace{}%
\AgdaBound{d₁}\AgdaSymbol{)}\AgdaSpace{}%
\AgdaSymbol{(}\AgdaBound{x₀}\AgdaSpace{}%
\AgdaOperator{\AgdaInductiveConstructor{,}}\AgdaSpace{}%
\AgdaBound{x₁}\AgdaSymbol{)}\AgdaSpace{}%
\AgdaSymbol{=}\AgdaSpace{}%
\AgdaFunction{cong₂}\AgdaSpace{}%
\AgdaOperator{\AgdaInductiveConstructor{\AgdaUnderscore{},\AgdaUnderscore{}}}\AgdaSpace{}%
\AgdaSymbol{(}\AgdaFunction{fmap-id}\AgdaSpace{}%
\AgdaBound{d₀}\AgdaSpace{}%
\AgdaBound{x₀}\AgdaSymbol{)}\AgdaSpace{}%
\AgdaSymbol{(}\AgdaFunction{fmap-id}\AgdaSpace{}%
\AgdaBound{d₁}\AgdaSpace{}%
\AgdaBound{x₁}\AgdaSymbol{)}\<%
\\
\>[0]\AgdaFunction{fmap-id}\AgdaSpace{}%
\AgdaSymbol{(}\AgdaInductiveConstructor{∐}\AgdaSpace{}%
\AgdaBound{n}\AgdaSpace{}%
\AgdaBound{g}\AgdaSymbol{)}\AgdaSpace{}%
\AgdaSymbol{(}\AgdaBound{m}\AgdaSpace{}%
\AgdaOperator{\AgdaInductiveConstructor{,}}\AgdaSpace{}%
\AgdaBound{x}\AgdaSymbol{)}\AgdaSpace{}%
\AgdaSymbol{=}\AgdaSpace{}%
\AgdaFunction{cong}\AgdaSpace{}%
\AgdaSymbol{(\AgdaUnderscore{}}\AgdaSpace{}%
\AgdaOperator{\AgdaInductiveConstructor{,\AgdaUnderscore{}}}\AgdaSymbol{)}\AgdaSpace{}%
\AgdaSymbol{(}\AgdaFunction{fmap-id}\AgdaSpace{}%
\AgdaSymbol{(}\AgdaBound{g}\AgdaSpace{}%
\AgdaBound{m}\AgdaSymbol{)}\AgdaSpace{}%
\AgdaBound{x}\AgdaSymbol{)}\<%
\end{code}
And mapping a composed function is the same as composing the mapped functions.
\begin{code}%
\>[0]\AgdaFunction{fmap-∘}%
\>[215I]\AgdaSymbol{:}\AgdaSpace{}%
\AgdaSymbol{∀}\AgdaSpace{}%
\AgdaBound{d}\AgdaSpace{}%
\AgdaSymbol{\{}\AgdaBound{X}\AgdaSpace{}%
\AgdaBound{Y}\AgdaSpace{}%
\AgdaBound{Z}\AgdaSymbol{\}}\AgdaSpace{}%
\AgdaSymbol{(}\AgdaBound{f}\AgdaSpace{}%
\AgdaSymbol{:}\AgdaSpace{}%
\AgdaBound{X}\AgdaSpace{}%
\AgdaSymbol{→}\AgdaSpace{}%
\AgdaBound{Y}\AgdaSymbol{)}\AgdaSpace{}%
\AgdaSymbol{(}\AgdaBound{g}\AgdaSpace{}%
\AgdaSymbol{:}\AgdaSpace{}%
\AgdaBound{Y}\AgdaSpace{}%
\AgdaSymbol{→}\AgdaSpace{}%
\AgdaBound{Z}\AgdaSymbol{)}\AgdaSpace{}%
\AgdaSymbol{(}\AgdaBound{x}\AgdaSpace{}%
\AgdaSymbol{:}\AgdaSpace{}%
\AgdaOperator{\AgdaFunction{⟦}}\AgdaSpace{}%
\AgdaBound{d}\AgdaSpace{}%
\AgdaOperator{\AgdaFunction{⟧}}\AgdaSpace{}%
\AgdaBound{X}\AgdaSymbol{)}\<%
\\
\>[.][@{}l@{}]\<[215I]%
\>[7]\AgdaSymbol{→}\AgdaSpace{}%
\AgdaFunction{fmap}\AgdaSpace{}%
\AgdaSymbol{\{}\AgdaBound{d}\AgdaSymbol{\}}\AgdaSpace{}%
\AgdaSymbol{(}\AgdaBound{g}\AgdaSpace{}%
\AgdaOperator{\AgdaFunction{∘}}\AgdaSpace{}%
\AgdaBound{f}\AgdaSymbol{)}\AgdaSpace{}%
\AgdaBound{x}\AgdaSpace{}%
\AgdaOperator{\AgdaDatatype{≡}}\AgdaSpace{}%
\AgdaSymbol{(}\AgdaFunction{fmap}\AgdaSpace{}%
\AgdaSymbol{\{}\AgdaBound{d}\AgdaSymbol{\}}\AgdaSpace{}%
\AgdaBound{g}\AgdaSpace{}%
\AgdaOperator{\AgdaFunction{∘}}\AgdaSpace{}%
\AgdaFunction{fmap}\AgdaSpace{}%
\AgdaSymbol{\{}\AgdaBound{d}\AgdaSymbol{\}}\AgdaSpace{}%
\AgdaBound{f}\AgdaSymbol{)}\AgdaSpace{}%
\AgdaBound{x}\<%
\end{code}
\begin{code}[hide]%
\>[0]\AgdaFunction{fmap-∘}\AgdaSpace{}%
\AgdaInductiveConstructor{I}\AgdaSpace{}%
\AgdaBound{f}\AgdaSpace{}%
\AgdaBound{g}\AgdaSpace{}%
\AgdaBound{x}\AgdaSpace{}%
\AgdaSymbol{=}\AgdaSpace{}%
\AgdaInductiveConstructor{refl}\<%
\\
\>[0]\AgdaFunction{fmap-∘}\AgdaSpace{}%
\AgdaSymbol{(}\AgdaInductiveConstructor{K}\AgdaSpace{}%
\AgdaBound{x₁}\AgdaSymbol{)}\AgdaSpace{}%
\AgdaBound{f}\AgdaSpace{}%
\AgdaBound{g}\AgdaSpace{}%
\AgdaBound{x}\AgdaSpace{}%
\AgdaSymbol{=}\AgdaSpace{}%
\AgdaInductiveConstructor{refl}\<%
\\
\>[0]\AgdaFunction{fmap-∘}\AgdaSpace{}%
\AgdaSymbol{(}\AgdaBound{d₀}\AgdaSpace{}%
\AgdaOperator{\AgdaInductiveConstructor{ẋ}}\AgdaSpace{}%
\AgdaBound{d₁}\AgdaSymbol{)}\AgdaSpace{}%
\AgdaBound{f}\AgdaSpace{}%
\AgdaBound{g}\AgdaSpace{}%
\AgdaSymbol{(}\AgdaBound{x₀}\AgdaSpace{}%
\AgdaOperator{\AgdaInductiveConstructor{,}}\AgdaSpace{}%
\AgdaBound{x₁}\AgdaSymbol{)}\AgdaSpace{}%
\AgdaSymbol{=}\AgdaSpace{}%
\AgdaFunction{cong₂}\AgdaSpace{}%
\AgdaOperator{\AgdaInductiveConstructor{\AgdaUnderscore{},\AgdaUnderscore{}}}\AgdaSpace{}%
\AgdaSymbol{(}\AgdaFunction{fmap-∘}\AgdaSpace{}%
\AgdaBound{d₀}\AgdaSpace{}%
\AgdaBound{f}\AgdaSpace{}%
\AgdaBound{g}\AgdaSpace{}%
\AgdaBound{x₀}\AgdaSymbol{)}\AgdaSpace{}%
\AgdaSymbol{(}\AgdaFunction{fmap-∘}\AgdaSpace{}%
\AgdaBound{d₁}\AgdaSpace{}%
\AgdaBound{f}\AgdaSpace{}%
\AgdaBound{g}\AgdaSpace{}%
\AgdaBound{x₁}\AgdaSymbol{)}\<%
\\
\>[0]\AgdaFunction{fmap-∘}\AgdaSpace{}%
\AgdaSymbol{(}\AgdaInductiveConstructor{∐}\AgdaSpace{}%
\AgdaBound{n}\AgdaSpace{}%
\AgdaBound{h}\AgdaSymbol{)}\AgdaSpace{}%
\AgdaBound{f}\AgdaSpace{}%
\AgdaBound{g}\AgdaSpace{}%
\AgdaSymbol{(}\AgdaBound{m}\AgdaSpace{}%
\AgdaOperator{\AgdaInductiveConstructor{,}}\AgdaSpace{}%
\AgdaBound{x}\AgdaSymbol{)}\AgdaSpace{}%
\AgdaSymbol{=}\AgdaSpace{}%
\AgdaFunction{cong}\AgdaSpace{}%
\AgdaSymbol{(\AgdaUnderscore{}}\AgdaSpace{}%
\AgdaOperator{\AgdaInductiveConstructor{,\AgdaUnderscore{}}}\AgdaSymbol{)}\AgdaSpace{}%
\AgdaSymbol{(}\AgdaFunction{fmap-∘}\AgdaSpace{}%
\AgdaSymbol{(}\AgdaBound{h}\AgdaSpace{}%
\AgdaBound{m}\AgdaSymbol{)}\AgdaSpace{}%
\AgdaBound{f}\AgdaSpace{}%
\AgdaBound{g}\AgdaSpace{}%
\AgdaBound{x}\AgdaSymbol{)}\<%
\end{code}

\begin{example}
Returning to our earlier example, we can now interpret the syntax of natural numbers as a \emph{functor} on \ad{Set}.
For example, the value representing zero is an element of \af{⟦}~\af{NatDesc}~\af{⟧}~\ab{X}, for any type \ab{X}:
\begin{code}%
\>[0]\AgdaFunction{zeroN}\AgdaSpace{}%
\AgdaSymbol{:}\AgdaSpace{}%
\AgdaSymbol{\{}\AgdaBound{X}\AgdaSpace{}%
\AgdaSymbol{:}\AgdaSpace{}%
\AgdaPrimitive{Set}\AgdaSymbol{\}}\AgdaSpace{}%
\AgdaSymbol{→}\AgdaSpace{}%
\AgdaOperator{\AgdaFunction{⟦}}\AgdaSpace{}%
\AgdaFunction{NatDesc}\AgdaSpace{}%
\AgdaOperator{\AgdaFunction{⟧}}\AgdaSpace{}%
\AgdaBound{X}\<%
\\
\>[0]\AgdaFunction{zeroN}\AgdaSpace{}%
\AgdaSymbol{=}\AgdaSpace{}%
\AgdaInductiveConstructor{zero}\AgdaSpace{}%
\AgdaOperator{\AgdaInductiveConstructor{,}}\AgdaSpace{}%
\AgdaInductiveConstructor{tt}\<%
\end{code}
Given an element \ab{x} of type \ab{X}, we can also construct a successor value:
\begin{code}%
\>[0]\AgdaFunction{sucN}\AgdaSpace{}%
\AgdaSymbol{:}\AgdaSpace{}%
\AgdaSymbol{\{}\AgdaBound{X}\AgdaSpace{}%
\AgdaSymbol{:}\AgdaSpace{}%
\AgdaPrimitive{Set}\AgdaSymbol{\}}\AgdaSpace{}%
\AgdaSymbol{→}\AgdaSpace{}%
\AgdaBound{X}\AgdaSpace{}%
\AgdaSymbol{→}\AgdaSpace{}%
\AgdaOperator{\AgdaFunction{⟦}}\AgdaSpace{}%
\AgdaFunction{NatDesc}\AgdaSpace{}%
\AgdaOperator{\AgdaFunction{⟧}}\AgdaSpace{}%
\AgdaBound{X}\<%
\\
\>[0]\AgdaFunction{sucN}\AgdaSpace{}%
\AgdaBound{x}\AgdaSpace{}%
\AgdaSymbol{=}\AgdaSpace{}%
\AgdaInductiveConstructor{suc}\AgdaSpace{}%
\AgdaInductiveConstructor{zero}\AgdaSpace{}%
\AgdaOperator{\AgdaInductiveConstructor{,}}\AgdaSpace{}%
\AgdaBound{x}\<%
\end{code}
\end{example}

The examples above suggest a recursive structure: we are building values of type \af{⟦}~\af{NatDesc}~\af{⟧}~\ab{X} from elements of \ab{X}, and would like to find a type \ab{X} that is \emph{isomorphic} to its image under the functor \af{⟦}~\af{NatDesc}~\af{⟧}. That is, we are looking for a solution to the recursive domain equation:
\[
\ab{X} \simeq \af{⟦}~\af{NatDesc}~\af{⟧}~\ab{X}
\]
In category theory, such a solution is given by the \emph{initial algebra} of the functor \af{⟦}~\ab{d}~\af{⟧}, which corresponds to its \emph{least fixed point}.
We define this in Agda by introducing a type \ad{μ[}~\ab{d}~\ad{]} representing the initial algebra of the functor \af{⟦}~\ab{d}~\af{⟧}.
\begin{code}%
\>[0]\AgdaKeyword{data}\AgdaSpace{}%
\AgdaOperator{\AgdaDatatype{μ[\AgdaUnderscore{}]}}\AgdaSpace{}%
\AgdaSymbol{(}\AgdaBound{d}\AgdaSpace{}%
\AgdaSymbol{:}\AgdaSpace{}%
\AgdaDatatype{Desc}\AgdaSymbol{)}\AgdaSpace{}%
\AgdaSymbol{:}\AgdaSpace{}%
\AgdaPrimitive{Set}\AgdaSpace{}%
\AgdaKeyword{where}\<%
\\
\>[0][@{}l@{\AgdaIndent{0}}]%
\>[2]\AgdaOperator{\AgdaInductiveConstructor{⟨\AgdaUnderscore{}⟩}}\AgdaSpace{}%
\AgdaSymbol{:}\AgdaSpace{}%
\AgdaOperator{\AgdaFunction{⟦}}\AgdaSpace{}%
\AgdaBound{d}\AgdaSpace{}%
\AgdaOperator{\AgdaFunction{⟧}}\AgdaSpace{}%
\AgdaSymbol{(}\AgdaOperator{\AgdaDatatype{μ[}}\AgdaSpace{}%
\AgdaBound{d}\AgdaSpace{}%
\AgdaOperator{\AgdaDatatype{]}}\AgdaSymbol{)}\AgdaSpace{}%
\AgdaSymbol{→}\AgdaSpace{}%
\AgdaOperator{\AgdaDatatype{μ[}}\AgdaSpace{}%
\AgdaBound{d}\AgdaSpace{}%
\AgdaOperator{\AgdaDatatype{]}}\<%
\end{code}
This type provides a canonical solution to the recursive domain equation $\ab{X} \simeq \af{⟦}~\ab{d}~\af{⟧}~\ab{X}$ in the sense that the constructor \ac{⟨\_⟩}~\as{:}~\af{⟦}~\ab{d}~\af{⟧}~\as{(}~\ad{μ[}~\ab{d}~\ad{]}\as{)~→}~\ad{μ[}~\ab{d}~\ad{]} and the function
\begin{code}%
\>[0]\AgdaFunction{⟨⟩ᵒ}\AgdaSpace{}%
\AgdaSymbol{:}\AgdaSpace{}%
\AgdaSymbol{\{}\AgdaBound{d}\AgdaSpace{}%
\AgdaSymbol{:}\AgdaSpace{}%
\AgdaDatatype{Desc}\AgdaSymbol{\}}\AgdaSpace{}%
\AgdaSymbol{→}\AgdaSpace{}%
\AgdaOperator{\AgdaDatatype{μ[}}\AgdaSpace{}%
\AgdaBound{d}\AgdaSpace{}%
\AgdaOperator{\AgdaDatatype{]}}\AgdaSpace{}%
\AgdaSymbol{→}\AgdaSpace{}%
\AgdaOperator{\AgdaFunction{⟦}}\AgdaSpace{}%
\AgdaBound{d}\AgdaSpace{}%
\AgdaOperator{\AgdaFunction{⟧}}\AgdaSpace{}%
\AgdaOperator{\AgdaDatatype{μ[}}\AgdaSpace{}%
\AgdaBound{d}\AgdaSpace{}%
\AgdaOperator{\AgdaDatatype{]}}\<%
\\
\>[0]\AgdaFunction{⟨⟩ᵒ}\AgdaSpace{}%
\AgdaOperator{\AgdaInductiveConstructor{⟨}}\AgdaSpace{}%
\AgdaBound{x}\AgdaSpace{}%
\AgdaOperator{\AgdaInductiveConstructor{⟩}}\AgdaSpace{}%
\AgdaSymbol{=}\AgdaSpace{}%
\AgdaBound{x}\<%
\end{code}
each witness one direction of the isomorphism $\ad{μ[}~\ab{d}~\ad{]} \simeq \af{⟦}~\ab{d}~\af{⟧}~\as{(}\ad{μ[}~\ab{d}~\ad{]}\as{)}$ in \ad{Set}.
This tells us that \ad{μ[}~\ab{d}~\ad{]} is a fixed point of the functor \af{⟦}~\ab{d}~\af{⟧}.

\begin{example}
The simple arithmetic language from the previous section can also be encoded as a description:
\begin{code}%
\>[0]\AgdaFunction{HR}\AgdaSpace{}%
\AgdaSymbol{:}\AgdaSpace{}%
\AgdaDatatype{Desc}\<%
\\
\>[0]\AgdaFunction{HR}\AgdaSpace{}%
\AgdaSymbol{=}\AgdaSpace{}%
\AgdaInductiveConstructor{∐}\AgdaSpace{}%
\AgdaNumber{2}\AgdaSpace{}%
\AgdaSymbol{(λ}\AgdaSpace{}%
\AgdaKeyword{where}\<%
\\
\>[0][@{}l@{\AgdaIndent{0}}]%
\>[2]\AgdaInductiveConstructor{zero}%
\>[14]\AgdaSymbol{→}\AgdaSpace{}%
\AgdaInductiveConstructor{K}\AgdaSpace{}%
\AgdaDatatype{ℕ}\<%
\\
\>[2]\AgdaSymbol{(}\AgdaInductiveConstructor{suc}\AgdaSpace{}%
\AgdaInductiveConstructor{zero}\AgdaSymbol{)}%
\>[14]\AgdaSymbol{→}\AgdaSpace{}%
\AgdaInductiveConstructor{I}\AgdaSpace{}%
\AgdaOperator{\AgdaInductiveConstructor{ẋ}}\AgdaSpace{}%
\AgdaInductiveConstructor{I}\AgdaSymbol{)}\<%
\end{code}
The following declare syntactic shorthands for constructing number and addition expressions, respectively:\footnote{We use Agda's pattern synonyms: \url{https://agda.readthedocs.io/en/latest/language/pattern-synonyms.html}}
\begin{code}%
\>[0]\AgdaKeyword{pattern}\AgdaSpace{}%
\AgdaInductiveConstructor{`numF}\AgdaSpace{}%
\AgdaBound{n}%
\>[19]\AgdaSymbol{=}\AgdaSpace{}%
\AgdaInductiveConstructor{zero}\AgdaSpace{}%
\AgdaOperator{\AgdaInductiveConstructor{,}}\AgdaSpace{}%
\AgdaBound{n}\<%
\\
\>[0]\AgdaKeyword{pattern}\AgdaSpace{}%
\AgdaInductiveConstructor{`num}\AgdaSpace{}%
\AgdaBound{n}%
\>[19]\AgdaSymbol{=}\AgdaSpace{}%
\AgdaOperator{\AgdaInductiveConstructor{⟨}}\AgdaSpace{}%
\AgdaInductiveConstructor{`numF}\AgdaSpace{}%
\AgdaBound{n}\AgdaSpace{}%
\AgdaOperator{\AgdaInductiveConstructor{⟩}}\<%
\\
\>[0]\AgdaKeyword{pattern}\AgdaSpace{}%
\AgdaInductiveConstructor{`addF}\AgdaSpace{}%
\AgdaBound{x}\AgdaSpace{}%
\AgdaBound{y}%
\>[19]\AgdaSymbol{=}\AgdaSpace{}%
\AgdaInductiveConstructor{suc}\AgdaSpace{}%
\AgdaInductiveConstructor{zero}\AgdaSpace{}%
\AgdaOperator{\AgdaInductiveConstructor{,}}\AgdaSpace{}%
\AgdaBound{x}\AgdaSpace{}%
\AgdaOperator{\AgdaInductiveConstructor{,}}\AgdaSpace{}%
\AgdaBound{y}\<%
\\
\>[0]\AgdaKeyword{pattern}\AgdaSpace{}%
\AgdaInductiveConstructor{`add}\AgdaSpace{}%
\AgdaBound{x}\AgdaSpace{}%
\AgdaBound{y}%
\>[19]\AgdaSymbol{=}\AgdaSpace{}%
\AgdaOperator{\AgdaInductiveConstructor{⟨}}\AgdaSpace{}%
\AgdaInductiveConstructor{`addF}\AgdaSpace{}%
\AgdaBound{x}\AgdaSpace{}%
\AgdaBound{y}\AgdaSpace{}%
\AgdaOperator{\AgdaInductiveConstructor{⟩}}\<%
\end{code}%
\begin{code}[hide]%
\>[0]\<%
\end{code}
Using these, the following term encodes the expression $1+2$:
\begin{code}%
\>[0]\AgdaFunction{1+2}\AgdaSpace{}%
\AgdaSymbol{:}\AgdaSpace{}%
\AgdaOperator{\AgdaDatatype{μ[}}\AgdaSpace{}%
\AgdaFunction{HR}\AgdaSpace{}%
\AgdaOperator{\AgdaDatatype{]}}\<%
\\
\>[0]\AgdaFunction{1+2}\AgdaSpace{}%
\AgdaSymbol{=}\AgdaSpace{}%
\AgdaInductiveConstructor{`add}\AgdaSpace{}%
\AgdaSymbol{(}\AgdaInductiveConstructor{`num}\AgdaSpace{}%
\AgdaNumber{1}\AgdaSymbol{)}\AgdaSpace{}%
\AgdaSymbol{(}\AgdaInductiveConstructor{`num}\AgdaSpace{}%
\AgdaNumber{2}\AgdaSymbol{)}\<%
\end{code}
\label{ex:hr}
\end{example}

\subsection{Algebras of Functors}

Given a functor \af{⟦}~\ab{d}~\af{⟧}~:~\ad{Set}~\as{→}~\ad{Set}, we can define what it means to be an \emph{algebra} for this functor.
An \af{⟦}~\ab{d}~\af{⟧}-algebra consists of (1) a carrier set \ab{A}, and (2) a function \ab{α}~\as{:}~\af{⟦}~\ab{d}~\af{⟧}~\ab{A}~\as{→}~\ab{A}.

Intuitively, the functor \af{⟦}~\ab{d}~\af{⟧} describes the shape of one layer of structure, and the algebra map \ab{α} tells us how to reduce or interpret that structure into a value of type \ab{A}.

We can define this in Agda as a record:\footnote{The carrier set of the algebra is a record type argument rather than record field.  Having it as an argument is more convenient for working with algebras with specific carrier sets in Agda.}
\begin{code}%
\>[0]\AgdaKeyword{record}\AgdaSpace{}%
\AgdaRecord{Alg}\AgdaSpace{}%
\AgdaSymbol{(}\AgdaBound{F}\AgdaSpace{}%
\AgdaSymbol{:}\AgdaSpace{}%
\AgdaPrimitive{Set}\AgdaSpace{}%
\AgdaSymbol{→}\AgdaSpace{}%
\AgdaPrimitive{Set}\AgdaSymbol{)}\AgdaSpace{}%
\AgdaSymbol{(}\AgdaBound{A}\AgdaSpace{}%
\AgdaSymbol{:}\AgdaSpace{}%
\AgdaPrimitive{Set}\AgdaSymbol{)}\AgdaSpace{}%
\AgdaSymbol{:}\AgdaSpace{}%
\AgdaPrimitive{Set}\AgdaSpace{}%
\AgdaKeyword{where}\<%
\\
\>[0][@{}l@{\AgdaIndent{0}}]%
\>[2]\AgdaKeyword{field}\AgdaSpace{}%
\AgdaField{alg}\AgdaSpace{}%
\AgdaSymbol{:}\AgdaSpace{}%
\AgdaBound{F}\AgdaSpace{}%
\AgdaBound{A}\AgdaSpace{}%
\AgdaSymbol{→}\AgdaSpace{}%
\AgdaBound{A}\<%
\end{code}
\begin{code}[hide]%
\>[0]\AgdaKeyword{open}\AgdaSpace{}%
\AgdaModule{Alg}\<%
\end{code}

Given an \af{⟦}~\ab{d}~\af{⟧}-algebra with carrier \ab{A} and structure map \ab{α}~\as{:}~\af{⟦}~\ab{d}~\af{⟧}~\ab{A}~\as{→}~\ab{A}, the \emph{catamorphism} induced by \ab{α} is a function that recursively collapses a value of type \ad{μ[}~\ab{d}~\ad{]} into a value of type \ab{A}, by interpreting each layer using \ab{α}.

In Agda, we define the catamorphism as follows (our notation is inspired by~\citet{MeijerFP91}):
\begin{code}%
\>[0]\AgdaSymbol{\{-\#}\AgdaSpace{}%
\AgdaKeyword{TERMINATING}\AgdaSpace{}%
\AgdaSymbol{\#-\}}\<%
\\
\>[0]\AgdaOperator{\AgdaFunction{⦉\AgdaUnderscore{}∼\AgdaUnderscore{}⦊}}\AgdaSpace{}%
\AgdaSymbol{:}\AgdaSpace{}%
\AgdaSymbol{\{}\AgdaBound{A}\AgdaSpace{}%
\AgdaSymbol{:}\AgdaSpace{}%
\AgdaPrimitive{Set}\AgdaSymbol{\}}\AgdaSpace{}%
\AgdaSymbol{(}\AgdaBound{d}\AgdaSpace{}%
\AgdaSymbol{:}\AgdaSpace{}%
\AgdaDatatype{Desc}\AgdaSymbol{)}\AgdaSpace{}%
\AgdaSymbol{→}\AgdaSpace{}%
\AgdaRecord{Alg}\AgdaSpace{}%
\AgdaOperator{\AgdaFunction{⟦}}\AgdaSpace{}%
\AgdaBound{d}\AgdaSpace{}%
\AgdaOperator{\AgdaFunction{⟧}}\AgdaSpace{}%
\AgdaBound{A}\AgdaSpace{}%
\AgdaSymbol{→}\AgdaSpace{}%
\AgdaOperator{\AgdaDatatype{μ[}}\AgdaSpace{}%
\AgdaBound{d}\AgdaSpace{}%
\AgdaOperator{\AgdaDatatype{]}}\AgdaSpace{}%
\AgdaSymbol{→}\AgdaSpace{}%
\AgdaBound{A}\<%
\\
\>[0]\AgdaOperator{\AgdaFunction{⦉}}\AgdaSpace{}%
\AgdaBound{d}\AgdaSpace{}%
\AgdaOperator{\AgdaFunction{∼}}\AgdaSpace{}%
\AgdaBound{a}\AgdaSpace{}%
\AgdaOperator{\AgdaFunction{⦊}}\AgdaSpace{}%
\AgdaSymbol{=}\AgdaSpace{}%
\AgdaField{alg}\AgdaSpace{}%
\AgdaBound{a}\AgdaSpace{}%
\AgdaOperator{\AgdaFunction{∘}}\AgdaSpace{}%
\AgdaFunction{fmap}\AgdaSpace{}%
\AgdaSymbol{\{}\AgdaBound{d}\AgdaSymbol{\}}\AgdaSpace{}%
\AgdaOperator{\AgdaFunction{⦉}}\AgdaSpace{}%
\AgdaBound{d}\AgdaSpace{}%
\AgdaOperator{\AgdaFunction{∼}}\AgdaSpace{}%
\AgdaBound{a}\AgdaSpace{}%
\AgdaOperator{\AgdaFunction{⦊}}\AgdaSpace{}%
\AgdaOperator{\AgdaFunction{∘}}\AgdaSpace{}%
\AgdaFunction{⟨⟩ᵒ}\<%
\end{code}
Agda cannot see that this function is terminating.
We could work around this issue by inlining the definition of \af{fmap}, following McBride~\cite[\S{3}]{McBride2011ornamental}.
We will not belabor this point here.

Catamorphisms let us define simple recursive traversals over a structure.
However, for defining reduction semantics, we will make use of a slightly more general recursion scheme---namely \emph{paramorphisms}~\cite{Meertens92}---which lets us recursively traverse a structure where each recursive case may depend on the structure of the term being traversed.
That is, whereas catamorphic algebras have type \ad{⟦}~\ab{d}~\ad{⟧}~\ab{A}~\as{→}~\ab{A}, paramorphic algebras have type \ad{⟦}~\ab{d}~\ad{⟧}~\as{(}\ad{μ[}~\ab{d}~\ad{]}~\ad{×}~\ab{A}\as{)}~\as{→}~\ab{A}.
The principle is otherwise the same as with catamorphisms.
\begin{code}%
\>[0]\AgdaSymbol{\{-\#}\AgdaSpace{}%
\AgdaKeyword{TERMINATING}\AgdaSpace{}%
\AgdaSymbol{\#-\}}\<%
\\
\>[0]\AgdaOperator{\AgdaFunction{⅋【\AgdaUnderscore{}∼\AgdaUnderscore{}】}}\AgdaSpace{}%
\AgdaSymbol{:}%
\>[483I]\AgdaSymbol{\{}\AgdaBound{A}\AgdaSpace{}%
\AgdaSymbol{:}\AgdaSpace{}%
\AgdaPrimitive{Set}\AgdaSymbol{\}}\AgdaSpace{}%
\AgdaSymbol{(}\AgdaBound{d}\AgdaSpace{}%
\AgdaSymbol{:}\AgdaSpace{}%
\AgdaDatatype{Desc}\AgdaSymbol{)}\<%
\\
\>[.][@{}l@{}]\<[483I]%
\>[9]\AgdaSymbol{→}\AgdaSpace{}%
\AgdaRecord{Alg}\AgdaSpace{}%
\AgdaSymbol{(λ}\AgdaSpace{}%
\AgdaBound{X}\AgdaSpace{}%
\AgdaSymbol{→}\AgdaSpace{}%
\AgdaOperator{\AgdaFunction{⟦}}\AgdaSpace{}%
\AgdaBound{d}\AgdaSpace{}%
\AgdaOperator{\AgdaFunction{⟧}}\AgdaSpace{}%
\AgdaSymbol{(}\AgdaOperator{\AgdaDatatype{μ[}}\AgdaSpace{}%
\AgdaBound{d}\AgdaSpace{}%
\AgdaOperator{\AgdaDatatype{]}}\AgdaSpace{}%
\AgdaOperator{\AgdaFunction{×}}\AgdaSpace{}%
\AgdaBound{X}\AgdaSymbol{))}\AgdaSpace{}%
\AgdaBound{A}\AgdaSpace{}%
\AgdaSymbol{→}\AgdaSpace{}%
\AgdaOperator{\AgdaDatatype{μ[}}\AgdaSpace{}%
\AgdaBound{d}\AgdaSpace{}%
\AgdaOperator{\AgdaDatatype{]}}\AgdaSpace{}%
\AgdaSymbol{→}\AgdaSpace{}%
\AgdaBound{A}\<%
\\
\>[0]\AgdaOperator{\AgdaFunction{⅋【}}\AgdaSpace{}%
\AgdaBound{d}\AgdaSpace{}%
\AgdaOperator{\AgdaFunction{∼}}\AgdaSpace{}%
\AgdaBound{a}\AgdaSpace{}%
\AgdaOperator{\AgdaFunction{】}}\AgdaSpace{}%
\AgdaSymbol{=}\AgdaSpace{}%
\AgdaField{alg}\AgdaSpace{}%
\AgdaBound{a}\AgdaSpace{}%
\AgdaOperator{\AgdaFunction{∘}}\AgdaSpace{}%
\AgdaFunction{fmap}\AgdaSpace{}%
\AgdaSymbol{\{}\AgdaBound{d}\AgdaSymbol{\}}\AgdaSpace{}%
\AgdaSymbol{(λ}\AgdaSpace{}%
\AgdaBound{x}\AgdaSpace{}%
\AgdaSymbol{→}\AgdaSpace{}%
\AgdaSymbol{(}\AgdaBound{x}\AgdaSpace{}%
\AgdaOperator{\AgdaInductiveConstructor{,}}\AgdaSpace{}%
\AgdaOperator{\AgdaFunction{⅋【}}\AgdaSpace{}%
\AgdaBound{d}\AgdaSpace{}%
\AgdaOperator{\AgdaFunction{∼}}\AgdaSpace{}%
\AgdaBound{a}\AgdaSpace{}%
\AgdaOperator{\AgdaFunction{】}}\AgdaSpace{}%
\AgdaBound{x}\AgdaSymbol{))}\AgdaSpace{}%
\AgdaOperator{\AgdaFunction{∘}}\AgdaSpace{}%
\AgdaFunction{⟨⟩ᵒ}\<%
\end{code}
\begin{code}[hide]%
\>[0]\AgdaComment{--\ By\ pattern\ matching\ explicitly\ rather\ than\ implicitly,\ Agda\ unfolds\ the}\<%
\\
\>[0]\AgdaComment{--\ function\ more\ carefully\ than\ the\ definition\ above.\ \ Without\ the\ pattern}\<%
\\
\>[0]\AgdaComment{--\ match,\ Agda\ will\ sometimes\ try\ to\ unfold\ the\ paramorphism\ too\ eagerly\ and}\<%
\\
\>[0]\AgdaComment{--\ (seemingly)\ non-terminate...}\<%
\\
\>[0]\AgdaComment{--}\<%
\\
\>[0]\AgdaComment{--\ So\ the\ pragma\ is\ safe\ in\ theory,\ but\ in\ practice\ it\ affects\ how\ long\ it\ takes}\<%
\\
\>[0]\AgdaComment{--\ to\ type\ check\ programs.}\<%
\\
\>[0]\AgdaSymbol{\{-\#}\AgdaSpace{}%
\AgdaKeyword{TERMINATING}\AgdaSpace{}%
\AgdaSymbol{\#-\}}\<%
\\
\>[0]\AgdaOperator{\AgdaFunction{【\AgdaUnderscore{}∼\AgdaUnderscore{}】}}\AgdaSpace{}%
\AgdaSymbol{:}\AgdaSpace{}%
\AgdaSymbol{\{}\AgdaBound{A}\AgdaSpace{}%
\AgdaSymbol{:}\AgdaSpace{}%
\AgdaPrimitive{Set}\AgdaSymbol{\}}\AgdaSpace{}%
\AgdaSymbol{(}\AgdaBound{d}\AgdaSpace{}%
\AgdaSymbol{:}\AgdaSpace{}%
\AgdaDatatype{Desc}\AgdaSymbol{)}\AgdaSpace{}%
\AgdaSymbol{→}\AgdaSpace{}%
\AgdaRecord{Alg}\AgdaSpace{}%
\AgdaSymbol{(}\AgdaOperator{\AgdaFunction{⟦}}\AgdaSpace{}%
\AgdaBound{d}\AgdaSpace{}%
\AgdaOperator{\AgdaFunction{⟧}}\AgdaSpace{}%
\AgdaOperator{\AgdaFunction{∘}}\AgdaSpace{}%
\AgdaSymbol{(}\AgdaFunction{const}\AgdaSpace{}%
\AgdaOperator{\AgdaDatatype{μ[}}\AgdaSpace{}%
\AgdaBound{d}\AgdaSpace{}%
\AgdaOperator{\AgdaDatatype{]}}\AgdaSpace{}%
\AgdaOperator{\AgdaFunction{∩}}\AgdaSpace{}%
\AgdaFunction{id}\AgdaSymbol{))}\AgdaSpace{}%
\AgdaBound{A}\AgdaSpace{}%
\AgdaSymbol{→}\AgdaSpace{}%
\AgdaOperator{\AgdaDatatype{μ[}}\AgdaSpace{}%
\AgdaBound{d}\AgdaSpace{}%
\AgdaOperator{\AgdaDatatype{]}}\AgdaSpace{}%
\AgdaSymbol{→}\AgdaSpace{}%
\AgdaBound{A}\<%
\\
\>[0]\AgdaOperator{\AgdaFunction{【}}\AgdaSpace{}%
\AgdaBound{d}\AgdaSpace{}%
\AgdaOperator{\AgdaFunction{∼}}\AgdaSpace{}%
\AgdaBound{a}\AgdaSpace{}%
\AgdaOperator{\AgdaFunction{】}}\AgdaSpace{}%
\AgdaOperator{\AgdaInductiveConstructor{⟨}}\AgdaSpace{}%
\AgdaBound{x}\AgdaSpace{}%
\AgdaOperator{\AgdaInductiveConstructor{⟩}}\AgdaSpace{}%
\AgdaSymbol{=}\AgdaSpace{}%
\AgdaField{alg}\AgdaSpace{}%
\AgdaBound{a}\AgdaSpace{}%
\AgdaSymbol{(}\AgdaFunction{fmap}\AgdaSpace{}%
\AgdaSymbol{\{}\AgdaBound{d}\AgdaSymbol{\}}\AgdaSpace{}%
\AgdaSymbol{(λ}\AgdaSpace{}%
\AgdaBound{x}\AgdaSpace{}%
\AgdaSymbol{→}\AgdaSpace{}%
\AgdaSymbol{(}\AgdaBound{x}\AgdaSpace{}%
\AgdaOperator{\AgdaInductiveConstructor{,}}\AgdaSpace{}%
\AgdaOperator{\AgdaFunction{【}}\AgdaSpace{}%
\AgdaBound{d}\AgdaSpace{}%
\AgdaOperator{\AgdaFunction{∼}}\AgdaSpace{}%
\AgdaBound{a}\AgdaSpace{}%
\AgdaOperator{\AgdaFunction{】}}\AgdaSpace{}%
\AgdaBound{x}\AgdaSymbol{))}\AgdaSpace{}%
\AgdaBound{x}\AgdaSymbol{)}\<%
\end{code}%
In \cref{sec:04-redsem-gen} we show how this notion of paramorphism lets us define decomposition functions generically.
Before doing so, we first need a generic notion of context, which is given by the \emph{derivative} of a syntax.

\subsection{Contexts as the Derivative of a Syntax}
\label{sec:derivatives}

Following McBride~\cite{mcbride2001derivative,McBride08}, one-hole contexts (or \emph{zippers}~\cite{Huet97}) are equivalent to the derivative of a syntax, in the sense that it obeys similar rules as the well-known concept of derivative from calculus.

In calculus, the derivative tells us the slope of the tangent at a given point.  That is, the best linear predictor of how a function $f\ x$ changes at a given point.
In data types, the derivative gives us a one-hole context for a functor.  That is, the best linear predictor of how a container $F\ A$ changes when you tweak it at a given position.
For data types, this position is determined by a \emph{context}.
The analogy between derivatives from calculus and derivatives of data types is reflected in how we define derivatives.
For example, consider the rules of differentiation from calculus.
\begin{mathpar}
  \partial_x\ x = 1
  \and
  \partial_x\ c = 0
  \and
  \partial_x\ (a \times b) = \left(\partial_x\ a\right) \times b + a \times (\partial_x\ b)
  \and
  \partial_x\ (a + b) = \partial_x\ a + \partial_x\ b
\end{mathpar}
These rules are mirrored in the definition of the derivative of a described data type:\footnote{The types \af{\_∩\_} and \af{\_∪\_} are products and sums on type \ad{Set}~\as{→}~\ad{Set}.}
\begin{code}%
\>[0]\AgdaOperator{\AgdaFunction{∂[\AgdaUnderscore{}]}}\AgdaSpace{}%
\AgdaSymbol{:}\AgdaSpace{}%
\AgdaDatatype{Desc}\AgdaSpace{}%
\AgdaSymbol{→}\AgdaSpace{}%
\AgdaPrimitive{Set}\AgdaSpace{}%
\AgdaSymbol{→}\AgdaSpace{}%
\AgdaPrimitive{Set}\<%
\\
\>[0]\AgdaOperator{\AgdaFunction{∂[}}\AgdaSpace{}%
\AgdaInductiveConstructor{I}%
\>[12]\AgdaOperator{\AgdaFunction{]}}%
\>[16]\AgdaSymbol{\AgdaUnderscore{}}%
\>[19]\AgdaSymbol{=}\AgdaSpace{}%
\AgdaRecord{⊤}\<%
\\
\>[0]\AgdaOperator{\AgdaFunction{∂[}}\AgdaSpace{}%
\AgdaInductiveConstructor{K}\AgdaSpace{}%
\AgdaBound{Y}%
\>[12]\AgdaOperator{\AgdaFunction{]}}%
\>[16]\AgdaSymbol{\AgdaUnderscore{}}%
\>[19]\AgdaSymbol{=}\AgdaSpace{}%
\AgdaFunction{⊥}\<%
\\
\>[0]\AgdaOperator{\AgdaFunction{∂[}}\AgdaSpace{}%
\AgdaBound{d₀}\AgdaSpace{}%
\AgdaOperator{\AgdaInductiveConstructor{ẋ}}\AgdaSpace{}%
\AgdaBound{d₁}%
\>[12]\AgdaOperator{\AgdaFunction{]}}%
\>[19]\AgdaSymbol{=}\AgdaSpace{}%
\AgdaSymbol{(}\AgdaOperator{\AgdaFunction{∂[}}\AgdaSpace{}%
\AgdaBound{d₀}\AgdaSpace{}%
\AgdaOperator{\AgdaFunction{]}}\AgdaSpace{}%
\AgdaOperator{\AgdaFunction{∩}}\AgdaSpace{}%
\AgdaOperator{\AgdaFunction{⟦}}\AgdaSpace{}%
\AgdaBound{d₁}\AgdaSpace{}%
\AgdaOperator{\AgdaFunction{⟧}}\AgdaSymbol{)}\AgdaSpace{}%
\AgdaOperator{\AgdaFunction{∪}}\AgdaSpace{}%
\AgdaSymbol{(}\AgdaOperator{\AgdaFunction{⟦}}\AgdaSpace{}%
\AgdaBound{d₀}\AgdaSpace{}%
\AgdaOperator{\AgdaFunction{⟧}}\AgdaSpace{}%
\AgdaOperator{\AgdaFunction{∩}}\AgdaSpace{}%
\AgdaOperator{\AgdaFunction{∂[}}\AgdaSpace{}%
\AgdaBound{d₁}\AgdaSpace{}%
\AgdaOperator{\AgdaFunction{]}}\AgdaSymbol{)}\<%
\\
\>[0]\AgdaOperator{\AgdaFunction{∂[}}\AgdaSpace{}%
\AgdaInductiveConstructor{∐}\AgdaSpace{}%
\AgdaBound{n}\AgdaSpace{}%
\AgdaBound{f}%
\>[12]\AgdaOperator{\AgdaFunction{]}}%
\>[16]\AgdaBound{X}%
\>[19]\AgdaSymbol{=}\AgdaSpace{}%
\AgdaFunction{Σ[}\AgdaSpace{}%
\AgdaBound{i}\AgdaSpace{}%
\AgdaFunction{∈}\AgdaSpace{}%
\AgdaDatatype{Fin}\AgdaSpace{}%
\AgdaBound{n}\AgdaSpace{}%
\AgdaFunction{]}\AgdaSpace{}%
\AgdaSymbol{(}\AgdaOperator{\AgdaFunction{∂[}}\AgdaSpace{}%
\AgdaBound{f}\AgdaSpace{}%
\AgdaBound{i}\AgdaSpace{}%
\AgdaOperator{\AgdaFunction{]}}\AgdaSpace{}%
\AgdaBound{X}\AgdaSymbol{)}\<%
\end{code}
As the type suggests we define the derivative of a data type as an endofunctor on \ad{Set}.
We can think of these functors as defining the context of a single AST node---a \emph{context frame}.
The case for \ac{I} says that the context frame of a non-terminal is given by a single hole.
The case for \ac{K} says that terminals cannot have holes.
For adjacent syntax tree nodes (\ab{d₀}~\ac{ẋ}~\ab{d₁}), we either have a hole on the left and a syntax tree on the right (\af{∂[}~\ab{d₀}~\af{]~∩~⟦}~\ab{d₁}~\af{⟧}); or we have syntax tree on the left and a hole on the right (\af{⟦}~\ab{d₀}~\af{⟧~∩~∂[}~\ab{d₁}~\af{]}).
For sums of syntax trees, we have a choice of contexts.

Derivatives are also functors, with the following action on morphisms.
\begin{code}%
\>[0]\AgdaFunction{∂map}\AgdaSpace{}%
\AgdaSymbol{:}\AgdaSpace{}%
\AgdaSymbol{∀}\AgdaSpace{}%
\AgdaSymbol{\{}\AgdaBound{d}\AgdaSpace{}%
\AgdaBound{X}\AgdaSpace{}%
\AgdaBound{Y}\AgdaSymbol{\}}\AgdaSpace{}%
\AgdaSymbol{→}\AgdaSpace{}%
\AgdaSymbol{(}\AgdaBound{X}\AgdaSpace{}%
\AgdaSymbol{→}\AgdaSpace{}%
\AgdaBound{Y}\AgdaSymbol{)}\AgdaSpace{}%
\AgdaSymbol{→}\AgdaSpace{}%
\AgdaOperator{\AgdaFunction{∂[}}\AgdaSpace{}%
\AgdaBound{d}\AgdaSpace{}%
\AgdaOperator{\AgdaFunction{]}}\AgdaSpace{}%
\AgdaBound{X}\AgdaSpace{}%
\AgdaSymbol{→}\AgdaSpace{}%
\AgdaOperator{\AgdaFunction{∂[}}\AgdaSpace{}%
\AgdaBound{d}\AgdaSpace{}%
\AgdaOperator{\AgdaFunction{]}}\AgdaSpace{}%
\AgdaBound{Y}\<%
\end{code}
\begin{code}[hide]%
\>[0]\AgdaFunction{∂map}\AgdaSpace{}%
\AgdaSymbol{\{}\AgdaInductiveConstructor{I}\AgdaSymbol{\}}\AgdaSpace{}%
\AgdaBound{f}\AgdaSpace{}%
\AgdaBound{x}\AgdaSpace{}%
\AgdaSymbol{=}\AgdaSpace{}%
\AgdaInductiveConstructor{tt}\<%
\\
\>[0]\AgdaFunction{∂map}\AgdaSpace{}%
\AgdaSymbol{\{}\AgdaInductiveConstructor{K}\AgdaSpace{}%
\AgdaBound{A}\AgdaSymbol{\}}\AgdaSpace{}%
\AgdaBound{f}\AgdaSpace{}%
\AgdaSymbol{()}\<%
\\
\>[0]\AgdaFunction{∂map}\AgdaSpace{}%
\AgdaSymbol{\{}\AgdaBound{d}\AgdaSpace{}%
\AgdaOperator{\AgdaInductiveConstructor{ẋ}}\AgdaSpace{}%
\AgdaBound{d₁}\AgdaSymbol{\}}\AgdaSpace{}%
\AgdaBound{f}\AgdaSpace{}%
\AgdaSymbol{(}\AgdaInductiveConstructor{inj₁}\AgdaSpace{}%
\AgdaSymbol{(}\AgdaBound{δ}\AgdaSpace{}%
\AgdaOperator{\AgdaInductiveConstructor{,}}\AgdaSpace{}%
\AgdaBound{x}\AgdaSymbol{))}\AgdaSpace{}%
\AgdaSymbol{=}\AgdaSpace{}%
\AgdaInductiveConstructor{inj₁}\AgdaSpace{}%
\AgdaSymbol{(}\AgdaFunction{∂map}\AgdaSpace{}%
\AgdaSymbol{\{}\AgdaBound{d}\AgdaSymbol{\}}\AgdaSpace{}%
\AgdaBound{f}\AgdaSpace{}%
\AgdaBound{δ}\AgdaSpace{}%
\AgdaOperator{\AgdaInductiveConstructor{,}}\AgdaSpace{}%
\AgdaFunction{fmap}\AgdaSpace{}%
\AgdaSymbol{\{}\AgdaBound{d₁}\AgdaSymbol{\}}\AgdaSpace{}%
\AgdaBound{f}\AgdaSpace{}%
\AgdaBound{x}\AgdaSymbol{)}\<%
\\
\>[0]\AgdaFunction{∂map}\AgdaSpace{}%
\AgdaSymbol{\{}\AgdaBound{d}\AgdaSpace{}%
\AgdaOperator{\AgdaInductiveConstructor{ẋ}}\AgdaSpace{}%
\AgdaBound{d₁}\AgdaSymbol{\}}\AgdaSpace{}%
\AgdaBound{f}\AgdaSpace{}%
\AgdaSymbol{(}\AgdaInductiveConstructor{inj₂}\AgdaSpace{}%
\AgdaSymbol{(}\AgdaBound{y}\AgdaSpace{}%
\AgdaOperator{\AgdaInductiveConstructor{,}}\AgdaSpace{}%
\AgdaBound{δ}\AgdaSymbol{))}\AgdaSpace{}%
\AgdaSymbol{=}\AgdaSpace{}%
\AgdaInductiveConstructor{inj₂}\AgdaSpace{}%
\AgdaSymbol{(}\AgdaFunction{fmap}\AgdaSpace{}%
\AgdaSymbol{\{}\AgdaBound{d}\AgdaSymbol{\}}\AgdaSpace{}%
\AgdaBound{f}\AgdaSpace{}%
\AgdaBound{y}\AgdaSpace{}%
\AgdaOperator{\AgdaInductiveConstructor{,}}\AgdaSpace{}%
\AgdaFunction{∂map}\AgdaSpace{}%
\AgdaSymbol{\{}\AgdaBound{d₁}\AgdaSymbol{\}}\AgdaSpace{}%
\AgdaBound{f}\AgdaSpace{}%
\AgdaBound{δ}\AgdaSymbol{)}\<%
\\
\>[0]\AgdaFunction{∂map}\AgdaSpace{}%
\AgdaSymbol{\{}\AgdaInductiveConstructor{∐}\AgdaSpace{}%
\AgdaBound{n}\AgdaSpace{}%
\AgdaBound{f}\AgdaSymbol{\}}\AgdaSpace{}%
\AgdaBound{g}\AgdaSpace{}%
\AgdaSymbol{(}\AgdaBound{m}\AgdaSpace{}%
\AgdaOperator{\AgdaInductiveConstructor{,}}\AgdaSpace{}%
\AgdaBound{δ}\AgdaSymbol{)}\AgdaSpace{}%
\AgdaSymbol{=}\AgdaSpace{}%
\AgdaBound{m}\AgdaSpace{}%
\AgdaOperator{\AgdaInductiveConstructor{,}}\AgdaSpace{}%
\AgdaFunction{∂map}\AgdaSpace{}%
\AgdaSymbol{\{}\AgdaBound{f}\AgdaSpace{}%
\AgdaBound{m}\AgdaSymbol{\}}\AgdaSpace{}%
\AgdaBound{g}\AgdaSpace{}%
\AgdaBound{δ}\<%
\end{code}%
\begin{code}[hide]%
\>[0]\AgdaFunction{∂map-id}%
\>[707I]\AgdaSymbol{:}\AgdaSpace{}%
\AgdaSymbol{∀}\AgdaSpace{}%
\AgdaBound{d}\AgdaSpace{}%
\AgdaSymbol{\{}\AgdaBound{X}\AgdaSymbol{\}}\AgdaSpace{}%
\AgdaSymbol{(}\AgdaBound{x}\AgdaSpace{}%
\AgdaSymbol{:}\AgdaSpace{}%
\AgdaOperator{\AgdaFunction{∂[}}\AgdaSpace{}%
\AgdaBound{d}\AgdaSpace{}%
\AgdaOperator{\AgdaFunction{]}}\AgdaSpace{}%
\AgdaBound{X}\AgdaSymbol{)}\<%
\\
\>[.][@{}l@{}]\<[707I]%
\>[8]\AgdaSymbol{→}\AgdaSpace{}%
\AgdaFunction{∂map}\AgdaSpace{}%
\AgdaSymbol{\{}\AgdaBound{d}\AgdaSymbol{\}}\AgdaSpace{}%
\AgdaFunction{id}\AgdaSpace{}%
\AgdaBound{x}\AgdaSpace{}%
\AgdaOperator{\AgdaDatatype{≡}}\AgdaSpace{}%
\AgdaFunction{id}\AgdaSpace{}%
\AgdaBound{x}\<%
\\
\>[0]\AgdaFunction{∂map-id}\AgdaSpace{}%
\AgdaInductiveConstructor{I}\AgdaSpace{}%
\AgdaInductiveConstructor{tt}\AgdaSpace{}%
\AgdaSymbol{=}\AgdaSpace{}%
\AgdaInductiveConstructor{refl}\<%
\\
\>[0]\AgdaFunction{∂map-id}\AgdaSpace{}%
\AgdaSymbol{(}\AgdaBound{d₀}\AgdaSpace{}%
\AgdaOperator{\AgdaInductiveConstructor{ẋ}}\AgdaSpace{}%
\AgdaBound{d₁}\AgdaSymbol{)}\AgdaSpace{}%
\AgdaSymbol{(}\AgdaInductiveConstructor{inj₁}\AgdaSpace{}%
\AgdaSymbol{(}\AgdaBound{x₀̂}\AgdaSpace{}%
\AgdaOperator{\AgdaInductiveConstructor{,}}\AgdaSpace{}%
\AgdaBound{x₁}\AgdaSymbol{))}\AgdaSpace{}%
\AgdaSymbol{=}\AgdaSpace{}%
\AgdaFunction{cong}\AgdaSpace{}%
\AgdaInductiveConstructor{inj₁}\<%
\\
\>[0][@{}l@{\AgdaIndent{0}}]%
\>[2]\AgdaSymbol{(}\AgdaFunction{cong₂}\AgdaSpace{}%
\AgdaOperator{\AgdaInductiveConstructor{\AgdaUnderscore{},\AgdaUnderscore{}}}\AgdaSpace{}%
\AgdaSymbol{(}\AgdaFunction{∂map-id}\AgdaSpace{}%
\AgdaBound{d₀}\AgdaSpace{}%
\AgdaBound{x₀̂}\AgdaSymbol{)}\AgdaSpace{}%
\AgdaSymbol{(}\AgdaFunction{fmap-id}\AgdaSpace{}%
\AgdaBound{d₁}\AgdaSpace{}%
\AgdaBound{x₁}\AgdaSymbol{))}\<%
\\
\>[0]\AgdaFunction{∂map-id}\AgdaSpace{}%
\AgdaSymbol{(}\AgdaBound{d₀}\AgdaSpace{}%
\AgdaOperator{\AgdaInductiveConstructor{ẋ}}\AgdaSpace{}%
\AgdaBound{d₁}\AgdaSymbol{)}\AgdaSpace{}%
\AgdaSymbol{(}\AgdaInductiveConstructor{inj₂}\AgdaSpace{}%
\AgdaSymbol{(}\AgdaBound{x₀}\AgdaSpace{}%
\AgdaOperator{\AgdaInductiveConstructor{,}}\AgdaSpace{}%
\AgdaBound{x₁̂}\AgdaSymbol{))}\AgdaSpace{}%
\AgdaSymbol{=}\AgdaSpace{}%
\AgdaFunction{cong}\AgdaSpace{}%
\AgdaInductiveConstructor{inj₂}\<%
\\
\>[0][@{}l@{\AgdaIndent{0}}]%
\>[2]\AgdaSymbol{(}\AgdaFunction{cong₂}\AgdaSpace{}%
\AgdaOperator{\AgdaInductiveConstructor{\AgdaUnderscore{},\AgdaUnderscore{}}}\AgdaSpace{}%
\AgdaSymbol{(}\AgdaFunction{fmap-id}\AgdaSpace{}%
\AgdaBound{d₀}\AgdaSpace{}%
\AgdaBound{x₀}\AgdaSymbol{)}\AgdaSpace{}%
\AgdaSymbol{(}\AgdaFunction{∂map-id}\AgdaSpace{}%
\AgdaBound{d₁}\AgdaSpace{}%
\AgdaBound{x₁̂}\AgdaSymbol{))}\<%
\\
\>[0]\AgdaFunction{∂map-id}\AgdaSpace{}%
\AgdaSymbol{(}\AgdaInductiveConstructor{∐}\AgdaSpace{}%
\AgdaBound{n}\AgdaSpace{}%
\AgdaBound{g}\AgdaSymbol{)}\AgdaSpace{}%
\AgdaSymbol{(}\AgdaBound{m}\AgdaSpace{}%
\AgdaOperator{\AgdaInductiveConstructor{,}}\AgdaSpace{}%
\AgdaBound{x}\AgdaSymbol{)}\AgdaSpace{}%
\AgdaSymbol{=}\AgdaSpace{}%
\AgdaFunction{cong}\AgdaSpace{}%
\AgdaSymbol{(\AgdaUnderscore{}}\AgdaSpace{}%
\AgdaOperator{\AgdaInductiveConstructor{,\AgdaUnderscore{}}}\AgdaSymbol{)}\AgdaSpace{}%
\AgdaSymbol{(}\AgdaFunction{∂map-id}\AgdaSpace{}%
\AgdaSymbol{(}\AgdaBound{g}\AgdaSpace{}%
\AgdaBound{m}\AgdaSymbol{)}\AgdaSpace{}%
\AgdaBound{x}\AgdaSymbol{)}\<%
\end{code}
\begin{code}[hide]%
\>[0]\AgdaFunction{∂map-∘}%
\>[776I]\AgdaSymbol{:}\AgdaSpace{}%
\AgdaSymbol{∀}\AgdaSpace{}%
\AgdaBound{d}\AgdaSpace{}%
\AgdaSymbol{\{}\AgdaBound{X}\AgdaSpace{}%
\AgdaBound{Y}\AgdaSpace{}%
\AgdaBound{Z}\AgdaSymbol{\}}\AgdaSpace{}%
\AgdaSymbol{(}\AgdaBound{f}\AgdaSpace{}%
\AgdaSymbol{:}\AgdaSpace{}%
\AgdaBound{X}\AgdaSpace{}%
\AgdaSymbol{→}\AgdaSpace{}%
\AgdaBound{Y}\AgdaSymbol{)}\AgdaSpace{}%
\AgdaSymbol{(}\AgdaBound{g}\AgdaSpace{}%
\AgdaSymbol{:}\AgdaSpace{}%
\AgdaBound{Y}\AgdaSpace{}%
\AgdaSymbol{→}\AgdaSpace{}%
\AgdaBound{Z}\AgdaSymbol{)}\AgdaSpace{}%
\AgdaSymbol{(}\AgdaBound{x}\AgdaSpace{}%
\AgdaSymbol{:}\AgdaSpace{}%
\AgdaOperator{\AgdaFunction{∂[}}\AgdaSpace{}%
\AgdaBound{d}\AgdaSpace{}%
\AgdaOperator{\AgdaFunction{]}}\AgdaSpace{}%
\AgdaBound{X}\AgdaSymbol{)}\<%
\\
\>[.][@{}l@{}]\<[776I]%
\>[7]\AgdaSymbol{→}\AgdaSpace{}%
\AgdaFunction{∂map}\AgdaSpace{}%
\AgdaSymbol{\{}\AgdaBound{d}\AgdaSymbol{\}}\AgdaSpace{}%
\AgdaSymbol{(}\AgdaBound{g}\AgdaSpace{}%
\AgdaOperator{\AgdaFunction{∘}}\AgdaSpace{}%
\AgdaBound{f}\AgdaSymbol{)}\AgdaSpace{}%
\AgdaBound{x}\AgdaSpace{}%
\AgdaOperator{\AgdaDatatype{≡}}\AgdaSpace{}%
\AgdaSymbol{(}\AgdaFunction{∂map}\AgdaSpace{}%
\AgdaSymbol{\{}\AgdaBound{d}\AgdaSymbol{\}}\AgdaSpace{}%
\AgdaBound{g}\AgdaSpace{}%
\AgdaOperator{\AgdaFunction{∘}}\AgdaSpace{}%
\AgdaFunction{∂map}\AgdaSpace{}%
\AgdaSymbol{\{}\AgdaBound{d}\AgdaSymbol{\}}\AgdaSpace{}%
\AgdaBound{f}\AgdaSymbol{)}\AgdaSpace{}%
\AgdaBound{x}\<%
\\
\>[0]\AgdaFunction{∂map-∘}\AgdaSpace{}%
\AgdaInductiveConstructor{I}\AgdaSpace{}%
\AgdaBound{f}\AgdaSpace{}%
\AgdaBound{g}\AgdaSpace{}%
\AgdaInductiveConstructor{tt}\AgdaSpace{}%
\AgdaSymbol{=}\AgdaSpace{}%
\AgdaInductiveConstructor{refl}\<%
\\
\>[0]\AgdaFunction{∂map-∘}\AgdaSpace{}%
\AgdaSymbol{(}\AgdaBound{d₀}\AgdaSpace{}%
\AgdaOperator{\AgdaInductiveConstructor{ẋ}}\AgdaSpace{}%
\AgdaBound{d₁}\AgdaSymbol{)}\AgdaSpace{}%
\AgdaBound{f}\AgdaSpace{}%
\AgdaBound{g}\AgdaSpace{}%
\AgdaSymbol{(}\AgdaInductiveConstructor{inj₁}\AgdaSpace{}%
\AgdaSymbol{(}\AgdaBound{x₀̂}\AgdaSpace{}%
\AgdaOperator{\AgdaInductiveConstructor{,}}\AgdaSpace{}%
\AgdaBound{x₁}\AgdaSymbol{))}\AgdaSpace{}%
\AgdaSymbol{=}\AgdaSpace{}%
\AgdaFunction{cong}\AgdaSpace{}%
\AgdaInductiveConstructor{inj₁}\<%
\\
\>[0][@{}l@{\AgdaIndent{0}}]%
\>[2]\AgdaSymbol{(}\AgdaFunction{cong₂}\AgdaSpace{}%
\AgdaOperator{\AgdaInductiveConstructor{\AgdaUnderscore{},\AgdaUnderscore{}}}\AgdaSpace{}%
\AgdaSymbol{(}\AgdaFunction{∂map-∘}\AgdaSpace{}%
\AgdaBound{d₀}\AgdaSpace{}%
\AgdaBound{f}\AgdaSpace{}%
\AgdaBound{g}\AgdaSpace{}%
\AgdaBound{x₀̂}\AgdaSymbol{)}\AgdaSpace{}%
\AgdaSymbol{(}\AgdaFunction{fmap-∘}\AgdaSpace{}%
\AgdaBound{d₁}\AgdaSpace{}%
\AgdaBound{f}\AgdaSpace{}%
\AgdaBound{g}\AgdaSpace{}%
\AgdaBound{x₁}\AgdaSymbol{))}\<%
\\
\>[0]\AgdaFunction{∂map-∘}\AgdaSpace{}%
\AgdaSymbol{(}\AgdaBound{d₀}\AgdaSpace{}%
\AgdaOperator{\AgdaInductiveConstructor{ẋ}}\AgdaSpace{}%
\AgdaBound{d₁}\AgdaSymbol{)}\AgdaSpace{}%
\AgdaBound{f}\AgdaSpace{}%
\AgdaBound{g}\AgdaSpace{}%
\AgdaSymbol{(}\AgdaInductiveConstructor{inj₂}\AgdaSpace{}%
\AgdaSymbol{(}\AgdaBound{x₀}\AgdaSpace{}%
\AgdaOperator{\AgdaInductiveConstructor{,}}\AgdaSpace{}%
\AgdaBound{x₁̂}\AgdaSymbol{))}\AgdaSpace{}%
\AgdaSymbol{=}\AgdaSpace{}%
\AgdaFunction{cong}\AgdaSpace{}%
\AgdaInductiveConstructor{inj₂}\<%
\\
\>[0][@{}l@{\AgdaIndent{0}}]%
\>[2]\AgdaSymbol{(}\AgdaFunction{cong₂}\AgdaSpace{}%
\AgdaOperator{\AgdaInductiveConstructor{\AgdaUnderscore{},\AgdaUnderscore{}}}\AgdaSpace{}%
\AgdaSymbol{(}\AgdaFunction{fmap-∘}\AgdaSpace{}%
\AgdaBound{d₀}\AgdaSpace{}%
\AgdaBound{f}\AgdaSpace{}%
\AgdaBound{g}\AgdaSpace{}%
\AgdaBound{x₀}\AgdaSymbol{)}\AgdaSpace{}%
\AgdaSymbol{(}\AgdaFunction{∂map-∘}\AgdaSpace{}%
\AgdaBound{d₁}\AgdaSpace{}%
\AgdaBound{f}\AgdaSpace{}%
\AgdaBound{g}\AgdaSpace{}%
\AgdaBound{x₁̂}\AgdaSymbol{))}\<%
\\
\>[0]\AgdaFunction{∂map-∘}\AgdaSpace{}%
\AgdaSymbol{(}\AgdaInductiveConstructor{∐}\AgdaSpace{}%
\AgdaBound{n}\AgdaSpace{}%
\AgdaBound{h}\AgdaSymbol{)}\AgdaSpace{}%
\AgdaBound{f}\AgdaSpace{}%
\AgdaBound{g}\AgdaSpace{}%
\AgdaSymbol{(}\AgdaBound{m}\AgdaSpace{}%
\AgdaOperator{\AgdaInductiveConstructor{,}}\AgdaSpace{}%
\AgdaBound{x}\AgdaSymbol{)}\AgdaSpace{}%
\AgdaSymbol{=}\AgdaSpace{}%
\AgdaFunction{cong}\AgdaSpace{}%
\AgdaSymbol{(\AgdaUnderscore{}}\AgdaSpace{}%
\AgdaOperator{\AgdaInductiveConstructor{,\AgdaUnderscore{}}}\AgdaSymbol{)}\AgdaSpace{}%
\AgdaSymbol{(}\AgdaFunction{∂map-∘}\AgdaSpace{}%
\AgdaSymbol{(}\AgdaBound{h}\AgdaSpace{}%
\AgdaBound{m}\AgdaSymbol{)}\AgdaSpace{}%
\AgdaBound{f}\AgdaSpace{}%
\AgdaBound{g}\AgdaSpace{}%
\AgdaBound{x}\AgdaSymbol{)}\<%
\end{code}

Just like the notion of recomposition for contexts and context frames from \cref{sec:02-redsem}, generic context frames have a generic notion of recomposition (or `plugging'):
\begin{code}%
\>[0]\AgdaFunction{plug}\AgdaSpace{}%
\AgdaSymbol{:}\AgdaSpace{}%
\AgdaSymbol{∀}\AgdaSpace{}%
\AgdaSymbol{\{}\AgdaBound{d}\AgdaSpace{}%
\AgdaBound{X}\AgdaSymbol{\}}\AgdaSpace{}%
\AgdaSymbol{→}\AgdaSpace{}%
\AgdaOperator{\AgdaFunction{∂[}}\AgdaSpace{}%
\AgdaBound{d}\AgdaSpace{}%
\AgdaOperator{\AgdaFunction{]}}\AgdaSpace{}%
\AgdaBound{X}\AgdaSpace{}%
\AgdaSymbol{→}\AgdaSpace{}%
\AgdaBound{X}\AgdaSpace{}%
\AgdaSymbol{→}\AgdaSpace{}%
\AgdaOperator{\AgdaFunction{⟦}}\AgdaSpace{}%
\AgdaBound{d}\AgdaSpace{}%
\AgdaOperator{\AgdaFunction{⟧}}\AgdaSpace{}%
\AgdaBound{X}\<%
\\
\>[0]\AgdaFunction{plug}\AgdaSpace{}%
\AgdaSymbol{\{}\AgdaInductiveConstructor{I}%
\>[14]\AgdaSymbol{\}}%
\>[17]\AgdaBound{δ}%
\>[33]\AgdaBound{x}%
\>[36]\AgdaSymbol{=}\AgdaSpace{}%
\AgdaBound{x}\<%
\\
\>[0]\AgdaFunction{plug}\AgdaSpace{}%
\AgdaSymbol{\{}\AgdaBound{d}\AgdaSpace{}%
\AgdaOperator{\AgdaInductiveConstructor{ẋ}}\AgdaSpace{}%
\AgdaBound{d₁}%
\>[14]\AgdaSymbol{\}}%
\>[17]\AgdaSymbol{(}\AgdaInductiveConstructor{inj₁}\AgdaSpace{}%
\AgdaSymbol{(}\AgdaBound{δ}\AgdaSpace{}%
\AgdaOperator{\AgdaInductiveConstructor{,}}\AgdaSpace{}%
\AgdaBound{t}\AgdaSymbol{))}%
\>[33]\AgdaBound{x}%
\>[36]\AgdaSymbol{=}\AgdaSpace{}%
\AgdaFunction{plug}\AgdaSpace{}%
\AgdaSymbol{\{}\AgdaBound{d}\AgdaSymbol{\}}\AgdaSpace{}%
\AgdaBound{δ}\AgdaSpace{}%
\AgdaBound{x}\AgdaSpace{}%
\AgdaOperator{\AgdaInductiveConstructor{,}}\AgdaSpace{}%
\AgdaBound{t}\<%
\\
\>[0]\AgdaFunction{plug}\AgdaSpace{}%
\AgdaSymbol{\{}\AgdaBound{d}\AgdaSpace{}%
\AgdaOperator{\AgdaInductiveConstructor{ẋ}}\AgdaSpace{}%
\AgdaBound{d₁}%
\>[14]\AgdaSymbol{\}}%
\>[17]\AgdaSymbol{(}\AgdaInductiveConstructor{inj₂}\AgdaSpace{}%
\AgdaSymbol{(}\AgdaBound{t}\AgdaSpace{}%
\AgdaOperator{\AgdaInductiveConstructor{,}}\AgdaSpace{}%
\AgdaBound{δ}\AgdaSymbol{))}%
\>[33]\AgdaBound{x}%
\>[36]\AgdaSymbol{=}\AgdaSpace{}%
\AgdaBound{t}\AgdaSpace{}%
\AgdaOperator{\AgdaInductiveConstructor{,}}\AgdaSpace{}%
\AgdaFunction{plug}\AgdaSpace{}%
\AgdaSymbol{\{}\AgdaBound{d₁}\AgdaSymbol{\}}\AgdaSpace{}%
\AgdaBound{δ}\AgdaSpace{}%
\AgdaBound{x}\<%
\\
\>[0]\AgdaFunction{plug}\AgdaSpace{}%
\AgdaSymbol{\{}\AgdaInductiveConstructor{∐}\AgdaSpace{}%
\AgdaBound{n}\AgdaSpace{}%
\AgdaBound{f}%
\>[14]\AgdaSymbol{\}}%
\>[17]\AgdaSymbol{(}\AgdaBound{m}\AgdaSpace{}%
\AgdaOperator{\AgdaInductiveConstructor{,}}\AgdaSpace{}%
\AgdaBound{δ}\AgdaSymbol{)}%
\>[33]\AgdaBound{x}%
\>[36]\AgdaSymbol{=}\AgdaSpace{}%
\AgdaBound{m}\AgdaSpace{}%
\AgdaOperator{\AgdaInductiveConstructor{,}}\AgdaSpace{}%
\AgdaFunction{plug}\AgdaSpace{}%
\AgdaSymbol{\{}\AgdaBound{f}\AgdaSpace{}%
\AgdaBound{m}\AgdaSymbol{\}}\AgdaSpace{}%
\AgdaBound{δ}\AgdaSpace{}%
\AgdaBound{x}\<%
\end{code}
Furthermore, similarly to how we represented contexts in \cref{sec:02-redsem} as a sequence of nested context frames, generic contexts are given by a nested (pluggable) generic context frames:
\begin{code}%
\>[0]\AgdaOperator{\AgdaFunction{Ctx[\AgdaUnderscore{}]}}\AgdaSpace{}%
\AgdaSymbol{:}\AgdaSpace{}%
\AgdaDatatype{Desc}\AgdaSpace{}%
\AgdaSymbol{→}\AgdaSpace{}%
\AgdaPrimitive{Set}\<%
\\
\>[0]\AgdaOperator{\AgdaFunction{Ctx[}}\AgdaSpace{}%
\AgdaBound{d}\AgdaSpace{}%
\AgdaOperator{\AgdaFunction{]}}\AgdaSpace{}%
\AgdaSymbol{=}\AgdaSpace{}%
\AgdaDatatype{List}\AgdaSpace{}%
\AgdaSymbol{(}\AgdaOperator{\AgdaFunction{∂[}}\AgdaSpace{}%
\AgdaBound{d}\AgdaSpace{}%
\AgdaOperator{\AgdaFunction{]}}\AgdaSpace{}%
\AgdaOperator{\AgdaDatatype{μ[}}\AgdaSpace{}%
\AgdaBound{d}\AgdaSpace{}%
\AgdaOperator{\AgdaDatatype{]}}\AgdaSymbol{)}\<%
\\
\\[\AgdaEmptyExtraSkip]%
\>[0]\AgdaFunction{plug-ctx}\AgdaSpace{}%
\AgdaSymbol{:}\AgdaSpace{}%
\AgdaSymbol{\{}\AgdaBound{d}\AgdaSpace{}%
\AgdaSymbol{:}\AgdaSpace{}%
\AgdaDatatype{Desc}\AgdaSymbol{\}}\AgdaSpace{}%
\AgdaSymbol{→}\AgdaSpace{}%
\AgdaOperator{\AgdaFunction{Ctx[}}\AgdaSpace{}%
\AgdaBound{d}\AgdaSpace{}%
\AgdaOperator{\AgdaFunction{]}}\AgdaSpace{}%
\AgdaSymbol{→}\AgdaSpace{}%
\AgdaOperator{\AgdaDatatype{μ[}}\AgdaSpace{}%
\AgdaBound{d}\AgdaSpace{}%
\AgdaOperator{\AgdaDatatype{]}}\AgdaSpace{}%
\AgdaSymbol{→}\AgdaSpace{}%
\AgdaOperator{\AgdaDatatype{μ[}}\AgdaSpace{}%
\AgdaBound{d}\AgdaSpace{}%
\AgdaOperator{\AgdaDatatype{]}}\<%
\\
\>[0]\AgdaFunction{plug-ctx}%
\>[14]\AgdaInductiveConstructor{[]}%
\>[25]\AgdaBound{x}%
\>[28]\AgdaSymbol{=}\AgdaSpace{}%
\AgdaBound{x}\<%
\\
\>[0]\AgdaFunction{plug-ctx}\AgdaSpace{}%
\AgdaSymbol{\{}\AgdaBound{d}\AgdaSymbol{\}}%
\>[14]\AgdaSymbol{(}\AgdaBound{frm}\AgdaSpace{}%
\AgdaOperator{\AgdaInductiveConstructor{∷}}\AgdaSpace{}%
\AgdaBound{c}\AgdaSymbol{)}%
\>[25]\AgdaBound{x}%
\>[28]\AgdaSymbol{=}\AgdaSpace{}%
\AgdaFunction{plug-ctx}\AgdaSpace{}%
\AgdaSymbol{\{}\AgdaBound{d}\AgdaSymbol{\}}\AgdaSpace{}%
\AgdaBound{c}\AgdaSpace{}%
\AgdaOperator{\AgdaInductiveConstructor{⟨}}\AgdaSpace{}%
\AgdaFunction{plug}\AgdaSpace{}%
\AgdaSymbol{\{}\AgdaBound{d}\AgdaSymbol{\}}\AgdaSpace{}%
\AgdaBound{frm}\AgdaSpace{}%
\AgdaBound{x}\AgdaSpace{}%
\AgdaOperator{\AgdaInductiveConstructor{⟩}}\<%
\end{code}
Contexts also compose, and composition preserves pluggability:
\begin{code}%
\>[0]\AgdaFunction{compose-plug}%
\>[14]\AgdaSymbol{:}\AgdaSpace{}%
\AgdaSymbol{∀}\AgdaSpace{}%
\AgdaSymbol{\{}\AgdaBound{d}\AgdaSpace{}%
\AgdaBound{x}\AgdaSymbol{\}}\AgdaSpace{}%
\AgdaSymbol{(}\AgdaBound{c₀}\AgdaSpace{}%
\AgdaBound{c₁}\AgdaSpace{}%
\AgdaSymbol{:}\AgdaSpace{}%
\AgdaOperator{\AgdaFunction{Ctx[}}\AgdaSpace{}%
\AgdaBound{d}\AgdaSpace{}%
\AgdaOperator{\AgdaFunction{]}}\AgdaSymbol{)}\<%
\\
\>[14]\AgdaSymbol{→}%
\>[17]\AgdaFunction{plug-ctx}\AgdaSpace{}%
\AgdaSymbol{\{}\AgdaBound{d}\AgdaSymbol{\}}\AgdaSpace{}%
\AgdaSymbol{(}\AgdaBound{c₀}\AgdaSpace{}%
\AgdaOperator{\AgdaFunction{++}}\AgdaSpace{}%
\AgdaBound{c₁}\AgdaSymbol{)}\AgdaSpace{}%
\AgdaBound{x}\<%
\\
\>[17]\AgdaOperator{\AgdaDatatype{≡}}\AgdaSpace{}%
\AgdaFunction{plug-ctx}\AgdaSpace{}%
\AgdaBound{c₁}\AgdaSpace{}%
\AgdaSymbol{(}\AgdaFunction{plug-ctx}\AgdaSpace{}%
\AgdaBound{c₀}\AgdaSpace{}%
\AgdaBound{x}\AgdaSymbol{)}\<%
\\
\>[0]\AgdaFunction{compose-plug}%
\>[18]\AgdaInductiveConstructor{[]}%
\>[30]\AgdaBound{c₁}\AgdaSpace{}%
\AgdaSymbol{=}\AgdaSpace{}%
\AgdaInductiveConstructor{refl}\<%
\\
\>[0]\AgdaFunction{compose-plug}\AgdaSpace{}%
\AgdaSymbol{\{}\AgdaBound{d}\AgdaSymbol{\}}%
\>[18]\AgdaSymbol{(}\AgdaBound{frm}\AgdaSpace{}%
\AgdaOperator{\AgdaInductiveConstructor{∷}}\AgdaSpace{}%
\AgdaBound{c₀}\AgdaSymbol{)}%
\>[30]\AgdaBound{c₁}\AgdaSpace{}%
\AgdaSymbol{=}\AgdaSpace{}%
\AgdaFunction{compose-plug}\AgdaSpace{}%
\AgdaBound{c₀}\AgdaSpace{}%
\AgdaBound{c₁}\<%
\end{code}

\begin{example}
Let us return to the syntax for the simple arithmetic language from \cref{ex:hr}.
The following defines the context frame $1 + []$:
\begin{code}%
\>[0]\AgdaFunction{1+[]}\AgdaSpace{}%
\AgdaSymbol{:}\AgdaSpace{}%
\AgdaOperator{\AgdaFunction{∂[}}\AgdaSpace{}%
\AgdaFunction{HR}\AgdaSpace{}%
\AgdaOperator{\AgdaFunction{]}}\AgdaSpace{}%
\AgdaOperator{\AgdaDatatype{μ[}}\AgdaSpace{}%
\AgdaFunction{HR}\AgdaSpace{}%
\AgdaOperator{\AgdaDatatype{]}}\<%
\\
\>[0]\AgdaFunction{1+[]}\AgdaSpace{}%
\AgdaSymbol{=}\AgdaSpace{}%
\AgdaInductiveConstructor{suc}\AgdaSpace{}%
\AgdaInductiveConstructor{zero}\AgdaSpace{}%
\AgdaOperator{\AgdaInductiveConstructor{,}}\AgdaSpace{}%
\AgdaInductiveConstructor{inj₂}\AgdaSpace{}%
\AgdaSymbol{(}\AgdaInductiveConstructor{`num}\AgdaSpace{}%
\AgdaNumber{1}\AgdaSpace{}%
\AgdaOperator{\AgdaInductiveConstructor{,}}\AgdaSpace{}%
\AgdaInductiveConstructor{tt}\AgdaSymbol{)}\<%
\end{code}
Now
\begin{code}[hide]%
\>[0]\AgdaFunction{plug-1+[]-with-2≡1+2}\AgdaSpace{}%
\AgdaSymbol{:}\<%
\end{code}%
\begin{code}[inline]%
\>[0][@{}l@{\AgdaIndent{1}}]%
\>[2]\AgdaFunction{plug}\AgdaSpace{}%
\AgdaSymbol{\{}\AgdaFunction{HR}\AgdaSymbol{\}}\AgdaSpace{}%
\AgdaFunction{1+[]}\AgdaSpace{}%
\AgdaSymbol{(}\AgdaInductiveConstructor{`num}\AgdaSpace{}%
\AgdaNumber{2}\AgdaSymbol{)}\AgdaSpace{}%
\AgdaOperator{\AgdaDatatype{≡}}\AgdaSpace{}%
\AgdaInductiveConstructor{`addF}\AgdaSpace{}%
\AgdaSymbol{(}\AgdaInductiveConstructor{`num}\AgdaSpace{}%
\AgdaNumber{1}\AgdaSymbol{)}\AgdaSpace{}%
\AgdaSymbol{(}\AgdaInductiveConstructor{`num}\AgdaSpace{}%
\AgdaNumber{2}\AgdaSymbol{)}\<%
\end{code}%
\begin{code}[hide]%
\>[0]\AgdaFunction{plug-1+[]-with-2≡1+2}\AgdaSpace{}%
\AgdaSymbol{=}\AgdaSpace{}%
\AgdaInductiveConstructor{refl}\<%
\end{code}%
. Contexts can be represented and plugged analogously.
%
\end{example}

The next section uses the summarized generic techniques to define reduction semantics.

\begin{code}[hide]%
\>[0]\AgdaSymbol{\{-\#}\AgdaSpace{}%
\AgdaKeyword{OPTIONS}\AgdaSpace{}%
\AgdaPragma{--guardedness}\AgdaSpace{}%
\AgdaSymbol{\#-\}}\<%
\\
\\[\AgdaEmptyExtraSkip]%
\>[0]\AgdaKeyword{open}\AgdaSpace{}%
\AgdaKeyword{import}\AgdaSpace{}%
\AgdaModule{Function}\<%
\\
\\[\AgdaEmptyExtraSkip]%
\>[0]\AgdaKeyword{open}\AgdaSpace{}%
\AgdaKeyword{import}\AgdaSpace{}%
\AgdaModule{Data.Empty}\<%
\\
\>[0]\AgdaKeyword{open}\AgdaSpace{}%
\AgdaKeyword{import}\AgdaSpace{}%
\AgdaModule{Data.Unit}\<%
\\
\>[0]\AgdaKeyword{open}\AgdaSpace{}%
\AgdaKeyword{import}\AgdaSpace{}%
\AgdaModule{Data.Bool}\<%
\\
\>[0]\AgdaKeyword{open}\AgdaSpace{}%
\AgdaKeyword{import}\AgdaSpace{}%
\AgdaModule{Data.Maybe}\<%
\\
\>[0]\AgdaKeyword{open}\AgdaSpace{}%
\AgdaKeyword{import}\AgdaSpace{}%
\AgdaModule{Data.Sum}\<%
\\
\>[0]\AgdaKeyword{open}\AgdaSpace{}%
\AgdaKeyword{import}\AgdaSpace{}%
\AgdaModule{Data.Product}\<%
\\
\>[0]\AgdaKeyword{open}\AgdaSpace{}%
\AgdaKeyword{import}\AgdaSpace{}%
\AgdaModule{Data.Nat}\<%
\\
\>[0]\AgdaKeyword{open}\AgdaSpace{}%
\AgdaKeyword{import}\AgdaSpace{}%
\AgdaModule{Data.Fin}\AgdaSpace{}%
\AgdaKeyword{hiding}\AgdaSpace{}%
\AgdaSymbol{(}\AgdaOperator{\AgdaFunction{\AgdaUnderscore{}+\AgdaUnderscore{}}}\AgdaSymbol{)}\<%
\\
\>[0]\AgdaKeyword{open}\AgdaSpace{}%
\AgdaKeyword{import}\AgdaSpace{}%
\AgdaModule{Data.List}\AgdaSpace{}%
\AgdaSymbol{as}\AgdaSpace{}%
\AgdaModule{List}\AgdaSpace{}%
\AgdaKeyword{renaming}\AgdaSpace{}%
\AgdaSymbol{(}\AgdaFunction{map}\AgdaSpace{}%
\AgdaSymbol{to}\AgdaSpace{}%
\AgdaFunction{⅋map}\AgdaSymbol{)}\<%
\\
\>[0]\AgdaKeyword{open}\AgdaSpace{}%
\AgdaKeyword{import}\AgdaSpace{}%
\AgdaModule{Data.List.Properties}\<%
\\
\>[0]\AgdaKeyword{open}\AgdaSpace{}%
\AgdaKeyword{import}\AgdaSpace{}%
\AgdaModule{Data.List.Membership.Propositional}\<%
\\
\>[0]\AgdaKeyword{open}\AgdaSpace{}%
\AgdaKeyword{import}\AgdaSpace{}%
\AgdaModule{Data.List.Membership.Propositional.Properties}\<%
\\
\>[0]\AgdaKeyword{open}\AgdaSpace{}%
\AgdaKeyword{import}\AgdaSpace{}%
\AgdaModule{Data.List.Relation.Unary.Any}\<%
\\
\>[0]\AgdaKeyword{open}\AgdaSpace{}%
\AgdaKeyword{import}\AgdaSpace{}%
\AgdaModule{Data.List.Relation.Unary.All}\AgdaSpace{}%
\AgdaSymbol{as}\AgdaSpace{}%
\AgdaModule{All}\<%
\\
\>[0]\AgdaKeyword{open}\AgdaSpace{}%
\AgdaKeyword{import}\AgdaSpace{}%
\AgdaModule{Data.List.Relation.Unary.All.Properties}\<%
\\
\\[\AgdaEmptyExtraSkip]%
\>[0]\AgdaKeyword{open}\AgdaSpace{}%
\AgdaKeyword{import}\AgdaSpace{}%
\AgdaModule{Relation.Nullary}\<%
\\
\>[0]\AgdaKeyword{open}\AgdaSpace{}%
\AgdaKeyword{import}\AgdaSpace{}%
\AgdaModule{Relation.Unary}\AgdaSpace{}%
\AgdaKeyword{renaming}\AgdaSpace{}%
\AgdaSymbol{(}\AgdaOperator{\AgdaFunction{\AgdaUnderscore{}∈\AgdaUnderscore{}}}\AgdaSpace{}%
\AgdaSymbol{to}\AgdaSpace{}%
\AgdaOperator{\AgdaFunction{\AgdaUnderscore{}∈′\AgdaUnderscore{}}}\AgdaSymbol{)}\<%
\\
\>[0]\AgdaKeyword{open}\AgdaSpace{}%
\AgdaKeyword{import}\AgdaSpace{}%
\AgdaModule{Relation.Binary.PropositionalEquality}\<%
\\
\\[\AgdaEmptyExtraSkip]%
\>[0]\AgdaKeyword{open}\AgdaSpace{}%
\AgdaKeyword{import}\AgdaSpace{}%
\AgdaModule{sections.03-prelude}\<%
\\
\\[\AgdaEmptyExtraSkip]%
\>[0]\AgdaKeyword{open}\AgdaSpace{}%
\AgdaModule{Alg}\<%
\\
\\[\AgdaEmptyExtraSkip]%
\>[0]\AgdaKeyword{module}\AgdaSpace{}%
\AgdaModule{sections.04-redsem-gen}\AgdaSpace{}%
\AgdaKeyword{where}\<%
\end{code}

\section{Reduction Semantics, Generically}
\label{sec:04-redsem-gen}

In this section we first (\cref{sec:algebraic-red-sem}) present a generic specification of standard reduction strategies.
Next (\cref{sec:generic-paras}), we present generic paramorphisms corresponding to this specification.
Proving that the paramorphisms respect the specification is left as future work.
Finally (\cref{sec:generic-cases}), we demonstrate how the generic definitions from \cref{sec:02-redsem,sec:algebraic-red-sem,sec:generic-paras} let us implement reduction semantics where the only code we have to write is a syntax specification, a contraction function, and code for choosing the right standard reduction strategy.

\subsection{An Algebraic Approach to Reduction Strategies}
\label{sec:algebraic-red-sem}

The following reduction strategic taxonomy characterizes what we understand by standard, deterministic reduction strategies.
\\[0.5em]
\emph{Horizontal reduction strategy.}  Most semantics prefer left-to-right or right-to-left order of evaluation, typically dubbed left-most and right-most reduction strategies.  In theory, other orders of evaluation are possible.  We deem such other evaluation orders non-standard, and restrict our attention to left-most and right-most order of reduction.
\\[0.5em]
\emph{Vertical reduction strategy.}  While many languages use an inner redexes over outer redexes, some languages prefer outer over inner, such as the Boolean normalization functions studied in Johannsen's thesis \cite{johannsen-thesis}.
\\[0.5em]
\indent To help us make the taxonomy above more precise, and to aid us in defining a generic decomposition function in \cref{sec:generic-paras}, we define a function which returns the list of all sub-terms and their context frames for a given term.
\begin{code}%
\>[0]\AgdaOperator{\AgdaFunction{S[\AgdaUnderscore{}∼\AgdaUnderscore{}]}}%
\>[8]\AgdaSymbol{:}\AgdaSpace{}%
\AgdaSymbol{∀}\AgdaSpace{}%
\AgdaSymbol{\{}\AgdaBound{X}\AgdaSymbol{\}}\AgdaSpace{}%
\AgdaSymbol{(}\AgdaBound{d}\AgdaSpace{}%
\AgdaSymbol{:}\AgdaSpace{}%
\AgdaDatatype{Desc}\AgdaSymbol{)}\<%
\\
\>[8]\AgdaSymbol{→}\AgdaSpace{}%
\AgdaOperator{\AgdaFunction{⟦}}\AgdaSpace{}%
\AgdaBound{d}\AgdaSpace{}%
\AgdaOperator{\AgdaFunction{⟧}}\AgdaSpace{}%
\AgdaBound{X}\AgdaSpace{}%
\AgdaSymbol{→}\AgdaSpace{}%
\AgdaDatatype{List}\AgdaSpace{}%
\AgdaSymbol{(}\AgdaOperator{\AgdaFunction{∂[}}\AgdaSpace{}%
\AgdaBound{d}\AgdaSpace{}%
\AgdaOperator{\AgdaFunction{]}}\AgdaSpace{}%
\AgdaBound{X}\AgdaSpace{}%
\AgdaOperator{\AgdaFunction{×}}\AgdaSpace{}%
\AgdaBound{X}\AgdaSymbol{)}\<%
\end{code}
\begin{code}[hide]%
\>[0]\AgdaOperator{\AgdaFunction{S[}}\AgdaSpace{}%
\AgdaInductiveConstructor{I}%
\>[11]\AgdaOperator{\AgdaFunction{∼}}\AgdaSpace{}%
\AgdaBound{t}%
\>[22]\AgdaOperator{\AgdaFunction{]}}%
\>[25]\AgdaSymbol{=}\AgdaSpace{}%
\AgdaSymbol{(}\AgdaInductiveConstructor{tt}\AgdaSpace{}%
\AgdaOperator{\AgdaInductiveConstructor{,}}\AgdaSpace{}%
\AgdaBound{t}\AgdaSymbol{)}\AgdaSpace{}%
\AgdaOperator{\AgdaInductiveConstructor{∷}}\AgdaSpace{}%
\AgdaInductiveConstructor{[]}\<%
\\
\>[0]\AgdaOperator{\AgdaFunction{S[}}\AgdaSpace{}%
\AgdaInductiveConstructor{K}\AgdaSpace{}%
\AgdaBound{x}%
\>[11]\AgdaOperator{\AgdaFunction{∼}}\AgdaSpace{}%
\AgdaBound{t}%
\>[22]\AgdaOperator{\AgdaFunction{]}}%
\>[25]\AgdaSymbol{=}\AgdaSpace{}%
\AgdaInductiveConstructor{[]}\<%
\\
\>[0]\AgdaOperator{\AgdaFunction{S[}}\AgdaSpace{}%
\AgdaBound{d₀}\AgdaSpace{}%
\AgdaOperator{\AgdaInductiveConstructor{ẋ}}\AgdaSpace{}%
\AgdaBound{d₁}%
\>[12]\AgdaOperator{\AgdaFunction{∼}}\AgdaSpace{}%
\AgdaSymbol{(}\AgdaBound{x₀}\AgdaSpace{}%
\AgdaOperator{\AgdaInductiveConstructor{,}}\AgdaSpace{}%
\AgdaBound{x₁}\AgdaSymbol{)}%
\>[25]\AgdaOperator{\AgdaFunction{]}}%
\>[28]\AgdaSymbol{=}\<%
\\
\>[0][@{}l@{\AgdaIndent{0}}]%
\>[2]\AgdaFunction{⅋map}\AgdaSpace{}%
\AgdaSymbol{(λ}\AgdaSpace{}%
\AgdaSymbol{(}\AgdaBound{x₀̂}\AgdaSpace{}%
\AgdaOperator{\AgdaInductiveConstructor{,}}\AgdaSpace{}%
\AgdaBound{x₀₀}\AgdaSymbol{)}\AgdaSpace{}%
\AgdaSymbol{→}\AgdaSpace{}%
\AgdaInductiveConstructor{inj₁}\AgdaSpace{}%
\AgdaSymbol{(}\AgdaBound{x₀̂}\AgdaSpace{}%
\AgdaOperator{\AgdaInductiveConstructor{,}}\AgdaSpace{}%
\AgdaBound{x₁}\AgdaSymbol{)}\AgdaSpace{}%
\AgdaOperator{\AgdaInductiveConstructor{,}}\AgdaSpace{}%
\AgdaBound{x₀₀}\AgdaSymbol{)}\AgdaSpace{}%
\AgdaOperator{\AgdaFunction{S[}}\AgdaSpace{}%
\AgdaBound{d₀}\AgdaSpace{}%
\AgdaOperator{\AgdaFunction{∼}}\AgdaSpace{}%
\AgdaBound{x₀}\AgdaSpace{}%
\AgdaOperator{\AgdaFunction{]}}\<%
\\
\>[2]\AgdaOperator{\AgdaFunction{++}}\AgdaSpace{}%
\AgdaFunction{⅋map}\AgdaSpace{}%
\AgdaSymbol{(λ}\AgdaSpace{}%
\AgdaSymbol{(}\AgdaBound{x₁̂}\AgdaSpace{}%
\AgdaOperator{\AgdaInductiveConstructor{,}}\AgdaSpace{}%
\AgdaBound{x₁₁}\AgdaSymbol{)}\AgdaSpace{}%
\AgdaSymbol{→}\AgdaSpace{}%
\AgdaInductiveConstructor{inj₂}\AgdaSpace{}%
\AgdaSymbol{(}\AgdaBound{x₀}\AgdaSpace{}%
\AgdaOperator{\AgdaInductiveConstructor{,}}\AgdaSpace{}%
\AgdaBound{x₁̂}\AgdaSymbol{)}\AgdaSpace{}%
\AgdaOperator{\AgdaInductiveConstructor{,}}\AgdaSpace{}%
\AgdaBound{x₁₁}\AgdaSymbol{)}\AgdaSpace{}%
\AgdaOperator{\AgdaFunction{S[}}\AgdaSpace{}%
\AgdaBound{d₁}\AgdaSpace{}%
\AgdaOperator{\AgdaFunction{∼}}\AgdaSpace{}%
\AgdaBound{x₁}\AgdaSpace{}%
\AgdaOperator{\AgdaFunction{]}}\<%
\\
\>[0]\AgdaOperator{\AgdaFunction{S[}}\AgdaSpace{}%
\AgdaInductiveConstructor{∐}\AgdaSpace{}%
\AgdaBound{n}\AgdaSpace{}%
\AgdaBound{f}%
\>[11]\AgdaOperator{\AgdaFunction{∼}}\AgdaSpace{}%
\AgdaSymbol{(}\AgdaBound{m}\AgdaSpace{}%
\AgdaOperator{\AgdaInductiveConstructor{,}}\AgdaSpace{}%
\AgdaBound{x}\AgdaSymbol{)}%
\>[22]\AgdaOperator{\AgdaFunction{]}}%
\>[25]\AgdaSymbol{=}\<%
\\
\>[0][@{}l@{\AgdaIndent{0}}]%
\>[2]\AgdaFunction{⅋map}\AgdaSpace{}%
\AgdaSymbol{(λ}\AgdaSpace{}%
\AgdaSymbol{(}\AgdaBound{x̂}\AgdaSpace{}%
\AgdaOperator{\AgdaInductiveConstructor{,}}\AgdaSpace{}%
\AgdaBound{x}\AgdaSymbol{)}\AgdaSpace{}%
\AgdaSymbol{→}\AgdaSpace{}%
\AgdaSymbol{(}\AgdaBound{m}\AgdaSpace{}%
\AgdaOperator{\AgdaInductiveConstructor{,}}\AgdaSpace{}%
\AgdaBound{x̂}\AgdaSymbol{)}\AgdaSpace{}%
\AgdaOperator{\AgdaInductiveConstructor{,}}\AgdaSpace{}%
\AgdaBound{x}\AgdaSymbol{)}\AgdaSpace{}%
\AgdaOperator{\AgdaFunction{S[}}\AgdaSpace{}%
\AgdaBound{f}\AgdaSpace{}%
\AgdaBound{m}\AgdaSpace{}%
\AgdaOperator{\AgdaFunction{∼}}\AgdaSpace{}%
\AgdaBound{x}\AgdaSpace{}%
\AgdaOperator{\AgdaFunction{]}}\<%
\end{code}
%
%
\begin{code}[hide]%
\>[0]\AgdaFunction{S-sound}%
\>[149I]\AgdaSymbol{:}\AgdaSpace{}%
\AgdaSymbol{∀}\AgdaSpace{}%
\AgdaSymbol{\{}\AgdaBound{d}\AgdaSpace{}%
\AgdaBound{X}\AgdaSymbol{\}}\AgdaSpace{}%
\AgdaSymbol{(}\AgdaBound{f}\AgdaSpace{}%
\AgdaSymbol{:}\AgdaSpace{}%
\AgdaOperator{\AgdaFunction{⟦}}\AgdaSpace{}%
\AgdaBound{d}\AgdaSpace{}%
\AgdaOperator{\AgdaFunction{⟧}}\AgdaSpace{}%
\AgdaBound{X}\AgdaSymbol{)}\<%
\\
\>[.][@{}l@{}]\<[149I]%
\>[8]\AgdaSymbol{→}\AgdaSpace{}%
\AgdaDatatype{All}\AgdaSpace{}%
\AgdaSymbol{((}\AgdaBound{f}\AgdaSpace{}%
\AgdaOperator{\AgdaDatatype{≡\AgdaUnderscore{}}}\AgdaSymbol{)}\AgdaSpace{}%
\AgdaOperator{\AgdaFunction{∘}}\AgdaSpace{}%
\AgdaFunction{uncurry}\AgdaSpace{}%
\AgdaSymbol{(}\AgdaFunction{plug}\AgdaSpace{}%
\AgdaSymbol{\{}\AgdaBound{d}\AgdaSymbol{\}))}\AgdaSpace{}%
\AgdaSymbol{(}\AgdaOperator{\AgdaFunction{S[}}\AgdaSpace{}%
\AgdaBound{d}\AgdaSpace{}%
\AgdaOperator{\AgdaFunction{∼}}\AgdaSpace{}%
\AgdaBound{f}\AgdaSpace{}%
\AgdaOperator{\AgdaFunction{]}}\AgdaSymbol{)}\<%
\end{code}%
\begin{code}[hide]%
\>[0]\AgdaFunction{S-sound}\AgdaSpace{}%
\AgdaSymbol{\{}\AgdaInductiveConstructor{I}\AgdaSymbol{\}}\AgdaSpace{}%
\AgdaBound{f}\AgdaSpace{}%
\AgdaSymbol{=}\AgdaSpace{}%
\AgdaInductiveConstructor{refl}\AgdaSpace{}%
\AgdaOperator{\AgdaInductiveConstructor{∷}}\AgdaSpace{}%
\AgdaInductiveConstructor{[]}\<%
\\
\>[0]\AgdaFunction{S-sound}\AgdaSpace{}%
\AgdaSymbol{\{}\AgdaInductiveConstructor{K}\AgdaSpace{}%
\AgdaBound{x}\AgdaSymbol{\}}\AgdaSpace{}%
\AgdaBound{f}\AgdaSpace{}%
\AgdaSymbol{=}\AgdaSpace{}%
\AgdaInductiveConstructor{[]}\<%
\\
\>[0]\AgdaFunction{S-sound}\AgdaSpace{}%
\AgdaSymbol{\{}\AgdaBound{d}\AgdaSpace{}%
\AgdaOperator{\AgdaInductiveConstructor{ẋ}}\AgdaSpace{}%
\AgdaBound{d₁}\AgdaSymbol{\}}\AgdaSpace{}%
\AgdaSymbol{(}\AgdaBound{f}\AgdaSpace{}%
\AgdaOperator{\AgdaInductiveConstructor{,}}\AgdaSpace{}%
\AgdaBound{f₁}\AgdaSymbol{)}\AgdaSpace{}%
\AgdaKeyword{with}\AgdaSpace{}%
\AgdaFunction{S-sound}\AgdaSpace{}%
\AgdaSymbol{\{}\AgdaBound{d}\AgdaSymbol{\}}\AgdaSpace{}%
\AgdaBound{f}\AgdaSpace{}%
\AgdaSymbol{|}\AgdaSpace{}%
\AgdaFunction{S-sound}\AgdaSpace{}%
\AgdaSymbol{\{}\AgdaBound{d₁}\AgdaSymbol{\}}\AgdaSpace{}%
\AgdaBound{f₁}\<%
\\
\>[0]\AgdaSymbol{...}\AgdaSpace{}%
\AgdaSymbol{|}\AgdaSpace{}%
\AgdaBound{a₁}\AgdaSpace{}%
\AgdaSymbol{|}\AgdaSpace{}%
\AgdaBound{a₂}\AgdaSpace{}%
\AgdaSymbol{=}\AgdaSpace{}%
\AgdaFunction{++⁺}\AgdaSpace{}%
\AgdaSymbol{\{}\AgdaArgument{xs}\AgdaSpace{}%
\AgdaSymbol{=}\AgdaSpace{}%
\AgdaFunction{⅋map}\AgdaSpace{}%
\AgdaSymbol{(λ}\AgdaSpace{}%
\AgdaSymbol{(}\AgdaBound{x}\AgdaSpace{}%
\AgdaOperator{\AgdaInductiveConstructor{,}}\AgdaSpace{}%
\AgdaBound{y}\AgdaSymbol{)}\AgdaSpace{}%
\AgdaSymbol{→}\AgdaSpace{}%
\AgdaInductiveConstructor{inj₁}\AgdaSpace{}%
\AgdaSymbol{(}\AgdaBound{x}\AgdaSpace{}%
\AgdaOperator{\AgdaInductiveConstructor{,}}\AgdaSpace{}%
\AgdaBound{f₁}\AgdaSymbol{)}\AgdaSpace{}%
\AgdaOperator{\AgdaInductiveConstructor{,}}\AgdaSpace{}%
\AgdaBound{y}\AgdaSymbol{)}\AgdaSpace{}%
\AgdaOperator{\AgdaFunction{S[}}\AgdaSpace{}%
\AgdaBound{d}\AgdaSpace{}%
\AgdaOperator{\AgdaFunction{∼}}\AgdaSpace{}%
\AgdaBound{f}\AgdaSpace{}%
\AgdaOperator{\AgdaFunction{]}}\AgdaSymbol{\}}\<%
\\
\>[0][@{}l@{\AgdaIndent{0}}]%
\>[2]\AgdaSymbol{(}\AgdaFunction{gmap⁺}\AgdaSpace{}%
\AgdaSymbol{(λ}\AgdaSpace{}%
\AgdaKeyword{where}\AgdaSpace{}%
\AgdaInductiveConstructor{refl}\AgdaSpace{}%
\AgdaSymbol{→}\AgdaSpace{}%
\AgdaInductiveConstructor{refl}\AgdaSymbol{)}\AgdaSpace{}%
\AgdaBound{a₁}\AgdaSymbol{)}\<%
\\
\>[2]\AgdaSymbol{(}\AgdaFunction{gmap⁺}\AgdaSpace{}%
\AgdaSymbol{(λ}\AgdaSpace{}%
\AgdaKeyword{where}\AgdaSpace{}%
\AgdaInductiveConstructor{refl}\AgdaSpace{}%
\AgdaSymbol{→}\AgdaSpace{}%
\AgdaInductiveConstructor{refl}\AgdaSymbol{)}\AgdaSpace{}%
\AgdaBound{a₂}\AgdaSymbol{)}\<%
\\
\>[0]\AgdaFunction{S-sound}\AgdaSpace{}%
\AgdaSymbol{\{}\AgdaInductiveConstructor{∐}\AgdaSpace{}%
\AgdaBound{n}\AgdaSpace{}%
\AgdaBound{f}\AgdaSymbol{\}}\AgdaSpace{}%
\AgdaSymbol{(}\AgdaBound{m}\AgdaSpace{}%
\AgdaOperator{\AgdaInductiveConstructor{,}}\AgdaSpace{}%
\AgdaBound{g}\AgdaSymbol{)}\AgdaSpace{}%
\AgdaSymbol{=}\AgdaSpace{}%
\AgdaFunction{gmap⁺}\AgdaSpace{}%
\AgdaSymbol{(λ}\AgdaSpace{}%
\AgdaKeyword{where}\AgdaSpace{}%
\AgdaInductiveConstructor{refl}\AgdaSpace{}%
\AgdaSymbol{→}\AgdaSpace{}%
\AgdaInductiveConstructor{refl}\AgdaSymbol{)}\AgdaSpace{}%
\AgdaSymbol{(}\AgdaFunction{S-sound}\AgdaSpace{}%
\AgdaSymbol{\{}\AgdaBound{f}\AgdaSpace{}%
\AgdaBound{m}\AgdaSymbol{\}}\AgdaSpace{}%
\AgdaBound{g}\AgdaSymbol{)}\<%
\end{code}%
%
%
\begin{code}[hide]%
\>[0]\AgdaFunction{S-complete}%
\>[12]\AgdaSymbol{:}\AgdaSpace{}%
\AgdaSymbol{∀}\AgdaSpace{}%
\AgdaSymbol{\{}\AgdaBound{d}\AgdaSpace{}%
\AgdaBound{X}\AgdaSymbol{\}}\AgdaSpace{}%
\AgdaSymbol{(}\AgdaBound{frm}\AgdaSpace{}%
\AgdaSymbol{:}\AgdaSpace{}%
\AgdaOperator{\AgdaFunction{∂[}}\AgdaSpace{}%
\AgdaBound{d}\AgdaSpace{}%
\AgdaOperator{\AgdaFunction{]}}\AgdaSpace{}%
\AgdaBound{X}\AgdaSymbol{)}\AgdaSpace{}%
\AgdaSymbol{(}\AgdaBound{x}\AgdaSpace{}%
\AgdaSymbol{:}\AgdaSpace{}%
\AgdaBound{X}\AgdaSymbol{)}\<%
\\
\>[12]\AgdaSymbol{→}\AgdaSpace{}%
\AgdaSymbol{(}\AgdaBound{frm}\AgdaSpace{}%
\AgdaOperator{\AgdaInductiveConstructor{,}}\AgdaSpace{}%
\AgdaBound{x}\AgdaSymbol{)}\AgdaSpace{}%
\AgdaOperator{\AgdaFunction{∈}}\AgdaSpace{}%
\AgdaOperator{\AgdaFunction{S[}}\AgdaSpace{}%
\AgdaBound{d}\AgdaSpace{}%
\AgdaOperator{\AgdaFunction{∼}}\AgdaSpace{}%
\AgdaFunction{plug}\AgdaSpace{}%
\AgdaSymbol{\{}\AgdaBound{d}\AgdaSymbol{\}}\AgdaSpace{}%
\AgdaBound{frm}\AgdaSpace{}%
\AgdaBound{x}\AgdaSpace{}%
\AgdaOperator{\AgdaFunction{]}}\<%
\end{code}
\begin{code}[hide]%
\>[0]\AgdaFunction{S-complete}\AgdaSpace{}%
\AgdaSymbol{\{}\AgdaInductiveConstructor{I}\AgdaSymbol{\}}\AgdaSpace{}%
\AgdaInductiveConstructor{tt}\AgdaSpace{}%
\AgdaBound{x}\AgdaSpace{}%
\AgdaSymbol{=}\AgdaSpace{}%
\AgdaInductiveConstructor{here}\AgdaSpace{}%
\AgdaInductiveConstructor{refl}\<%
\\
\>[0]\AgdaFunction{S-complete}\AgdaSpace{}%
\AgdaSymbol{\{}\AgdaBound{d}\AgdaSpace{}%
\AgdaOperator{\AgdaInductiveConstructor{ẋ}}\AgdaSpace{}%
\AgdaBound{d₁}\AgdaSymbol{\}}\AgdaSpace{}%
\AgdaSymbol{(}\AgdaInductiveConstructor{inj₁}\AgdaSpace{}%
\AgdaSymbol{(}\AgdaBound{a}\AgdaSpace{}%
\AgdaOperator{\AgdaInductiveConstructor{,}}\AgdaSpace{}%
\AgdaBound{t}\AgdaSymbol{))}\AgdaSpace{}%
\AgdaBound{x}\AgdaSpace{}%
\AgdaSymbol{=}\AgdaSpace{}%
\AgdaFunction{∈-++⁺ˡ}\<%
\\
\>[0][@{}l@{\AgdaIndent{0}}]%
\>[2]\AgdaSymbol{(}\AgdaFunction{∈-map⁺}\<%
\\
\>[2][@{}l@{\AgdaIndent{0}}]%
\>[4]\AgdaSymbol{(λ}\AgdaSpace{}%
\AgdaSymbol{(}\AgdaBound{a}\AgdaSpace{}%
\AgdaOperator{\AgdaInductiveConstructor{,}}\AgdaSpace{}%
\AgdaBound{y}\AgdaSymbol{)}\AgdaSpace{}%
\AgdaSymbol{→}\AgdaSpace{}%
\AgdaInductiveConstructor{inj₁}\AgdaSpace{}%
\AgdaSymbol{(}\AgdaBound{a}\AgdaSpace{}%
\AgdaOperator{\AgdaInductiveConstructor{,}}\AgdaSpace{}%
\AgdaBound{t}\AgdaSymbol{)}\AgdaSpace{}%
\AgdaOperator{\AgdaInductiveConstructor{,}}\AgdaSpace{}%
\AgdaBound{y}\AgdaSymbol{)}\<%
\\
\>[4]\AgdaSymbol{(}\AgdaFunction{S-complete}\AgdaSpace{}%
\AgdaSymbol{\{}\AgdaBound{d}\AgdaSymbol{\}}\AgdaSpace{}%
\AgdaBound{a}\AgdaSpace{}%
\AgdaBound{x}\AgdaSymbol{))}\<%
\\
\>[0]\AgdaFunction{S-complete}\AgdaSpace{}%
\AgdaSymbol{\{}\AgdaBound{d}\AgdaSpace{}%
\AgdaOperator{\AgdaInductiveConstructor{ẋ}}\AgdaSpace{}%
\AgdaBound{d₁}\AgdaSymbol{\}}\AgdaSpace{}%
\AgdaSymbol{(}\AgdaInductiveConstructor{inj₂}\AgdaSpace{}%
\AgdaSymbol{(}\AgdaBound{t}\AgdaSpace{}%
\AgdaOperator{\AgdaInductiveConstructor{,}}\AgdaSpace{}%
\AgdaBound{b}\AgdaSymbol{))}\AgdaSpace{}%
\AgdaBound{x}\AgdaSpace{}%
\AgdaSymbol{=}\AgdaSpace{}%
\AgdaFunction{∈-++⁺ʳ}\AgdaSpace{}%
\AgdaSymbol{\AgdaUnderscore{}}\<%
\\
\>[0][@{}l@{\AgdaIndent{0}}]%
\>[2]\AgdaSymbol{(}\AgdaFunction{∈-map⁺}\<%
\\
\>[2][@{}l@{\AgdaIndent{0}}]%
\>[4]\AgdaSymbol{(λ}\AgdaSpace{}%
\AgdaSymbol{(}\AgdaBound{b}\AgdaSpace{}%
\AgdaOperator{\AgdaInductiveConstructor{,}}\AgdaSpace{}%
\AgdaBound{y}\AgdaSymbol{)}\AgdaSpace{}%
\AgdaSymbol{→}\AgdaSpace{}%
\AgdaInductiveConstructor{inj₂}\AgdaSpace{}%
\AgdaSymbol{(}\AgdaBound{t}\AgdaSpace{}%
\AgdaOperator{\AgdaInductiveConstructor{,}}\AgdaSpace{}%
\AgdaBound{b}\AgdaSymbol{)}\AgdaSpace{}%
\AgdaOperator{\AgdaInductiveConstructor{,}}\AgdaSpace{}%
\AgdaBound{y}\AgdaSymbol{)}\<%
\\
\>[4]\AgdaSymbol{(}\AgdaFunction{S-complete}\AgdaSpace{}%
\AgdaSymbol{\{}\AgdaBound{d₁}\AgdaSymbol{\}}\AgdaSpace{}%
\AgdaBound{b}\AgdaSpace{}%
\AgdaBound{x}\AgdaSymbol{))}\<%
\\
\>[0]\AgdaFunction{S-complete}\AgdaSpace{}%
\AgdaSymbol{\{}\AgdaInductiveConstructor{∐}\AgdaSpace{}%
\AgdaBound{n}\AgdaSpace{}%
\AgdaBound{f}\AgdaSymbol{\}}\AgdaSpace{}%
\AgdaSymbol{(}\AgdaBound{m}\AgdaSpace{}%
\AgdaOperator{\AgdaInductiveConstructor{,}}\AgdaSpace{}%
\AgdaBound{fr}\AgdaSymbol{)}\AgdaSpace{}%
\AgdaBound{x}\AgdaSpace{}%
\AgdaSymbol{=}\AgdaSpace{}%
\AgdaFunction{∈-map⁺}\<%
\\
\>[0][@{}l@{\AgdaIndent{0}}]%
\>[2]\AgdaSymbol{(λ}\AgdaSpace{}%
\AgdaSymbol{(}\AgdaBound{fr}\AgdaSpace{}%
\AgdaOperator{\AgdaInductiveConstructor{,}}\AgdaSpace{}%
\AgdaBound{x}\AgdaSymbol{)}\AgdaSpace{}%
\AgdaSymbol{→}\AgdaSpace{}%
\AgdaSymbol{(}\AgdaBound{m}\AgdaSpace{}%
\AgdaOperator{\AgdaInductiveConstructor{,}}\AgdaSpace{}%
\AgdaBound{fr}\AgdaSymbol{)}\AgdaSpace{}%
\AgdaOperator{\AgdaInductiveConstructor{,}}\AgdaSpace{}%
\AgdaBound{x}\AgdaSymbol{)}\<%
\\
\>[2]\AgdaSymbol{(}\AgdaFunction{S-complete}\AgdaSpace{}%
\AgdaSymbol{\{}\AgdaBound{f}\AgdaSpace{}%
\AgdaBound{m}\AgdaSymbol{\}}\AgdaSpace{}%
\AgdaBound{fr}\AgdaSpace{}%
\AgdaBound{x}\AgdaSymbol{)}\<%
\\
\\[\AgdaEmptyExtraSkip]%
\\[\AgdaEmptyExtraSkip]%
\>[0]\AgdaFunction{DecStrat}\AgdaSpace{}%
\AgdaSymbol{:}\AgdaSpace{}%
\AgdaDatatype{Desc}\AgdaSpace{}%
\AgdaSymbol{→}\AgdaSpace{}%
\AgdaPrimitive{Set}\<%
\\
\>[0]\AgdaFunction{DecStrat}\AgdaSpace{}%
\AgdaBound{d}\AgdaSpace{}%
\AgdaSymbol{=}\AgdaSpace{}%
\AgdaOperator{\AgdaFunction{Ctx[}}\AgdaSpace{}%
\AgdaBound{d}\AgdaSpace{}%
\AgdaOperator{\AgdaFunction{]}}\AgdaSpace{}%
\AgdaSymbol{→}\AgdaSpace{}%
\AgdaDatatype{Maybe}\AgdaSpace{}%
\AgdaSymbol{(}\AgdaOperator{\AgdaFunction{Ctx[}}\AgdaSpace{}%
\AgdaBound{d}\AgdaSpace{}%
\AgdaOperator{\AgdaFunction{]}}\AgdaSpace{}%
\AgdaOperator{\AgdaFunction{×}}\AgdaSpace{}%
\AgdaOperator{\AgdaDatatype{μ[}}\AgdaSpace{}%
\AgdaBound{d}\AgdaSpace{}%
\AgdaOperator{\AgdaDatatype{]}}\AgdaSymbol{)}\<%
\end{code}
In \cref{app:characterization} we use this definition to define the horizontal/vertical/strength taxonomy more precisely in Agda.

\subsection{Generic Decomposition}
\label{sec:generic-paras}

We can define decomposition generically using algebras and paramorphisms.
To simplify the definition, we use a parameterized Agda module.
\begin{code}%
\>[0]\AgdaKeyword{module}%
\>[8]\AgdaModule{GenericReduction}\<%
\\
\>[0][@{}l@{\AgdaIndent{0}}]%
\>[4]\AgdaSymbol{\{}\AgdaBound{d}\AgdaSpace{}%
\AgdaSymbol{:}\AgdaSpace{}%
\AgdaDatatype{Desc}\AgdaSymbol{\}}\<%
\\
\>[4]\AgdaSymbol{(}\AgdaBound{is-redex}\AgdaSpace{}%
\AgdaBound{is-val}\AgdaSpace{}%
\AgdaSymbol{:}\AgdaSpace{}%
\AgdaOperator{\AgdaDatatype{μ[}}\AgdaSpace{}%
\AgdaBound{d}\AgdaSpace{}%
\AgdaOperator{\AgdaDatatype{]}}\AgdaSpace{}%
\AgdaSymbol{→}\AgdaSpace{}%
\AgdaDatatype{Bool}\AgdaSymbol{)}\<%
\\
\>[4]\AgdaSymbol{(}\AgdaBound{redex-pos}\AgdaSpace{}%
\AgdaSymbol{:}\AgdaSpace{}%
\AgdaSymbol{∀}\AgdaSpace{}%
\AgdaSymbol{\{}\AgdaBound{X}\AgdaSpace{}%
\AgdaBound{e}\AgdaSymbol{\}}\AgdaSpace{}%
\AgdaSymbol{\{}\AgdaBound{x}\AgdaSpace{}%
\AgdaSymbol{:}\AgdaSpace{}%
\AgdaOperator{\AgdaFunction{⟦}}\AgdaSpace{}%
\AgdaBound{d}\AgdaSpace{}%
\AgdaOperator{\AgdaFunction{⟧}}\AgdaSpace{}%
\AgdaBound{X}\AgdaSymbol{\}}\AgdaSpace{}%
\AgdaSymbol{→}\AgdaSpace{}%
\AgdaBound{e}\AgdaSpace{}%
\AgdaOperator{\AgdaFunction{∈}}\AgdaSpace{}%
\AgdaOperator{\AgdaFunction{S[}}\AgdaSpace{}%
\AgdaBound{d}\AgdaSpace{}%
\AgdaOperator{\AgdaFunction{∼}}\AgdaSpace{}%
\AgdaBound{x}\AgdaSpace{}%
\AgdaOperator{\AgdaFunction{]}}\AgdaSpace{}%
\AgdaSymbol{→}\AgdaSpace{}%
\AgdaDatatype{Bool}\AgdaSymbol{)}\<%
\\
\>[0][@{}l@{\AgdaIndent{0}}]%
\>[2]\AgdaKeyword{where}\<%
\end{code}
When importing this module we must provide a concrete description and three concrete functions characterizing redexes, values, and \emph{redex positions}.

The redex position function \ab{redex-pos} associates each sub-term position (computed by the \af{S[\_∼\_]} function) of a term \ab{x} with a Boolean.
This Boolean tells us, for each sub-term position, whether we should search for a redex in that sub-term (\ac{true}) or not (\ac{false}).
The purpose of \ab{redex-pos} is to distinguish sub-term positions which should not be reducible.
For example, \emph{weak reduction strategies}~\cite{LagoM08,ForsterKR20} for the $\lambda$ calculus do not admit reduction under $\lambda$s.
Similarly, it is common for reduction semantics to not search for redexes in branches of if-expressions.
\Cref{sec:decomp-lambda-shift-reset} contains an example semantics with a weak reduction strategy.

The following algebra uses the module parameters to define a generic, left-most, inner-most reduction strategy.
\begin{code}%
\>[2]\AgdaFunction{algLMIMW}\AgdaSpace{}%
\AgdaSymbol{:}\AgdaSpace{}%
\AgdaRecord{Alg}\AgdaSpace{}%
\AgdaSymbol{(λ}\AgdaSpace{}%
\AgdaBound{X}\AgdaSpace{}%
\AgdaSymbol{→}\AgdaSpace{}%
\AgdaOperator{\AgdaFunction{⟦}}\AgdaSpace{}%
\AgdaBound{d}\AgdaSpace{}%
\AgdaOperator{\AgdaFunction{⟧}}\AgdaSpace{}%
\AgdaSymbol{(}\AgdaOperator{\AgdaDatatype{μ[}}\AgdaSpace{}%
\AgdaBound{d}\AgdaSpace{}%
\AgdaOperator{\AgdaDatatype{]}}\AgdaSpace{}%
\AgdaOperator{\AgdaFunction{×}}\AgdaSpace{}%
\AgdaBound{X}\AgdaSymbol{))}\AgdaSpace{}%
\AgdaSymbol{(}\AgdaFunction{DecStrat}\AgdaSpace{}%
\AgdaBound{d}\AgdaSymbol{)}\<%
\\
\>[2]\AgdaField{alg}\AgdaSpace{}%
\AgdaFunction{algLMIMW}\AgdaSpace{}%
\AgdaBound{x}\AgdaSpace{}%
\AgdaBound{c}\AgdaSpace{}%
\AgdaSymbol{=}\<%
\\
\>[2][@{}l@{\AgdaIndent{0}}]%
\>[4]\AgdaKeyword{let}\AgdaSpace{}%
\AgdaBound{t}\AgdaSpace{}%
\AgdaSymbol{=}\AgdaSpace{}%
\AgdaOperator{\AgdaInductiveConstructor{⟨}}\AgdaSpace{}%
\AgdaFunction{fmap}\AgdaSpace{}%
\AgdaSymbol{\{}\AgdaBound{d}\AgdaSymbol{\}}\AgdaSpace{}%
\AgdaField{proj₁}\AgdaSpace{}%
\AgdaBound{x}\AgdaSpace{}%
\AgdaOperator{\AgdaInductiveConstructor{⟩}}\AgdaSpace{}%
\AgdaKeyword{in}\<%
\\
\>[4]\AgdaFunction{maybe}\AgdaSpace{}%
\AgdaInductiveConstructor{just}\<%
\\
\>[4][@{}l@{\AgdaIndent{0}}]%
\>[6]\AgdaSymbol{(}\AgdaOperator{\AgdaFunction{if}}\AgdaSpace{}%
\AgdaBound{is-redex}\AgdaSpace{}%
\AgdaBound{t}\AgdaSpace{}%
\AgdaOperator{\AgdaFunction{then}}\AgdaSpace{}%
\AgdaInductiveConstructor{just}\AgdaSpace{}%
\AgdaSymbol{(}\AgdaBound{c}\AgdaSpace{}%
\AgdaOperator{\AgdaInductiveConstructor{,}}\AgdaSpace{}%
\AgdaBound{t}\AgdaSymbol{)}\AgdaSpace{}%
\AgdaOperator{\AgdaFunction{else}}\AgdaSpace{}%
\AgdaInductiveConstructor{nothing}\AgdaSymbol{)}\<%
\\
\>[6]\AgdaSymbol{(}\AgdaFunction{find-left-most}\AgdaSpace{}%
\AgdaOperator{\AgdaFunction{S[}}\AgdaSpace{}%
\AgdaBound{d}\AgdaSpace{}%
\AgdaOperator{\AgdaFunction{∼}}\AgdaSpace{}%
\AgdaBound{x}\AgdaSpace{}%
\AgdaOperator{\AgdaFunction{]}}\AgdaSpace{}%
\AgdaBound{redex-pos}\AgdaSymbol{)}\<%
\\
\>[4]\AgdaKeyword{where}\<%
\\
\>[4][@{}l@{\AgdaIndent{0}}]%
\>[6]\AgdaFunction{find-left-most}%
\>[440I]\AgdaSymbol{:}\AgdaSpace{}%
\AgdaSymbol{(}\AgdaBound{xs}\AgdaSpace{}%
\AgdaSymbol{:}\AgdaSpace{}%
\AgdaDatatype{List}%
\>[444I]\AgdaSymbol{(}%
\>[37]\AgdaOperator{\AgdaFunction{∂[}}\AgdaSpace{}%
\AgdaBound{d}\AgdaSpace{}%
\AgdaOperator{\AgdaFunction{]}}\AgdaSpace{}%
\AgdaSymbol{(}\AgdaOperator{\AgdaDatatype{μ[}}\AgdaSpace{}%
\AgdaBound{d}\AgdaSpace{}%
\AgdaOperator{\AgdaDatatype{]}}\AgdaSpace{}%
\AgdaOperator{\AgdaFunction{×}}\AgdaSpace{}%
\AgdaFunction{DecStrat}\AgdaSpace{}%
\AgdaBound{d}\AgdaSymbol{)}\<%
\\
\>[.][@{}l@{}]\<[444I]%
\>[34]\AgdaOperator{\AgdaFunction{×}}%
\>[37]\AgdaOperator{\AgdaDatatype{μ[}}\AgdaSpace{}%
\AgdaBound{d}\AgdaSpace{}%
\AgdaOperator{\AgdaDatatype{]}}\AgdaSpace{}%
\AgdaOperator{\AgdaFunction{×}}\AgdaSpace{}%
\AgdaFunction{DecStrat}\AgdaSpace{}%
\AgdaBound{d}\AgdaSpace{}%
\AgdaSymbol{))}\<%
\\
\>[.][@{}l@{}]\<[440I]%
\>[21]\AgdaSymbol{→}\AgdaSpace{}%
\AgdaSymbol{(∀}\AgdaSpace{}%
\AgdaSymbol{\{}\AgdaBound{e}\AgdaSymbol{\}}\AgdaSpace{}%
\AgdaSymbol{→}\AgdaSpace{}%
\AgdaBound{e}\AgdaSpace{}%
\AgdaOperator{\AgdaFunction{∈}}\AgdaSpace{}%
\AgdaBound{xs}\AgdaSpace{}%
\AgdaSymbol{→}\AgdaSpace{}%
\AgdaDatatype{Bool}\AgdaSymbol{)}\<%
\\
\>[21]\AgdaSymbol{→}\AgdaSpace{}%
\AgdaDatatype{Maybe}\AgdaSpace{}%
\AgdaSymbol{(}\AgdaOperator{\AgdaFunction{Ctx[}}\AgdaSpace{}%
\AgdaBound{d}\AgdaSpace{}%
\AgdaOperator{\AgdaFunction{]}}\AgdaSpace{}%
\AgdaOperator{\AgdaFunction{×}}\AgdaSpace{}%
\AgdaOperator{\AgdaDatatype{μ[}}\AgdaSpace{}%
\AgdaBound{d}\AgdaSpace{}%
\AgdaOperator{\AgdaDatatype{]}}\AgdaSymbol{)}\<%
\\
\>[6]\AgdaFunction{find-left-most}\AgdaSpace{}%
\AgdaInductiveConstructor{[]}\AgdaSpace{}%
\AgdaBound{rp}\AgdaSpace{}%
\AgdaSymbol{=}\AgdaSpace{}%
\AgdaInductiveConstructor{nothing}\<%
\\
\>[6]\AgdaFunction{find-left-most}\AgdaSpace{}%
\AgdaSymbol{((}\AgdaBound{frm}\AgdaSpace{}%
\AgdaOperator{\AgdaInductiveConstructor{,}}\AgdaSpace{}%
\AgdaBound{t}\AgdaSpace{}%
\AgdaOperator{\AgdaInductiveConstructor{,}}\AgdaSpace{}%
\AgdaBound{f}\AgdaSymbol{)}\AgdaSpace{}%
\AgdaOperator{\AgdaInductiveConstructor{∷}}\AgdaSpace{}%
\AgdaBound{c₀}\AgdaSymbol{)}\AgdaSpace{}%
\AgdaBound{rp}\AgdaSpace{}%
\AgdaKeyword{with}\AgdaSpace{}%
\AgdaBound{rp}\AgdaSpace{}%
\AgdaSymbol{(}\AgdaInductiveConstructor{here}\AgdaSpace{}%
\AgdaInductiveConstructor{refl}\AgdaSymbol{)}\<%
\\
\>[6]\AgdaSymbol{...}\AgdaSpace{}%
\AgdaSymbol{|}\AgdaSpace{}%
\AgdaInductiveConstructor{true}\AgdaSpace{}%
\AgdaSymbol{=}\AgdaSpace{}%
\AgdaFunction{maybe}\AgdaSpace{}%
\AgdaInductiveConstructor{just}\<%
\\
\>[6][@{}l@{\AgdaIndent{0}}]%
\>[8]\AgdaSymbol{(}\AgdaOperator{\AgdaFunction{if}}\AgdaSpace{}%
\AgdaBound{is-val}\AgdaSpace{}%
\AgdaBound{t}%
\>[22]\AgdaOperator{\AgdaFunction{then}}\AgdaSpace{}%
\AgdaFunction{find-left-most}\AgdaSpace{}%
\AgdaBound{c₀}\AgdaSpace{}%
\AgdaSymbol{(}\AgdaBound{rp}\AgdaSpace{}%
\AgdaOperator{\AgdaFunction{∘}}\AgdaSpace{}%
\AgdaInductiveConstructor{there}\AgdaSymbol{)}\<%
\\
\>[22]\AgdaOperator{\AgdaFunction{else}}\AgdaSpace{}%
\AgdaInductiveConstructor{nothing}\AgdaSymbol{)}\<%
\\
\>[8]\AgdaSymbol{(}\AgdaBound{f}\AgdaSpace{}%
\AgdaSymbol{(}\AgdaFunction{∂map}\AgdaSpace{}%
\AgdaSymbol{\{}\AgdaBound{d}\AgdaSymbol{\}}\AgdaSpace{}%
\AgdaField{proj₁}\AgdaSpace{}%
\AgdaBound{frm}\AgdaSpace{}%
\AgdaOperator{\AgdaInductiveConstructor{∷}}\AgdaSpace{}%
\AgdaBound{c}\AgdaSymbol{))}\<%
\\
\>[6]\AgdaSymbol{...}\AgdaSpace{}%
\AgdaSymbol{|}\AgdaSpace{}%
\AgdaInductiveConstructor{false}\AgdaSpace{}%
\AgdaSymbol{=}\AgdaSpace{}%
\AgdaInductiveConstructor{nothing}\<%
\end{code}
This algebra uses \af{find-left-most} to first check if a left-most-inner-most redex exists.
If it does, we \ac{just} use that.
If it does not, we check whether the current term is a redex, and return it if it is.
If it is not, we return \ac{nothing}.
The \af{find-left-most} function ensures that all terms to the left of the found redex are values, and returns \ac{nothing} if they are not.

Using a paramorphism to fold the algebra over a syntax tree yields a decomposition function:
\begin{code}%
\>[2]\AgdaFunction{decomposeLMIMW}\AgdaSpace{}%
\AgdaSymbol{:}\AgdaSpace{}%
\AgdaOperator{\AgdaDatatype{μ[}}\AgdaSpace{}%
\AgdaBound{d}\AgdaSpace{}%
\AgdaOperator{\AgdaDatatype{]}}\AgdaSpace{}%
\AgdaSymbol{→}\AgdaSpace{}%
\AgdaDatatype{Maybe}\AgdaSpace{}%
\AgdaSymbol{(}\AgdaOperator{\AgdaFunction{Ctx[}}\AgdaSpace{}%
\AgdaBound{d}\AgdaSpace{}%
\AgdaOperator{\AgdaFunction{]}}\AgdaSpace{}%
\AgdaOperator{\AgdaFunction{×}}\AgdaSpace{}%
\AgdaOperator{\AgdaDatatype{μ[}}\AgdaSpace{}%
\AgdaBound{d}\AgdaSpace{}%
\AgdaOperator{\AgdaDatatype{]}}\AgdaSymbol{)}\<%
\\
\>[2]\AgdaFunction{decomposeLMIMW}\AgdaSpace{}%
\AgdaSymbol{=}\AgdaSpace{}%
\AgdaFunction{flip}\AgdaSpace{}%
\AgdaOperator{\AgdaFunction{【}}\AgdaSpace{}%
\AgdaBound{d}\AgdaSpace{}%
\AgdaOperator{\AgdaFunction{∼}}\AgdaSpace{}%
\AgdaFunction{algLMIMW}\AgdaSpace{}%
\AgdaOperator{\AgdaFunction{】}}\AgdaSpace{}%
\AgdaInductiveConstructor{[]}\<%
\end{code}
Other variants of standard reduction strategy variants can be provided analogously.

As we illustrate next, the standard strategy above suffices to define reduction for both the arithmetic language from~\cref{sec:simple-arith} and the language with $\lambda$s and shift/reset from~\cref{sec:lambda-shift-reset}.

\subsubsection{Decomposition for Arithmetic Expressions}

To make use of our generic reduction strategy, we define the following functions.
\begin{code}%
\>[0]\AgdaFunction{is-redex-HR}\AgdaSpace{}%
\AgdaFunction{is-val-HR}\AgdaSpace{}%
\AgdaSymbol{:}\AgdaSpace{}%
\AgdaOperator{\AgdaDatatype{μ[}}\AgdaSpace{}%
\AgdaFunction{HR}\AgdaSpace{}%
\AgdaOperator{\AgdaDatatype{]}}\AgdaSpace{}%
\AgdaSymbol{→}\AgdaSpace{}%
\AgdaDatatype{Bool}\<%
\\
\>[0]\AgdaFunction{redex-pos-HR}%
\>[14]\AgdaSymbol{:}\AgdaSpace{}%
\AgdaSymbol{∀}\AgdaSpace{}%
\AgdaSymbol{\{}\AgdaBound{X}\AgdaSpace{}%
\AgdaBound{e}\AgdaSymbol{\}}\AgdaSpace{}%
\AgdaSymbol{\{}\AgdaBound{x}\AgdaSpace{}%
\AgdaSymbol{:}\AgdaSpace{}%
\AgdaOperator{\AgdaFunction{⟦}}\AgdaSpace{}%
\AgdaFunction{HR}\AgdaSpace{}%
\AgdaOperator{\AgdaFunction{⟧}}\AgdaSpace{}%
\AgdaBound{X}\AgdaSymbol{\}}\AgdaSpace{}%
\AgdaSymbol{→}\AgdaSpace{}%
\AgdaBound{e}\AgdaSpace{}%
\AgdaOperator{\AgdaFunction{∈}}\AgdaSpace{}%
\AgdaOperator{\AgdaFunction{S[}}\AgdaSpace{}%
\AgdaFunction{HR}\AgdaSpace{}%
\AgdaOperator{\AgdaFunction{∼}}\AgdaSpace{}%
\AgdaBound{x}\AgdaSpace{}%
\AgdaOperator{\AgdaFunction{]}}\<%
\\
\>[14]\AgdaSymbol{→}\AgdaSpace{}%
\AgdaDatatype{Bool}\<%
\\
\\[\AgdaEmptyExtraSkip]%
\>[0]\AgdaFunction{is-redex-HR}\AgdaSpace{}%
\AgdaSymbol{(}\AgdaInductiveConstructor{`add}\AgdaSpace{}%
\AgdaSymbol{(}\AgdaInductiveConstructor{`num}\AgdaSpace{}%
\AgdaBound{n₀}\AgdaSymbol{)}\AgdaSpace{}%
\AgdaSymbol{(}\AgdaInductiveConstructor{`num}\AgdaSpace{}%
\AgdaBound{n₁}\AgdaSymbol{))}%
\>[40]\AgdaSymbol{=}\AgdaSpace{}%
\AgdaInductiveConstructor{true}\<%
\\
\>[0]\AgdaCatchallClause{\AgdaFunction{is-redex-HR}}\AgdaSpace{}%
\AgdaCatchallClause{\AgdaSymbol{\AgdaUnderscore{}}}%
\>[40]\AgdaSymbol{=}\AgdaSpace{}%
\AgdaInductiveConstructor{false}\<%
\\
\\[\AgdaEmptyExtraSkip]%
\>[0]\AgdaFunction{is-val-HR}\AgdaSpace{}%
\AgdaSymbol{(}\AgdaInductiveConstructor{`num}\AgdaSpace{}%
\AgdaBound{n}\AgdaSymbol{)}%
\>[20]\AgdaSymbol{=}\AgdaSpace{}%
\AgdaInductiveConstructor{true}\<%
\\
\>[0]\AgdaCatchallClause{\AgdaFunction{is-val-HR}}\AgdaSpace{}%
\AgdaCatchallClause{\AgdaSymbol{\AgdaUnderscore{}}}%
\>[20]\AgdaSymbol{=}\AgdaSpace{}%
\AgdaInductiveConstructor{false}\<%
\\
\\[\AgdaEmptyExtraSkip]%
\>[0]\AgdaFunction{redex-pos-HR}\AgdaSpace{}%
\AgdaSymbol{\AgdaUnderscore{}}\AgdaSpace{}%
\AgdaSymbol{=}\AgdaSpace{}%
\AgdaInductiveConstructor{true}\AgdaSpace{}%
\AgdaComment{--\ All\ sub-terms\ reducible}\<%
\end{code}
We can now import the \am{GenericReduction} module,
\begin{code}[hide]%
\>[0]\AgdaKeyword{pattern}\AgdaSpace{}%
\AgdaInductiveConstructor{`add-l}\AgdaSpace{}%
\AgdaBound{e}%
\>[19]\AgdaSymbol{=}\AgdaSpace{}%
\AgdaInductiveConstructor{suc}\AgdaSpace{}%
\AgdaInductiveConstructor{zero}\AgdaSpace{}%
\AgdaOperator{\AgdaInductiveConstructor{,}}\AgdaSpace{}%
\AgdaInductiveConstructor{inj₁}\AgdaSpace{}%
\AgdaSymbol{(}\AgdaInductiveConstructor{tt}\AgdaSpace{}%
\AgdaOperator{\AgdaInductiveConstructor{,}}\AgdaSpace{}%
\AgdaBound{e}\AgdaSymbol{)}\<%
\\
\>[0]\AgdaKeyword{pattern}\AgdaSpace{}%
\AgdaInductiveConstructor{`add-r}\AgdaSpace{}%
\AgdaBound{v}%
\>[19]\AgdaSymbol{=}\AgdaSpace{}%
\AgdaInductiveConstructor{suc}\AgdaSpace{}%
\AgdaInductiveConstructor{zero}\AgdaSpace{}%
\AgdaOperator{\AgdaInductiveConstructor{,}}\AgdaSpace{}%
\AgdaInductiveConstructor{inj₂}\AgdaSpace{}%
\AgdaSymbol{(}\AgdaBound{v}\AgdaSpace{}%
\AgdaOperator{\AgdaInductiveConstructor{,}}\AgdaSpace{}%
\AgdaInductiveConstructor{tt}\AgdaSymbol{)}\<%
\\
\>[0]\AgdaKeyword{module}\AgdaSpace{}%
\AgdaModule{HRReduction}\AgdaSpace{}%
\AgdaKeyword{where}\<%
\\
\>[0][@{}l@{\AgdaIndent{0}}]%
\>[2]\AgdaKeyword{open}%
\>[597I]\AgdaModule{GenericReduction}\<%
\\
\>[597I][@{}l@{\AgdaIndent{0}}]%
\>[9]\AgdaSymbol{\{}\AgdaFunction{HR}\AgdaSymbol{\}}\<%
\\
\>[9]\AgdaFunction{is-redex-HR}\<%
\\
\>[9]\AgdaFunction{is-val-HR}\<%
\\
\>[9]\AgdaFunction{redex-pos-HR}\<%
\end{code}%
which in turn lets us decompose terms similarly to \cref{sec:simple-arith}; e.g.:
\begin{code}%
\>[2]\AgdaFunction{decomp-[1+2]+[3+4]}\AgdaSpace{}%
\AgdaSymbol{:}\<%
\\
\>[2][@{}l@{\AgdaIndent{0}}]%
\>[4]\AgdaFunction{decomposeLMIMW}\AgdaSpace{}%
\AgdaSymbol{(}\AgdaInductiveConstructor{`add}%
\>[26]\AgdaSymbol{(}\AgdaInductiveConstructor{`add}\AgdaSpace{}%
\AgdaSymbol{(}\AgdaInductiveConstructor{`num}\AgdaSpace{}%
\AgdaNumber{1}\AgdaSymbol{)}\AgdaSpace{}%
\AgdaSymbol{(}\AgdaInductiveConstructor{`num}\AgdaSpace{}%
\AgdaNumber{2}\AgdaSymbol{))}\<%
\\
\>[26]\AgdaSymbol{(}\AgdaInductiveConstructor{`add}\AgdaSpace{}%
\AgdaSymbol{(}\AgdaInductiveConstructor{`num}\AgdaSpace{}%
\AgdaNumber{3}\AgdaSymbol{)}\AgdaSpace{}%
\AgdaSymbol{(}\AgdaInductiveConstructor{`num}\AgdaSpace{}%
\AgdaNumber{4}\AgdaSymbol{)))}\<%
\\
\>[4]\AgdaOperator{\AgdaDatatype{≡}}\AgdaSpace{}%
\AgdaInductiveConstructor{just}%
\>[609I]\AgdaSymbol{(}\AgdaSpace{}%
\AgdaInductiveConstructor{`add-l}\AgdaSpace{}%
\AgdaSymbol{(}\AgdaInductiveConstructor{`add}\AgdaSpace{}%
\AgdaSymbol{(}\AgdaInductiveConstructor{`num}\AgdaSpace{}%
\AgdaNumber{3}\AgdaSymbol{)}\AgdaSpace{}%
\AgdaSymbol{(}\AgdaInductiveConstructor{`num}\AgdaSpace{}%
\AgdaNumber{4}\AgdaSymbol{))}\AgdaSpace{}%
\AgdaOperator{\AgdaInductiveConstructor{∷}}\AgdaSpace{}%
\AgdaInductiveConstructor{[]}\<%
\\
\>[.][@{}l@{}]\<[609I]%
\>[11]\AgdaOperator{\AgdaInductiveConstructor{,}}\AgdaSpace{}%
\AgdaInductiveConstructor{`add}\AgdaSpace{}%
\AgdaSymbol{(}\AgdaInductiveConstructor{`num}\AgdaSpace{}%
\AgdaNumber{1}\AgdaSymbol{)}\AgdaSpace{}%
\AgdaSymbol{(}\AgdaInductiveConstructor{`num}\AgdaSpace{}%
\AgdaNumber{2}\AgdaSymbol{)}\AgdaSpace{}%
\AgdaSymbol{)}\<%
\\
\>[2]\AgdaFunction{decomp-[1+2]+[3+4]}\AgdaSpace{}%
\AgdaSymbol{=}\AgdaSpace{}%
\AgdaInductiveConstructor{refl}\<%
\end{code}
The key difference from \cref{sec:simple-arith} is that contexts and decomposition is now defined generically.
In other words, all we need to do to obtain a decomposition function for a language, is to define a syntax, its redexes, values, a function identifying (ir)reducible positions, and import the \am{GenericReduction} module.
\begin{code}[hide]%
\>[2]\AgdaFunction{decomp-[1+2]+3}\AgdaSpace{}%
\AgdaSymbol{:}%
\>[627I]\AgdaFunction{decomposeLMIMW}\<%
\\
\>[627I][@{}l@{\AgdaIndent{0}}]%
\>[21]\AgdaSymbol{(}\AgdaInductiveConstructor{`add}\AgdaSpace{}%
\AgdaSymbol{(}\AgdaInductiveConstructor{`add}\AgdaSpace{}%
\AgdaSymbol{(}\AgdaInductiveConstructor{`num}\AgdaSpace{}%
\AgdaNumber{1}\AgdaSymbol{)}\AgdaSpace{}%
\AgdaSymbol{(}\AgdaInductiveConstructor{`num}\AgdaSpace{}%
\AgdaNumber{2}\AgdaSymbol{))}\AgdaSpace{}%
\AgdaSymbol{(}\AgdaInductiveConstructor{`num}\AgdaSpace{}%
\AgdaNumber{3}\AgdaSymbol{))}\<%
\\
\>[.][@{}l@{}]\<[627I]%
\>[19]\AgdaOperator{\AgdaDatatype{≡}}\AgdaSpace{}%
\AgdaInductiveConstructor{just}%
\>[636I]\AgdaSymbol{(}\AgdaSpace{}%
\AgdaInductiveConstructor{`add-l}\AgdaSpace{}%
\AgdaSymbol{(}\AgdaInductiveConstructor{`num}\AgdaSpace{}%
\AgdaNumber{3}\AgdaSymbol{)}\AgdaSpace{}%
\AgdaOperator{\AgdaInductiveConstructor{∷}}\AgdaSpace{}%
\AgdaInductiveConstructor{[]}\<%
\\
\>[.][@{}l@{}]\<[636I]%
\>[26]\AgdaOperator{\AgdaInductiveConstructor{,}}\AgdaSpace{}%
\AgdaInductiveConstructor{`add}\AgdaSpace{}%
\AgdaSymbol{(}\AgdaInductiveConstructor{`num}\AgdaSpace{}%
\AgdaNumber{1}\AgdaSymbol{)}\AgdaSpace{}%
\AgdaSymbol{(}\AgdaInductiveConstructor{`num}\AgdaSpace{}%
\AgdaNumber{2}\AgdaSymbol{))}\<%
\\
\>[2]\AgdaFunction{decomp-[1+2]+3}\AgdaSpace{}%
\AgdaSymbol{=}\AgdaSpace{}%
\AgdaInductiveConstructor{refl}\<%
\\
\\[\AgdaEmptyExtraSkip]%
\>[2]\AgdaFunction{decomp-1+[2+3]}\AgdaSpace{}%
\AgdaSymbol{:}%
\>[20]\AgdaFunction{decomposeLMIMW}\<%
\\
\>[20][@{}l@{\AgdaIndent{0}}]%
\>[22]\AgdaSymbol{(}\AgdaInductiveConstructor{`add}\AgdaSpace{}%
\AgdaSymbol{(}\AgdaInductiveConstructor{`num}\AgdaSpace{}%
\AgdaNumber{1}\AgdaSymbol{)}\AgdaSpace{}%
\AgdaSymbol{(}\AgdaInductiveConstructor{`add}\AgdaSpace{}%
\AgdaSymbol{(}\AgdaInductiveConstructor{`num}\AgdaSpace{}%
\AgdaNumber{2}\AgdaSymbol{)}\AgdaSpace{}%
\AgdaSymbol{(}\AgdaInductiveConstructor{`num}\AgdaSpace{}%
\AgdaNumber{3}\AgdaSymbol{)))}\<%
\\
\>[20]\AgdaOperator{\AgdaDatatype{≡}}\AgdaSpace{}%
\AgdaInductiveConstructor{just}%
\>[658I]\AgdaSymbol{(}\AgdaSpace{}%
\AgdaInductiveConstructor{`add-r}\AgdaSpace{}%
\AgdaSymbol{(}\AgdaInductiveConstructor{`num}\AgdaSpace{}%
\AgdaNumber{1}\AgdaSymbol{)}\AgdaSpace{}%
\AgdaOperator{\AgdaInductiveConstructor{∷}}\AgdaSpace{}%
\AgdaInductiveConstructor{[]}\<%
\\
\>[.][@{}l@{}]\<[658I]%
\>[27]\AgdaOperator{\AgdaInductiveConstructor{,}}\AgdaSpace{}%
\AgdaInductiveConstructor{`add}\AgdaSpace{}%
\AgdaSymbol{(}\AgdaInductiveConstructor{`num}\AgdaSpace{}%
\AgdaNumber{2}\AgdaSymbol{)}\AgdaSpace{}%
\AgdaSymbol{(}\AgdaInductiveConstructor{`num}\AgdaSpace{}%
\AgdaNumber{3}\AgdaSymbol{)}\AgdaSpace{}%
\AgdaSymbol{)}\<%
\\
\>[2]\AgdaFunction{decomp-1+[2+3]}\AgdaSpace{}%
\AgdaSymbol{=}\AgdaSpace{}%
\AgdaInductiveConstructor{refl}\<%
\\
\\[\AgdaEmptyExtraSkip]%
\>[2]\AgdaFunction{decomp-[1+2]}\AgdaSpace{}%
\AgdaSymbol{:}%
\>[18]\AgdaFunction{decomposeLMIMW}\<%
\\
\>[18][@{}l@{\AgdaIndent{0}}]%
\>[20]\AgdaSymbol{(}\AgdaInductiveConstructor{`add}\AgdaSpace{}%
\AgdaSymbol{(}\AgdaInductiveConstructor{`num}\AgdaSpace{}%
\AgdaNumber{1}\AgdaSymbol{)}\AgdaSpace{}%
\AgdaSymbol{(}\AgdaInductiveConstructor{`num}\AgdaSpace{}%
\AgdaNumber{2}\AgdaSymbol{))}\<%
\\
\>[18]\AgdaOperator{\AgdaDatatype{≡}}\AgdaSpace{}%
\AgdaInductiveConstructor{just}%
\>[678I]\AgdaSymbol{(}\AgdaSpace{}%
\AgdaInductiveConstructor{[]}\<%
\\
\>[.][@{}l@{}]\<[678I]%
\>[25]\AgdaOperator{\AgdaInductiveConstructor{,}}\AgdaSpace{}%
\AgdaInductiveConstructor{`add}\AgdaSpace{}%
\AgdaSymbol{(}\AgdaInductiveConstructor{`num}\AgdaSpace{}%
\AgdaNumber{1}\AgdaSymbol{)}\AgdaSpace{}%
\AgdaSymbol{(}\AgdaInductiveConstructor{`num}\AgdaSpace{}%
\AgdaNumber{2}\AgdaSymbol{)}\AgdaSpace{}%
\AgdaSymbol{)}\<%
\\
\>[2]\AgdaFunction{decomp-[1+2]}\AgdaSpace{}%
\AgdaSymbol{=}\AgdaSpace{}%
\AgdaInductiveConstructor{refl}\<%
\end{code}

We can obtain a full reduction semantics by providing a contraction function and using a generic driver loop.
We do so in \cref{sec:generic-cases}.

\subsubsection{Decomposition for $\lambda$s and Shift/Reset}
\label{sec:decomp-lambda-shift-reset}

First, we demonstrate how we can obtain a decomposition function for $\lambda$s, shifts, and resets in a similarly generic manner.
First, we need a syntax.
\begin{code}[hide]%
\>[0]\AgdaKeyword{pattern}\AgdaSpace{}%
\AgdaInductiveConstructor{\#0}\AgdaSpace{}%
\AgdaSymbol{=}\AgdaSpace{}%
\AgdaInductiveConstructor{zero}\<%
\\
\>[0]\AgdaKeyword{pattern}\AgdaSpace{}%
\AgdaInductiveConstructor{\#1}\AgdaSpace{}%
\AgdaSymbol{=}\AgdaSpace{}%
\AgdaInductiveConstructor{suc}\AgdaSpace{}%
\AgdaInductiveConstructor{zero}\<%
\\
\>[0]\AgdaKeyword{pattern}\AgdaSpace{}%
\AgdaInductiveConstructor{\#2}\AgdaSpace{}%
\AgdaSymbol{=}\AgdaSpace{}%
\AgdaInductiveConstructor{suc}\AgdaSpace{}%
\AgdaSymbol{(}\AgdaInductiveConstructor{suc}\AgdaSpace{}%
\AgdaInductiveConstructor{zero}\AgdaSymbol{)}\<%
\\
\>[0]\AgdaKeyword{pattern}\AgdaSpace{}%
\AgdaInductiveConstructor{\#3}\AgdaSpace{}%
\AgdaSymbol{=}\AgdaSpace{}%
\AgdaInductiveConstructor{suc}\AgdaSpace{}%
\AgdaSymbol{(}\AgdaInductiveConstructor{suc}\AgdaSpace{}%
\AgdaSymbol{(}\AgdaInductiveConstructor{suc}\AgdaSpace{}%
\AgdaInductiveConstructor{zero}\AgdaSymbol{))}\<%
\\
\>[0]\AgdaKeyword{pattern}\AgdaSpace{}%
\AgdaInductiveConstructor{\#4}\AgdaSpace{}%
\AgdaSymbol{=}\AgdaSpace{}%
\AgdaInductiveConstructor{suc}\AgdaSpace{}%
\AgdaSymbol{(}\AgdaInductiveConstructor{suc}\AgdaSpace{}%
\AgdaSymbol{(}\AgdaInductiveConstructor{suc}\AgdaSpace{}%
\AgdaSymbol{(}\AgdaInductiveConstructor{suc}\AgdaSpace{}%
\AgdaInductiveConstructor{zero}\AgdaSymbol{)))}\<%
\\
\>[0]\AgdaKeyword{pattern}\AgdaSpace{}%
\AgdaInductiveConstructor{\#5}\AgdaSpace{}%
\AgdaSymbol{=}\AgdaSpace{}%
\AgdaInductiveConstructor{suc}\AgdaSpace{}%
\AgdaSymbol{(}\AgdaInductiveConstructor{suc}\AgdaSpace{}%
\AgdaSymbol{(}\AgdaInductiveConstructor{suc}\AgdaSpace{}%
\AgdaSymbol{(}\AgdaInductiveConstructor{suc}\AgdaSpace{}%
\AgdaSymbol{(}\AgdaInductiveConstructor{suc}\AgdaSpace{}%
\AgdaInductiveConstructor{zero}\AgdaSymbol{))))}\<%
\\
\>[0]\AgdaKeyword{pattern}\AgdaSpace{}%
\AgdaInductiveConstructor{\#6}\AgdaSpace{}%
\AgdaSymbol{=}\AgdaSpace{}%
\AgdaInductiveConstructor{suc}\AgdaSpace{}%
\AgdaSymbol{(}\AgdaInductiveConstructor{suc}\AgdaSpace{}%
\AgdaSymbol{(}\AgdaInductiveConstructor{suc}\AgdaSpace{}%
\AgdaSymbol{(}\AgdaInductiveConstructor{suc}\AgdaSpace{}%
\AgdaSymbol{(}\AgdaInductiveConstructor{suc}\AgdaSpace{}%
\AgdaSymbol{(}\AgdaInductiveConstructor{suc}\AgdaSpace{}%
\AgdaInductiveConstructor{zero}\AgdaSymbol{)))))}\<%
\\
\>[0]\AgdaKeyword{pattern}\AgdaSpace{}%
\AgdaInductiveConstructor{\#7}\AgdaSpace{}%
\AgdaSymbol{=}\AgdaSpace{}%
\AgdaInductiveConstructor{suc}\AgdaSpace{}%
\AgdaSymbol{(}\AgdaInductiveConstructor{suc}\AgdaSpace{}%
\AgdaSymbol{(}\AgdaInductiveConstructor{suc}\AgdaSpace{}%
\AgdaSymbol{(}\AgdaInductiveConstructor{suc}\AgdaSpace{}%
\AgdaSymbol{(}\AgdaInductiveConstructor{suc}\AgdaSpace{}%
\AgdaSymbol{(}\AgdaInductiveConstructor{suc}\AgdaSpace{}%
\AgdaSymbol{(}\AgdaInductiveConstructor{suc}\AgdaSpace{}%
\AgdaInductiveConstructor{zero}\AgdaSymbol{))))))}\<%
\\
\>[0]\AgdaKeyword{pattern}\AgdaSpace{}%
\AgdaInductiveConstructor{\#8}\AgdaSpace{}%
\AgdaSymbol{=}\AgdaSpace{}%
\AgdaInductiveConstructor{suc}\AgdaSpace{}%
\AgdaSymbol{(}\AgdaInductiveConstructor{suc}\AgdaSpace{}%
\AgdaSymbol{(}\AgdaInductiveConstructor{suc}\AgdaSpace{}%
\AgdaSymbol{(}\AgdaInductiveConstructor{suc}\AgdaSpace{}%
\AgdaSymbol{(}\AgdaInductiveConstructor{suc}\AgdaSpace{}%
\AgdaSymbol{(}\AgdaInductiveConstructor{suc}\AgdaSpace{}%
\AgdaSymbol{(}\AgdaInductiveConstructor{suc}\AgdaSpace{}%
\AgdaSymbol{(}\AgdaInductiveConstructor{suc}\AgdaSpace{}%
\AgdaInductiveConstructor{zero}\AgdaSymbol{)))))))}\<%
\\
\>[0]\AgdaKeyword{pattern}\AgdaSpace{}%
\AgdaInductiveConstructor{\#9}\AgdaSpace{}%
\AgdaSymbol{=}\AgdaSpace{}%
\AgdaInductiveConstructor{suc}\AgdaSpace{}%
\AgdaSymbol{(}\AgdaInductiveConstructor{suc}\AgdaSpace{}%
\AgdaSymbol{(}\AgdaInductiveConstructor{suc}\AgdaSpace{}%
\AgdaSymbol{(}\AgdaInductiveConstructor{suc}\AgdaSpace{}%
\AgdaSymbol{(}\AgdaInductiveConstructor{suc}\AgdaSpace{}%
\AgdaSymbol{(}\AgdaInductiveConstructor{suc}\AgdaSpace{}%
\AgdaSymbol{(}\AgdaInductiveConstructor{suc}\AgdaSpace{}%
\AgdaSymbol{(}\AgdaInductiveConstructor{suc}\AgdaSpace{}%
\AgdaSymbol{(}\AgdaInductiveConstructor{suc}\AgdaSpace{}%
\AgdaInductiveConstructor{zero}\AgdaSymbol{))))))))}\<%
\end{code}
\begin{code}%
\>[0]\AgdaFunction{LSR}\AgdaSpace{}%
\AgdaSymbol{=}\AgdaSpace{}%
\AgdaInductiveConstructor{∐}\AgdaSpace{}%
\AgdaNumber{7}\AgdaSpace{}%
\AgdaSymbol{(λ}\AgdaSpace{}%
\AgdaKeyword{where}\<%
\\
\>[0][@{}l@{\AgdaIndent{0}}]%
\>[2]\AgdaInductiveConstructor{\#0}\AgdaSpace{}%
\AgdaSymbol{→}\AgdaSpace{}%
\AgdaInductiveConstructor{K}\AgdaSpace{}%
\AgdaDatatype{ℕ}%
\>[17]\AgdaComment{--\ num\ ℕ}\<%
\\
\>[2]\AgdaInductiveConstructor{\#1}\AgdaSpace{}%
\AgdaSymbol{→}\AgdaSpace{}%
\AgdaInductiveConstructor{I}\AgdaSpace{}%
\AgdaOperator{\AgdaInductiveConstructor{ẋ}}\AgdaSpace{}%
\AgdaInductiveConstructor{I}%
\>[17]\AgdaComment{--\ add\ Term\ Term}\<%
\\
\>[2]\AgdaInductiveConstructor{\#2}\AgdaSpace{}%
\AgdaSymbol{→}\AgdaSpace{}%
\AgdaInductiveConstructor{I}%
\>[17]\AgdaComment{--\ lam\ Term}\<%
\\
\>[2]\AgdaInductiveConstructor{\#3}\AgdaSpace{}%
\AgdaSymbol{→}\AgdaSpace{}%
\AgdaInductiveConstructor{K}\AgdaSpace{}%
\AgdaDatatype{ℕ}%
\>[17]\AgdaComment{--\ var\ ℕ}\<%
\\
\>[2]\AgdaInductiveConstructor{\#4}\AgdaSpace{}%
\AgdaSymbol{→}\AgdaSpace{}%
\AgdaInductiveConstructor{I}\AgdaSpace{}%
\AgdaOperator{\AgdaInductiveConstructor{ẋ}}\AgdaSpace{}%
\AgdaInductiveConstructor{I}%
\>[17]\AgdaComment{--\ app\ Term\ Term}\<%
\\
\>[2]\AgdaInductiveConstructor{\#5}\AgdaSpace{}%
\AgdaSymbol{→}\AgdaSpace{}%
\AgdaInductiveConstructor{I}%
\>[17]\AgdaComment{--\ reset\ Term}\<%
\\
\>[2]\AgdaInductiveConstructor{\#6}\AgdaSpace{}%
\AgdaSymbol{→}\AgdaSpace{}%
\AgdaInductiveConstructor{I}\AgdaSymbol{)}%
\>[17]\AgdaComment{--\ shift\ Term}\<%
\end{code}
We use pattern synonyms for convenient short-hands.
\begin{code}%
\>[0]\AgdaKeyword{pattern}\AgdaSpace{}%
\AgdaInductiveConstructor{`lam}\AgdaSpace{}%
\AgdaBound{t}%
\>[20]\AgdaSymbol{=}\AgdaSpace{}%
\AgdaOperator{\AgdaInductiveConstructor{⟨}}\AgdaSpace{}%
\AgdaInductiveConstructor{\#2}\AgdaSpace{}%
\AgdaOperator{\AgdaInductiveConstructor{,}}\AgdaSpace{}%
\AgdaBound{t}\AgdaSpace{}%
\AgdaOperator{\AgdaInductiveConstructor{⟩}}%
\>[34]\AgdaSymbol{;}\AgdaSpace{}%
\AgdaKeyword{pattern}\AgdaSpace{}%
\AgdaInductiveConstructor{`var}\AgdaSpace{}%
\AgdaBound{n}%
\>[56]\AgdaSymbol{=}\AgdaSpace{}%
\AgdaOperator{\AgdaInductiveConstructor{⟨}}\AgdaSpace{}%
\AgdaInductiveConstructor{\#3}\AgdaSpace{}%
\AgdaOperator{\AgdaInductiveConstructor{,}}\AgdaSpace{}%
\AgdaBound{n}\AgdaSpace{}%
\AgdaOperator{\AgdaInductiveConstructor{⟩}}\<%
\\
\>[0]\AgdaKeyword{pattern}\AgdaSpace{}%
\AgdaInductiveConstructor{`app}\AgdaSpace{}%
\AgdaBound{t₀}\AgdaSpace{}%
\AgdaBound{t₁}%
\>[20]\AgdaSymbol{=}\AgdaSpace{}%
\AgdaOperator{\AgdaInductiveConstructor{⟨}}\AgdaSpace{}%
\AgdaInductiveConstructor{\#4}\AgdaSpace{}%
\AgdaOperator{\AgdaInductiveConstructor{,}}\AgdaSpace{}%
\AgdaBound{t₀}\AgdaSpace{}%
\AgdaOperator{\AgdaInductiveConstructor{,}}\AgdaSpace{}%
\AgdaBound{t₁}\AgdaSpace{}%
\AgdaOperator{\AgdaInductiveConstructor{⟩}}\<%
\\
\>[0]\AgdaKeyword{pattern}\AgdaSpace{}%
\AgdaInductiveConstructor{`reset}\AgdaSpace{}%
\AgdaBound{t}%
\>[20]\AgdaSymbol{=}\AgdaSpace{}%
\AgdaOperator{\AgdaInductiveConstructor{⟨}}\AgdaSpace{}%
\AgdaInductiveConstructor{\#5}\AgdaSpace{}%
\AgdaOperator{\AgdaInductiveConstructor{,}}\AgdaSpace{}%
\AgdaBound{t}\AgdaSpace{}%
\AgdaOperator{\AgdaInductiveConstructor{⟩}}%
\>[34]\AgdaSymbol{;}\AgdaSpace{}%
\AgdaKeyword{pattern}\AgdaSpace{}%
\AgdaInductiveConstructor{`shift}\AgdaSpace{}%
\AgdaBound{t}%
\>[56]\AgdaSymbol{=}\AgdaSpace{}%
\AgdaOperator{\AgdaInductiveConstructor{⟨}}\AgdaSpace{}%
\AgdaInductiveConstructor{\#6}\AgdaSpace{}%
\AgdaOperator{\AgdaInductiveConstructor{,}}\AgdaSpace{}%
\AgdaBound{t}\AgdaSpace{}%
\AgdaOperator{\AgdaInductiveConstructor{⟩}}\<%
\end{code}%
We also define three functions that characterize numbers and $\lambda$s as values; $\lambda$s and shift expressions as having irreducible sub-terms; and additions of numbers, applications of a $\lambda$, resets with a value sub-term, or shift expressions as redexes.
\begin{code}%
\>[0]\AgdaFunction{is-val-LSR}\AgdaSpace{}%
\AgdaFunction{is-redex-LSR}\AgdaSpace{}%
\AgdaSymbol{:}\AgdaSpace{}%
\AgdaOperator{\AgdaDatatype{μ[}}\AgdaSpace{}%
\AgdaFunction{LSR}\AgdaSpace{}%
\AgdaOperator{\AgdaDatatype{]}}\AgdaSpace{}%
\AgdaSymbol{→}\AgdaSpace{}%
\AgdaDatatype{Bool}\<%
\\
\>[0]\AgdaFunction{redex-pos-LSR}%
\>[15]\AgdaSymbol{:}\AgdaSpace{}%
\AgdaSymbol{∀}\AgdaSpace{}%
\AgdaSymbol{\{}\AgdaBound{X}\AgdaSpace{}%
\AgdaBound{e}\AgdaSymbol{\}}\AgdaSpace{}%
\AgdaSymbol{\{}\AgdaBound{x}\AgdaSpace{}%
\AgdaSymbol{:}\AgdaSpace{}%
\AgdaOperator{\AgdaFunction{⟦}}\AgdaSpace{}%
\AgdaFunction{LSR}\AgdaSpace{}%
\AgdaOperator{\AgdaFunction{⟧}}\AgdaSpace{}%
\AgdaBound{X}\AgdaSymbol{\}}\AgdaSpace{}%
\AgdaSymbol{→}\AgdaSpace{}%
\AgdaBound{e}\AgdaSpace{}%
\AgdaOperator{\AgdaFunction{∈}}\AgdaSpace{}%
\AgdaOperator{\AgdaFunction{S[}}\AgdaSpace{}%
\AgdaFunction{LSR}\AgdaSpace{}%
\AgdaOperator{\AgdaFunction{∼}}\AgdaSpace{}%
\AgdaBound{x}\AgdaSpace{}%
\AgdaOperator{\AgdaFunction{]}}\<%
\\
\>[15]\AgdaSymbol{→}\AgdaSpace{}%
\AgdaDatatype{Bool}\<%
\end{code}
\begin{code}[hide]%
\>[0]\AgdaKeyword{pattern}\AgdaSpace{}%
\AgdaInductiveConstructor{`lamF}\AgdaSpace{}%
\AgdaBound{t}%
\>[20]\AgdaSymbol{=}\AgdaSpace{}%
\AgdaInductiveConstructor{\#2}\AgdaSpace{}%
\AgdaOperator{\AgdaInductiveConstructor{,}}\AgdaSpace{}%
\AgdaBound{t}%
\>[30]\AgdaSymbol{;}\AgdaSpace{}%
\AgdaKeyword{pattern}\AgdaSpace{}%
\AgdaInductiveConstructor{`varF}\AgdaSpace{}%
\AgdaBound{n}%
\>[53]\AgdaSymbol{=}\AgdaSpace{}%
\AgdaInductiveConstructor{\#3}\AgdaSpace{}%
\AgdaOperator{\AgdaInductiveConstructor{,}}\AgdaSpace{}%
\AgdaBound{n}\<%
\\
\>[0]\AgdaKeyword{pattern}\AgdaSpace{}%
\AgdaInductiveConstructor{`appF}\AgdaSpace{}%
\AgdaBound{t₀}\AgdaSpace{}%
\AgdaBound{t₁}%
\>[21]\AgdaSymbol{=}\AgdaSpace{}%
\AgdaInductiveConstructor{\#4}\AgdaSpace{}%
\AgdaOperator{\AgdaInductiveConstructor{,}}\AgdaSpace{}%
\AgdaBound{t₀}\AgdaSpace{}%
\AgdaOperator{\AgdaInductiveConstructor{,}}\AgdaSpace{}%
\AgdaBound{t₁}\<%
\\
\>[0]\AgdaKeyword{pattern}\AgdaSpace{}%
\AgdaInductiveConstructor{`resetF}\AgdaSpace{}%
\AgdaBound{t}%
\>[21]\AgdaSymbol{=}\AgdaSpace{}%
\AgdaInductiveConstructor{\#5}\AgdaSpace{}%
\AgdaOperator{\AgdaInductiveConstructor{,}}\AgdaSpace{}%
\AgdaBound{t}%
\>[31]\AgdaSymbol{;}\AgdaSpace{}%
\AgdaKeyword{pattern}\AgdaSpace{}%
\AgdaInductiveConstructor{`shiftF}\AgdaSpace{}%
\AgdaBound{t}%
\>[54]\AgdaSymbol{=}\AgdaSpace{}%
\AgdaInductiveConstructor{\#6}\AgdaSpace{}%
\AgdaOperator{\AgdaInductiveConstructor{,}}\AgdaSpace{}%
\AgdaBound{t}\<%
\\
\\[\AgdaEmptyExtraSkip]%
\>[0]\AgdaFunction{is-val-LSR}\AgdaSpace{}%
\AgdaSymbol{(}\AgdaInductiveConstructor{`num}\AgdaSpace{}%
\AgdaBound{n}\AgdaSymbol{)}%
\>[21]\AgdaSymbol{=}\AgdaSpace{}%
\AgdaInductiveConstructor{true}\<%
\\
\>[0]\AgdaFunction{is-val-LSR}\AgdaSpace{}%
\AgdaSymbol{(}\AgdaInductiveConstructor{`lam}\AgdaSpace{}%
\AgdaBound{t}\AgdaSymbol{)}%
\>[21]\AgdaSymbol{=}\AgdaSpace{}%
\AgdaInductiveConstructor{true}\<%
\\
\>[0]\AgdaCatchallClause{\AgdaFunction{is-val-LSR}}\AgdaSpace{}%
\AgdaCatchallClause{\AgdaSymbol{\AgdaUnderscore{}}}%
\>[21]\AgdaSymbol{=}\AgdaSpace{}%
\AgdaInductiveConstructor{false}\<%
\\
\\[\AgdaEmptyExtraSkip]%
\>[0]\AgdaFunction{is-redex-LSR}\AgdaSpace{}%
\AgdaSymbol{(}\AgdaInductiveConstructor{`add}\AgdaSpace{}%
\AgdaSymbol{(}\AgdaInductiveConstructor{`num}\AgdaSpace{}%
\AgdaBound{n₀}\AgdaSymbol{)}\AgdaSpace{}%
\AgdaSymbol{(}\AgdaInductiveConstructor{`num}\AgdaSpace{}%
\AgdaBound{n₁}\AgdaSymbol{))}%
\>[41]\AgdaSymbol{=}\AgdaSpace{}%
\AgdaInductiveConstructor{true}\<%
\\
\>[0]\AgdaFunction{is-redex-LSR}\AgdaSpace{}%
\AgdaSymbol{(}\AgdaInductiveConstructor{`app}\AgdaSpace{}%
\AgdaSymbol{(}\AgdaInductiveConstructor{`lam}\AgdaSpace{}%
\AgdaBound{t}\AgdaSymbol{)}\AgdaSpace{}%
\AgdaBound{v}\AgdaSymbol{)}%
\>[41]\AgdaSymbol{=}\AgdaSpace{}%
\AgdaFunction{is-val-LSR}\AgdaSpace{}%
\AgdaBound{v}\<%
\\
\>[0]\AgdaFunction{is-redex-LSR}\AgdaSpace{}%
\AgdaSymbol{(}\AgdaInductiveConstructor{`reset}\AgdaSpace{}%
\AgdaBound{v}\AgdaSymbol{)}%
\>[41]\AgdaSymbol{=}\AgdaSpace{}%
\AgdaFunction{is-val-LSR}\AgdaSpace{}%
\AgdaBound{v}\<%
\\
\>[0]\AgdaFunction{is-redex-LSR}\AgdaSpace{}%
\AgdaSymbol{(}\AgdaInductiveConstructor{`shift}\AgdaSpace{}%
\AgdaBound{t}\AgdaSymbol{)}%
\>[41]\AgdaSymbol{=}\AgdaSpace{}%
\AgdaInductiveConstructor{true}\<%
\\
\>[0]\AgdaCatchallClause{\AgdaFunction{is-redex-LSR}}\AgdaSpace{}%
\AgdaCatchallClause{\AgdaSymbol{\AgdaUnderscore{}}}%
\>[41]\AgdaSymbol{=}\AgdaSpace{}%
\AgdaInductiveConstructor{false}\<%
\\
\\[\AgdaEmptyExtraSkip]%
\>[0]\AgdaFunction{redex-pos-LSR}\AgdaSpace{}%
\AgdaSymbol{\{}\AgdaArgument{x}\AgdaSpace{}%
\AgdaSymbol{=}\AgdaSpace{}%
\AgdaInductiveConstructor{`lamF}\AgdaSpace{}%
\AgdaSymbol{\AgdaUnderscore{}\}}\AgdaSpace{}%
\AgdaSymbol{\AgdaUnderscore{}}\AgdaSpace{}%
\AgdaSymbol{=}\AgdaSpace{}%
\AgdaInductiveConstructor{false}\<%
\\
\>[0]\AgdaFunction{redex-pos-LSR}\AgdaSpace{}%
\AgdaSymbol{\{}\AgdaArgument{x}\AgdaSpace{}%
\AgdaSymbol{=}\AgdaSpace{}%
\AgdaInductiveConstructor{`shiftF}\AgdaSpace{}%
\AgdaSymbol{\AgdaUnderscore{}\}}\AgdaSpace{}%
\AgdaSymbol{\AgdaUnderscore{}}\AgdaSpace{}%
\AgdaSymbol{=}\AgdaSpace{}%
\AgdaInductiveConstructor{false}\<%
\\
\>[0]\AgdaCatchallClause{\AgdaFunction{redex-pos-LSR}}\AgdaSpace{}%
\AgdaCatchallClause{\AgdaSymbol{\AgdaUnderscore{}}}\AgdaSpace{}%
\AgdaSymbol{=}\AgdaSpace{}%
\AgdaInductiveConstructor{true}\<%
\end{code}
\begin{code}[hide]%
\>[0]\AgdaKeyword{pattern}\AgdaSpace{}%
\AgdaInductiveConstructor{`lam-c}%
\>[17]\AgdaSymbol{=}\AgdaSpace{}%
\AgdaInductiveConstructor{\#2}\AgdaSpace{}%
\AgdaOperator{\AgdaInductiveConstructor{,}}\AgdaSpace{}%
\AgdaInductiveConstructor{tt}\<%
\\
\>[0]\AgdaKeyword{pattern}\AgdaSpace{}%
\AgdaInductiveConstructor{`app-l}\AgdaSpace{}%
\AgdaBound{e}\AgdaSpace{}%
\AgdaSymbol{=}\AgdaSpace{}%
\AgdaInductiveConstructor{\#4}\AgdaSpace{}%
\AgdaOperator{\AgdaInductiveConstructor{,}}\AgdaSpace{}%
\AgdaInductiveConstructor{inj₁}\AgdaSpace{}%
\AgdaSymbol{(}\AgdaInductiveConstructor{tt}\AgdaSpace{}%
\AgdaOperator{\AgdaInductiveConstructor{,}}\AgdaSpace{}%
\AgdaBound{e}\AgdaSymbol{)}\<%
\\
\>[0]\AgdaKeyword{pattern}\AgdaSpace{}%
\AgdaInductiveConstructor{`app-r}\AgdaSpace{}%
\AgdaBound{v}\AgdaSpace{}%
\AgdaSymbol{=}\AgdaSpace{}%
\AgdaInductiveConstructor{\#4}\AgdaSpace{}%
\AgdaOperator{\AgdaInductiveConstructor{,}}\AgdaSpace{}%
\AgdaInductiveConstructor{inj₂}\AgdaSpace{}%
\AgdaSymbol{(}\AgdaBound{v}\AgdaSpace{}%
\AgdaOperator{\AgdaInductiveConstructor{,}}\AgdaSpace{}%
\AgdaInductiveConstructor{tt}\AgdaSymbol{)}\<%
\\
\>[0]\AgdaKeyword{pattern}\AgdaSpace{}%
\AgdaInductiveConstructor{`reset-c}\AgdaSpace{}%
\AgdaSymbol{=}\AgdaSpace{}%
\AgdaInductiveConstructor{\#5}\AgdaSpace{}%
\AgdaOperator{\AgdaInductiveConstructor{,}}\AgdaSpace{}%
\AgdaInductiveConstructor{tt}\<%
\\
\\[\AgdaEmptyExtraSkip]%
\>[0]\AgdaKeyword{module}\AgdaSpace{}%
\AgdaModule{LSRReduction}\AgdaSpace{}%
\AgdaKeyword{where}\<%
\\
\>[0][@{}l@{\AgdaIndent{0}}]%
\>[2]\AgdaKeyword{open}%
\>[958I]\AgdaModule{GenericReduction}\<%
\\
\>[958I][@{}l@{\AgdaIndent{0}}]%
\>[9]\AgdaSymbol{\{}\AgdaFunction{LSR}\AgdaSymbol{\}}\<%
\\
\>[9]\AgdaFunction{is-redex-LSR}\<%
\\
\>[9]\AgdaFunction{is-val-LSR}\<%
\\
\>[9]\AgdaFunction{redex-pos-LSR}\AgdaSpace{}%
\AgdaKeyword{public}\<%
\end{code}
Importing the generic reduction module yields a decomposition function with the desired reduction strategy.
%
\begin{code}[hide]%
\>[2]\AgdaFunction{decomp-1+reset[2+shift[\textbackslash{}[0-[0-[3+4]]]]]}\AgdaSpace{}%
\AgdaSymbol{:}\<%
\\
\>[2][@{}l@{\AgdaIndent{0}}]%
\>[4]\AgdaFunction{decomposeLMIMW}\<%
\\
\>[4][@{}l@{\AgdaIndent{0}}]%
\>[6]\AgdaSymbol{(}\AgdaInductiveConstructor{`add}\AgdaSpace{}%
\AgdaSymbol{(}\AgdaInductiveConstructor{`num}\AgdaSpace{}%
\AgdaNumber{1}\AgdaSymbol{)}\<%
\\
\>[6][@{}l@{\AgdaIndent{0}}]%
\>[8]\AgdaSymbol{(}\AgdaInductiveConstructor{`reset}\<%
\\
\>[8][@{}l@{\AgdaIndent{0}}]%
\>[10]\AgdaSymbol{(}\AgdaInductiveConstructor{`add}\AgdaSpace{}%
\AgdaSymbol{(}\AgdaInductiveConstructor{`num}\AgdaSpace{}%
\AgdaNumber{2}\AgdaSymbol{)}\<%
\\
\>[10][@{}l@{\AgdaIndent{0}}]%
\>[12]\AgdaSymbol{(}\AgdaInductiveConstructor{`shift}\<%
\\
\>[12][@{}l@{\AgdaIndent{0}}]%
\>[14]\AgdaSymbol{(}\AgdaInductiveConstructor{`app}\AgdaSpace{}%
\AgdaSymbol{(}\AgdaInductiveConstructor{`var}\AgdaSpace{}%
\AgdaNumber{0}\AgdaSymbol{)}\<%
\\
\>[14][@{}l@{\AgdaIndent{0}}]%
\>[16]\AgdaSymbol{(}\AgdaInductiveConstructor{`app}\AgdaSpace{}%
\AgdaSymbol{(}\AgdaInductiveConstructor{`var}\AgdaSpace{}%
\AgdaNumber{0}\AgdaSymbol{)}\<%
\\
\>[16][@{}l@{\AgdaIndent{0}}]%
\>[18]\AgdaSymbol{(}\AgdaInductiveConstructor{`add}\<%
\\
\>[18][@{}l@{\AgdaIndent{0}}]%
\>[20]\AgdaSymbol{(}\AgdaInductiveConstructor{`num}\AgdaSpace{}%
\AgdaNumber{3}\AgdaSymbol{)}\<%
\\
\>[20]\AgdaSymbol{(}\AgdaInductiveConstructor{`num}\AgdaSpace{}%
\AgdaNumber{4}\AgdaSymbol{))))))))}\<%
\\
\>[4]\AgdaOperator{\AgdaDatatype{≡}}\AgdaSpace{}%
\AgdaInductiveConstructor{just}%
\>[972I]\AgdaSymbol{(}\AgdaSpace{}%
\AgdaInductiveConstructor{`add-r}\AgdaSpace{}%
\AgdaSymbol{(}\AgdaInductiveConstructor{`num}\AgdaSpace{}%
\AgdaNumber{2}\AgdaSymbol{)}\AgdaSpace{}%
\AgdaOperator{\AgdaInductiveConstructor{∷}}\AgdaSpace{}%
\AgdaInductiveConstructor{`reset-c}\AgdaSpace{}%
\AgdaOperator{\AgdaInductiveConstructor{∷}}\AgdaSpace{}%
\AgdaInductiveConstructor{`add-r}\AgdaSpace{}%
\AgdaSymbol{(}\AgdaInductiveConstructor{`num}\AgdaSpace{}%
\AgdaNumber{1}\AgdaSymbol{)}\AgdaSpace{}%
\AgdaOperator{\AgdaInductiveConstructor{∷}}\AgdaSpace{}%
\AgdaInductiveConstructor{[]}\<%
\\
\>[.][@{}l@{}]\<[972I]%
\>[11]\AgdaOperator{\AgdaInductiveConstructor{,}}%
\>[984I]\AgdaInductiveConstructor{`shift}\<%
\\
\>[984I][@{}l@{\AgdaIndent{0}}]%
\>[15]\AgdaSymbol{(}\AgdaInductiveConstructor{`app}\AgdaSpace{}%
\AgdaSymbol{(}\AgdaInductiveConstructor{`var}\AgdaSpace{}%
\AgdaNumber{0}\AgdaSymbol{)}\<%
\\
\>[15][@{}l@{\AgdaIndent{0}}]%
\>[17]\AgdaSymbol{(}\AgdaInductiveConstructor{`app}\AgdaSpace{}%
\AgdaSymbol{(}\AgdaInductiveConstructor{`var}\AgdaSpace{}%
\AgdaNumber{0}\AgdaSymbol{)}\<%
\\
\>[17][@{}l@{\AgdaIndent{0}}]%
\>[19]\AgdaSymbol{(}\AgdaInductiveConstructor{`add}\AgdaSpace{}%
\AgdaSymbol{(}\AgdaInductiveConstructor{`num}\AgdaSpace{}%
\AgdaNumber{3}\AgdaSymbol{)}\AgdaSpace{}%
\AgdaSymbol{(}\AgdaInductiveConstructor{`num}\AgdaSpace{}%
\AgdaNumber{4}\AgdaSymbol{))))}\AgdaSpace{}%
\AgdaSymbol{)}\<%
\\
\>[2]\AgdaFunction{decomp-1+reset[2+shift[\textbackslash{}[0-[0-[3+4]]]]]}\AgdaSpace{}%
\AgdaSymbol{=}\AgdaSpace{}%
\AgdaInductiveConstructor{refl}\<%
\end{code}
\begin{code}[hide]%
\>[2]\AgdaComment{--\ More\ tests.}\<%
\\
\>[2]\AgdaFunction{decomp-lam[1+2]}\AgdaSpace{}%
\AgdaSymbol{:}\<%
\\
\>[2][@{}l@{\AgdaIndent{0}}]%
\>[4]\AgdaFunction{decomposeLMIMW}\<%
\\
\>[4][@{}l@{\AgdaIndent{0}}]%
\>[6]\AgdaSymbol{(}\AgdaInductiveConstructor{`lam}\AgdaSpace{}%
\AgdaSymbol{(}\AgdaInductiveConstructor{`add}\AgdaSpace{}%
\AgdaSymbol{(}\AgdaInductiveConstructor{`num}\AgdaSpace{}%
\AgdaNumber{1}\AgdaSymbol{)}\AgdaSpace{}%
\AgdaSymbol{(}\AgdaInductiveConstructor{`num}\AgdaSpace{}%
\AgdaNumber{2}\AgdaSymbol{)))}\<%
\\
\>[4]\AgdaOperator{\AgdaDatatype{≡}}\AgdaSpace{}%
\AgdaInductiveConstructor{nothing}\<%
\\
\>[2]\AgdaFunction{decomp-lam[1+2]}\AgdaSpace{}%
\AgdaSymbol{=}\AgdaSpace{}%
\AgdaInductiveConstructor{refl}\<%
\\
\\[\AgdaEmptyExtraSkip]%
\>[2]\AgdaFunction{decomp-app-lam}\AgdaSpace{}%
\AgdaSymbol{:}\<%
\\
\>[2][@{}l@{\AgdaIndent{0}}]%
\>[4]\AgdaFunction{decomposeLMIMW}\<%
\\
\>[4][@{}l@{\AgdaIndent{0}}]%
\>[6]\AgdaSymbol{(}\AgdaInductiveConstructor{`app}\AgdaSpace{}%
\AgdaSymbol{(}\AgdaInductiveConstructor{`lam}\AgdaSpace{}%
\AgdaSymbol{(}\AgdaInductiveConstructor{`var}\AgdaSpace{}%
\AgdaNumber{0}\AgdaSymbol{))}\AgdaSpace{}%
\AgdaSymbol{(}\AgdaInductiveConstructor{`num}\AgdaSpace{}%
\AgdaNumber{1}\AgdaSymbol{))}\<%
\\
\>[4]\AgdaOperator{\AgdaDatatype{≡}}\AgdaSpace{}%
\AgdaInductiveConstructor{just}\AgdaSpace{}%
\AgdaSymbol{(}\AgdaInductiveConstructor{[]}\AgdaSpace{}%
\AgdaOperator{\AgdaInductiveConstructor{,}}\AgdaSpace{}%
\AgdaInductiveConstructor{`app}\AgdaSpace{}%
\AgdaSymbol{(}\AgdaInductiveConstructor{`lam}\AgdaSpace{}%
\AgdaSymbol{(}\AgdaInductiveConstructor{`var}\AgdaSpace{}%
\AgdaNumber{0}\AgdaSymbol{))}\AgdaSpace{}%
\AgdaSymbol{(}\AgdaInductiveConstructor{`num}\AgdaSpace{}%
\AgdaNumber{1}\AgdaSymbol{))}\<%
\\
\>[2]\AgdaFunction{decomp-app-lam}\AgdaSpace{}%
\AgdaSymbol{=}\AgdaSpace{}%
\AgdaInductiveConstructor{refl}\<%
\\
\\[\AgdaEmptyExtraSkip]%
\>[2]\AgdaFunction{decomp-app-lam-add}\AgdaSpace{}%
\AgdaSymbol{:}\<%
\\
\>[2][@{}l@{\AgdaIndent{0}}]%
\>[4]\AgdaFunction{decomposeLMIMW}\<%
\\
\>[4][@{}l@{\AgdaIndent{0}}]%
\>[6]\AgdaSymbol{(}\AgdaInductiveConstructor{`app}\AgdaSpace{}%
\AgdaSymbol{(}\AgdaInductiveConstructor{`lam}\AgdaSpace{}%
\AgdaSymbol{(}\AgdaInductiveConstructor{`var}\AgdaSpace{}%
\AgdaNumber{0}\AgdaSymbol{))}\AgdaSpace{}%
\AgdaSymbol{(}\AgdaInductiveConstructor{`add}\AgdaSpace{}%
\AgdaSymbol{(}\AgdaInductiveConstructor{`num}\AgdaSpace{}%
\AgdaNumber{1}\AgdaSymbol{)}\AgdaSpace{}%
\AgdaSymbol{(}\AgdaInductiveConstructor{`num}\AgdaSpace{}%
\AgdaNumber{2}\AgdaSymbol{)))}\<%
\\
\>[4]\AgdaOperator{\AgdaDatatype{≡}}\AgdaSpace{}%
\AgdaInductiveConstructor{just}\AgdaSpace{}%
\AgdaSymbol{(}\AgdaInductiveConstructor{`app-r}\AgdaSpace{}%
\AgdaSymbol{(}\AgdaInductiveConstructor{`lam}\AgdaSpace{}%
\AgdaSymbol{(}\AgdaInductiveConstructor{`var}\AgdaSpace{}%
\AgdaNumber{0}\AgdaSymbol{))}\AgdaSpace{}%
\AgdaOperator{\AgdaInductiveConstructor{∷}}\AgdaSpace{}%
\AgdaInductiveConstructor{[]}\AgdaSpace{}%
\AgdaOperator{\AgdaInductiveConstructor{,}}\AgdaSpace{}%
\AgdaInductiveConstructor{`add}\AgdaSpace{}%
\AgdaSymbol{(}\AgdaInductiveConstructor{`num}\AgdaSpace{}%
\AgdaNumber{1}\AgdaSymbol{)}\AgdaSpace{}%
\AgdaSymbol{(}\AgdaInductiveConstructor{`num}\AgdaSpace{}%
\AgdaNumber{2}\AgdaSymbol{))}\<%
\\
\>[2]\AgdaFunction{decomp-app-lam-add}\AgdaSpace{}%
\AgdaSymbol{=}\AgdaSpace{}%
\AgdaInductiveConstructor{refl}\<%
\\
\>[0]\<%
\\
\>[2]\AgdaFunction{decomp-app-app-lam-add}\AgdaSpace{}%
\AgdaSymbol{:}\<%
\\
\>[2][@{}l@{\AgdaIndent{0}}]%
\>[4]\AgdaFunction{decomposeLMIMW}\<%
\\
\>[4][@{}l@{\AgdaIndent{0}}]%
\>[6]\AgdaSymbol{(}\AgdaInductiveConstructor{`app}\AgdaSpace{}%
\AgdaSymbol{(}\AgdaInductiveConstructor{`app}\AgdaSpace{}%
\AgdaSymbol{(}\AgdaInductiveConstructor{`lam}\AgdaSpace{}%
\AgdaSymbol{(}\AgdaInductiveConstructor{`var}\AgdaSpace{}%
\AgdaNumber{0}\AgdaSymbol{))}\AgdaSpace{}%
\AgdaSymbol{(}\AgdaInductiveConstructor{`num}\AgdaSpace{}%
\AgdaNumber{1}\AgdaSymbol{))}\AgdaSpace{}%
\AgdaSymbol{(}\AgdaInductiveConstructor{`add}\AgdaSpace{}%
\AgdaSymbol{(}\AgdaInductiveConstructor{`num}\AgdaSpace{}%
\AgdaNumber{2}\AgdaSymbol{)}\AgdaSpace{}%
\AgdaSymbol{(}\AgdaInductiveConstructor{`num}\AgdaSpace{}%
\AgdaNumber{3}\AgdaSymbol{)))}\<%
\\
\>[4]\AgdaOperator{\AgdaDatatype{≡}}\AgdaSpace{}%
\AgdaInductiveConstructor{just}%
\>[1059I]\AgdaSymbol{(}\AgdaSpace{}%
\AgdaInductiveConstructor{`app-l}\AgdaSpace{}%
\AgdaSymbol{(}\AgdaInductiveConstructor{`add}\AgdaSpace{}%
\AgdaSymbol{(}\AgdaInductiveConstructor{`num}\AgdaSpace{}%
\AgdaNumber{2}\AgdaSymbol{)}\AgdaSpace{}%
\AgdaSymbol{(}\AgdaInductiveConstructor{`num}\AgdaSpace{}%
\AgdaNumber{3}\AgdaSymbol{))}\AgdaSpace{}%
\AgdaOperator{\AgdaInductiveConstructor{∷}}\AgdaSpace{}%
\AgdaInductiveConstructor{[]}\<%
\\
\>[.][@{}l@{}]\<[1059I]%
\>[11]\AgdaOperator{\AgdaInductiveConstructor{,}}\AgdaSpace{}%
\AgdaInductiveConstructor{`app}\AgdaSpace{}%
\AgdaSymbol{(}\AgdaInductiveConstructor{`lam}\AgdaSpace{}%
\AgdaSymbol{(}\AgdaInductiveConstructor{`var}\AgdaSpace{}%
\AgdaNumber{0}\AgdaSymbol{))}\AgdaSpace{}%
\AgdaSymbol{(}\AgdaInductiveConstructor{`num}\AgdaSpace{}%
\AgdaNumber{1}\AgdaSymbol{)}\AgdaSpace{}%
\AgdaSymbol{)}\<%
\\
\>[2]\AgdaFunction{decomp-app-app-lam-add}\AgdaSpace{}%
\AgdaSymbol{=}\AgdaSpace{}%
\AgdaInductiveConstructor{refl}\<%
\\
\>[0]\<%
\\
\>[2]\AgdaFunction{decomp-add-shift-v}\AgdaSpace{}%
\AgdaSymbol{:}\<%
\\
\>[2][@{}l@{\AgdaIndent{0}}]%
\>[4]\AgdaFunction{decomposeLMIMW}\<%
\\
\>[4][@{}l@{\AgdaIndent{0}}]%
\>[6]\AgdaSymbol{(}\AgdaInductiveConstructor{`add}\AgdaSpace{}%
\AgdaSymbol{(}\AgdaInductiveConstructor{`shift}\AgdaSpace{}%
\AgdaSymbol{(}\AgdaInductiveConstructor{`num}\AgdaSpace{}%
\AgdaNumber{0}\AgdaSymbol{))}\AgdaSpace{}%
\AgdaSymbol{(}\AgdaInductiveConstructor{`num}\AgdaSpace{}%
\AgdaNumber{1}\AgdaSymbol{))}\<%
\\
\>[4]\AgdaOperator{\AgdaDatatype{≡}}\AgdaSpace{}%
\AgdaInductiveConstructor{just}%
\>[1084I]\AgdaSymbol{(}\AgdaSpace{}%
\AgdaInductiveConstructor{`add-l}\AgdaSpace{}%
\AgdaSymbol{(}\AgdaInductiveConstructor{`num}\AgdaSpace{}%
\AgdaNumber{1}\AgdaSymbol{)}\AgdaSpace{}%
\AgdaOperator{\AgdaInductiveConstructor{∷}}\AgdaSpace{}%
\AgdaInductiveConstructor{[]}\<%
\\
\>[.][@{}l@{}]\<[1084I]%
\>[11]\AgdaOperator{\AgdaInductiveConstructor{,}}\AgdaSpace{}%
\AgdaInductiveConstructor{`shift}\AgdaSpace{}%
\AgdaSymbol{(}\AgdaInductiveConstructor{`num}\AgdaSpace{}%
\AgdaNumber{0}\AgdaSymbol{)}\AgdaSpace{}%
\AgdaSymbol{)}\<%
\\
\>[2]\AgdaFunction{decomp-add-shift-v}\AgdaSpace{}%
\AgdaSymbol{=}\AgdaSpace{}%
\AgdaInductiveConstructor{refl}\<%
\\
\>[0]\<%
\\
\>[2]\AgdaFunction{decomp-lam-1+2}\AgdaSpace{}%
\AgdaSymbol{:}%
\>[1097I]\AgdaFunction{decomposeLMIMW}\<%
\\
\>[1097I][@{}l@{\AgdaIndent{0}}]%
\>[21]\AgdaSymbol{(}\AgdaInductiveConstructor{`lam}\AgdaSpace{}%
\AgdaSymbol{(}\AgdaInductiveConstructor{`add}\AgdaSpace{}%
\AgdaSymbol{(}\AgdaInductiveConstructor{`num}\AgdaSpace{}%
\AgdaNumber{1}\AgdaSymbol{)}\AgdaSpace{}%
\AgdaSymbol{(}\AgdaInductiveConstructor{`num}\AgdaSpace{}%
\AgdaNumber{2}\AgdaSymbol{)))}\<%
\\
\>[.][@{}l@{}]\<[1097I]%
\>[19]\AgdaOperator{\AgdaDatatype{≡}}\AgdaSpace{}%
\AgdaInductiveConstructor{nothing}\<%
\\
\>[2]\AgdaFunction{decomp-lam-1+2}\AgdaSpace{}%
\AgdaSymbol{=}\AgdaSpace{}%
\AgdaInductiveConstructor{refl}\<%
\\
\>[0]\<%
\\
\>[2]\AgdaFunction{decomp-shift}\AgdaSpace{}%
\AgdaSymbol{:}%
\>[1107I]\AgdaFunction{decomposeLMIMW}\<%
\\
\>[1107I][@{}l@{\AgdaIndent{0}}]%
\>[19]\AgdaSymbol{(}\AgdaInductiveConstructor{`shift}\AgdaSpace{}%
\AgdaSymbol{(}\AgdaInductiveConstructor{`add}\AgdaSpace{}%
\AgdaSymbol{(}\AgdaInductiveConstructor{`num}\AgdaSpace{}%
\AgdaNumber{1}\AgdaSymbol{)}\AgdaSpace{}%
\AgdaSymbol{(}\AgdaInductiveConstructor{`num}\AgdaSpace{}%
\AgdaNumber{2}\AgdaSymbol{)))}\<%
\\
\>[.][@{}l@{}]\<[1107I]%
\>[17]\AgdaOperator{\AgdaDatatype{≡}}\AgdaSpace{}%
\AgdaInductiveConstructor{just}%
\>[1114I]\AgdaSymbol{(}\AgdaSpace{}%
\AgdaInductiveConstructor{[]}\<%
\\
\>[.][@{}l@{}]\<[1114I]%
\>[24]\AgdaOperator{\AgdaInductiveConstructor{,}}\AgdaSpace{}%
\AgdaInductiveConstructor{`shift}\AgdaSpace{}%
\AgdaSymbol{(}\AgdaInductiveConstructor{`add}\AgdaSpace{}%
\AgdaSymbol{(}\AgdaInductiveConstructor{`num}\AgdaSpace{}%
\AgdaNumber{1}\AgdaSymbol{)}\AgdaSpace{}%
\AgdaSymbol{(}\AgdaInductiveConstructor{`num}\AgdaSpace{}%
\AgdaNumber{2}\AgdaSymbol{))}\AgdaSpace{}%
\AgdaSymbol{)}\<%
\\
\>[2]\AgdaFunction{decomp-shift}\AgdaSpace{}%
\AgdaSymbol{=}\AgdaSpace{}%
\AgdaInductiveConstructor{refl}\<%
\end{code}

A caveat of the generic framework presented here is that the generic contexts we get from the machinery from \cref{sec:derivatives} can represent decompositions at \emph{any} recurisve position.
This means that we get a syntax of contexts with redexes occurring inside of a $\lambda$ expressions, for example.
As discussed in \cref{sec:generic-paras}, weak reduction strategies for $\lambda$ calculi should not admit such decompositions.
The \ab{redex-pos} function that \af{algLMIMW} ensures that such contexts are never yielded as decompositions in practice.

A potential approach to avoiding this caveat could be to refine our definition of derivative \af{∂[\_]} by a specification of which sub-terms we get a zipper for, and which we do not.
This would render invalid reduction contexts, such as a context which represents a decomposition past a $\lambda$ binder, unrepresentable in practice.
We leave exploring this approach to future work.

\subsubsection{Decomposition Soundness}

It is future work to mechanize in Agda that the decomposition functions yielded by the algebras discussed here respects the taxonomy in~\cref{app:characterization}.
It is also future work to mechanize soundness of plugging in Agda; i.e.:
\begin{code}%
\>[2]\AgdaKeyword{postulate}\<%
\\
\>[2][@{}l@{\AgdaIndent{0}}]%
\>[4]\AgdaPostulate{decompose-sound}%
\>[21]\AgdaSymbol{:}\AgdaSpace{}%
\AgdaSymbol{∀}\AgdaSpace{}%
\AgdaSymbol{\{}\AgdaBound{x}\AgdaSpace{}%
\AgdaBound{c}\AgdaSpace{}%
\AgdaBound{t}\AgdaSymbol{\}}\<%
\\
\>[21]\AgdaSymbol{→}\AgdaSpace{}%
\AgdaFunction{decomposeLMIMW}\AgdaSpace{}%
\AgdaBound{x}\AgdaSpace{}%
\AgdaOperator{\AgdaDatatype{≡}}\AgdaSpace{}%
\AgdaInductiveConstructor{just}\AgdaSpace{}%
\AgdaSymbol{(}\AgdaBound{c}\AgdaSpace{}%
\AgdaOperator{\AgdaInductiveConstructor{,}}\AgdaSpace{}%
\AgdaBound{t}\AgdaSymbol{)}\<%
\\
\>[21]\AgdaSymbol{→}\AgdaSpace{}%
\AgdaFunction{plug-ctx}\AgdaSpace{}%
\AgdaBound{c}\AgdaSpace{}%
\AgdaBound{t}\AgdaSpace{}%
\AgdaOperator{\AgdaDatatype{≡}}\AgdaSpace{}%
\AgdaBound{x}\<%
\end{code}
We have a proof sketch on pen-and-paper that the property holds.
Proving it in Agda requires mechanizing paramorphic induction, and carefully applying this principle in a manner that is consistent with the generic sub-term function \af{S[\_∼\_]}.
While mechanizing this is future work, a key selling point of our generic framework is exactly that we get a notion of decomposition that is \emph{sound by construction}; i.e., that \af{decompose-sound} holds.

\subsection[Lambdas and Shift/Reset, Revisited]{$\lambda$s and Shift/Reset, Revisited}
\label{sec:generic-cases}

Having defined a generic notion of decomposition, and demonstrated how it is derived from a syntax definition, all that remains to obtain a full reduction semantics for a language is to define (1) a contraction function, and (2) a generic driver loop for one-step reduction.

\subsubsection{Contraction for $\lambda$s and Shift/Reset}

Contraction is defined analogously to the definition in \cref{sec:lambda-shift-reset}.
To this end, we need a notion of substitution.
We could define substitution generically, using Allais' generic techniques~\cite{AllaisACMM21}.
For simplicity, we use a direct substitution function for the syntax described by \af{LSR}, rather than a generic substitution function.
\begin{code}%
\>[0]\AgdaFunction{Subst}\AgdaSpace{}%
\AgdaSymbol{:}\AgdaSpace{}%
\AgdaPrimitive{Set}\<%
\\
\>[0]\AgdaFunction{Subst}\AgdaSpace{}%
\AgdaSymbol{=}\AgdaSpace{}%
\AgdaDatatype{ℕ}\AgdaSpace{}%
\AgdaSymbol{→}\AgdaSpace{}%
\AgdaOperator{\AgdaDatatype{μ[}}\AgdaSpace{}%
\AgdaFunction{LSR}\AgdaSpace{}%
\AgdaOperator{\AgdaDatatype{]}}\<%
\\
\\[\AgdaEmptyExtraSkip]%
\>[0]\AgdaFunction{sb0}\AgdaSpace{}%
\AgdaSymbol{:}\AgdaSpace{}%
\AgdaOperator{\AgdaDatatype{μ[}}\AgdaSpace{}%
\AgdaFunction{LSR}\AgdaSpace{}%
\AgdaOperator{\AgdaDatatype{]}}\AgdaSpace{}%
\AgdaSymbol{→}\AgdaSpace{}%
\AgdaFunction{Subst}\AgdaSpace{}%
\AgdaComment{--\ Substitute\ 0th\ De\ Bruijn\ index}\<%
\\
\\[\AgdaEmptyExtraSkip]%
\>[0]\AgdaFunction{sb}\AgdaSpace{}%
\AgdaSymbol{:}\AgdaSpace{}%
\AgdaOperator{\AgdaDatatype{μ[}}\AgdaSpace{}%
\AgdaFunction{LSR}\AgdaSpace{}%
\AgdaOperator{\AgdaDatatype{]}}\AgdaSpace{}%
\AgdaSymbol{→}\AgdaSpace{}%
\AgdaFunction{Subst}\AgdaSpace{}%
\AgdaSymbol{→}\AgdaSpace{}%
\AgdaOperator{\AgdaDatatype{μ[}}\AgdaSpace{}%
\AgdaFunction{LSR}\AgdaSpace{}%
\AgdaOperator{\AgdaDatatype{]}}\AgdaSpace{}%
\AgdaComment{--\ Apply\ substitution}\<%
\end{code}
\begin{code}[hide]%
\>[0]\AgdaFunction{sb0}\AgdaSpace{}%
\AgdaBound{t}\AgdaSpace{}%
\AgdaNumber{0}%
\>[15]\AgdaSymbol{=}\AgdaSpace{}%
\AgdaBound{t}\<%
\\
\>[0]\AgdaFunction{sb0}\AgdaSpace{}%
\AgdaSymbol{\AgdaUnderscore{}}\AgdaSpace{}%
\AgdaSymbol{(}\AgdaInductiveConstructor{suc}\AgdaSpace{}%
\AgdaBound{n}\AgdaSymbol{)}%
\>[15]\AgdaSymbol{=}\AgdaSpace{}%
\AgdaInductiveConstructor{`var}\AgdaSpace{}%
\AgdaBound{n}\<%
\\
\\[\AgdaEmptyExtraSkip]%
\>[0]\AgdaFunction{Ren}\AgdaSpace{}%
\AgdaSymbol{:}\AgdaSpace{}%
\AgdaPrimitive{Set}\<%
\\
\>[0]\AgdaFunction{Ren}\AgdaSpace{}%
\AgdaSymbol{=}\AgdaSpace{}%
\AgdaDatatype{ℕ}\AgdaSpace{}%
\AgdaSymbol{→}\AgdaSpace{}%
\AgdaDatatype{ℕ}\<%
\\
\\[\AgdaEmptyExtraSkip]%
\>[0]\AgdaFunction{rn+}\AgdaSpace{}%
\AgdaSymbol{:}\AgdaSpace{}%
\AgdaFunction{Ren}\AgdaSpace{}%
\AgdaSymbol{→}\AgdaSpace{}%
\AgdaFunction{Ren}\<%
\\
\>[0]\AgdaFunction{rn+}\AgdaSpace{}%
\AgdaBound{r}\AgdaSpace{}%
\AgdaInductiveConstructor{zero}\AgdaSpace{}%
\AgdaSymbol{=}\AgdaSpace{}%
\AgdaInductiveConstructor{zero}\<%
\\
\>[0]\AgdaFunction{rn+}\AgdaSpace{}%
\AgdaBound{r}\AgdaSpace{}%
\AgdaSymbol{(}\AgdaInductiveConstructor{suc}\AgdaSpace{}%
\AgdaBound{n}\AgdaSymbol{)}\AgdaSpace{}%
\AgdaSymbol{=}\AgdaSpace{}%
\AgdaBound{r}\AgdaSpace{}%
\AgdaBound{n}\<%
\\
\\[\AgdaEmptyExtraSkip]%
\>[0]\AgdaFunction{rn}\AgdaSpace{}%
\AgdaSymbol{:}\AgdaSpace{}%
\AgdaOperator{\AgdaDatatype{μ[}}\AgdaSpace{}%
\AgdaFunction{LSR}\AgdaSpace{}%
\AgdaOperator{\AgdaDatatype{]}}\AgdaSpace{}%
\AgdaSymbol{→}\AgdaSpace{}%
\AgdaFunction{Ren}\AgdaSpace{}%
\AgdaSymbol{→}\AgdaSpace{}%
\AgdaOperator{\AgdaDatatype{μ[}}\AgdaSpace{}%
\AgdaFunction{LSR}\AgdaSpace{}%
\AgdaOperator{\AgdaDatatype{]}}\<%
\\
\>[0]\AgdaFunction{rn}\AgdaSpace{}%
\AgdaSymbol{(}\AgdaInductiveConstructor{`add}\AgdaSpace{}%
\AgdaBound{t₀}\AgdaSpace{}%
\AgdaBound{t₁}\AgdaSymbol{)}\AgdaSpace{}%
\AgdaBound{r}\AgdaSpace{}%
\AgdaSymbol{=}\AgdaSpace{}%
\AgdaInductiveConstructor{`add}\AgdaSpace{}%
\AgdaSymbol{(}\AgdaFunction{rn}\AgdaSpace{}%
\AgdaBound{t₀}\AgdaSpace{}%
\AgdaBound{r}\AgdaSymbol{)}\AgdaSpace{}%
\AgdaSymbol{(}\AgdaFunction{rn}\AgdaSpace{}%
\AgdaBound{t₁}\AgdaSpace{}%
\AgdaBound{r}\AgdaSymbol{)}\<%
\\
\>[0]\AgdaFunction{rn}\AgdaSpace{}%
\AgdaSymbol{(}\AgdaInductiveConstructor{`num}\AgdaSpace{}%
\AgdaBound{n}\AgdaSymbol{)}\AgdaSpace{}%
\AgdaBound{r}\AgdaSpace{}%
\AgdaSymbol{=}\AgdaSpace{}%
\AgdaInductiveConstructor{`num}\AgdaSpace{}%
\AgdaBound{n}\<%
\\
\>[0]\AgdaFunction{rn}\AgdaSpace{}%
\AgdaSymbol{(}\AgdaInductiveConstructor{`lam}\AgdaSpace{}%
\AgdaBound{t₀}\AgdaSymbol{)}\AgdaSpace{}%
\AgdaBound{r}\AgdaSpace{}%
\AgdaSymbol{=}\AgdaSpace{}%
\AgdaInductiveConstructor{`lam}\AgdaSpace{}%
\AgdaSymbol{(}\AgdaFunction{rn}\AgdaSpace{}%
\AgdaBound{t₀}\AgdaSpace{}%
\AgdaSymbol{(}\AgdaFunction{rn+}\AgdaSpace{}%
\AgdaBound{r}\AgdaSymbol{))}\<%
\\
\>[0]\AgdaFunction{rn}\AgdaSpace{}%
\AgdaSymbol{(}\AgdaInductiveConstructor{`var}\AgdaSpace{}%
\AgdaBound{n}\AgdaSymbol{)}\AgdaSpace{}%
\AgdaBound{r}\AgdaSpace{}%
\AgdaSymbol{=}\AgdaSpace{}%
\AgdaInductiveConstructor{`var}\AgdaSpace{}%
\AgdaSymbol{(}\AgdaBound{r}\AgdaSpace{}%
\AgdaBound{n}\AgdaSymbol{)}\<%
\\
\>[0]\AgdaFunction{rn}\AgdaSpace{}%
\AgdaSymbol{(}\AgdaInductiveConstructor{`app}\AgdaSpace{}%
\AgdaBound{t₀}\AgdaSpace{}%
\AgdaBound{t₁}\AgdaSymbol{)}\AgdaSpace{}%
\AgdaBound{r}\AgdaSpace{}%
\AgdaSymbol{=}\AgdaSpace{}%
\AgdaInductiveConstructor{`app}\AgdaSpace{}%
\AgdaSymbol{(}\AgdaFunction{rn}\AgdaSpace{}%
\AgdaBound{t₀}\AgdaSpace{}%
\AgdaBound{r}\AgdaSymbol{)}\AgdaSpace{}%
\AgdaSymbol{(}\AgdaFunction{rn}\AgdaSpace{}%
\AgdaBound{t₁}\AgdaSpace{}%
\AgdaBound{r}\AgdaSymbol{)}\<%
\\
\>[0]\AgdaFunction{rn}\AgdaSpace{}%
\AgdaSymbol{(}\AgdaInductiveConstructor{`reset}\AgdaSpace{}%
\AgdaBound{t₀}\AgdaSymbol{)}\AgdaSpace{}%
\AgdaBound{r}\AgdaSpace{}%
\AgdaSymbol{=}\AgdaSpace{}%
\AgdaInductiveConstructor{`reset}\AgdaSpace{}%
\AgdaSymbol{(}\AgdaFunction{rn}\AgdaSpace{}%
\AgdaBound{t₀}\AgdaSpace{}%
\AgdaBound{r}\AgdaSymbol{)}\<%
\\
\>[0]\AgdaFunction{rn}\AgdaSpace{}%
\AgdaSymbol{(}\AgdaInductiveConstructor{`shift}\AgdaSpace{}%
\AgdaBound{t₀}\AgdaSymbol{)}\AgdaSpace{}%
\AgdaBound{r}\AgdaSpace{}%
\AgdaSymbol{=}\AgdaSpace{}%
\AgdaInductiveConstructor{`shift}\AgdaSpace{}%
\AgdaSymbol{(}\AgdaFunction{rn}\AgdaSpace{}%
\AgdaBound{t₀}\AgdaSpace{}%
\AgdaSymbol{(}\AgdaFunction{rn+}\AgdaSpace{}%
\AgdaBound{r}\AgdaSymbol{))}\<%
\\
\\[\AgdaEmptyExtraSkip]%
\>[0]\AgdaFunction{sb+}\AgdaSpace{}%
\AgdaSymbol{:}\AgdaSpace{}%
\AgdaFunction{Subst}\AgdaSpace{}%
\AgdaSymbol{→}\AgdaSpace{}%
\AgdaFunction{Subst}\<%
\\
\>[0]\AgdaFunction{sb+}\AgdaSpace{}%
\AgdaBound{s}\AgdaSpace{}%
\AgdaInductiveConstructor{zero}%
\>[14]\AgdaSymbol{=}\AgdaSpace{}%
\AgdaInductiveConstructor{`var}\AgdaSpace{}%
\AgdaInductiveConstructor{zero}\<%
\\
\>[0]\AgdaFunction{sb+}\AgdaSpace{}%
\AgdaBound{s}\AgdaSpace{}%
\AgdaSymbol{(}\AgdaInductiveConstructor{suc}\AgdaSpace{}%
\AgdaBound{n}\AgdaSymbol{)}\AgdaSpace{}%
\AgdaSymbol{=}\AgdaSpace{}%
\AgdaFunction{rn}\AgdaSpace{}%
\AgdaSymbol{(}\AgdaBound{s}\AgdaSpace{}%
\AgdaBound{n}\AgdaSymbol{)}\AgdaSpace{}%
\AgdaInductiveConstructor{suc}\<%
\\
\\[\AgdaEmptyExtraSkip]%
\>[0]\AgdaFunction{sb}\AgdaSpace{}%
\AgdaSymbol{(}\AgdaInductiveConstructor{`add}\AgdaSpace{}%
\AgdaBound{t₀}\AgdaSpace{}%
\AgdaBound{t₁}\AgdaSymbol{)}\AgdaSpace{}%
\AgdaBound{s}\AgdaSpace{}%
\AgdaSymbol{=}\AgdaSpace{}%
\AgdaInductiveConstructor{`add}\AgdaSpace{}%
\AgdaSymbol{(}\AgdaFunction{sb}\AgdaSpace{}%
\AgdaBound{t₀}\AgdaSpace{}%
\AgdaBound{s}\AgdaSymbol{)}\AgdaSpace{}%
\AgdaSymbol{(}\AgdaFunction{sb}\AgdaSpace{}%
\AgdaBound{t₁}\AgdaSpace{}%
\AgdaBound{s}\AgdaSymbol{)}\<%
\\
\>[0]\AgdaFunction{sb}\AgdaSpace{}%
\AgdaSymbol{(}\AgdaInductiveConstructor{`num}\AgdaSpace{}%
\AgdaBound{n}\AgdaSymbol{)}%
\>[16]\AgdaSymbol{\AgdaUnderscore{}}\AgdaSpace{}%
\AgdaSymbol{=}\AgdaSpace{}%
\AgdaInductiveConstructor{`num}\AgdaSpace{}%
\AgdaBound{n}\<%
\\
\>[0]\AgdaFunction{sb}\AgdaSpace{}%
\AgdaSymbol{(}\AgdaInductiveConstructor{`lam}\AgdaSpace{}%
\AgdaBound{t₀}\AgdaSymbol{)}%
\>[16]\AgdaBound{s}\AgdaSpace{}%
\AgdaSymbol{=}\AgdaSpace{}%
\AgdaInductiveConstructor{`lam}\AgdaSpace{}%
\AgdaSymbol{(}\AgdaFunction{sb}\AgdaSpace{}%
\AgdaBound{t₀}\AgdaSpace{}%
\AgdaSymbol{(}\AgdaFunction{sb+}\AgdaSpace{}%
\AgdaBound{s}\AgdaSymbol{))}\<%
\\
\>[0]\AgdaFunction{sb}\AgdaSpace{}%
\AgdaSymbol{(}\AgdaInductiveConstructor{`var}\AgdaSpace{}%
\AgdaBound{x}\AgdaSymbol{)}%
\>[16]\AgdaBound{s}\AgdaSpace{}%
\AgdaSymbol{=}\AgdaSpace{}%
\AgdaBound{s}\AgdaSpace{}%
\AgdaBound{x}\<%
\\
\>[0]\AgdaFunction{sb}\AgdaSpace{}%
\AgdaSymbol{(}\AgdaInductiveConstructor{`app}\AgdaSpace{}%
\AgdaBound{t₀}\AgdaSpace{}%
\AgdaBound{t₁}\AgdaSymbol{)}\AgdaSpace{}%
\AgdaBound{s}\AgdaSpace{}%
\AgdaSymbol{=}\AgdaSpace{}%
\AgdaInductiveConstructor{`app}\AgdaSpace{}%
\AgdaSymbol{(}\AgdaFunction{sb}\AgdaSpace{}%
\AgdaBound{t₀}\AgdaSpace{}%
\AgdaBound{s}\AgdaSymbol{)}\AgdaSpace{}%
\AgdaSymbol{(}\AgdaFunction{sb}\AgdaSpace{}%
\AgdaBound{t₁}\AgdaSpace{}%
\AgdaBound{s}\AgdaSymbol{)}\<%
\\
\>[0]\AgdaFunction{sb}\AgdaSpace{}%
\AgdaSymbol{(}\AgdaInductiveConstructor{`reset}\AgdaSpace{}%
\AgdaBound{t₀}\AgdaSymbol{)}%
\>[16]\AgdaBound{s}\AgdaSpace{}%
\AgdaSymbol{=}\AgdaSpace{}%
\AgdaInductiveConstructor{`reset}\AgdaSpace{}%
\AgdaSymbol{(}\AgdaFunction{sb}\AgdaSpace{}%
\AgdaBound{t₀}\AgdaSpace{}%
\AgdaBound{s}\AgdaSymbol{)}\<%
\\
\>[0]\AgdaFunction{sb}\AgdaSpace{}%
\AgdaSymbol{(}\AgdaInductiveConstructor{`shift}\AgdaSpace{}%
\AgdaBound{t₀}\AgdaSymbol{)}%
\>[16]\AgdaBound{s}\AgdaSpace{}%
\AgdaSymbol{=}\AgdaSpace{}%
\AgdaInductiveConstructor{`shift}\AgdaSpace{}%
\AgdaSymbol{(}\AgdaFunction{sb}\AgdaSpace{}%
\AgdaBound{t₀}\AgdaSpace{}%
\AgdaSymbol{(}\AgdaFunction{sb+}\AgdaSpace{}%
\AgdaBound{s}\AgdaSymbol{))}\<%
\\
\\[\AgdaEmptyExtraSkip]%
\>[0]\AgdaFunction{↑}\AgdaSpace{}%
\AgdaSymbol{:}\AgdaSpace{}%
\AgdaOperator{\AgdaDatatype{μ[}}\AgdaSpace{}%
\AgdaFunction{LSR}\AgdaSpace{}%
\AgdaOperator{\AgdaDatatype{]}}\AgdaSpace{}%
\AgdaSymbol{→}\AgdaSpace{}%
\AgdaOperator{\AgdaDatatype{μ[}}\AgdaSpace{}%
\AgdaFunction{LSR}\AgdaSpace{}%
\AgdaOperator{\AgdaDatatype{]}}\<%
\\
\>[0]\AgdaFunction{↑}\AgdaSpace{}%
\AgdaBound{t}\AgdaSpace{}%
\AgdaSymbol{=}\AgdaSpace{}%
\AgdaFunction{rn}\AgdaSpace{}%
\AgdaBound{t}\AgdaSpace{}%
\AgdaInductiveConstructor{suc}\<%
\end{code}
Using this function, contraction is defined analogously to \cref{sec:lambda-shift-reset}.
We only discuss the most interesting case:
\begin{code}%
\>[0]\AgdaFunction{contract}%
\>[10]\AgdaSymbol{:}\AgdaSpace{}%
\AgdaOperator{\AgdaDatatype{μ[}}\AgdaSpace{}%
\AgdaFunction{LSR}\AgdaSpace{}%
\AgdaOperator{\AgdaDatatype{]}}\AgdaSpace{}%
\AgdaSymbol{→}\AgdaSpace{}%
\AgdaOperator{\AgdaFunction{Ctx[}}\AgdaSpace{}%
\AgdaFunction{LSR}\AgdaSpace{}%
\AgdaOperator{\AgdaFunction{]}}\<%
\\
\>[10]\AgdaSymbol{→}\AgdaSpace{}%
\AgdaDatatype{Maybe}\AgdaSpace{}%
\AgdaSymbol{(}\AgdaOperator{\AgdaDatatype{μ[}}\AgdaSpace{}%
\AgdaFunction{LSR}\AgdaSpace{}%
\AgdaOperator{\AgdaDatatype{]}}\AgdaSpace{}%
\AgdaOperator{\AgdaFunction{×}}\AgdaSpace{}%
\AgdaOperator{\AgdaFunction{Ctx[}}\AgdaSpace{}%
\AgdaFunction{LSR}\AgdaSpace{}%
\AgdaOperator{\AgdaFunction{]}}\AgdaSymbol{)}\<%
\\
\>[0]\AgdaFunction{contract}\AgdaSpace{}%
\AgdaSymbol{(}\AgdaInductiveConstructor{`shift}\AgdaSpace{}%
\AgdaBound{t₀}\AgdaSymbol{)}\AgdaSpace{}%
\AgdaBound{c}\AgdaSpace{}%
\AgdaSymbol{=}\AgdaSpace{}%
\AgdaKeyword{do}\<%
\\
\>[0][@{}l@{\AgdaIndent{0}}]%
\>[2]\AgdaSymbol{(}\AgdaBound{c₁}\AgdaSpace{}%
\AgdaOperator{\AgdaInductiveConstructor{,}}\AgdaSpace{}%
\AgdaBound{t₁}\AgdaSymbol{)}\AgdaSpace{}%
\AgdaOperator{\AgdaFunction{←}}\AgdaSpace{}%
\AgdaFunction{unwind}%
\>[22]\AgdaSymbol{(}\AgdaFunction{⅋map}\AgdaSpace{}%
\AgdaSymbol{(}\AgdaFunction{∂map}\AgdaSpace{}%
\AgdaSymbol{\{}\AgdaFunction{LSR}\AgdaSymbol{\}}\AgdaSpace{}%
\AgdaFunction{↑}\AgdaSymbol{)}\AgdaSpace{}%
\AgdaBound{c}\AgdaSymbol{)}\AgdaSpace{}%
\AgdaSymbol{(}\AgdaInductiveConstructor{`var}\AgdaSpace{}%
\AgdaNumber{0}\AgdaSymbol{)}\<%
\\
\>[2]\AgdaInductiveConstructor{just}\AgdaSpace{}%
\AgdaSymbol{(}\AgdaInductiveConstructor{`app}\AgdaSpace{}%
\AgdaSymbol{(}\AgdaInductiveConstructor{`lam}\AgdaSpace{}%
\AgdaBound{t₀}\AgdaSymbol{)}\AgdaSpace{}%
\AgdaSymbol{(}\AgdaInductiveConstructor{`lam}\AgdaSpace{}%
\AgdaSymbol{(}\AgdaInductiveConstructor{`reset}\AgdaSpace{}%
\AgdaBound{t₁}\AgdaSymbol{))}\AgdaSpace{}%
\AgdaOperator{\AgdaInductiveConstructor{,}}\AgdaSpace{}%
\AgdaBound{c₁}\AgdaSymbol{)}\<%
\\
\>[2]\AgdaKeyword{where}\<%
\\
\>[2][@{}l@{\AgdaIndent{0}}]%
\>[4]\AgdaFunction{unwind}%
\>[12]\AgdaSymbol{:}\AgdaSpace{}%
\AgdaOperator{\AgdaFunction{Ctx[}}\AgdaSpace{}%
\AgdaFunction{LSR}\AgdaSpace{}%
\AgdaOperator{\AgdaFunction{]}}\AgdaSpace{}%
\AgdaSymbol{→}\AgdaSpace{}%
\AgdaOperator{\AgdaDatatype{μ[}}\AgdaSpace{}%
\AgdaFunction{LSR}\AgdaSpace{}%
\AgdaOperator{\AgdaDatatype{]}}\<%
\\
\>[12]\AgdaSymbol{→}\AgdaSpace{}%
\AgdaDatatype{Maybe}\AgdaSpace{}%
\AgdaSymbol{(}\AgdaOperator{\AgdaFunction{Ctx[}}\AgdaSpace{}%
\AgdaFunction{LSR}\AgdaSpace{}%
\AgdaOperator{\AgdaFunction{]}}\AgdaSpace{}%
\AgdaOperator{\AgdaFunction{×}}\AgdaSpace{}%
\AgdaOperator{\AgdaDatatype{μ[}}\AgdaSpace{}%
\AgdaFunction{LSR}\AgdaSpace{}%
\AgdaOperator{\AgdaDatatype{]}}\AgdaSymbol{)}\<%
\\
\>[4]\AgdaFunction{unwind}\AgdaSpace{}%
\AgdaInductiveConstructor{[]}%
\>[28]\AgdaSymbol{\AgdaUnderscore{}}%
\>[31]\AgdaSymbol{=}\AgdaSpace{}%
\AgdaInductiveConstructor{nothing}\<%
\\
\>[4]\AgdaFunction{unwind}\AgdaSpace{}%
\AgdaSymbol{(}\AgdaInductiveConstructor{`reset-c}%
\>[22]\AgdaOperator{\AgdaInductiveConstructor{∷}}\AgdaSpace{}%
\AgdaBound{c}\AgdaSymbol{)}%
\>[28]\AgdaBound{t}%
\>[31]\AgdaSymbol{=}\AgdaSpace{}%
\AgdaInductiveConstructor{just}\AgdaSpace{}%
\AgdaSymbol{(}\AgdaInductiveConstructor{`reset-c}\AgdaSpace{}%
\AgdaOperator{\AgdaInductiveConstructor{∷}}\AgdaSpace{}%
\AgdaBound{c}\AgdaSpace{}%
\AgdaOperator{\AgdaInductiveConstructor{,}}\AgdaSpace{}%
\AgdaBound{t}\AgdaSymbol{)}\<%
\\
\>[4]\AgdaCatchallClause{\AgdaFunction{unwind}}\AgdaSpace{}%
\AgdaCatchallClause{\AgdaSymbol{(}}\AgdaCatchallClause{\AgdaBound{frm}}%
\>[22]\AgdaCatchallClause{\AgdaOperator{\AgdaInductiveConstructor{∷}}}\AgdaSpace{}%
\AgdaCatchallClause{\AgdaBound{c}}\AgdaCatchallClause{\AgdaSymbol{)}}%
\>[28]\AgdaCatchallClause{\AgdaBound{t}}%
\>[31]\AgdaSymbol{=}\AgdaSpace{}%
\AgdaFunction{unwind}\AgdaSpace{}%
\AgdaBound{c}\AgdaSpace{}%
\AgdaOperator{\AgdaInductiveConstructor{⟨}}\AgdaSpace{}%
\AgdaFunction{plug}\AgdaSpace{}%
\AgdaSymbol{\{}\AgdaFunction{LSR}\AgdaSymbol{\}}\AgdaSpace{}%
\AgdaBound{frm}\AgdaSpace{}%
\AgdaBound{t}\AgdaSpace{}%
\AgdaOperator{\AgdaInductiveConstructor{⟩}}\<%
\end{code}
\begin{code}[hide]%
\>[0]\AgdaFunction{contract}\AgdaSpace{}%
\AgdaSymbol{(}\AgdaInductiveConstructor{`add}\AgdaSpace{}%
\AgdaSymbol{(}\AgdaInductiveConstructor{`num}\AgdaSpace{}%
\AgdaBound{n₀}\AgdaSymbol{)}\AgdaSpace{}%
\AgdaSymbol{(}\AgdaInductiveConstructor{`num}\AgdaSpace{}%
\AgdaBound{n₁}\AgdaSymbol{))}\AgdaSpace{}%
\AgdaBound{c}\AgdaSpace{}%
\AgdaSymbol{=}\<%
\\
\>[0][@{}l@{\AgdaIndent{0}}]%
\>[2]\AgdaInductiveConstructor{just}\AgdaSpace{}%
\AgdaSymbol{(}\AgdaInductiveConstructor{`num}\AgdaSpace{}%
\AgdaSymbol{(}\AgdaBound{n₀}\AgdaSpace{}%
\AgdaOperator{\AgdaPrimitive{+}}\AgdaSpace{}%
\AgdaBound{n₁}\AgdaSymbol{)}\AgdaSpace{}%
\AgdaOperator{\AgdaInductiveConstructor{,}}\AgdaSpace{}%
\AgdaBound{c}\AgdaSymbol{)}\<%
\\
\>[0]\AgdaFunction{contract}\AgdaSpace{}%
\AgdaSymbol{(}\AgdaInductiveConstructor{`app}\AgdaSpace{}%
\AgdaSymbol{(}\AgdaInductiveConstructor{`lam}\AgdaSpace{}%
\AgdaBound{t₀}\AgdaSymbol{)}\AgdaSpace{}%
\AgdaBound{t₁}\AgdaSymbol{)}\AgdaSpace{}%
\AgdaBound{c}\AgdaSpace{}%
\AgdaKeyword{with}\AgdaSpace{}%
\AgdaFunction{is-val-LSR}\AgdaSpace{}%
\AgdaBound{t₁}\<%
\\
\>[0]\AgdaSymbol{...}\AgdaSpace{}%
\AgdaSymbol{|}\AgdaSpace{}%
\AgdaInductiveConstructor{true}\AgdaSpace{}%
\AgdaSymbol{=}\AgdaSpace{}%
\AgdaInductiveConstructor{just}\AgdaSpace{}%
\AgdaSymbol{(}\AgdaFunction{sb}\AgdaSpace{}%
\AgdaBound{t₀}\AgdaSpace{}%
\AgdaSymbol{(}\AgdaFunction{sb0}\AgdaSpace{}%
\AgdaBound{t₁}\AgdaSymbol{)}\AgdaSpace{}%
\AgdaOperator{\AgdaInductiveConstructor{,}}\AgdaSpace{}%
\AgdaBound{c}\AgdaSymbol{)}\<%
\\
\>[0]\AgdaSymbol{...}\AgdaSpace{}%
\AgdaSymbol{|}\AgdaSpace{}%
\AgdaInductiveConstructor{false}\AgdaSpace{}%
\AgdaSymbol{=}\AgdaSpace{}%
\AgdaInductiveConstructor{nothing}\<%
\\
\>[0]\AgdaFunction{contract}\AgdaSpace{}%
\AgdaSymbol{(}\AgdaInductiveConstructor{`reset}\AgdaSpace{}%
\AgdaBound{t₀}\AgdaSymbol{)}\AgdaSpace{}%
\AgdaBound{c}\AgdaSpace{}%
\AgdaKeyword{with}\AgdaSpace{}%
\AgdaFunction{is-val-LSR}\AgdaSpace{}%
\AgdaBound{t₀}\<%
\\
\>[0]\AgdaSymbol{...}\AgdaSpace{}%
\AgdaSymbol{|}\AgdaSpace{}%
\AgdaInductiveConstructor{true}\AgdaSpace{}%
\AgdaSymbol{=}\AgdaSpace{}%
\AgdaInductiveConstructor{just}\AgdaSpace{}%
\AgdaSymbol{(}\AgdaBound{t₀}\AgdaSpace{}%
\AgdaOperator{\AgdaInductiveConstructor{,}}\AgdaSpace{}%
\AgdaBound{c}\AgdaSymbol{)}\<%
\\
\>[0]\AgdaSymbol{...}\AgdaSpace{}%
\AgdaSymbol{|}\AgdaSpace{}%
\AgdaInductiveConstructor{false}\AgdaSpace{}%
\AgdaSymbol{=}\AgdaSpace{}%
\AgdaInductiveConstructor{nothing}\<%
\\
\>[0]\AgdaCatchallClause{\AgdaFunction{contract}}\AgdaSpace{}%
\AgdaCatchallClause{\AgdaSymbol{\AgdaUnderscore{}}}\AgdaSpace{}%
\AgdaCatchallClause{\AgdaSymbol{\AgdaUnderscore{}}}\AgdaSpace{}%
\AgdaSymbol{=}\AgdaSpace{}%
\AgdaInductiveConstructor{nothing}\<%
\end{code}
Two differences from earlier is that (1) the \af{unwind} function here uses the generic \af{map} and \af{∂map} functions to increase De Bruijn indices by one inside the context, and (2) we use \af{plug} to generically unwind the context.

\begin{code}[hide]%
\>[0]\AgdaKeyword{record}\AgdaSpace{}%
\AgdaRecord{Delay}\AgdaSpace{}%
\AgdaSymbol{(}\AgdaBound{A}\AgdaSpace{}%
\AgdaSymbol{:}\AgdaSpace{}%
\AgdaPrimitive{Set}\AgdaSymbol{)}\AgdaSpace{}%
\AgdaSymbol{:}\AgdaSpace{}%
\AgdaPrimitive{Set}\AgdaSpace{}%
\AgdaKeyword{where}\<%
\\
\>[0][@{}l@{\AgdaIndent{0}}]%
\>[2]\AgdaKeyword{constructor}\AgdaSpace{}%
\AgdaOperator{\AgdaCoinductiveConstructor{D⟨\AgdaUnderscore{}⟩}}\<%
\\
\>[2]\AgdaKeyword{coinductive}\<%
\\
\>[2]\AgdaKeyword{field}\AgdaSpace{}%
\AgdaField{insp}\AgdaSpace{}%
\AgdaSymbol{:}\AgdaSpace{}%
\AgdaBound{A}\AgdaSpace{}%
\AgdaOperator{\AgdaDatatype{⊎}}\AgdaSpace{}%
\AgdaRecord{Delay}\AgdaSpace{}%
\AgdaBound{A}\<%
\\
\\[\AgdaEmptyExtraSkip]%
\>[0]\AgdaKeyword{open}\AgdaSpace{}%
\AgdaModule{Delay}\<%
\\
\\[\AgdaEmptyExtraSkip]%
\>[0]\AgdaKeyword{mutual}\<%
\\
\>[0][@{}l@{\AgdaIndent{0}}]%
\>[2]\AgdaKeyword{data}\AgdaSpace{}%
\AgdaOperator{\AgdaDatatype{\AgdaUnderscore{}≈₀\AgdaUnderscore{}}}\AgdaSpace{}%
\AgdaSymbol{\{}\AgdaBound{A}\AgdaSymbol{\}}\AgdaSpace{}%
\AgdaSymbol{:}\AgdaSpace{}%
\AgdaRecord{Delay}\AgdaSpace{}%
\AgdaBound{A}\AgdaSpace{}%
\AgdaSymbol{→}\AgdaSpace{}%
\AgdaRecord{Delay}\AgdaSpace{}%
\AgdaBound{A}\AgdaSpace{}%
\AgdaSymbol{→}\AgdaSpace{}%
\AgdaPrimitive{Set}\AgdaSpace{}%
\AgdaKeyword{where}\<%
\\
\>[2][@{}l@{\AgdaIndent{0}}]%
\>[4]\AgdaInductiveConstructor{val}%
\>[1509I]\AgdaSymbol{:}\AgdaSpace{}%
\AgdaSymbol{\{}\AgdaBound{d₀}\AgdaSpace{}%
\AgdaBound{d₁}\AgdaSpace{}%
\AgdaSymbol{:}\AgdaSpace{}%
\AgdaRecord{Delay}\AgdaSpace{}%
\AgdaBound{A}\AgdaSymbol{\}}\AgdaSpace{}%
\AgdaSymbol{\{}\AgdaBound{x}\AgdaSpace{}%
\AgdaSymbol{:}\AgdaSpace{}%
\AgdaBound{A}\AgdaSymbol{\}}\<%
\\
\>[.][@{}l@{}]\<[1509I]%
\>[8]\AgdaSymbol{→}\AgdaSpace{}%
\AgdaField{insp}\AgdaSpace{}%
\AgdaBound{d₀}\AgdaSpace{}%
\AgdaOperator{\AgdaDatatype{≡}}\AgdaSpace{}%
\AgdaInductiveConstructor{inj₁}\AgdaSpace{}%
\AgdaBound{x}\AgdaSpace{}%
\AgdaSymbol{→}\AgdaSpace{}%
\AgdaField{insp}\AgdaSpace{}%
\AgdaBound{d₁}\AgdaSpace{}%
\AgdaOperator{\AgdaDatatype{≡}}\AgdaSpace{}%
\AgdaInductiveConstructor{inj₁}\AgdaSpace{}%
\AgdaBound{x}\<%
\\
\>[8]\AgdaSymbol{→}\AgdaSpace{}%
\AgdaBound{d₀}\AgdaSpace{}%
\AgdaOperator{\AgdaDatatype{≈₀}}\AgdaSpace{}%
\AgdaBound{d₁}\<%
\\
\>[4]\AgdaInductiveConstructor{tau}%
\>[1532I]\AgdaSymbol{:}\AgdaSpace{}%
\AgdaSymbol{\{}\AgdaBound{d₀}\AgdaSpace{}%
\AgdaBound{d₀₁}\AgdaSpace{}%
\AgdaBound{d₁}\AgdaSpace{}%
\AgdaBound{d₁₁}\AgdaSpace{}%
\AgdaSymbol{:}\AgdaSpace{}%
\AgdaRecord{Delay}\AgdaSpace{}%
\AgdaBound{A}\AgdaSymbol{\}}\<%
\\
\>[.][@{}l@{}]\<[1532I]%
\>[8]\AgdaSymbol{→}\AgdaSpace{}%
\AgdaField{insp}\AgdaSpace{}%
\AgdaBound{d₀}\AgdaSpace{}%
\AgdaOperator{\AgdaDatatype{≡}}\AgdaSpace{}%
\AgdaInductiveConstructor{inj₂}\AgdaSpace{}%
\AgdaBound{d₀₁}\AgdaSpace{}%
\AgdaSymbol{→}\AgdaSpace{}%
\AgdaField{insp}\AgdaSpace{}%
\AgdaBound{d₁}\AgdaSpace{}%
\AgdaOperator{\AgdaDatatype{≡}}\AgdaSpace{}%
\AgdaInductiveConstructor{inj₂}\AgdaSpace{}%
\AgdaBound{d₁₁}\<%
\\
\>[8]\AgdaSymbol{→}\AgdaSpace{}%
\AgdaBound{d₀₁}\AgdaSpace{}%
\AgdaOperator{\AgdaRecord{≈}}\AgdaSpace{}%
\AgdaBound{d₁₁}\AgdaSpace{}%
\AgdaSymbol{→}\AgdaSpace{}%
\AgdaBound{d₀}\AgdaSpace{}%
\AgdaOperator{\AgdaDatatype{≈₀}}\AgdaSpace{}%
\AgdaBound{d₁}\<%
\\
\\[\AgdaEmptyExtraSkip]%
\>[2]\AgdaKeyword{record}\AgdaSpace{}%
\AgdaOperator{\AgdaRecord{\AgdaUnderscore{}≈\AgdaUnderscore{}}}\AgdaSpace{}%
\AgdaSymbol{\{}\AgdaBound{A}\AgdaSymbol{\}}\AgdaSpace{}%
\AgdaSymbol{(}\AgdaBound{d₀}\AgdaSpace{}%
\AgdaBound{d₁}\AgdaSpace{}%
\AgdaSymbol{:}\AgdaSpace{}%
\AgdaRecord{Delay}\AgdaSpace{}%
\AgdaBound{A}\AgdaSymbol{)}\AgdaSpace{}%
\AgdaSymbol{:}\AgdaSpace{}%
\AgdaPrimitive{Set}\AgdaSpace{}%
\AgdaKeyword{where}\<%
\\
\>[2][@{}l@{\AgdaIndent{0}}]%
\>[4]\AgdaKeyword{coinductive}\<%
\\
\>[4]\AgdaKeyword{constructor}\AgdaSpace{}%
\AgdaOperator{\AgdaCoinductiveConstructor{≈⟨\AgdaUnderscore{}⟩}}\<%
\\
\>[4]\AgdaKeyword{field}\AgdaSpace{}%
\AgdaField{bisim}\AgdaSpace{}%
\AgdaSymbol{:}\AgdaSpace{}%
\AgdaBound{d₀}\AgdaSpace{}%
\AgdaOperator{\AgdaDatatype{≈₀}}\AgdaSpace{}%
\AgdaBound{d₁}\<%
\\
\\[\AgdaEmptyExtraSkip]%
\>[0]\AgdaFunction{delay}\AgdaSpace{}%
\AgdaSymbol{:}\AgdaSpace{}%
\AgdaSymbol{∀}\AgdaSpace{}%
\AgdaSymbol{\{}\AgdaBound{A}\AgdaSymbol{\}}\AgdaSpace{}%
\AgdaSymbol{→}\AgdaSpace{}%
\AgdaDatatype{ℕ}\AgdaSpace{}%
\AgdaSymbol{→}\AgdaSpace{}%
\AgdaBound{A}\AgdaSpace{}%
\AgdaSymbol{→}\AgdaSpace{}%
\AgdaRecord{Delay}\AgdaSpace{}%
\AgdaBound{A}\<%
\\
\>[0]\AgdaField{insp}\AgdaSpace{}%
\AgdaSymbol{(}\AgdaFunction{delay}\AgdaSpace{}%
\AgdaInductiveConstructor{zero}\AgdaSpace{}%
\AgdaBound{x}\AgdaSymbol{)}%
\>[23]\AgdaSymbol{=}\AgdaSpace{}%
\AgdaInductiveConstructor{inj₁}\AgdaSpace{}%
\AgdaBound{x}\<%
\\
\>[0]\AgdaField{insp}\AgdaSpace{}%
\AgdaSymbol{(}\AgdaFunction{delay}\AgdaSpace{}%
\AgdaSymbol{(}\AgdaInductiveConstructor{suc}\AgdaSpace{}%
\AgdaBound{n}\AgdaSymbol{)}\AgdaSpace{}%
\AgdaBound{x}\AgdaSymbol{)}\AgdaSpace{}%
\AgdaSymbol{=}\AgdaSpace{}%
\AgdaInductiveConstructor{inj₂}\AgdaSpace{}%
\AgdaSymbol{(}\AgdaFunction{delay}\AgdaSpace{}%
\AgdaBound{n}\AgdaSpace{}%
\AgdaBound{x}\AgdaSymbol{)}\<%
\\
\\[\AgdaEmptyExtraSkip]%
\>[0]\AgdaKeyword{open}\AgdaSpace{}%
\AgdaOperator{\AgdaModule{\AgdaUnderscore{}≈\AgdaUnderscore{}}}\<%
\end{code}

\subsubsection{Generic Driver Loop}

We obtain a reduction semantic interpreter from our syntax definition and contraction function by calling a generic driver loop.
The following function implements this loop, and is generic in the notion of syntax description \ab{d}, value predicate \ab{is-val}, and contraction function \ab{contract}.
\begin{code}[hide]%
\>[0]\AgdaKeyword{module}\AgdaSpace{}%
\AgdaModule{DriverLoop}\<%
\\
\>[0][@{}l@{\AgdaIndent{0}}]%
\>[4]\AgdaSymbol{\{}\AgdaBound{d}\AgdaSpace{}%
\AgdaSymbol{:}\AgdaSpace{}%
\AgdaDatatype{Desc}\AgdaSymbol{\}}\<%
\\
\>[4]\AgdaSymbol{(}\AgdaBound{is-val}%
\>[15]\AgdaSymbol{:}\AgdaSpace{}%
\AgdaOperator{\AgdaDatatype{μ[}}\AgdaSpace{}%
\AgdaBound{d}\AgdaSpace{}%
\AgdaOperator{\AgdaDatatype{]}}\AgdaSpace{}%
\AgdaSymbol{→}\AgdaSpace{}%
\AgdaDatatype{Bool}\AgdaSymbol{)}\<%
\\
\>[4]\AgdaSymbol{(}\AgdaBound{decomp}%
\>[15]\AgdaSymbol{:}\AgdaSpace{}%
\AgdaOperator{\AgdaDatatype{μ[}}\AgdaSpace{}%
\AgdaBound{d}\AgdaSpace{}%
\AgdaOperator{\AgdaDatatype{]}}\AgdaSpace{}%
\AgdaSymbol{→}\AgdaSpace{}%
\AgdaDatatype{Maybe}\AgdaSpace{}%
\AgdaSymbol{(}\AgdaOperator{\AgdaFunction{Ctx[}}\AgdaSpace{}%
\AgdaBound{d}\AgdaSpace{}%
\AgdaOperator{\AgdaFunction{]}}\AgdaSpace{}%
\AgdaOperator{\AgdaFunction{×}}\AgdaSpace{}%
\AgdaOperator{\AgdaDatatype{μ[}}\AgdaSpace{}%
\AgdaBound{d}\AgdaSpace{}%
\AgdaOperator{\AgdaDatatype{]}}\AgdaSymbol{))}\<%
\\
\>[4]\AgdaSymbol{(}\AgdaBound{contract}%
\>[15]\AgdaSymbol{:}\AgdaSpace{}%
\AgdaOperator{\AgdaDatatype{μ[}}\AgdaSpace{}%
\AgdaBound{d}\AgdaSpace{}%
\AgdaOperator{\AgdaDatatype{]}}\AgdaSpace{}%
\AgdaSymbol{→}\AgdaSpace{}%
\AgdaOperator{\AgdaFunction{Ctx[}}\AgdaSpace{}%
\AgdaBound{d}\AgdaSpace{}%
\AgdaOperator{\AgdaFunction{]}}\<%
\\
\>[15]\AgdaSymbol{→}\AgdaSpace{}%
\AgdaDatatype{Maybe}\AgdaSpace{}%
\AgdaSymbol{(}\AgdaOperator{\AgdaDatatype{μ[}}\AgdaSpace{}%
\AgdaBound{d}\AgdaSpace{}%
\AgdaOperator{\AgdaDatatype{]}}\AgdaSpace{}%
\AgdaOperator{\AgdaFunction{×}}\AgdaSpace{}%
\AgdaOperator{\AgdaFunction{Ctx[}}\AgdaSpace{}%
\AgdaBound{d}\AgdaSpace{}%
\AgdaOperator{\AgdaFunction{]}}\AgdaSymbol{))}\AgdaSpace{}%
\AgdaKeyword{where}\<%
\end{code}
\begin{code}%
\>[0][@{}l@{\AgdaIndent{1}}]%
\>[2]\AgdaFunction{drive}\AgdaSpace{}%
\AgdaSymbol{:}\AgdaSpace{}%
\AgdaOperator{\AgdaDatatype{μ[}}\AgdaSpace{}%
\AgdaBound{d}\AgdaSpace{}%
\AgdaOperator{\AgdaDatatype{]}}\AgdaSpace{}%
\AgdaSymbol{→}\AgdaSpace{}%
\AgdaRecord{Delay}\AgdaSpace{}%
\AgdaSymbol{(}\AgdaDatatype{Maybe}\AgdaSpace{}%
\AgdaOperator{\AgdaDatatype{μ[}}\AgdaSpace{}%
\AgdaBound{d}\AgdaSpace{}%
\AgdaOperator{\AgdaDatatype{]}}\AgdaSymbol{)}\<%
\\
\>[2]\AgdaField{insp}\AgdaSpace{}%
\AgdaSymbol{(}\AgdaFunction{drive}\AgdaSpace{}%
\AgdaBound{t}\AgdaSymbol{)}\AgdaSpace{}%
\AgdaKeyword{with}\AgdaSpace{}%
\AgdaBound{is-val}\AgdaSpace{}%
\AgdaBound{t}\<%
\\
\>[2]\AgdaSymbol{...}\AgdaSpace{}%
\AgdaSymbol{|}%
\>[9]\AgdaInductiveConstructor{true}%
\>[16]\AgdaSymbol{=}\AgdaSpace{}%
\AgdaInductiveConstructor{inj₁}\AgdaSpace{}%
\AgdaSymbol{(}\AgdaInductiveConstructor{just}\AgdaSpace{}%
\AgdaBound{t}\AgdaSymbol{)}\<%
\\
\>[2]\AgdaSymbol{...}\AgdaSpace{}%
\AgdaSymbol{|}%
\>[9]\AgdaInductiveConstructor{false}\AgdaSpace{}%
\AgdaKeyword{with}\AgdaSpace{}%
\AgdaBound{decomp}\AgdaSpace{}%
\AgdaBound{t}\<%
\\
\>[2]\AgdaSymbol{...}%
\>[9]\AgdaSymbol{|}%
\>[12]\AgdaInductiveConstructor{nothing}\AgdaSpace{}%
\AgdaSymbol{=}\AgdaSpace{}%
\AgdaInductiveConstructor{inj₁}\AgdaSpace{}%
\AgdaInductiveConstructor{nothing}\<%
\\
\>[2]\AgdaSymbol{...}%
\>[9]\AgdaSymbol{|}%
\>[12]\AgdaInductiveConstructor{just}\AgdaSpace{}%
\AgdaSymbol{(}\AgdaBound{c₀}\AgdaSpace{}%
\AgdaOperator{\AgdaInductiveConstructor{,}}\AgdaSpace{}%
\AgdaBound{t₀}\AgdaSymbol{)}\AgdaSpace{}%
\AgdaKeyword{with}\AgdaSpace{}%
\AgdaBound{contract}\AgdaSpace{}%
\AgdaBound{t₀}\AgdaSpace{}%
\AgdaBound{c₀}\<%
\\
\>[2]\AgdaSymbol{...}%
\>[12]\AgdaSymbol{|}\AgdaSpace{}%
\AgdaInductiveConstructor{nothing}\AgdaSpace{}%
\AgdaSymbol{=}\AgdaSpace{}%
\AgdaInductiveConstructor{inj₁}\AgdaSpace{}%
\AgdaInductiveConstructor{nothing}\<%
\\
\>[2]\AgdaSymbol{...}%
\>[12]\AgdaSymbol{|}\AgdaSpace{}%
\AgdaInductiveConstructor{just}\AgdaSpace{}%
\AgdaSymbol{(}\AgdaBound{t₁}\AgdaSpace{}%
\AgdaOperator{\AgdaInductiveConstructor{,}}\AgdaSpace{}%
\AgdaBound{c₁}\AgdaSymbol{)}\AgdaSpace{}%
\AgdaSymbol{=}\AgdaSpace{}%
\AgdaInductiveConstructor{inj₂}\AgdaSpace{}%
\AgdaSymbol{(}\AgdaFunction{drive}\AgdaSpace{}%
\AgdaSymbol{(}\AgdaFunction{plug-ctx}\AgdaSpace{}%
\AgdaBound{c₁}\AgdaSpace{}%
\AgdaBound{t₁}\AgdaSymbol{))}\<%
\end{code}%
\begin{code}[hide]%
\>[0]\AgdaKeyword{module}\AgdaSpace{}%
\AgdaModule{LSRDriver}\AgdaSpace{}%
\AgdaKeyword{where}\<%
\\
\>[0][@{}l@{\AgdaIndent{0}}]%
\>[2]\AgdaKeyword{open}\AgdaSpace{}%
\AgdaModule{LSRReduction}\<%
\\
\>[2]\AgdaKeyword{open}\AgdaSpace{}%
\AgdaModule{DriverLoop}\AgdaSpace{}%
\AgdaSymbol{\{}\AgdaFunction{LSR}\AgdaSymbol{\}}\AgdaSpace{}%
\AgdaFunction{is-val-LSR}\AgdaSpace{}%
\AgdaFunction{decomposeLMIMW}\AgdaSpace{}%
\AgdaFunction{contract}\<%
\end{code}%
%
%
\begin{code}[hide]%
\>[2]\AgdaFunction{test-shift}\AgdaSpace{}%
\AgdaSymbol{:}%
\>[1691I]\AgdaFunction{drive}%
\>[1692I]\AgdaSymbol{(}\AgdaInductiveConstructor{`add}\AgdaSpace{}%
\AgdaSymbol{(}\AgdaInductiveConstructor{`num}\AgdaSpace{}%
\AgdaNumber{1}\AgdaSymbol{)}\<%
\\
\>[1692I][@{}l@{\AgdaIndent{0}}]%
\>[23]\AgdaSymbol{(}\AgdaInductiveConstructor{`reset}\<%
\\
\>[23][@{}l@{\AgdaIndent{0}}]%
\>[25]\AgdaSymbol{(}\AgdaInductiveConstructor{`add}\AgdaSpace{}%
\AgdaSymbol{(}\AgdaInductiveConstructor{`num}\AgdaSpace{}%
\AgdaNumber{2}\AgdaSymbol{)}\<%
\\
\>[25][@{}l@{\AgdaIndent{0}}]%
\>[27]\AgdaSymbol{(}\AgdaInductiveConstructor{`shift}\<%
\\
\>[27][@{}l@{\AgdaIndent{0}}]%
\>[29]\AgdaSymbol{(}\AgdaInductiveConstructor{`app}\AgdaSpace{}%
\AgdaSymbol{(}\AgdaInductiveConstructor{`var}\AgdaSpace{}%
\AgdaNumber{0}\AgdaSymbol{)}\<%
\\
\>[29][@{}l@{\AgdaIndent{0}}]%
\>[31]\AgdaSymbol{(}\AgdaInductiveConstructor{`app}\AgdaSpace{}%
\AgdaSymbol{(}\AgdaInductiveConstructor{`var}\AgdaSpace{}%
\AgdaNumber{0}\AgdaSymbol{)}\AgdaSpace{}%
\AgdaSymbol{(}\AgdaInductiveConstructor{`num}\AgdaSpace{}%
\AgdaNumber{3}\AgdaSymbol{)))))))}\<%
\\
\>[.][@{}l@{}]\<[1691I]%
\>[15]\AgdaOperator{\AgdaRecord{≈}}\AgdaSpace{}%
\AgdaFunction{delay}\AgdaSpace{}%
\AgdaNumber{10}\AgdaSpace{}%
\AgdaSymbol{(}\AgdaInductiveConstructor{just}\AgdaSpace{}%
\AgdaSymbol{(}\AgdaInductiveConstructor{`num}\AgdaSpace{}%
\AgdaNumber{8}\AgdaSymbol{))}\<%
\\
\>[2]\AgdaField{bisim}\AgdaSpace{}%
\AgdaFunction{test-shift}\AgdaSpace{}%
\AgdaSymbol{=}\<%
\\
\>[2][@{}l@{\AgdaIndent{0}}]%
\>[4]\AgdaInductiveConstructor{tau}\AgdaSpace{}%
\AgdaInductiveConstructor{refl}\AgdaSpace{}%
\AgdaInductiveConstructor{refl}\<%
\\
\>[4][@{}l@{\AgdaIndent{0}}]%
\>[6]\AgdaOperator{\AgdaCoinductiveConstructor{≈⟨}}\AgdaSpace{}%
\AgdaInductiveConstructor{tau}\AgdaSpace{}%
\AgdaInductiveConstructor{refl}\AgdaSpace{}%
\AgdaInductiveConstructor{refl}\<%
\\
\>[6][@{}l@{\AgdaIndent{0}}]%
\>[8]\AgdaOperator{\AgdaCoinductiveConstructor{≈⟨}}\AgdaSpace{}%
\AgdaInductiveConstructor{tau}\AgdaSpace{}%
\AgdaInductiveConstructor{refl}\AgdaSpace{}%
\AgdaInductiveConstructor{refl}\<%
\\
\>[8][@{}l@{\AgdaIndent{0}}]%
\>[10]\AgdaOperator{\AgdaCoinductiveConstructor{≈⟨}}\AgdaSpace{}%
\AgdaInductiveConstructor{tau}\AgdaSpace{}%
\AgdaInductiveConstructor{refl}\AgdaSpace{}%
\AgdaInductiveConstructor{refl}\<%
\\
\>[10][@{}l@{\AgdaIndent{0}}]%
\>[12]\AgdaOperator{\AgdaCoinductiveConstructor{≈⟨}}\AgdaSpace{}%
\AgdaInductiveConstructor{tau}\AgdaSpace{}%
\AgdaInductiveConstructor{refl}\AgdaSpace{}%
\AgdaInductiveConstructor{refl}\<%
\\
\>[12][@{}l@{\AgdaIndent{0}}]%
\>[14]\AgdaOperator{\AgdaCoinductiveConstructor{≈⟨}}\AgdaSpace{}%
\AgdaInductiveConstructor{tau}\AgdaSpace{}%
\AgdaInductiveConstructor{refl}\AgdaSpace{}%
\AgdaInductiveConstructor{refl}\<%
\\
\>[14][@{}l@{\AgdaIndent{0}}]%
\>[16]\AgdaOperator{\AgdaCoinductiveConstructor{≈⟨}}\AgdaSpace{}%
\AgdaInductiveConstructor{tau}\AgdaSpace{}%
\AgdaInductiveConstructor{refl}\AgdaSpace{}%
\AgdaInductiveConstructor{refl}\<%
\\
\>[16][@{}l@{\AgdaIndent{0}}]%
\>[18]\AgdaOperator{\AgdaCoinductiveConstructor{≈⟨}}\AgdaSpace{}%
\AgdaInductiveConstructor{tau}\AgdaSpace{}%
\AgdaInductiveConstructor{refl}\AgdaSpace{}%
\AgdaInductiveConstructor{refl}\<%
\\
\>[18][@{}l@{\AgdaIndent{0}}]%
\>[20]\AgdaOperator{\AgdaCoinductiveConstructor{≈⟨}}\AgdaSpace{}%
\AgdaInductiveConstructor{tau}\AgdaSpace{}%
\AgdaInductiveConstructor{refl}\AgdaSpace{}%
\AgdaInductiveConstructor{refl}\<%
\\
\>[20][@{}l@{\AgdaIndent{0}}]%
\>[22]\AgdaOperator{\AgdaCoinductiveConstructor{≈⟨}}\AgdaSpace{}%
\AgdaInductiveConstructor{tau}\AgdaSpace{}%
\AgdaInductiveConstructor{refl}\AgdaSpace{}%
\AgdaInductiveConstructor{refl}\<%
\\
\>[22][@{}l@{\AgdaIndent{0}}]%
\>[24]\AgdaOperator{\AgdaCoinductiveConstructor{≈⟨}}\AgdaSpace{}%
\AgdaInductiveConstructor{val}\AgdaSpace{}%
\AgdaInductiveConstructor{refl}\AgdaSpace{}%
\AgdaInductiveConstructor{refl}\AgdaSpace{}%
\AgdaOperator{\AgdaCoinductiveConstructor{⟩}}\AgdaSpace{}%
\AgdaOperator{\AgdaCoinductiveConstructor{⟩}}\AgdaSpace{}%
\AgdaOperator{\AgdaCoinductiveConstructor{⟩}}\AgdaSpace{}%
\AgdaOperator{\AgdaCoinductiveConstructor{⟩}}\AgdaSpace{}%
\AgdaOperator{\AgdaCoinductiveConstructor{⟩}}\AgdaSpace{}%
\AgdaOperator{\AgdaCoinductiveConstructor{⟩}}\AgdaSpace{}%
\AgdaOperator{\AgdaCoinductiveConstructor{⟩}}\AgdaSpace{}%
\AgdaOperator{\AgdaCoinductiveConstructor{⟩}}\AgdaSpace{}%
\AgdaOperator{\AgdaCoinductiveConstructor{⟩}}\AgdaSpace{}%
\AgdaOperator{\AgdaCoinductiveConstructor{⟩}}\<%
\end{code}

This concludes our generic framework which lets us obtain a reduction semantics from a syntax definition and a contraction function, without boilerplate codex.
It is safe to use this framework for languages that meet the criteria and characterizations summarized in \cref{sec:algebraic-red-sem}.

\begin{code}[hide]%
\>[0]\AgdaSymbol{\{-\#}\AgdaSpace{}%
\AgdaKeyword{OPTIONS}\AgdaSpace{}%
\AgdaPragma{--guardedness}\AgdaSpace{}%
\AgdaSymbol{\#-\}}\<%
\\
\\[\AgdaEmptyExtraSkip]%
\>[0]\AgdaKeyword{module}\AgdaSpace{}%
\AgdaModule{sections.05-rel-work}\AgdaSpace{}%
\AgdaKeyword{where}\<%
\\
\\[\AgdaEmptyExtraSkip]%
\>[0]\AgdaKeyword{open}\AgdaSpace{}%
\AgdaKeyword{import}\AgdaSpace{}%
\AgdaModule{Data.Product}\<%
\\
\>[0]\AgdaKeyword{open}\AgdaSpace{}%
\AgdaKeyword{import}\AgdaSpace{}%
\AgdaModule{Data.Sum}\<%
\\
\>[0]\AgdaKeyword{open}\AgdaSpace{}%
\AgdaKeyword{import}\AgdaSpace{}%
\AgdaModule{Data.Bool}\<%
\\
\>[0]\AgdaKeyword{open}\AgdaSpace{}%
\AgdaKeyword{import}\AgdaSpace{}%
\AgdaModule{Data.Maybe}\<%
\\
\>[0]\AgdaKeyword{open}\AgdaSpace{}%
\AgdaKeyword{import}\AgdaSpace{}%
\AgdaModule{Data.List}\<%
\\
\\[\AgdaEmptyExtraSkip]%
\>[0]\AgdaKeyword{open}\AgdaSpace{}%
\AgdaKeyword{import}\AgdaSpace{}%
\AgdaModule{Relation.Binary.PropositionalEquality}\<%
\\
\\[\AgdaEmptyExtraSkip]%
\>[0]\AgdaKeyword{open}\AgdaSpace{}%
\AgdaKeyword{import}\AgdaSpace{}%
\AgdaModule{sections.03-prelude}\<%
\\
\>[0]\AgdaKeyword{open}\AgdaSpace{}%
\AgdaModule{Alg}\<%
\\
\>[0]\AgdaKeyword{open}\AgdaSpace{}%
\AgdaKeyword{import}\AgdaSpace{}%
\AgdaModule{sections.04-redsem-gen}\<%
\\
\>[0]\AgdaKeyword{open}\AgdaSpace{}%
\AgdaModule{Delay}\<%
\\
\>[0]\AgdaKeyword{open}\AgdaSpace{}%
\AgdaOperator{\AgdaModule{\AgdaUnderscore{}≈\AgdaUnderscore{}}}\<%
\end{code}

\section{Related Work}
\label{sec:07-rel-work}

We discuss related and future lines of work.

\subsection{Interderiving Semantic Artifacts}

Reduction semantics~\cite{FelleisenH92} is a well-established approach to specifying operational semantics for programming language core calculi, including calculi involving control constructs.
Other foundational approaches include structural operational semantics (SOS)~\cite{Plotkin04a,mosses2004modular,TuriP97}; abstract machines~\cite{Landin64} which bridge semantics and implementation; denotational semantics~\cite{scott-strachey-toward} which provide mathematical meaning to programs; and natural semantics~\cite{Kahn87} which define evaluation in terms of big-step relations.
Danvy et al.'s correspondences~\cite{AgerDM05,BiernackaD07,Danvy08} demonstrates how these styles are inter-derivable.

This paper offers a complementary contribution: a \emph{generic} account of reduction semantics, drawing on generic programming---especially McBride's theory of ornaments and derivatives~\cite{mcbride2001derivative}.
Our goal is to isolate what is \emph{essential} to reduction semantics from what is \emph{structural boilerplate}.
As we demonstrate in this paper, this lets us eliminate boilerplate code in executable prototype implementations of reduction semantics.
Furthermore, these implementations are \emph{typed}, \emph{compositionally defined}, and \emph{correct by construction}.
In principle, the generic framework we present here could integrate with Danvy's transformations, extending them to the generic setting and providing mechanically verifiable pipelines between high-level and low-level semantic artifacts.

\subsection{Limitations}

We discuss some key limitations of our current framework.

We do not currently offer generic support for syntax with binding.
It is, in principle possible to extend our approach to handle binding generically.
In particular, we could model intrinsically well-bound syntax for lambda abstractions, let-expressions, etc. by building on Allais et al.'s~\cite{AllaisACMM21} generic approach to functorial syntax with binding.

We do not currently support mutually recursive syntactic categories, such as statements and expressions in imperative languages. While our use of a single Desc type streamlines the semantics of simple calculi, handling mutually recursive syntax would require a more general treatment of datatypes. McBride et al.'s work on dependent polynomial functors and containers~\cite{AbbottAGM03,mcbride2001derivative,chapman2010thegentle} provides the foundational tools to extend our approach in this direction, enabling a generic reduction semantics framework for richer language classes.

Third, we do not address concurrency or interleaving semantics.
These require non-deterministic operational semantics and notions of structural equivalence that fall outside the scope of deterministic reduction semantics as presented here.
Capturing such behavior generically would likely require a fundamentally different treatment of evaluation contexts and redex identification.

Finally, our decomposition function is based on a paramorphism, which explicitly pairs subterms with their decompositions.
While this approach is expressive and generic, one might explore alternative recursion schemes that natively support evaluation contexts and unwinding.
McBride's tail-recursive schemes~\cite{mcbride2001derivative} and related formulations of context-aware folds could offer a more structured or optimized means of computing decompositions.
Whether these schemes can support the full expressive needs of reduction semantics---including customizable context traversal and backtracking---remains an open question.

\subsubsection{Other Approaches to Executable and Generic Reduction Semantics}

Existing semantic specification tools such as PLT Redex~\cite{KleinCDEFFMRTF12}, the K Framework~\cite{Rosu2010anoverview,Rosu17}, and Mosses et al.'s tooling for funcons~\cite{BinsbergenMS19,Churchill2014reusable} support deriving interpreters from concise specifications that require minimal boilerplate code.
A main benefit of our approach over these tools is that, by working in a dependently typed host language, we can generically mechanize and verify correctness guarantees, once and for all.

Danvy notes that reduction contexts have a zipper-like structure~\cite{Danvy08afp}.
We build on this observation to derive generic reduction semantics based on datatype differentiation and functorial representations of syntax.
Several existing approaches build on similar or related ideas.

The work of Cortinas~and~Swierstra~\cite{CortinasS18} uses many of the same techniques as we do.
They use derivatives of datatypes (in the McBride sense) to define verified abstract machines.
While their goal is to connect source-level semantics with machine implementations, ours is to provide a modular and reusable specification of reduction semantics directly.
Nevertheless, their work offers useful inspiration: they define abstract machines by catamorphic algebras over syntax, a pattern that we reuse for defining reduction strategies.
A key distinction is that not all reduction strategies can be expressed catamorphically, especially in the presence of control constructs or leftmost-innermost traversals.
Our framework admits general recursion or corecursion for defining strategies, whereas they take a more principled (termination-ensuring) approach.

Biernacka et al.~\cite{BiernackaBLS22} develop a semantic framework based on zippers, geared toward non-deterministic semantics.
While their use of zippers to define semantics is similar in spirit to ours, their focus is on defining abstract machines and non-determinism rather than generic reduction semantics.

Several papers on refocusing~\cite{Danvy2004refocusingin} develop notions of reduction semantics and contexts that resemble zippers.
For example, Sieczkowski~\cite{SieczkowskiBB10} develops a generic specification of reduction semantics in Coq.
His generic contexts are not explicitly based on zippers but closely mirrors their structure.
Sieczkowski uses this generic formalization to prove the correctness of refocusing.
\emph{Generalized refocusing}~\cite{BiernackaCZ17} is also formalized in Coq using a generic notion of reduction context that resembles zippers.
They use this to formalize a generalized refocusing transformation can refocus a wider class of reduction semantics than previous.
The focus of the works of Sieczkowski and Biernacka et al. is on correctness and derivability of refocusing whereas we aim to factor the construction of reduction semantics itself into generic programming components.

\subsubsection{Refocusing}

It is, in principle, possible to refocus~\cite{Danvy2004refocusingin} our generic reduction semantic interpreters.
For example, the following loop implements refocusing:
\begin{code}[hide]%
\>[0]\AgdaKeyword{module}\AgdaSpace{}%
\AgdaModule{RefocusedLoop}\<%
\\
\>[0][@{}l@{\AgdaIndent{0}}]%
\>[4]\AgdaSymbol{\{}\AgdaBound{d}\AgdaSpace{}%
\AgdaSymbol{:}\AgdaSpace{}%
\AgdaDatatype{Desc}\AgdaSymbol{\}}\<%
\\
\>[4]\AgdaSymbol{(}\AgdaBound{is-val}%
\>[15]\AgdaSymbol{:}\AgdaSpace{}%
\AgdaOperator{\AgdaDatatype{μ[}}\AgdaSpace{}%
\AgdaBound{d}\AgdaSpace{}%
\AgdaOperator{\AgdaDatatype{]}}\AgdaSpace{}%
\AgdaSymbol{→}\AgdaSpace{}%
\AgdaDatatype{Bool}\AgdaSymbol{)}\<%
\\
\>[4]\AgdaSymbol{(}\AgdaBound{decomp}%
\>[15]\AgdaSymbol{:}\AgdaSpace{}%
\AgdaOperator{\AgdaDatatype{μ[}}\AgdaSpace{}%
\AgdaBound{d}\AgdaSpace{}%
\AgdaOperator{\AgdaDatatype{]}}\AgdaSpace{}%
\AgdaSymbol{→}\AgdaSpace{}%
\AgdaFunction{DecStrat}\AgdaSpace{}%
\AgdaBound{d}\AgdaSymbol{)}\<%
\\
\>[4]\AgdaSymbol{(}\AgdaBound{contract}%
\>[15]\AgdaSymbol{:}\AgdaSpace{}%
\AgdaOperator{\AgdaDatatype{μ[}}\AgdaSpace{}%
\AgdaBound{d}\AgdaSpace{}%
\AgdaOperator{\AgdaDatatype{]}}\AgdaSpace{}%
\AgdaSymbol{→}\AgdaSpace{}%
\AgdaOperator{\AgdaFunction{Ctx[}}\AgdaSpace{}%
\AgdaBound{d}\AgdaSpace{}%
\AgdaOperator{\AgdaFunction{]}}\<%
\\
\>[15]\AgdaSymbol{→}\AgdaSpace{}%
\AgdaDatatype{Maybe}\AgdaSpace{}%
\AgdaSymbol{(}\AgdaOperator{\AgdaDatatype{μ[}}\AgdaSpace{}%
\AgdaBound{d}\AgdaSpace{}%
\AgdaOperator{\AgdaDatatype{]}}\AgdaSpace{}%
\AgdaOperator{\AgdaFunction{×}}\AgdaSpace{}%
\AgdaOperator{\AgdaFunction{Ctx[}}\AgdaSpace{}%
\AgdaBound{d}\AgdaSpace{}%
\AgdaOperator{\AgdaFunction{]}}\AgdaSymbol{))}\AgdaSpace{}%
\AgdaKeyword{where}\<%
\\
\\[\AgdaEmptyExtraSkip]%
\>[0][@{}l@{\AgdaIndent{0}}]%
\>[2]\AgdaFunction{plug-val}\AgdaSpace{}%
\AgdaSymbol{:}\AgdaSpace{}%
\AgdaOperator{\AgdaDatatype{μ[}}\AgdaSpace{}%
\AgdaBound{d}\AgdaSpace{}%
\AgdaOperator{\AgdaDatatype{]}}\AgdaSpace{}%
\AgdaSymbol{→}\AgdaSpace{}%
\AgdaOperator{\AgdaFunction{Ctx[}}\AgdaSpace{}%
\AgdaBound{d}\AgdaSpace{}%
\AgdaOperator{\AgdaFunction{]}}\AgdaSpace{}%
\AgdaSymbol{→}\AgdaSpace{}%
\AgdaSymbol{(}\AgdaOperator{\AgdaDatatype{μ[}}\AgdaSpace{}%
\AgdaBound{d}\AgdaSpace{}%
\AgdaOperator{\AgdaDatatype{]}}\AgdaSpace{}%
\AgdaOperator{\AgdaFunction{×}}\AgdaSpace{}%
\AgdaOperator{\AgdaFunction{Ctx[}}\AgdaSpace{}%
\AgdaBound{d}\AgdaSpace{}%
\AgdaOperator{\AgdaFunction{]}}\AgdaSymbol{)}\<%
\\
\>[2]\AgdaFunction{plug-val}\AgdaSpace{}%
\AgdaBound{t}\AgdaSpace{}%
\AgdaBound{c}\AgdaSpace{}%
\AgdaKeyword{with}\AgdaSpace{}%
\AgdaBound{c}\AgdaSpace{}%
\AgdaSymbol{|}\AgdaSpace{}%
\AgdaBound{is-val}\AgdaSpace{}%
\AgdaBound{t}\<%
\\
\>[2]\AgdaSymbol{...}\AgdaSpace{}%
\AgdaSymbol{|}\AgdaSpace{}%
\AgdaInductiveConstructor{[]}%
\>[19]\AgdaSymbol{|}\AgdaSpace{}%
\AgdaInductiveConstructor{true}%
\>[28]\AgdaSymbol{=}\AgdaSpace{}%
\AgdaBound{t}\AgdaSpace{}%
\AgdaOperator{\AgdaInductiveConstructor{,}}\AgdaSpace{}%
\AgdaBound{c}\<%
\\
\>[2]\AgdaSymbol{...}\AgdaSpace{}%
\AgdaSymbol{|}\AgdaSpace{}%
\AgdaSymbol{(}\AgdaBound{frm}\AgdaSpace{}%
\AgdaOperator{\AgdaInductiveConstructor{∷}}\AgdaSpace{}%
\AgdaBound{c}\AgdaSymbol{)}%
\>[19]\AgdaSymbol{|}\AgdaSpace{}%
\AgdaInductiveConstructor{true}%
\>[28]\AgdaSymbol{=}\AgdaSpace{}%
\AgdaFunction{plug-val}\AgdaSpace{}%
\AgdaSymbol{(}\AgdaOperator{\AgdaInductiveConstructor{⟨}}\AgdaSpace{}%
\AgdaFunction{plug}\AgdaSpace{}%
\AgdaSymbol{\{}\AgdaBound{d}\AgdaSymbol{\}}\AgdaSpace{}%
\AgdaBound{frm}\AgdaSpace{}%
\AgdaBound{t}\AgdaSpace{}%
\AgdaOperator{\AgdaInductiveConstructor{⟩}}\AgdaSymbol{)}\AgdaSpace{}%
\AgdaBound{c}\<%
\\
\>[2]\AgdaCatchallClause{\AgdaSymbol{...}}\AgdaSpace{}%
\AgdaCatchallClause{\AgdaSymbol{|}}\AgdaSpace{}%
\AgdaCatchallClause{\AgdaSymbol{\AgdaUnderscore{}}}%
\>[19]\AgdaCatchallClause{\AgdaSymbol{|}}\AgdaSpace{}%
\AgdaCatchallClause{\AgdaInductiveConstructor{false}}%
\>[28]\AgdaSymbol{=}\AgdaSpace{}%
\AgdaBound{t}\AgdaSpace{}%
\AgdaOperator{\AgdaInductiveConstructor{,}}\AgdaSpace{}%
\AgdaBound{c}\<%
\end{code}
\begin{code}%
\>[2]\AgdaFunction{refocus}\AgdaSpace{}%
\AgdaSymbol{:}\AgdaSpace{}%
\AgdaOperator{\AgdaDatatype{μ[}}\AgdaSpace{}%
\AgdaBound{d}\AgdaSpace{}%
\AgdaOperator{\AgdaDatatype{]}}\AgdaSpace{}%
\AgdaSymbol{→}\AgdaSpace{}%
\AgdaOperator{\AgdaFunction{Ctx[}}\AgdaSpace{}%
\AgdaBound{d}\AgdaSpace{}%
\AgdaOperator{\AgdaFunction{]}}\AgdaSpace{}%
\AgdaSymbol{→}\AgdaSpace{}%
\AgdaRecord{Delay}\AgdaSpace{}%
\AgdaSymbol{(}\AgdaDatatype{Maybe}\AgdaSpace{}%
\AgdaOperator{\AgdaDatatype{μ[}}\AgdaSpace{}%
\AgdaBound{d}\AgdaSpace{}%
\AgdaOperator{\AgdaDatatype{]}}\AgdaSymbol{)}\<%
\\
\>[2]\AgdaField{insp}\AgdaSpace{}%
\AgdaSymbol{(}\AgdaFunction{refocus}\AgdaSpace{}%
\AgdaBound{t}\AgdaSpace{}%
\AgdaBound{c}\AgdaSymbol{)}\AgdaSpace{}%
\AgdaKeyword{with}\AgdaSpace{}%
\AgdaBound{c}\AgdaSpace{}%
\AgdaSymbol{|}\AgdaSpace{}%
\AgdaBound{is-val}\AgdaSpace{}%
\AgdaBound{t}\<%
\\
\>[2]\AgdaSymbol{...}\AgdaSpace{}%
\AgdaSymbol{|}%
\>[9]\AgdaInductiveConstructor{[]}%
\>[13]\AgdaSymbol{|}\AgdaSpace{}%
\AgdaInductiveConstructor{true}\AgdaSpace{}%
\AgdaSymbol{=}\AgdaSpace{}%
\AgdaInductiveConstructor{inj₁}\AgdaSpace{}%
\AgdaSymbol{(}\AgdaInductiveConstructor{just}\AgdaSpace{}%
\AgdaBound{t}\AgdaSymbol{)}\<%
\\
\>[2]\AgdaCatchallClause{\AgdaSymbol{...}}\AgdaSpace{}%
\AgdaCatchallClause{\AgdaSymbol{|}}%
\>[9]\AgdaCatchallClause{\AgdaSymbol{\AgdaUnderscore{}}}%
\>[13]\AgdaCatchallClause{\AgdaSymbol{|}}\AgdaSpace{}%
\AgdaCatchallClause{\AgdaInductiveConstructor{true}}\AgdaSpace{}%
\AgdaSymbol{=}\<%
\\
\>[2][@{}l@{\AgdaIndent{0}}]%
\>[4]\AgdaKeyword{let}\AgdaSpace{}%
\AgdaSymbol{(}\AgdaBound{t₀}\AgdaSpace{}%
\AgdaOperator{\AgdaInductiveConstructor{,}}\AgdaSpace{}%
\AgdaBound{c₀}\AgdaSymbol{)}\AgdaSpace{}%
\AgdaSymbol{=}\AgdaSpace{}%
\AgdaFunction{plug-val}\AgdaSpace{}%
\AgdaBound{t}\AgdaSpace{}%
\AgdaBound{c}\AgdaSpace{}%
\AgdaKeyword{in}\AgdaSpace{}%
\AgdaInductiveConstructor{inj₂}\AgdaSpace{}%
\AgdaSymbol{(}\AgdaFunction{refocus}\AgdaSpace{}%
\AgdaBound{t₀}\AgdaSpace{}%
\AgdaBound{c₀}\AgdaSymbol{)}\<%
\\
\>[2]\AgdaCatchallClause{\AgdaSymbol{...}}\AgdaSpace{}%
\AgdaCatchallClause{\AgdaSymbol{|}}%
\>[9]\AgdaCatchallClause{\AgdaSymbol{\AgdaUnderscore{}}}%
\>[13]\AgdaCatchallClause{\AgdaSymbol{|}}\AgdaSpace{}%
\AgdaCatchallClause{\AgdaInductiveConstructor{false}}\AgdaSpace{}%
\AgdaKeyword{with}\AgdaSpace{}%
\AgdaBound{decomp}\AgdaSpace{}%
\AgdaBound{t}\AgdaSpace{}%
\AgdaBound{c}\<%
\\
\>[2]\AgdaSymbol{...}%
\>[9]\AgdaSymbol{|}%
\>[12]\AgdaInductiveConstructor{nothing}\AgdaSpace{}%
\AgdaSymbol{=}\AgdaSpace{}%
\AgdaInductiveConstructor{inj₁}\AgdaSpace{}%
\AgdaInductiveConstructor{nothing}\<%
\\
\>[2]\AgdaSymbol{...}%
\>[9]\AgdaSymbol{|}%
\>[12]\AgdaInductiveConstructor{just}\AgdaSpace{}%
\AgdaSymbol{(}\AgdaBound{c₀}\AgdaSpace{}%
\AgdaOperator{\AgdaInductiveConstructor{,}}\AgdaSpace{}%
\AgdaBound{t₀}\AgdaSymbol{)}\AgdaSpace{}%
\AgdaKeyword{with}\AgdaSpace{}%
\AgdaBound{contract}\AgdaSpace{}%
\AgdaBound{t₀}\AgdaSpace{}%
\AgdaBound{c₀}\<%
\\
\>[2]\AgdaSymbol{...}%
\>[12]\AgdaSymbol{|}\AgdaSpace{}%
\AgdaInductiveConstructor{nothing}\AgdaSpace{}%
\AgdaSymbol{=}\AgdaSpace{}%
\AgdaInductiveConstructor{inj₁}\AgdaSpace{}%
\AgdaInductiveConstructor{nothing}\<%
\\
\>[2]\AgdaSymbol{...}%
\>[12]\AgdaSymbol{|}\AgdaSpace{}%
\AgdaInductiveConstructor{just}\AgdaSpace{}%
\AgdaSymbol{(}\AgdaBound{t₁}\AgdaSpace{}%
\AgdaOperator{\AgdaInductiveConstructor{,}}\AgdaSpace{}%
\AgdaBound{c₁}\AgdaSymbol{)}\AgdaSpace{}%
\AgdaSymbol{=}\AgdaSpace{}%
\AgdaInductiveConstructor{inj₂}\AgdaSpace{}%
\AgdaSymbol{(}\AgdaFunction{refocus}\AgdaSpace{}%
\AgdaBound{t₁}\AgdaSpace{}%
\AgdaBound{c₁}\AgdaSymbol{)}\AgdaSpace{}%
\AgdaComment{--\ continue\ locally}\<%
\end{code}
Here \af{plug-val} plugs a value into a reduction context in a locality-preserving manner.
Whereas our earlier \af{drive} function recomposed the entire term after each reduction, \af{refocus} continues reducing locally, thereby avoiding unnecessary decomposition and recomposition.

Two key questions here are: (1) how should \af{plug-val} be defined; and (2) when is it safe to refocus, in the sense that \af{drive}~\ab{t}~=~\af{refocus}~\ab{t}~\ac{[]}?
Following Danvy and Nielsen~\cite{Danvy2004refocusingin}, for left-/right-most inner-most reduction strategies (such as the one used in the examples in this paper), \af{plug-val} can simply plug a value into a context to yield a new term, which will either be a value or contain the next redex.
Following Danvy and Johannsen~\cite{johannsen-thesis,DanvyJ13}, \af{plug-val} must yield terms that are guaranteed to decompose to the left-/right-most (depending on strategy), outer-most redex.

\begin{code}[hide]%
\>[0]\AgdaKeyword{module}\AgdaSpace{}%
\AgdaModule{LSRRefocused}\AgdaSpace{}%
\AgdaKeyword{where}\<%
\\
\>[0][@{}l@{\AgdaIndent{0}}]%
\>[2]\AgdaKeyword{open}\AgdaSpace{}%
\AgdaModule{LSRReduction}\<%
\\
\>[2]\AgdaKeyword{open}\AgdaSpace{}%
\AgdaModule{RefocusedLoop}\AgdaSpace{}%
\AgdaSymbol{\{}\AgdaFunction{LSR}\AgdaSymbol{\}}\AgdaSpace{}%
\AgdaFunction{is-val-LSR}\AgdaSpace{}%
\AgdaOperator{\AgdaFunction{【}}\AgdaSpace{}%
\AgdaFunction{LSR}\AgdaSpace{}%
\AgdaOperator{\AgdaFunction{∼}}\AgdaSpace{}%
\AgdaFunction{algLMIMW}\AgdaSpace{}%
\AgdaOperator{\AgdaFunction{】}}\AgdaSpace{}%
\AgdaFunction{contract}\<%
\\
\\[\AgdaEmptyExtraSkip]%
\>[2]\AgdaFunction{test-shift}\AgdaSpace{}%
\AgdaSymbol{:}%
\>[188I]\AgdaFunction{refocus}%
\>[189I]\AgdaSymbol{(}\AgdaInductiveConstructor{`add}\AgdaSpace{}%
\AgdaSymbol{(}\AgdaInductiveConstructor{`num}\AgdaSpace{}%
\AgdaNumber{1}\AgdaSymbol{)}\<%
\\
\>[.][@{}l@{}]\<[189I]%
\>[23]\AgdaSymbol{(}\AgdaInductiveConstructor{`reset}\<%
\\
\>[23][@{}l@{\AgdaIndent{0}}]%
\>[25]\AgdaSymbol{(}\AgdaInductiveConstructor{`add}\AgdaSpace{}%
\AgdaSymbol{(}\AgdaInductiveConstructor{`num}\AgdaSpace{}%
\AgdaNumber{2}\AgdaSymbol{)}\<%
\\
\>[25][@{}l@{\AgdaIndent{0}}]%
\>[27]\AgdaSymbol{(}\AgdaInductiveConstructor{`shift}\<%
\\
\>[27][@{}l@{\AgdaIndent{0}}]%
\>[29]\AgdaSymbol{(}\AgdaInductiveConstructor{`app}\AgdaSpace{}%
\AgdaSymbol{(}\AgdaInductiveConstructor{`var}\AgdaSpace{}%
\AgdaNumber{0}\AgdaSymbol{)}\<%
\\
\>[29][@{}l@{\AgdaIndent{0}}]%
\>[31]\AgdaSymbol{(}\AgdaInductiveConstructor{`app}\AgdaSpace{}%
\AgdaSymbol{(}\AgdaInductiveConstructor{`var}\AgdaSpace{}%
\AgdaNumber{0}\AgdaSymbol{)}\AgdaSpace{}%
\AgdaSymbol{(}\AgdaInductiveConstructor{`num}\AgdaSpace{}%
\AgdaNumber{3}\AgdaSymbol{)))))))}\AgdaSpace{}%
\AgdaInductiveConstructor{[]}\<%
\\
\>[.][@{}l@{}]\<[188I]%
\>[15]\AgdaOperator{\AgdaRecord{≈}}\AgdaSpace{}%
\AgdaFunction{delay}\AgdaSpace{}%
\AgdaNumber{15}\AgdaSpace{}%
\AgdaSymbol{(}\AgdaInductiveConstructor{just}\AgdaSpace{}%
\AgdaSymbol{(}\AgdaInductiveConstructor{`num}\AgdaSpace{}%
\AgdaNumber{8}\AgdaSymbol{))}\<%
\\
\>[2]\AgdaField{bisim}\AgdaSpace{}%
\AgdaFunction{test-shift}\AgdaSpace{}%
\AgdaSymbol{=}\<%
\\
\>[2][@{}l@{\AgdaIndent{0}}]%
\>[4]\AgdaInductiveConstructor{tau}\AgdaSpace{}%
\AgdaInductiveConstructor{refl}\AgdaSpace{}%
\AgdaInductiveConstructor{refl}\<%
\\
\>[4][@{}l@{\AgdaIndent{0}}]%
\>[6]\AgdaOperator{\AgdaCoinductiveConstructor{≈⟨}}\AgdaSpace{}%
\AgdaInductiveConstructor{tau}\AgdaSpace{}%
\AgdaInductiveConstructor{refl}\AgdaSpace{}%
\AgdaInductiveConstructor{refl}\<%
\\
\>[6][@{}l@{\AgdaIndent{0}}]%
\>[8]\AgdaOperator{\AgdaCoinductiveConstructor{≈⟨}}\AgdaSpace{}%
\AgdaInductiveConstructor{tau}\AgdaSpace{}%
\AgdaInductiveConstructor{refl}\AgdaSpace{}%
\AgdaInductiveConstructor{refl}\<%
\\
\>[8][@{}l@{\AgdaIndent{0}}]%
\>[10]\AgdaOperator{\AgdaCoinductiveConstructor{≈⟨}}\AgdaSpace{}%
\AgdaInductiveConstructor{tau}\AgdaSpace{}%
\AgdaInductiveConstructor{refl}\AgdaSpace{}%
\AgdaInductiveConstructor{refl}\<%
\\
\>[10][@{}l@{\AgdaIndent{0}}]%
\>[12]\AgdaOperator{\AgdaCoinductiveConstructor{≈⟨}}\AgdaSpace{}%
\AgdaInductiveConstructor{tau}\AgdaSpace{}%
\AgdaInductiveConstructor{refl}\AgdaSpace{}%
\AgdaInductiveConstructor{refl}\<%
\\
\>[12][@{}l@{\AgdaIndent{0}}]%
\>[14]\AgdaOperator{\AgdaCoinductiveConstructor{≈⟨}}\AgdaSpace{}%
\AgdaInductiveConstructor{tau}\AgdaSpace{}%
\AgdaInductiveConstructor{refl}\AgdaSpace{}%
\AgdaInductiveConstructor{refl}\<%
\\
\>[14][@{}l@{\AgdaIndent{0}}]%
\>[16]\AgdaOperator{\AgdaCoinductiveConstructor{≈⟨}}\AgdaSpace{}%
\AgdaInductiveConstructor{tau}\AgdaSpace{}%
\AgdaInductiveConstructor{refl}\AgdaSpace{}%
\AgdaInductiveConstructor{refl}\<%
\\
\>[16][@{}l@{\AgdaIndent{0}}]%
\>[18]\AgdaOperator{\AgdaCoinductiveConstructor{≈⟨}}\AgdaSpace{}%
\AgdaInductiveConstructor{tau}\AgdaSpace{}%
\AgdaInductiveConstructor{refl}\AgdaSpace{}%
\AgdaInductiveConstructor{refl}\<%
\\
\>[18][@{}l@{\AgdaIndent{0}}]%
\>[20]\AgdaOperator{\AgdaCoinductiveConstructor{≈⟨}}\AgdaSpace{}%
\AgdaInductiveConstructor{tau}\AgdaSpace{}%
\AgdaInductiveConstructor{refl}\AgdaSpace{}%
\AgdaInductiveConstructor{refl}\<%
\\
\>[20][@{}l@{\AgdaIndent{0}}]%
\>[22]\AgdaOperator{\AgdaCoinductiveConstructor{≈⟨}}\AgdaSpace{}%
\AgdaInductiveConstructor{tau}\AgdaSpace{}%
\AgdaInductiveConstructor{refl}\AgdaSpace{}%
\AgdaInductiveConstructor{refl}\<%
\\
\>[22][@{}l@{\AgdaIndent{0}}]%
\>[24]\AgdaOperator{\AgdaCoinductiveConstructor{≈⟨}}\AgdaSpace{}%
\AgdaInductiveConstructor{tau}\AgdaSpace{}%
\AgdaInductiveConstructor{refl}\AgdaSpace{}%
\AgdaInductiveConstructor{refl}\<%
\\
\>[24][@{}l@{\AgdaIndent{0}}]%
\>[26]\AgdaOperator{\AgdaCoinductiveConstructor{≈⟨}}\AgdaSpace{}%
\AgdaInductiveConstructor{tau}\AgdaSpace{}%
\AgdaInductiveConstructor{refl}\AgdaSpace{}%
\AgdaInductiveConstructor{refl}\<%
\\
\>[26][@{}l@{\AgdaIndent{0}}]%
\>[28]\AgdaOperator{\AgdaCoinductiveConstructor{≈⟨}}\AgdaSpace{}%
\AgdaInductiveConstructor{tau}\AgdaSpace{}%
\AgdaInductiveConstructor{refl}\AgdaSpace{}%
\AgdaInductiveConstructor{refl}\<%
\\
\>[28][@{}l@{\AgdaIndent{0}}]%
\>[30]\AgdaOperator{\AgdaCoinductiveConstructor{≈⟨}}\AgdaSpace{}%
\AgdaInductiveConstructor{tau}\AgdaSpace{}%
\AgdaInductiveConstructor{refl}\AgdaSpace{}%
\AgdaInductiveConstructor{refl}\<%
\\
\>[30][@{}l@{\AgdaIndent{0}}]%
\>[32]\AgdaOperator{\AgdaCoinductiveConstructor{≈⟨}}\AgdaSpace{}%
\AgdaInductiveConstructor{tau}\AgdaSpace{}%
\AgdaInductiveConstructor{refl}\AgdaSpace{}%
\AgdaInductiveConstructor{refl}\<%
\\
\>[32][@{}l@{\AgdaIndent{0}}]%
\>[34]\AgdaOperator{\AgdaCoinductiveConstructor{≈⟨}}\AgdaSpace{}%
\AgdaInductiveConstructor{val}\AgdaSpace{}%
\AgdaInductiveConstructor{refl}\AgdaSpace{}%
\AgdaInductiveConstructor{refl}\AgdaSpace{}%
\AgdaOperator{\AgdaCoinductiveConstructor{⟩}}\AgdaSpace{}%
\AgdaOperator{\AgdaCoinductiveConstructor{⟩}}\AgdaSpace{}%
\AgdaOperator{\AgdaCoinductiveConstructor{⟩}}\AgdaSpace{}%
\AgdaOperator{\AgdaCoinductiveConstructor{⟩}}\AgdaSpace{}%
\AgdaOperator{\AgdaCoinductiveConstructor{⟩}}\AgdaSpace{}%
\AgdaOperator{\AgdaCoinductiveConstructor{⟩}}\AgdaSpace{}%
\AgdaOperator{\AgdaCoinductiveConstructor{⟩}}\AgdaSpace{}%
\AgdaOperator{\AgdaCoinductiveConstructor{⟩}}\AgdaSpace{}%
\AgdaOperator{\AgdaCoinductiveConstructor{⟩}}\AgdaSpace{}%
\AgdaOperator{\AgdaCoinductiveConstructor{⟩}}\AgdaSpace{}%
\AgdaOperator{\AgdaCoinductiveConstructor{⟩}}\AgdaSpace{}%
\AgdaOperator{\AgdaCoinductiveConstructor{⟩}}\AgdaSpace{}%
\AgdaOperator{\AgdaCoinductiveConstructor{⟩}}\AgdaSpace{}%
\AgdaOperator{\AgdaCoinductiveConstructor{⟩}}\AgdaSpace{}%
\AgdaOperator{\AgdaCoinductiveConstructor{⟩}}\<%
\end{code}

\begin{code}[hide]%
\>[0]\AgdaKeyword{module}\AgdaSpace{}%
\AgdaModule{sections.06-conclusion}\AgdaSpace{}%
\AgdaKeyword{where}\<%
\end{code}

\section{Conclusion}
\label{sec:08-conclusion}

We have presented a generic framework for specifying reduction semantics based on datatype derivatives~\cite{mcbride2001derivative}.
Our approach separates essential reduction behavior from the syntactic boilerplate typically required to implement reduction-based interpreters.
By representing syntax generically and contexts as derivatives, we capture the core structure of reduction semantics in a way that is reusable, compositional, typed, and correct by construction.

The framework in theory enables correctness proofs, exploiting the compositional structure afforded by our use of a paramorphism.
Mechanizing this compositional framework in Agda is future work.

In other future work, we propose to explore how the framework could serve as a basis for a generic semantics toolkit that integrates with Danvy's methodology for inter-derivable semantics artifacts.

\begin{acks}
The author thanks the anonymous reviewers for their insightful comments which improved the paper.
\end{acks}

\bibliographystyle{ACM-Reference-Format}
\bibliography{references.bib}


\begin{thebibliography}{44}


\ifx \showCODEN    \undefined \def \showCODEN     #1{\unskip}     \fi
\ifx \showISBNx    \undefined \def \showISBNx     #1{\unskip}     \fi
\ifx \showISBNxiii \undefined \def \showISBNxiii  #1{\unskip}     \fi
\ifx \showISSN     \undefined \def \showISSN      #1{\unskip}     \fi
\ifx \showLCCN     \undefined \def \showLCCN      #1{\unskip}     \fi
\ifx \shownote     \undefined \def \shownote      #1{#1}          \fi
\ifx \showarticletitle \undefined \def \showarticletitle #1{#1}   \fi
\ifx \showURL      \undefined \def \showURL       {\relax}        \fi
\providecommand\bibfield[2]{#2}
\providecommand\bibinfo[2]{#2}
\providecommand\natexlab[1]{#1}
\providecommand\showeprint[2][]{arXiv:#2}

\bibitem[joh(2015)]%
        {johannsen-thesis}
 \bibinfo{year}{2015}\natexlab{}.
\newblock \emph{\bibinfo{title}{On Computational Small Steps and Big Steps:
  Refocusing for Outermost Reduction}}.
\newblock \bibinfo{thesistype}{Ph.\,D. Dissertation}.
  \bibinfo{school}{Department of Computer Science, Aarhus University}.
\newblock


\bibitem[Abbott et~al\mbox{.}(2003)]%
        {AbbottAGM03}
\bibfield{author}{\bibinfo{person}{Michael~Gordon Abbott},
  \bibinfo{person}{Thorsten Altenkirch}, \bibinfo{person}{Neil Ghani}, {and}
  \bibinfo{person}{Conor McBride}.} \bibinfo{year}{2003}\natexlab{}.
\newblock \showarticletitle{Derivatives of Containers}. In
  \bibinfo{booktitle}{\emph{Typed Lambda Calculi and Applications, 6th
  International Conference, {TLCA} 2003, Valencia, Spain, June 10-12, 2003,
  Proceedings}} \emph{(\bibinfo{series}{Lecture Notes in Computer Science},
  Vol.~\bibinfo{volume}{2701})}, \bibfield{editor}{\bibinfo{person}{Martin
  Hofmann}} (Ed.). \bibinfo{publisher}{Springer}, \bibinfo{pages}{16--30}.
\newblock
\showISBNx{3-540-40332-9}
\href{https://doi.org/10.1007/3-540-44904-3\_2}{doi:\nolinkurl{10.1007/3-540-44904-3\_2}}


\bibitem[Abel and Chapman(2014)]%
        {AbelC14}
\bibfield{author}{\bibinfo{person}{Andreas Abel} {and} \bibinfo{person}{James
  Chapman}.} \bibinfo{year}{2014}\natexlab{}.
\newblock \showarticletitle{Normalization by Evaluation in the Delay Monad: {A}
  Case Study for Coinduction via Copatterns and Sized Types}. In
  \bibinfo{booktitle}{\emph{Proceedings 5th Workshop on Mathematically
  Structured Functional Programming, MSFP@ETAPS 2014, Grenoble, France, 12
  April 2014}} \emph{(\bibinfo{series}{{EPTCS}}, Vol.~\bibinfo{volume}{153})},
  \bibfield{editor}{\bibinfo{person}{Paul~Blain Levy} {and}
  \bibinfo{person}{Neel Krishnaswami}} (Eds.). \bibinfo{pages}{51--67}.
\newblock
\href{https://doi.org/10.4204/EPTCS.153.4}{doi:\nolinkurl{10.4204/EPTCS.153.4}}


\bibitem[Ager et~al\mbox{.}(2005)]%
        {AgerDM05}
\bibfield{author}{\bibinfo{person}{Mads~Sig Ager}, \bibinfo{person}{Olivier
  Danvy}, {and} \bibinfo{person}{Jan Midtgaard}.}
  \bibinfo{year}{2005}\natexlab{}.
\newblock \showarticletitle{A functional correspondence between monadic
  evaluators and abstract machines for languages with computational effects}.
\newblock \bibinfo{journal}{\emph{Theor. Comput. Sci.}} \bibinfo{volume}{342},
  \bibinfo{number}{1} (\bibinfo{year}{2005}), \bibinfo{pages}{149--172}.
\newblock
\href{https://doi.org/10.1016/J.TCS.2005.06.008}{doi:\nolinkurl{10.1016/J.TCS.2005.06.008}}


\bibitem[Allais et~al\mbox{.}(2021)]%
        {AllaisACMM21}
\bibfield{author}{\bibinfo{person}{Guillaume Allais}, \bibinfo{person}{Robert
  Atkey}, \bibinfo{person}{James Chapman}, \bibinfo{person}{Conor McBride},
  {and} \bibinfo{person}{James McKinna}.} \bibinfo{year}{2021}\natexlab{}.
\newblock \showarticletitle{A type- and scope-safe universe of syntaxes with
  binding: their semantics and proofs}.
\newblock \bibinfo{journal}{\emph{J. Funct. Program.}}  \bibinfo{volume}{31}
  (\bibinfo{year}{2021}), \bibinfo{pages}{e22}.
\newblock
\href{https://doi.org/10.1017/S0956796820000076}{doi:\nolinkurl{10.1017/S0956796820000076}}


\bibitem[Biernacka et~al\mbox{.}(2022)]%
        {BiernackaBLS22}
\bibfield{author}{\bibinfo{person}{Malgorzata Biernacka},
  \bibinfo{person}{Dariusz Biernacki}, \bibinfo{person}{Sergue{\"{\i}}
  Lenglet}, {and} \bibinfo{person}{Alan Schmitt}.}
  \bibinfo{year}{2022}\natexlab{}.
\newblock \showarticletitle{Non-Deterministic Abstract Machines}. In
  \bibinfo{booktitle}{\emph{33rd International Conference on Concurrency
  Theory, {CONCUR} 2022, September 12-16, 2022, Warsaw, Poland}}
  \emph{(\bibinfo{series}{LIPIcs}, Vol.~\bibinfo{volume}{243})},
  \bibfield{editor}{\bibinfo{person}{Bartek Klin}, \bibinfo{person}{Slawomir
  Lasota}, {and} \bibinfo{person}{Anca Muscholl}} (Eds.).
  \bibinfo{publisher}{Schloss Dagstuhl - Leibniz-Zentrum f{\"{u}}r Informatik},
  \bibinfo{pages}{7:1--7:24}.
\newblock
\showISBNx{978-3-95977-246-4}
\href{https://doi.org/10.4230/LIPICS.CONCUR.2022.7}{doi:\nolinkurl{10.4230/LIPICS.CONCUR.2022.7}}


\bibitem[Biernacka et~al\mbox{.}(2017)]%
        {BiernackaCZ17}
\bibfield{author}{\bibinfo{person}{Malgorzata Biernacka},
  \bibinfo{person}{Witold Charatonik}, {and} \bibinfo{person}{Klara
  Zielinska}.} \bibinfo{year}{2017}\natexlab{}.
\newblock \showarticletitle{Generalized Refocusing: From Hybrid Strategies to
  Abstract Machines}. In \bibinfo{booktitle}{\emph{2nd International Conference
  on Formal Structures for Computation and Deduction, {FSCD} 2017, September
  3-9, 2017, Oxford, {UK}}} \emph{(\bibinfo{series}{LIPIcs},
  Vol.~\bibinfo{volume}{84})}, \bibfield{editor}{\bibinfo{person}{Dale Miller}}
  (Ed.). \bibinfo{publisher}{Schloss Dagstuhl - Leibniz-Zentrum f{\"{u}}r
  Informatik}, \bibinfo{pages}{10:1--10:17}.
\newblock
\showISBNx{978-3-95977-047-7}
\href{https://doi.org/10.4230/LIPICS.FSCD.2017.10}{doi:\nolinkurl{10.4230/LIPICS.FSCD.2017.10}}


\bibitem[Biernacka and Danvy(2007)]%
        {BiernackaD07}
\bibfield{author}{\bibinfo{person}{Malgorzata Biernacka} {and}
  \bibinfo{person}{Olivier Danvy}.} \bibinfo{year}{2007}\natexlab{}.
\newblock \showarticletitle{A syntactic correspondence between
  context-sensitive calculi and abstract machines}.
\newblock \bibinfo{journal}{\emph{Theor. Comput. Sci.}} \bibinfo{volume}{375},
  \bibinfo{number}{1-3} (\bibinfo{year}{2007}), \bibinfo{pages}{76--108}.
\newblock
\href{https://doi.org/10.1016/J.TCS.2006.12.028}{doi:\nolinkurl{10.1016/J.TCS.2006.12.028}}


\bibitem[Capretta(2005)]%
        {Capretta05}
\bibfield{author}{\bibinfo{person}{Venanzio Capretta}.}
  \bibinfo{year}{2005}\natexlab{}.
\newblock \showarticletitle{General recursion via coinductive types}.
\newblock \bibinfo{journal}{\emph{Log. Methods Comput. Sci.}}
  \bibinfo{volume}{1}, \bibinfo{number}{2} (\bibinfo{year}{2005}).
\newblock
\href{https://doi.org/10.2168/LMCS-1(2:1)2005}{doi:\nolinkurl{10.2168/LMCS-1(2:1)2005}}


\bibitem[Chapman et~al\mbox{.}(2010)]%
        {chapman2010thegentle}
\bibfield{author}{\bibinfo{person}{James Chapman},
  \bibinfo{person}{Pierre{-}{\'{E}}variste Dagand}, \bibinfo{person}{Conor
  McBride}, {and} \bibinfo{person}{Peter Morris}.}
  \bibinfo{year}{2010}\natexlab{}.
\newblock \showarticletitle{The gentle art of levitation}. In
  \bibinfo{booktitle}{\emph{Proceeding of the 15th {ACM} {SIGPLAN}
  international conference on Functional programming, {ICFP} 2010, Baltimore,
  Maryland, USA, September 27-29, 2010}},
  \bibfield{editor}{\bibinfo{person}{Paul Hudak} {and}
  \bibinfo{person}{Stephanie Weirich}} (Eds.). \bibinfo{publisher}{{ACM}},
  \bibinfo{pages}{3--14}.
\newblock
\showISBNx{978-1-60558-794-3}
\href{https://doi.org/10.1145/1863543.1863547}{doi:\nolinkurl{10.1145/1863543.1863547}}


\bibitem[Churchill et~al\mbox{.}(2015)]%
        {Churchill2014reusable}
\bibfield{author}{\bibinfo{person}{Martin Churchill}, \bibinfo{person}{Peter~D.
  Mosses}, \bibinfo{person}{Neil Sculthorpe}, {and} \bibinfo{person}{Paolo
  Torrini}.} \bibinfo{year}{2015}\natexlab{}.
\newblock \showarticletitle{Reusable Components of Semantic Specifications}.
\newblock \bibinfo{journal}{\emph{{LNCS} Trans. Aspect Oriented Softw. Dev.}}
  \bibinfo{volume}{12} (\bibinfo{year}{2015}), \bibinfo{pages}{132--179}.
\newblock
\showISBNx{978-3-662-46733-6}
\href{https://doi.org/10.1007/978-3-662-46734-3\_4}{doi:\nolinkurl{10.1007/978-3-662-46734-3\_4}}


\bibitem[Community({[n.\,d.]})]%
        {agda-stdlib}
\bibfield{author}{\bibinfo{person}{The~Agda Community}.}
  \bibinfo{year}{[n.\,d.]}\natexlab{}.
\newblock \bibinfo{booktitle}{\emph{{Agda Standard Library}}}.
\newblock
\urldef\tempurl%
\url{https://github.com/agda/agda-stdlib}
\showURL{%
\tempurl}


\bibitem[Danielsson(2012)]%
        {Danielsson12}
\bibfield{author}{\bibinfo{person}{Nils~Anders Danielsson}.}
  \bibinfo{year}{2012}\natexlab{}.
\newblock \showarticletitle{Operational semantics using the partiality monad}.
  In \bibinfo{booktitle}{\emph{{ACM} {SIGPLAN} International Conference on
  Functional Programming, ICFP'12, Copenhagen, Denmark, September 9-15, 2012}},
  \bibfield{editor}{\bibinfo{person}{Peter Thiemann} {and}
  \bibinfo{person}{Robby~Bruce Findler}} (Eds.). \bibinfo{publisher}{{ACM}},
  \bibinfo{pages}{127--138}.
\newblock
\showISBNx{978-1-4503-1054-3}
\href{https://doi.org/10.1145/2364527.2364546}{doi:\nolinkurl{10.1145/2364527.2364546}}


\bibitem[Danvy(2008a)]%
        {Danvy08}
\bibfield{author}{\bibinfo{person}{Olivier Danvy}.}
  \bibinfo{year}{2008}\natexlab{a}.
\newblock \showarticletitle{Defunctionalized interpreters for programming
  languages}. In \bibinfo{booktitle}{\emph{Proceeding of the 13th {ACM}
  {SIGPLAN} international conference on Functional programming, {ICFP} 2008,
  Victoria, BC, Canada, September 20-28, 2008}},
  \bibfield{editor}{\bibinfo{person}{James Hook} {and} \bibinfo{person}{Peter
  Thiemann}} (Eds.). \bibinfo{publisher}{{ACM}}, \bibinfo{pages}{131--142}.
\newblock
\showISBNx{978-1-59593-919-7}
\href{https://doi.org/10.1145/1411204.1411206}{doi:\nolinkurl{10.1145/1411204.1411206}}


\bibitem[Danvy(2008b)]%
        {Danvy08afp}
\bibfield{author}{\bibinfo{person}{Olivier Danvy}.}
  \bibinfo{year}{2008}\natexlab{b}.
\newblock \showarticletitle{From Reduction-Based to Reduction-Free
  Normalization}. In \bibinfo{booktitle}{\emph{Advanced Functional Programming,
  6th International School, {AFP} 2008, Heijen, The Netherlands, May 2008,
  Revised Lectures}} \emph{(\bibinfo{series}{Lecture Notes in Computer
  Science}, Vol.~\bibinfo{volume}{5832})},
  \bibfield{editor}{\bibinfo{person}{Pieter W.~M. Koopman},
  \bibinfo{person}{Rinus Plasmeijer}, {and} \bibinfo{person}{S.~Doaitse
  Swierstra}} (Eds.). \bibinfo{publisher}{Springer}, \bibinfo{pages}{66--164}.
\newblock
\showISBNx{978-3-642-04651-3}
\href{https://doi.org/10.1007/978-3-642-04652-0\_3}{doi:\nolinkurl{10.1007/978-3-642-04652-0\_3}}


\bibitem[Danvy and Filinski(1990)]%
        {DanvyF90}
\bibfield{author}{\bibinfo{person}{Olivier Danvy} {and}
  \bibinfo{person}{Andrzej Filinski}.} \bibinfo{year}{1990}\natexlab{}.
\newblock \showarticletitle{Abstracting Control}. In
  \bibinfo{booktitle}{\emph{Proceedings of the 1990 {ACM} Conference on {LISP}
  and Functional Programming, {LFP} 1990, Nice, France, 27-29 June 1990}},
  \bibfield{editor}{\bibinfo{person}{Gilles Kahn}} (Ed.).
  \bibinfo{publisher}{{ACM}}, \bibinfo{pages}{151--160}.
\newblock
\showISBNx{0-89791-368-X}
\href{https://doi.org/10.1145/91556.91622}{doi:\nolinkurl{10.1145/91556.91622}}


\bibitem[Danvy and Johannsen(2013)]%
        {DanvyJ13}
\bibfield{author}{\bibinfo{person}{Olivier Danvy} {and} \bibinfo{person}{Jacob
  Johannsen}.} \bibinfo{year}{2013}\natexlab{}.
\newblock \showarticletitle{From Outermost Reduction Semantics to Abstract
  Machine}. In \bibinfo{booktitle}{\emph{Logic-Based Program Synthesis and
  Transformation, 23rd International Symposium, {LOPSTR} 2013, Madrid, Spain,
  September 18-19, 2013, Revised Selected Papers}}
  \emph{(\bibinfo{series}{Lecture Notes in Computer Science},
  Vol.~\bibinfo{volume}{8901})}, \bibfield{editor}{\bibinfo{person}{Gopal
  Gupta} {and} \bibinfo{person}{Ricardo Pe{\~{n}}a}} (Eds.).
  \bibinfo{publisher}{Springer}, \bibinfo{pages}{91--108}.
\newblock
\href{https://doi.org/10.1007/978-3-319-14125-1\_6}{doi:\nolinkurl{10.1007/978-3-319-14125-1\_6}}


\bibitem[Danvy et~al\mbox{.}(2011)]%
        {DanvyJZ11}
\bibfield{author}{\bibinfo{person}{Olivier Danvy}, \bibinfo{person}{Jacob
  Johannsen}, {and} \bibinfo{person}{Ian Zerny}.}
  \bibinfo{year}{2011}\natexlab{}.
\newblock \showarticletitle{A walk in the semantic park}. In
  \bibinfo{booktitle}{\emph{Proceedings of the 2011 {ACM} {SIGPLAN} Workshop on
  Partial Evaluation and Program Manipulation, {PEPM} 2011, Austin, TX, USA,
  January 24-25, 2011}}, \bibfield{editor}{\bibinfo{person}{Siau{-}Cheng Khoo}
  {and} \bibinfo{person}{Jeremy~G. Siek}} (Eds.). \bibinfo{publisher}{{ACM}},
  \bibinfo{pages}{1--12}.
\newblock
\showISBNx{978-1-4503-0485-6}
\href{https://doi.org/10.1145/1929501.1929503}{doi:\nolinkurl{10.1145/1929501.1929503}}


\bibitem[Danvy and Nielsen(2004)]%
        {Danvy2004refocusingin}
\bibfield{author}{\bibinfo{person}{Olivier Danvy} {and}
  \bibinfo{person}{Lasse~R. Nielsen}.} \bibinfo{year}{2004}\natexlab{}.
\newblock \bibinfo{booktitle}{\emph{Refocusing in Reduction Semantics}}.
\newblock \bibinfo{type}{BRICS Research Series} RS-04-26.
  \bibinfo{institution}{Department of Computer Science, Aarhus University}.
\newblock
\urldef\tempurl%
\url{http://www.brics.dk/RS/04/26/}
\showURL{%
\tempurl}


\bibitem[Felleisen(1988)]%
        {Felleisen88}
\bibfield{author}{\bibinfo{person}{Matthias Felleisen}.}
  \bibinfo{year}{1988}\natexlab{}.
\newblock \showarticletitle{The Theory and Practice of First-Class Prompts}. In
  \bibinfo{booktitle}{\emph{Conference Record of the Fifteenth Annual {ACM}
  Symposium on Principles of Programming Languages, San Diego, California, USA,
  January 10-13, 1988}}, \bibfield{editor}{\bibinfo{person}{Jeanne Ferrante}
  {and} \bibinfo{person}{Peter Mager}} (Eds.). \bibinfo{publisher}{{ACM}
  Press}, \bibinfo{pages}{180--190}.
\newblock
\showISBNx{0-89791-252-7}
\href{https://doi.org/10.1145/73560.73576}{doi:\nolinkurl{10.1145/73560.73576}}


\bibitem[Felleisen and Hieb(1992)]%
        {FelleisenH92}
\bibfield{author}{\bibinfo{person}{Matthias Felleisen} {and}
  \bibinfo{person}{Robert Hieb}.} \bibinfo{year}{1992}\natexlab{}.
\newblock \showarticletitle{The Revised Report on the Syntactic Theories of
  Sequential Control and State}.
\newblock \bibinfo{journal}{\emph{Theor. Comput. Sci.}} \bibinfo{volume}{103},
  \bibinfo{number}{2} (\bibinfo{year}{1992}), \bibinfo{pages}{235--271}.
\newblock
\href{https://doi.org/10.1016/0304-3975(92)90014-7}{doi:\nolinkurl{10.1016/0304-3975(92)90014-7}}


\bibitem[Forster et~al\mbox{.}(2020)]%
        {ForsterKR20}
\bibfield{author}{\bibinfo{person}{Yannick Forster}, \bibinfo{person}{Fabian
  Kunze}, {and} \bibinfo{person}{Marc Roth}.} \bibinfo{year}{2020}\natexlab{}.
\newblock \showarticletitle{The weak call-by-value {\(\lambda\)}-calculus is
  reasonable for both time and space}.
\newblock \bibinfo{journal}{\emph{Proc. {ACM} Program. Lang.}}
  \bibinfo{volume}{4}, \bibinfo{number}{{POPL}} (\bibinfo{year}{2020}),
  \bibinfo{pages}{27:1--27:23}.
\newblock
\href{https://doi.org/10.1145/3371095}{doi:\nolinkurl{10.1145/3371095}}


\bibitem[Huet(1997)]%
        {Huet97}
\bibfield{author}{\bibinfo{person}{G{\'{e}}rard~P. Huet}.}
  \bibinfo{year}{1997}\natexlab{}.
\newblock \showarticletitle{The Zipper}.
\newblock \bibinfo{journal}{\emph{J. Funct. Program.}} \bibinfo{volume}{7},
  \bibinfo{number}{5} (\bibinfo{year}{1997}), \bibinfo{pages}{549--554}.
\newblock
\href{https://doi.org/10.1017/S0956796897002864}{doi:\nolinkurl{10.1017/S0956796897002864}}


\bibitem[Kahn(1987)]%
        {Kahn87}
\bibfield{author}{\bibinfo{person}{Gilles Kahn}.}
  \bibinfo{year}{1987}\natexlab{}.
\newblock \showarticletitle{Natural Semantics}. In
  \bibinfo{booktitle}{\emph{{STACS} 87, 4th Annual Symposium on Theoretical
  Aspects of Computer Science, Passau, Germany, February 19-21, 1987,
  Proceedings}} \emph{(\bibinfo{series}{Lecture Notes in Computer Science},
  Vol.~\bibinfo{volume}{247})},
  \bibfield{editor}{\bibinfo{person}{Franz{-}Josef Brandenburg},
  \bibinfo{person}{Guy Vidal{-}Naquet}, {and} \bibinfo{person}{Martin Wirsing}}
  (Eds.). \bibinfo{publisher}{Springer}, \bibinfo{pages}{22--39}.
\newblock
\showISBNx{3-540-17219-X}
\href{https://doi.org/10.1007/BFB0039592}{doi:\nolinkurl{10.1007/BFB0039592}}


\bibitem[Klein et~al\mbox{.}(2012)]%
        {KleinCDEFFMRTF12}
\bibfield{author}{\bibinfo{person}{Casey Klein}, \bibinfo{person}{John
  Clements}, \bibinfo{person}{Christos Dimoulas}, \bibinfo{person}{Carl
  Eastlund}, \bibinfo{person}{Matthias Felleisen}, \bibinfo{person}{Matthew
  Flatt}, \bibinfo{person}{Jay~A. McCarthy}, \bibinfo{person}{Jon Rafkind},
  \bibinfo{person}{Sam Tobin{-}Hochstadt}, {and} \bibinfo{person}{Robert~Bruce
  Findler}.} \bibinfo{year}{2012}\natexlab{}.
\newblock \showarticletitle{Run your research: on the effectiveness of
  lightweight mechanization}. In \bibinfo{booktitle}{\emph{Proceedings of the
  39th {ACM} {SIGPLAN-SIGACT} Symposium on Principles of Programming Languages,
  {POPL} 2012, Philadelphia, Pennsylvania, USA, January 22-28, 2012}},
  \bibfield{editor}{\bibinfo{person}{John Field} {and} \bibinfo{person}{Michael
  Hicks}} (Eds.). \bibinfo{publisher}{{ACM}}, \bibinfo{pages}{285--296}.
\newblock
\showISBNx{978-1-4503-1083-3}
\href{https://doi.org/10.1145/2103656.2103691}{doi:\nolinkurl{10.1145/2103656.2103691}}


\bibitem[Lago and Martini(2008)]%
        {LagoM08}
\bibfield{author}{\bibinfo{person}{Ugo~Dal Lago} {and} \bibinfo{person}{Simone
  Martini}.} \bibinfo{year}{2008}\natexlab{}.
\newblock \showarticletitle{The weak lambda calculus as a reasonable machine}.
\newblock \bibinfo{journal}{\emph{Theor. Comput. Sci.}} \bibinfo{volume}{398},
  \bibinfo{number}{1-3} (\bibinfo{year}{2008}), \bibinfo{pages}{32--50}.
\newblock
\href{https://doi.org/10.1016/J.TCS.2008.01.044}{doi:\nolinkurl{10.1016/J.TCS.2008.01.044}}


\bibitem[Landin(1964)]%
        {Landin64}
\bibfield{author}{\bibinfo{person}{P.~J. Landin}.}
  \bibinfo{year}{1964}\natexlab{}.
\newblock \showarticletitle{The Mechanical Evaluation of Expressions}.
\newblock \bibinfo{journal}{\emph{Comput. J.}} \bibinfo{volume}{6},
  \bibinfo{number}{4} (\bibinfo{year}{1964}), \bibinfo{pages}{308--320}.
\newblock
\href{https://doi.org/10.1093/COMJNL/6.4.308}{doi:\nolinkurl{10.1093/COMJNL/6.4.308}}


\bibitem[McBride(2001)]%
        {mcbride2001derivative}
\bibfield{author}{\bibinfo{person}{Conor McBride}.}
  \bibinfo{year}{2001}\natexlab{}.
\newblock \bibinfo{title}{The Derivative of a Regular Type is its Type of
  One-Hole Contexts}.  (\bibinfo{year}{2001}).
\newblock


\bibitem[McBride(2008)]%
        {McBride08}
\bibfield{author}{\bibinfo{person}{Conor McBride}.}
  \bibinfo{year}{2008}\natexlab{}.
\newblock \showarticletitle{Clowns to the left of me, jokers to the right
  (pearl): dissecting data structures}. In
  \bibinfo{booktitle}{\emph{Proceedings of the 35th {ACM} {SIGPLAN-SIGACT}
  Symposium on Principles of Programming Languages, {POPL} 2008, San Francisco,
  California, USA, January 7-12, 2008}},
  \bibfield{editor}{\bibinfo{person}{George~C. Necula} {and}
  \bibinfo{person}{Philip Wadler}} (Eds.). \bibinfo{publisher}{{ACM}},
  \bibinfo{pages}{287--295}.
\newblock
\showISBNx{978-1-59593-689-9}
\href{https://doi.org/10.1145/1328438.1328474}{doi:\nolinkurl{10.1145/1328438.1328474}}


\bibitem[McBride(2011)]%
        {McBride2011ornamental}
\bibfield{author}{\bibinfo{person}{Conor McBride}.}
  \bibinfo{year}{2011}\natexlab{}.
\newblock \bibinfo{title}{Ornamental Algebras, Algebraic Ornaments}.
  (\bibinfo{year}{2011}).
\newblock
\newblock
\shownote{Unpublished manuscript}.


\bibitem[Meertens(1992)]%
        {Meertens92}
\bibfield{author}{\bibinfo{person}{Lambert G. L.~T. Meertens}.}
  \bibinfo{year}{1992}\natexlab{}.
\newblock \showarticletitle{Paramorphisms}.
\newblock \bibinfo{journal}{\emph{Formal Aspects Comput.}} \bibinfo{volume}{4},
  \bibinfo{number}{5} (\bibinfo{year}{1992}), \bibinfo{pages}{413--424}.
\newblock
\href{https://doi.org/10.1007/BF01211391}{doi:\nolinkurl{10.1007/BF01211391}}


\bibitem[Meijer et~al\mbox{.}(1991)]%
        {MeijerFP91}
\bibfield{author}{\bibinfo{person}{Erik Meijer}, \bibinfo{person}{Maarten~M.
  Fokkinga}, {and} \bibinfo{person}{Ross Paterson}.}
  \bibinfo{year}{1991}\natexlab{}.
\newblock \showarticletitle{Functional Programming with Bananas, Lenses,
  Envelopes and Barbed Wire}. In \bibinfo{booktitle}{\emph{Functional
  Programming Languages and Computer Architecture, 5th {ACM} Conference,
  Cambridge, MA, USA, August 26-30, 1991, Proceedings}}
  \emph{(\bibinfo{series}{Lecture Notes in Computer Science},
  Vol.~\bibinfo{volume}{523})}, \bibfield{editor}{\bibinfo{person}{John
  Hughes}} (Ed.). \bibinfo{publisher}{Springer}, \bibinfo{pages}{124--144}.
\newblock
\showISBNx{3-540-54396-1}
\href{https://doi.org/10.1007/3540543961\_7}{doi:\nolinkurl{10.1007/3540543961\_7}}


\bibitem[Mosses(2004)]%
        {mosses2004modular}
\bibfield{author}{\bibinfo{person}{Peter~D. Mosses}.}
  \bibinfo{year}{2004}\natexlab{}.
\newblock \showarticletitle{Modular structural operational semantics}.
\newblock \bibinfo{journal}{\emph{J. Log. Algebraic Methods Program.}}
  \bibinfo{volume}{60-61} (\bibinfo{year}{2004}), \bibinfo{pages}{195--228}.
\newblock
\href{https://doi.org/10.1016/j.jlap.2004.03.008}{doi:\nolinkurl{10.1016/j.jlap.2004.03.008}}


\bibitem[Norell(2009)]%
        {norell2008dependently}
\bibfield{author}{\bibinfo{person}{Ulf Norell}.}
  \bibinfo{year}{2009}\natexlab{}.
\newblock \showarticletitle{Dependently typed programming in Agda}. In
  \bibinfo{booktitle}{\emph{Proceedings of TLDI'09: 2009 {ACM} {SIGPLAN}
  International Workshop on Types in Languages Design and Implementation,
  Savannah, GA, USA, January 24, 2009}},
  \bibfield{editor}{\bibinfo{person}{Andrew Kennedy} {and}
  \bibinfo{person}{Amal Ahmed}} (Eds.). \bibinfo{publisher}{{ACM}},
  \bibinfo{pages}{1--2}.
\newblock
\href{https://doi.org/10.1145/1481861.1481862}{doi:\nolinkurl{10.1145/1481861.1481862}}


\bibitem[Plotkin(2004)]%
        {Plotkin04a}
\bibfield{author}{\bibinfo{person}{Gordon~D. Plotkin}.}
  \bibinfo{year}{2004}\natexlab{}.
\newblock \showarticletitle{A structural approach to operational semantics}.
\newblock \bibinfo{journal}{\emph{J. Log. Algebraic Methods Program.}}
  \bibinfo{volume}{60-61} (\bibinfo{year}{2004}), \bibinfo{pages}{17--139}.
\newblock


\bibitem[Pretschner et~al\mbox{.}(2017)]%
        {DBLP:series/natosec/50}
\bibfield{editor}{\bibinfo{person}{Alexander Pretschner},
  \bibinfo{person}{Doron Peled}, {and} \bibinfo{person}{Thomas Hutzelmann}}
  (Eds.). \bibinfo{year}{2017}\natexlab{}.
\newblock \bibinfo{booktitle}{\emph{Dependable Software Systems Engineering}}.
\newblock Volume~50 of Pretschner et~al\mbox{.} \citeN{DBLP:series/natosec/50}.
\newblock
\showISBNx{978-1-61499-809-9}
\urldef\tempurl%
\url{http://ebooks.iospress.nl/volume/dependable-software-systems-engineering-3}
\showURL{%
\tempurl}


\bibitem[Reynolds(1998)]%
        {reynolds98definitional}
\bibfield{author}{\bibinfo{person}{John~C. Reynolds}.}
  \bibinfo{year}{1998}\natexlab{}.
\newblock \showarticletitle{Definitional Interpreters for Higher-Order
  Programming Languages}.
\newblock \bibinfo{journal}{\emph{High. Order Symb. Comput.}}
  \bibinfo{volume}{11}, \bibinfo{number}{4} (\bibinfo{year}{1998}),
  \bibinfo{pages}{363--397}.
\newblock
\href{https://doi.org/10.1023/A:1010027404223}{doi:\nolinkurl{10.1023/A:1010027404223}}


\bibitem[Rosu(2017)]%
        {Rosu17}
\bibfield{author}{\bibinfo{person}{Grigore Rosu}.}
  \bibinfo{year}{2017}\natexlab{}.
\newblock \showarticletitle{{\(\mathbb{K}\)}: {A} Semantic Framework for
  Programming Languages and Formal Analysis Tools}.
\newblock See \citeN{DBLP:series/natosec/50}, \bibinfo{pages}{186--206}.
\newblock
\showISBNx{978-1-61499-809-9}
\href{https://doi.org/10.3233/978-1-61499-810-5-186}{doi:\nolinkurl{10.3233/978-1-61499-810-5-186}}


\bibitem[Rosu and Serbanuta(2010)]%
        {Rosu2010anoverview}
\bibfield{author}{\bibinfo{person}{Grigore Rosu} {and}
  \bibinfo{person}{Traian{-}Florin Serbanuta}.}
  \bibinfo{year}{2010}\natexlab{}.
\newblock \showarticletitle{An overview of the {K} semantic framework}.
\newblock \bibinfo{journal}{\emph{J. Log. Algebraic Methods Program.}}
  \bibinfo{volume}{79}, \bibinfo{number}{6} (\bibinfo{year}{2010}),
  \bibinfo{pages}{397--434}.
\newblock
\href{https://doi.org/10.1016/j.jlap.2010.03.012}{doi:\nolinkurl{10.1016/j.jlap.2010.03.012}}


\bibitem[Scott and Strachey(1971)]%
        {scott-strachey-toward}
\bibfield{author}{\bibinfo{person}{Dana Scott} {and}
  \bibinfo{person}{Christopher Strachey}.} \bibinfo{year}{1971}\natexlab{}.
\newblock \showarticletitle{Toward a mathematical semantics for computer
  languages}. In \bibinfo{booktitle}{\emph{Proc. Symp. on Computers and
  Automata}} \emph{(\bibinfo{series}{Microwave Research Inst. Symposia Series},
  Vol.~\bibinfo{volume}{21})}. \bibinfo{publisher}{Polytechnic Inst. of
  Brooklyn}, \bibinfo{pages}{19–46}.
\newblock
\urldef\tempurl%
\url{https://www.cs.ox.ac.uk/files/3228/PRG06.pdf}
\showURL{%
\tempurl}
\newblock
\shownote{Also: Tech. Monograph PRG-6, Oxford Univ. Computing Lab., Programming
  Research Group (1971)}.


\bibitem[Sieczkowski et~al\mbox{.}(2010)]%
        {SieczkowskiBB10}
\bibfield{author}{\bibinfo{person}{Filip Sieczkowski},
  \bibinfo{person}{Malgorzata Biernacka}, {and} \bibinfo{person}{Dariusz
  Biernacki}.} \bibinfo{year}{2010}\natexlab{}.
\newblock \showarticletitle{Automating Derivations of Abstract Machines from
  Reduction Semantics: - {A} Generic Formalization of Refocusing in Coq}. In
  \bibinfo{booktitle}{\emph{Implementation and Application of Functional
  Languages - 22nd International Symposium, {IFL} 2010, Alphen aan den Rijn,
  The Netherlands, September 1-3, 2010, Revised Selected Papers}}
  \emph{(\bibinfo{series}{Lecture Notes in Computer Science},
  Vol.~\bibinfo{volume}{6647})}, \bibfield{editor}{\bibinfo{person}{Jurriaan
  Hage} {and} \bibinfo{person}{Marco~T. Moraz{\'{a}}n}} (Eds.).
  \bibinfo{publisher}{Springer}, \bibinfo{pages}{72--88}.
\newblock
\showISBNx{978-3-642-24275-5}
\href{https://doi.org/10.1007/978-3-642-24276-2\_5}{doi:\nolinkurl{10.1007/978-3-642-24276-2\_5}}


\bibitem[{Tom{\'{e}} Corti{\~{n}}as} and Swierstra(2018)]%
        {CortinasS18}
\bibfield{author}{\bibinfo{person}{Carlos {Tom{\'{e}} Corti{\~{n}}as}} {and}
  \bibinfo{person}{Wouter Swierstra}.} \bibinfo{year}{2018}\natexlab{}.
\newblock \showarticletitle{From algebra to abstract machine: a verified
  generic construction}. In \bibinfo{booktitle}{\emph{Proceedings of the 3rd
  {ACM} {SIGPLAN} International Workshop on Type-Driven Development, TyDe@ICFP
  2018, St. Louis, MO, USA, September 27, 2018}},
  \bibfield{editor}{\bibinfo{person}{Richard~A. Eisenberg} {and}
  \bibinfo{person}{Niki Vazou}} (Eds.). \bibinfo{publisher}{{ACM}},
  \bibinfo{pages}{78--90}.
\newblock
\href{https://doi.org/10.1145/3240719.3241787}{doi:\nolinkurl{10.1145/3240719.3241787}}


\bibitem[Turi and Plotkin(1997)]%
        {TuriP97}
\bibfield{author}{\bibinfo{person}{Daniele Turi} {and}
  \bibinfo{person}{Gordon~D. Plotkin}.} \bibinfo{year}{1997}\natexlab{}.
\newblock \showarticletitle{Towards a Mathematical Operational Semantics}. In
  \bibinfo{booktitle}{\emph{Proceedings, 12th Annual {IEEE} Symposium on Logic
  in Computer Science, Warsaw, Poland, June 29 - July 2, 1997}}.
  \bibinfo{publisher}{{IEEE} Computer Society}, \bibinfo{pages}{280--291}.
\newblock
\showISBNx{0-8186-7925-5}
\href{https://doi.org/10.1109/LICS.1997.614955}{doi:\nolinkurl{10.1109/LICS.1997.614955}}


\bibitem[van Binsbergen et~al\mbox{.}(2019)]%
        {BinsbergenMS19}
\bibfield{author}{\bibinfo{person}{L.~Thomas van Binsbergen},
  \bibinfo{person}{Peter~D. Mosses}, {and} \bibinfo{person}{Neil Sculthorpe}.}
  \bibinfo{year}{2019}\natexlab{}.
\newblock \showarticletitle{Executable component-based semantics}.
\newblock \bibinfo{journal}{\emph{J. Log. Algebraic Methods Program.}}
  \bibinfo{volume}{103} (\bibinfo{year}{2019}), \bibinfo{pages}{184--212}.
\newblock
\href{https://doi.org/10.1016/J.JLAMP.2018.12.004}{doi:\nolinkurl{10.1016/J.JLAMP.2018.12.004}}


\end{thebibliography}

\appendix

\begin{code}[hide]%
\>[0]\AgdaSymbol{\{-\#}\AgdaSpace{}%
\AgdaKeyword{OPTIONS}\AgdaSpace{}%
\AgdaPragma{--guardedness}\AgdaSpace{}%
\AgdaSymbol{\#-\}}\<%
\\
\\[\AgdaEmptyExtraSkip]%
\>[0]\AgdaKeyword{open}\AgdaSpace{}%
\AgdaKeyword{import}\AgdaSpace{}%
\AgdaModule{Function}\<%
\\
\\[\AgdaEmptyExtraSkip]%
\>[0]\AgdaKeyword{open}\AgdaSpace{}%
\AgdaKeyword{import}\AgdaSpace{}%
\AgdaModule{Data.Empty}\<%
\\
\>[0]\AgdaKeyword{open}\AgdaSpace{}%
\AgdaKeyword{import}\AgdaSpace{}%
\AgdaModule{Data.Unit}\<%
\\
\>[0]\AgdaKeyword{open}\AgdaSpace{}%
\AgdaKeyword{import}\AgdaSpace{}%
\AgdaModule{Data.Bool}\<%
\\
\>[0]\AgdaKeyword{open}\AgdaSpace{}%
\AgdaKeyword{import}\AgdaSpace{}%
\AgdaModule{Data.Maybe}\<%
\\
\>[0]\AgdaKeyword{open}\AgdaSpace{}%
\AgdaKeyword{import}\AgdaSpace{}%
\AgdaModule{Data.Sum}\<%
\\
\>[0]\AgdaKeyword{open}\AgdaSpace{}%
\AgdaKeyword{import}\AgdaSpace{}%
\AgdaModule{Data.Product}\<%
\\
\>[0]\AgdaKeyword{open}\AgdaSpace{}%
\AgdaKeyword{import}\AgdaSpace{}%
\AgdaModule{Data.Nat}\<%
\\
\>[0]\AgdaKeyword{open}\AgdaSpace{}%
\AgdaKeyword{import}\AgdaSpace{}%
\AgdaModule{Data.Fin}\AgdaSpace{}%
\AgdaKeyword{hiding}\AgdaSpace{}%
\AgdaSymbol{(}\AgdaOperator{\AgdaFunction{\AgdaUnderscore{}+\AgdaUnderscore{}}}\AgdaSymbol{)}\<%
\\
\>[0]\AgdaKeyword{open}\AgdaSpace{}%
\AgdaKeyword{import}\AgdaSpace{}%
\AgdaModule{Data.List}\AgdaSpace{}%
\AgdaSymbol{as}\AgdaSpace{}%
\AgdaModule{List}\AgdaSpace{}%
\AgdaKeyword{renaming}\AgdaSpace{}%
\AgdaSymbol{(}\AgdaFunction{map}\AgdaSpace{}%
\AgdaSymbol{to}\AgdaSpace{}%
\AgdaFunction{⅋map}\AgdaSymbol{)}\<%
\\
\>[0]\AgdaKeyword{open}\AgdaSpace{}%
\AgdaKeyword{import}\AgdaSpace{}%
\AgdaModule{Data.List.Properties}\<%
\\
\>[0]\AgdaKeyword{open}\AgdaSpace{}%
\AgdaKeyword{import}\AgdaSpace{}%
\AgdaModule{Data.List.Membership.Propositional}\<%
\\
\>[0]\AgdaKeyword{open}\AgdaSpace{}%
\AgdaKeyword{import}\AgdaSpace{}%
\AgdaModule{Data.List.Membership.Propositional.Properties}\<%
\\
\>[0]\AgdaKeyword{open}\AgdaSpace{}%
\AgdaKeyword{import}\AgdaSpace{}%
\AgdaModule{Data.List.Relation.Unary.Any}\<%
\\
\>[0]\AgdaKeyword{open}\AgdaSpace{}%
\AgdaKeyword{import}\AgdaSpace{}%
\AgdaModule{Data.List.Relation.Unary.All}\AgdaSpace{}%
\AgdaSymbol{as}\AgdaSpace{}%
\AgdaModule{All}\<%
\\
\>[0]\AgdaKeyword{open}\AgdaSpace{}%
\AgdaKeyword{import}\AgdaSpace{}%
\AgdaModule{Data.List.Relation.Unary.All.Properties}\<%
\\
\\[\AgdaEmptyExtraSkip]%
\>[0]\AgdaKeyword{open}\AgdaSpace{}%
\AgdaKeyword{import}\AgdaSpace{}%
\AgdaModule{Relation.Nullary}\<%
\\
\>[0]\AgdaKeyword{open}\AgdaSpace{}%
\AgdaKeyword{import}\AgdaSpace{}%
\AgdaModule{Relation.Unary}\AgdaSpace{}%
\AgdaKeyword{renaming}\AgdaSpace{}%
\AgdaSymbol{(}\AgdaOperator{\AgdaFunction{\AgdaUnderscore{}∈\AgdaUnderscore{}}}\AgdaSpace{}%
\AgdaSymbol{to}\AgdaSpace{}%
\AgdaOperator{\AgdaFunction{\AgdaUnderscore{}∈′\AgdaUnderscore{}}}\AgdaSymbol{)}\<%
\\
\>[0]\AgdaKeyword{open}\AgdaSpace{}%
\AgdaKeyword{import}\AgdaSpace{}%
\AgdaModule{Relation.Binary.PropositionalEquality}\<%
\\
\\[\AgdaEmptyExtraSkip]%
\>[0]\AgdaKeyword{open}\AgdaSpace{}%
\AgdaKeyword{import}\AgdaSpace{}%
\AgdaModule{sections.03-prelude}\<%
\\
\\[\AgdaEmptyExtraSkip]%
\>[0]\AgdaKeyword{open}\AgdaSpace{}%
\AgdaModule{Alg}\<%
\\
\\[\AgdaEmptyExtraSkip]%
\>[0]\AgdaKeyword{open}\AgdaSpace{}%
\AgdaKeyword{import}\AgdaSpace{}%
\AgdaModule{sections.04-redsem-gen}\<%
\\
\\[\AgdaEmptyExtraSkip]%
\>[0]\AgdaKeyword{module}\AgdaSpace{}%
\AgdaModule{sections.app-algebraic}\AgdaSpace{}%
\AgdaKeyword{where}\<%
\end{code}

\section{An Algebraic Characterization of Reduction Strategies in Agda}
\label{app:characterization}

We give an algebraic characterization of reduction strategies.
To this end, we will use an Agda module to parameterize our definitions over the algebra we characterize, as well as language-specific functions for characterizing redexes and values.
\begin{code}%
\>[0]\AgdaKeyword{module}\AgdaSpace{}%
\AgdaModule{ReductionAlgebra}\<%
\\
\>[0][@{}l@{\AgdaIndent{0}}]%
\>[4]\AgdaSymbol{\{}\AgdaBound{d}\AgdaSpace{}%
\AgdaSymbol{:}\AgdaSpace{}%
\AgdaDatatype{Desc}\AgdaSymbol{\}}\<%
\\
\>[4]\AgdaSymbol{(}\AgdaBound{a}\AgdaSpace{}%
\AgdaSymbol{:}\AgdaSpace{}%
\AgdaRecord{Alg}%
\>[14]\AgdaSymbol{(λ}\AgdaSpace{}%
\AgdaBound{X}\AgdaSpace{}%
\AgdaSymbol{→}\AgdaSpace{}%
\AgdaOperator{\AgdaFunction{⟦}}\AgdaSpace{}%
\AgdaBound{d}\AgdaSpace{}%
\AgdaOperator{\AgdaFunction{⟧}}\AgdaSpace{}%
\AgdaSymbol{(}\AgdaOperator{\AgdaDatatype{μ[}}\AgdaSpace{}%
\AgdaBound{d}\AgdaSpace{}%
\AgdaOperator{\AgdaDatatype{]}}\AgdaSpace{}%
\AgdaOperator{\AgdaFunction{×}}\AgdaSpace{}%
\AgdaBound{X}\AgdaSymbol{))}\<%
\\
\>[14]\AgdaSymbol{(}\AgdaOperator{\AgdaFunction{Ctx[}}\AgdaSpace{}%
\AgdaBound{d}\AgdaSpace{}%
\AgdaOperator{\AgdaFunction{]}}\AgdaSpace{}%
\AgdaSymbol{→}\AgdaSpace{}%
\AgdaDatatype{Maybe}\AgdaSpace{}%
\AgdaSymbol{(}\AgdaOperator{\AgdaFunction{Ctx[}}\AgdaSpace{}%
\AgdaBound{d}\AgdaSpace{}%
\AgdaOperator{\AgdaFunction{]}}\AgdaSpace{}%
\AgdaOperator{\AgdaFunction{×}}\AgdaSpace{}%
\AgdaOperator{\AgdaDatatype{μ[}}\AgdaSpace{}%
\AgdaBound{d}\AgdaSpace{}%
\AgdaOperator{\AgdaDatatype{]}}\AgdaSymbol{)))}\<%
\\
\>[4]\AgdaSymbol{(}\AgdaBound{is-redex}%
\>[15]\AgdaSymbol{:}\AgdaSpace{}%
\AgdaOperator{\AgdaDatatype{μ[}}\AgdaSpace{}%
\AgdaBound{d}\AgdaSpace{}%
\AgdaOperator{\AgdaDatatype{]}}\AgdaSpace{}%
\AgdaSymbol{→}\AgdaSpace{}%
\AgdaDatatype{Bool}\AgdaSymbol{)}\<%
\\
\>[4]\AgdaSymbol{(}\AgdaBound{is-val}%
\>[15]\AgdaSymbol{:}\AgdaSpace{}%
\AgdaOperator{\AgdaDatatype{μ[}}\AgdaSpace{}%
\AgdaBound{d}\AgdaSpace{}%
\AgdaOperator{\AgdaDatatype{]}}\AgdaSpace{}%
\AgdaSymbol{→}\AgdaSpace{}%
\AgdaDatatype{Bool}\AgdaSymbol{)}\<%
\\
\>[4]\AgdaSymbol{(}\AgdaBound{redex→¬val}%
\>[17]\AgdaSymbol{:}\AgdaSpace{}%
\AgdaSymbol{∀}\AgdaSpace{}%
\AgdaBound{x}\<%
\\
\>[17]\AgdaSymbol{→}%
\>[20]\AgdaSymbol{(}\AgdaSpace{}%
\AgdaBound{is-redex}\AgdaSpace{}%
\AgdaBound{x}\AgdaSpace{}%
\AgdaOperator{\AgdaDatatype{≡}}\AgdaSpace{}%
\AgdaInductiveConstructor{true}\<%
\\
\>[20]\AgdaSymbol{→}\AgdaSpace{}%
\AgdaBound{is-val}\AgdaSpace{}%
\AgdaBound{x}\AgdaSpace{}%
\AgdaOperator{\AgdaDatatype{≡}}\AgdaSpace{}%
\AgdaInductiveConstructor{false}\AgdaSpace{}%
\AgdaSymbol{))}\<%
\\
\>[0][@{}l@{\AgdaIndent{0}}]%
\>[2]\AgdaKeyword{where}\<%
\end{code}
The carrier (\ad{Ctx[}~\ab{d}~\ad{]}~\as{→}~\ad{Maybe}~\as{(}\af{Ctx[}~\ab{d}~\af{]}~\ad{×}~\ad{μ[}~\ab{d}~\ad{]}))) of the algebra \ab{a} represents a function which, when applied to the ``current'' context, returns a possible decomposition, given by a context and possible redex.
In other words, a decomposition strategy.
We introduce the following abbreviated type for such strategies:
\begin{code}%
\>[2]\AgdaFunction{⅋DecStrat}\AgdaSpace{}%
\AgdaSymbol{:}\AgdaSpace{}%
\AgdaDatatype{Desc}\AgdaSpace{}%
\AgdaSymbol{→}\AgdaSpace{}%
\AgdaPrimitive{Set}\<%
\\
\>[2]\AgdaFunction{⅋DecStrat}\AgdaSpace{}%
\AgdaBound{d}\AgdaSpace{}%
\AgdaSymbol{=}\AgdaSpace{}%
\AgdaOperator{\AgdaFunction{Ctx[}}\AgdaSpace{}%
\AgdaBound{d}\AgdaSpace{}%
\AgdaOperator{\AgdaFunction{]}}\AgdaSpace{}%
\AgdaSymbol{→}\AgdaSpace{}%
\AgdaDatatype{Maybe}\AgdaSpace{}%
\AgdaSymbol{(}\AgdaOperator{\AgdaFunction{Ctx[}}\AgdaSpace{}%
\AgdaBound{d}\AgdaSpace{}%
\AgdaOperator{\AgdaFunction{]}}\AgdaSpace{}%
\AgdaOperator{\AgdaFunction{×}}\AgdaSpace{}%
\AgdaOperator{\AgdaDatatype{μ[}}\AgdaSpace{}%
\AgdaBound{d}\AgdaSpace{}%
\AgdaOperator{\AgdaDatatype{]}}\AgdaSymbol{)}\<%
\end{code}
This algebra and carrier type corresponds to the recursive structure of the decomposition functions used in \cref{sec:02-redsem}.

\subsection{Stuckness}
A basic property that all reduction strategies satisfy is that reduction is stuck when no decomposition exists, and when the current term is not a redex either.
\begin{code}%
\>[2]\AgdaFunction{no-redex}%
\>[12]\AgdaSymbol{=}%
\>[15]\AgdaSymbol{∀}\AgdaSpace{}%
\AgdaSymbol{\{}\AgdaBound{x}\AgdaSpace{}%
\AgdaSymbol{:}\AgdaSpace{}%
\AgdaOperator{\AgdaFunction{⟦}}\AgdaSpace{}%
\AgdaBound{d}\AgdaSpace{}%
\AgdaOperator{\AgdaFunction{⟧}}\AgdaSpace{}%
\AgdaSymbol{(}\AgdaOperator{\AgdaDatatype{μ[}}\AgdaSpace{}%
\AgdaBound{d}\AgdaSpace{}%
\AgdaOperator{\AgdaDatatype{]}}\AgdaSpace{}%
\AgdaOperator{\AgdaFunction{×}}%
\>[39]\AgdaFunction{DecStrat}\AgdaSpace{}%
\AgdaBound{d}\AgdaSymbol{)\}}\AgdaSpace{}%
\AgdaSymbol{\{}\AgdaBound{c}\AgdaSymbol{\}}\<%
\\
\>[15]\AgdaSymbol{→}\AgdaSpace{}%
\AgdaBound{is-redex}\AgdaSpace{}%
\AgdaOperator{\AgdaInductiveConstructor{⟨}}\AgdaSpace{}%
\AgdaFunction{fmap}\AgdaSpace{}%
\AgdaSymbol{\{}\AgdaBound{d}\AgdaSymbol{\}}\AgdaSpace{}%
\AgdaField{proj₁}\AgdaSpace{}%
\AgdaBound{x}\AgdaSpace{}%
\AgdaOperator{\AgdaInductiveConstructor{⟩}}\AgdaSpace{}%
\AgdaOperator{\AgdaDatatype{≡}}\AgdaSpace{}%
\AgdaInductiveConstructor{false}\<%
\\
\>[15]\AgdaSymbol{→}\AgdaSpace{}%
\AgdaDatatype{All}\AgdaSpace{}%
\AgdaSymbol{(λ}\AgdaSpace{}%
\AgdaSymbol{(\AgdaUnderscore{}}\AgdaSpace{}%
\AgdaOperator{\AgdaInductiveConstructor{,}}\AgdaSpace{}%
\AgdaSymbol{\AgdaUnderscore{}}\AgdaSpace{}%
\AgdaOperator{\AgdaInductiveConstructor{,}}\AgdaSpace{}%
\AgdaBound{m}\AgdaSymbol{)}\AgdaSpace{}%
\AgdaSymbol{→}\AgdaSpace{}%
\AgdaBound{m}\AgdaSpace{}%
\AgdaBound{c}\AgdaSpace{}%
\AgdaOperator{\AgdaDatatype{≡}}\AgdaSpace{}%
\AgdaInductiveConstructor{nothing}\AgdaSymbol{)}\AgdaSpace{}%
\AgdaOperator{\AgdaFunction{S[}}\AgdaSpace{}%
\AgdaBound{d}\AgdaSpace{}%
\AgdaOperator{\AgdaFunction{∼}}\AgdaSpace{}%
\AgdaBound{x}\AgdaSpace{}%
\AgdaOperator{\AgdaFunction{]}}\<%
\\
\>[15]\AgdaSymbol{→}\AgdaSpace{}%
\AgdaField{alg}\AgdaSpace{}%
\AgdaBound{a}\AgdaSpace{}%
\AgdaBound{x}\AgdaSpace{}%
\AgdaBound{c}\AgdaSpace{}%
\AgdaOperator{\AgdaDatatype{≡}}\AgdaSpace{}%
\AgdaInductiveConstructor{nothing}\<%
\end{code}
Recall that a paramorphic algebra accepts as input a syntax node where each recursive sub-tree has been replaced by a pair comprising (1) the original recursive sub-tree, and (2) the recursively computed value (e.g., a possible decomposition) for that sub-tree.
Thus \af{fmap}~\as{\{}\ab{d}\as{\}}~\aF{proj₁}~\ab{x} discards the recursively computed result for the current term, yielding a plain AST node.
The type \ad{All}~\ab{P}~\ab{xs} ensures that the predicate \ab{P}~\ab{x} is true for all \ab{x}~\as{∈}~\ab{xs}.
The property above uses \ad{All} to ensure that each recursively computed decomposition for the current AST node is empty.

\subsection{Horizontal Reduction Strategy}

The following property says that an algebra \ab{a} implements a left-most reduction strategy.
\begin{code}%
\>[2]\AgdaFunction{left-most}%
\>[13]\AgdaSymbol{=}%
\>[16]\AgdaSymbol{∀}\AgdaSpace{}%
\AgdaSymbol{\{}\AgdaBound{x}\AgdaSpace{}%
\AgdaBound{xs}\AgdaSpace{}%
\AgdaBound{ys}\AgdaSpace{}%
\AgdaBound{frm}\AgdaSpace{}%
\AgdaBound{c}\AgdaSpace{}%
\AgdaBound{t₀}\AgdaSpace{}%
\AgdaBound{f}\AgdaSpace{}%
\AgdaBound{c₁}\AgdaSpace{}%
\AgdaBound{t₁}\AgdaSymbol{\}}\<%
\\
\>[16]\AgdaSymbol{→}\AgdaSpace{}%
\AgdaOperator{\AgdaFunction{S[}}\AgdaSpace{}%
\AgdaBound{d}\AgdaSpace{}%
\AgdaOperator{\AgdaFunction{∼}}\AgdaSpace{}%
\AgdaBound{x}\AgdaSpace{}%
\AgdaOperator{\AgdaFunction{]}}\AgdaSpace{}%
\AgdaOperator{\AgdaDatatype{≡}}\AgdaSpace{}%
\AgdaBound{xs}\AgdaSpace{}%
\AgdaOperator{\AgdaFunction{++}}\AgdaSpace{}%
\AgdaSymbol{(}\AgdaBound{frm}\AgdaSpace{}%
\AgdaOperator{\AgdaInductiveConstructor{,}}\AgdaSpace{}%
\AgdaBound{t₀}\AgdaSpace{}%
\AgdaOperator{\AgdaInductiveConstructor{,}}\AgdaSpace{}%
\AgdaBound{f}\AgdaSymbol{)}\AgdaSpace{}%
\AgdaOperator{\AgdaInductiveConstructor{∷}}\AgdaSpace{}%
\AgdaBound{ys}\<%
\\
\>[16]\AgdaSymbol{→}\AgdaSpace{}%
\AgdaBound{f}\AgdaSpace{}%
\AgdaBound{c}\AgdaSpace{}%
\AgdaOperator{\AgdaDatatype{≡}}\AgdaSpace{}%
\AgdaInductiveConstructor{just}\AgdaSpace{}%
\AgdaSymbol{(}\AgdaBound{c₁}\AgdaSpace{}%
\AgdaOperator{\AgdaInductiveConstructor{,}}\AgdaSpace{}%
\AgdaBound{t₁}\AgdaSymbol{)}\<%
\\
\>[16]\AgdaSymbol{→}\AgdaSpace{}%
\AgdaDatatype{All}\AgdaSpace{}%
\AgdaSymbol{(λ}\AgdaSpace{}%
\AgdaSymbol{(\AgdaUnderscore{}}\AgdaSpace{}%
\AgdaOperator{\AgdaInductiveConstructor{,}}\AgdaSpace{}%
\AgdaBound{y}\AgdaSpace{}%
\AgdaOperator{\AgdaInductiveConstructor{,}}\AgdaSpace{}%
\AgdaSymbol{\AgdaUnderscore{})}\AgdaSpace{}%
\AgdaSymbol{→}\AgdaSpace{}%
\AgdaBound{is-val}\AgdaSpace{}%
\AgdaBound{y}\AgdaSpace{}%
\AgdaOperator{\AgdaDatatype{≡}}\AgdaSpace{}%
\AgdaInductiveConstructor{true}\AgdaSymbol{)}\AgdaSpace{}%
\AgdaBound{xs}\<%
\\
\>[16]\AgdaSymbol{→}\AgdaSpace{}%
\AgdaField{alg}\AgdaSpace{}%
\AgdaBound{a}\AgdaSpace{}%
\AgdaBound{x}\AgdaSpace{}%
\AgdaBound{c}\AgdaSpace{}%
\AgdaOperator{\AgdaDatatype{≡}}\AgdaSpace{}%
\AgdaInductiveConstructor{just}\AgdaSpace{}%
\AgdaSymbol{(}\AgdaFunction{∂map}\AgdaSpace{}%
\AgdaSymbol{\{}\AgdaBound{d}\AgdaSymbol{\}}\AgdaSpace{}%
\AgdaField{proj₁}\AgdaSpace{}%
\AgdaBound{frm}\AgdaSpace{}%
\AgdaOperator{\AgdaInductiveConstructor{∷}}\AgdaSpace{}%
\AgdaBound{c₁}\AgdaSpace{}%
\AgdaOperator{\AgdaInductiveConstructor{,}}\AgdaSpace{}%
\AgdaBound{t₁}\AgdaSymbol{)}\<%
\end{code}
Specifically, when the algebra is applied to a term whose left-most sub-term which has a valid decomposition, such that all of the terms to the left of the term are values, the algebra yields the decomposition in that sub-term as result.
A property characterizing right-most reduction could be defined in a similar way.

\subsection{Vertical Reduction Strategy}

An algebra implements an inner-most reduction strategy when it prefers inner redexes over outer ones.\footnote{The type \ad{Any}~\ab{P}~\ab{xs} ensures that the predicate \ab{P}~\ab{x} is true for at least one \ab{x}~\as{∈}~\ab{xs}.
}
\begin{code}%
\>[2]\AgdaFunction{inner-most}\AgdaSpace{}%
\AgdaSymbol{=}\<%
\\
\>[2][@{}l@{\AgdaIndent{0}}]%
\>[4]\AgdaSymbol{∀}\AgdaSpace{}%
\AgdaSymbol{\{}\AgdaBound{x}\AgdaSpace{}%
\AgdaBound{c}\AgdaSpace{}%
\AgdaBound{xs}\AgdaSymbol{\}}\<%
\\
\>[4]\AgdaSymbol{→}\AgdaSpace{}%
\AgdaOperator{\AgdaFunction{S[}}\AgdaSpace{}%
\AgdaBound{d}\AgdaSpace{}%
\AgdaOperator{\AgdaFunction{∼}}\AgdaSpace{}%
\AgdaBound{x}\AgdaSpace{}%
\AgdaOperator{\AgdaFunction{]}}\AgdaSpace{}%
\AgdaOperator{\AgdaDatatype{≡}}\AgdaSpace{}%
\AgdaBound{xs}\<%
\\
\>[4]\AgdaSymbol{→}\AgdaSpace{}%
\AgdaBound{is-redex}\AgdaSpace{}%
\AgdaOperator{\AgdaInductiveConstructor{⟨}}\AgdaSpace{}%
\AgdaFunction{fmap}\AgdaSpace{}%
\AgdaSymbol{\{}\AgdaBound{d}\AgdaSymbol{\}}\AgdaSpace{}%
\AgdaField{proj₁}\AgdaSpace{}%
\AgdaBound{x}\AgdaSpace{}%
\AgdaOperator{\AgdaInductiveConstructor{⟩}}\AgdaSpace{}%
\AgdaOperator{\AgdaDatatype{≡}}\AgdaSpace{}%
\AgdaInductiveConstructor{true}\<%
\\
\>[4]\AgdaSymbol{→}\AgdaSpace{}%
\AgdaDatatype{Any}%
\>[249I]\AgdaSymbol{(λ}\AgdaSpace{}%
\AgdaSymbol{(}\AgdaBound{frm}\AgdaSpace{}%
\AgdaOperator{\AgdaInductiveConstructor{,}}\AgdaSpace{}%
\AgdaSymbol{\AgdaUnderscore{}}\AgdaSpace{}%
\AgdaOperator{\AgdaInductiveConstructor{,}}\AgdaSpace{}%
\AgdaBound{m}\AgdaSymbol{)}\AgdaSpace{}%
\AgdaSymbol{→}\<%
\\
\>[249I][@{}l@{\AgdaIndent{0}}]%
\>[12]\AgdaFunction{Is-just}\AgdaSpace{}%
\AgdaSymbol{(}\AgdaBound{m}\AgdaSpace{}%
\AgdaSymbol{(}\AgdaFunction{∂map}\AgdaSpace{}%
\AgdaSymbol{\{}\AgdaBound{d}\AgdaSymbol{\}}\AgdaSpace{}%
\AgdaField{proj₁}\AgdaSpace{}%
\AgdaBound{frm}\AgdaSpace{}%
\AgdaOperator{\AgdaInductiveConstructor{∷}}\AgdaSpace{}%
\AgdaBound{c}\AgdaSymbol{)))}\AgdaSpace{}%
\AgdaBound{xs}\<%
\\
\>[4]\AgdaSymbol{→}\AgdaSpace{}%
\AgdaDatatype{Any}%
\>[265I]\AgdaSymbol{(λ}\AgdaSpace{}%
\AgdaSymbol{(}\AgdaBound{frm}%
\>[19]\AgdaOperator{\AgdaInductiveConstructor{,}}\AgdaSpace{}%
\AgdaSymbol{\AgdaUnderscore{}}\AgdaSpace{}%
\AgdaOperator{\AgdaInductiveConstructor{,}}\AgdaSpace{}%
\AgdaBound{f}\AgdaSymbol{)}\AgdaSpace{}%
\AgdaSymbol{→}\<%
\\
\>[265I][@{}l@{\AgdaIndent{0}}]%
\>[12]\AgdaSymbol{∀}\AgdaSpace{}%
\AgdaSymbol{\{}\AgdaBound{c₀}\AgdaSpace{}%
\AgdaBound{t₀}\AgdaSymbol{\}}\<%
\\
\>[12]\AgdaSymbol{→}%
\>[15]\AgdaBound{f}\AgdaSpace{}%
\AgdaSymbol{(}\AgdaFunction{∂map}\AgdaSpace{}%
\AgdaSymbol{\{}\AgdaBound{d}\AgdaSymbol{\}}\AgdaSpace{}%
\AgdaField{proj₁}\AgdaSpace{}%
\AgdaBound{frm}\AgdaSpace{}%
\AgdaOperator{\AgdaInductiveConstructor{∷}}\AgdaSpace{}%
\AgdaBound{c}\AgdaSymbol{)}\AgdaSpace{}%
\AgdaOperator{\AgdaDatatype{≡}}\AgdaSpace{}%
\AgdaInductiveConstructor{just}\AgdaSpace{}%
\AgdaSymbol{(}\AgdaBound{c₀}\AgdaSpace{}%
\AgdaOperator{\AgdaInductiveConstructor{,}}\AgdaSpace{}%
\AgdaBound{t₀}\AgdaSymbol{)}\<%
\\
\>[12]\AgdaSymbol{→}%
\>[15]\AgdaField{alg}\AgdaSpace{}%
\AgdaBound{a}\AgdaSpace{}%
\AgdaBound{x}\AgdaSpace{}%
\AgdaBound{c}\AgdaSpace{}%
\AgdaOperator{\AgdaDatatype{≡}}\AgdaSpace{}%
\AgdaInductiveConstructor{just}\AgdaSpace{}%
\AgdaSymbol{(}\AgdaBound{c₀}\AgdaSpace{}%
\AgdaOperator{\AgdaInductiveConstructor{,}}\AgdaSpace{}%
\AgdaBound{t₀}\AgdaSymbol{))}\AgdaSpace{}%
\AgdaBound{xs}\<%
\end{code}
Outer-most reduction strategies prefer outer redexes unconditionally:
\begin{code}%
\>[2]\AgdaFunction{outer-most}\AgdaSpace{}%
\AgdaSymbol{=}%
\>[16]\AgdaSymbol{∀}\AgdaSpace{}%
\AgdaSymbol{\{}\AgdaBound{x}\AgdaSpace{}%
\AgdaBound{c}\AgdaSpace{}%
\AgdaBound{t}\AgdaSymbol{\}}\<%
\\
\>[16]\AgdaSymbol{→}\AgdaSpace{}%
\AgdaBound{t}\AgdaSpace{}%
\AgdaOperator{\AgdaDatatype{≡}}\AgdaSpace{}%
\AgdaOperator{\AgdaInductiveConstructor{⟨}}\AgdaSpace{}%
\AgdaFunction{fmap}\AgdaSpace{}%
\AgdaSymbol{\{}\AgdaBound{d}\AgdaSymbol{\}}\AgdaSpace{}%
\AgdaField{proj₁}\AgdaSpace{}%
\AgdaBound{x}\AgdaSpace{}%
\AgdaOperator{\AgdaInductiveConstructor{⟩}}\<%
\\
\>[16]\AgdaSymbol{→}\AgdaSpace{}%
\AgdaBound{is-redex}\AgdaSpace{}%
\AgdaBound{t}\AgdaSpace{}%
\AgdaOperator{\AgdaDatatype{≡}}\AgdaSpace{}%
\AgdaInductiveConstructor{true}\<%
\\
\>[16]\AgdaSymbol{→}\AgdaSpace{}%
\AgdaField{alg}\AgdaSpace{}%
\AgdaBound{a}\AgdaSpace{}%
\AgdaBound{x}\AgdaSpace{}%
\AgdaBound{c}\AgdaSpace{}%
\AgdaOperator{\AgdaDatatype{≡}}\AgdaSpace{}%
\AgdaInductiveConstructor{just}\AgdaSpace{}%
\AgdaSymbol{(}\AgdaBound{c}\AgdaSpace{}%
\AgdaOperator{\AgdaInductiveConstructor{,}}\AgdaSpace{}%
\AgdaBound{t}\AgdaSymbol{)}\<%
\end{code}


\end{document}